\documentclass[10pt]{memoir}

\usepackage[english]{babel}
\usepackage[T1]{fontenc}
\usepackage{alphabeta}
\usepackage{hyperref}
\hypersetup{
    colorlinks   = true, 
    urlcolor     = green, 
    linkcolor    = blue, 
    citecolor   = red 
}
\usepackage{dsfont}

\newcolumntype{x}[1]{>{\centering\arraybackslash}m{#1}}

\usepackage[section]{placeins}
\usepackage{stackrel}
\usepackage{multirow}
\usepackage{multicol}
\usepackage[dvipsnames]{xcolor}
\usepackage{tikz}
\usepackage{tikz-cd}
\usetikzlibrary{calc}
\usetikzlibrary{arrows,positioning,decorations.pathmorphing,trees}
\usepackage{etoolbox}
\usepackage{array}
\usepackage{textalpha}
\usepackage[osf]{mathpazo}
\usepackage[mathcal,mathbf]{euler}
\usepackage{amsmath,amssymb,amsthm}
\usepackage{mathrsfs}
\usepackage[utf8]{inputenc}
\usepackage{graphicx,sidecap}
\usepackage{subfiles}
\usepackage{siunitx} 
\SetMathAlphabet{\mathrm}{normal}{U}{eur}{m}{n}
\definecolor{dhc1}{RGB}{255, 255, 0}
\definecolor{dhc2}{RGB}{0, 180, 0}
\definecolor{dhc3}{RGB}{10, 0, 255}
\definecolor{dhc4}{RGB}{255, 179, 0}
\definecolor{dhc5}{RGB}{9, 255, 255}
\definecolor{dhc6}{RGB}{9, 179, 255}
\definecolor{dhc7}{RGB}{9, 179, 255}

\definecolor{rsa1}{RGB}{200,55,200} 
\definecolor{rsa2}{RGB}{55,200,55}
\definecolor{rsa3}{RGB}{255,200,55}
\definecolor{rsa4}{RGB}{199,255,255}
\definecolor{rsa5}{RGB}{255,200,55}
\definecolor{rsak1}{RGB}{255,0,0}
\definecolor{rsak2}{RGB}{0,255,255}

\newcommand{\twocollet}[3][A]{%
  \begin{tikzpicture}
    \begin{scope}
      \node[inner sep=0pt,outer sep=0pt] (a) {\phantom{#1}};
      \clip (a.south west) rectangle ($(a.north)-(0.5\pgflinewidth,0)$);
      \node[inner sep=0pt,outer sep=0pt,text=#2]  {#1};
    \end{scope}
    \clip (a.south east) rectangle ($(a.north)-(0.5\pgflinewidth,0)$);
      \node[inner sep=0pt,outer sep=0pt,text=#3]  {#1};
  \end{tikzpicture}
}

\newcommand{\circleft}{
\begin{tikzpicture}[scale=0.15]
\draw[->] (0,0) arc (0:350:1);
\end{tikzpicture}
}
\newcommand{\circright}{
\begin{tikzpicture}[scale=0.15]
\draw[<-] (0,0) arc (0:350:1);
\end{tikzpicture}
}

\newcommand{\udbase}{
        \begin{tikzpicture}[scale=0.2]
            \draw [<->] (-1, 0) -- (1, 0);           
            \draw [<->] (0, 1) -- (0, -1);        
        \end{tikzpicture}
}
\newcommand{\udarr}{
        \begin{tikzpicture}[scale=0.13]          
            \draw [<->] (0, 1) -- (0, -1);        
        \end{tikzpicture}
}
\newcommand{\lrarr}{
        \begin{tikzpicture}[scale=0.15]
            \draw [<->] (-1, 0) -- (1, 0);                
        \end{tikzpicture}
}
\newcommand{\lrbase}{
        \begin{tikzpicture}[scale=0.15]
            \draw [<->] (-1, -1) -- (1, 1);           
            \draw [<->] (1, -1) -- (-1, 1);        
        \end{tikzpicture}
}
\newcommand{\ruarr}{
        \begin{tikzpicture}[scale=0.13]
            \draw [<->] (-1, -1) -- (1, 1);           
      
        \end{tikzpicture}
}
\newcommand{\rdarr}{
        \begin{tikzpicture}[scale=0.13]         
            \draw [<->] (1, -1) -- (-1, 1);        
        \end{tikzpicture}
}

\setSingleSpace{1.2}
\SingleSpacing

\setstocksize{297mm}{210mm}
\settrimmedsize{\stockheight}{\stockwidth}{*}
\settrims{0cm}{0cm}

\settypeblocksize{45\baselineskip}{130mm}{*}
\setlrmargins{*}{20mm}{*}
\setulmargins{*}{30mm}{*}

\setheadfoot{\baselineskip}{\baselineskip} 
\setheaderspaces{20mm}{*}{*} 

\checkandfixthelayout[lines]

\setsecnumdepth{subsection}
\settocdepth{subsection}
\bibintoc

\newcommand{\swelledrule}{%
    \tikz \filldraw[scale=.018,thick]%
    (0,0) -- (300,0) -- cycle;}
\makeatletter
\chapterstyle{madsen} 
\aliaspagestyle{chapter}{empty}

\newcommand{\hsection}[1]{
    \refstepcounter{section}
    \section*{\arabic{chapter}.\arabic{section}\hspace{1em}{#1}}
}

\newcommand{\hsubsection}[1]{
    \refstepcounter{subsection}
    \subsection*{\arabic{chapter}.\arabic{section}.\arabic{subsection}\hspace{1em}{#1}}
}

\newcommand{\listanswername}{Side information and notes}
\newlistof{listofanswers}{ans}{\listanswername}
\newlistentry[chapter]{answers}{ans}{0}

\newcommand{\hnote}[1]
{%
  \refstepcounter{answers}%
  \addcontentsline{ans}{answers}{\protect\numberline{\theanswers}#1}\par
}

\captiondelim{\null\newline}
\captionnamefont{\small\scshape\bfseries}
\captiontitlefont{\small\scshape\linespread{1}\selectfont}
\captionstyle[\raggedright]{}
\normalcaptionwidth
\def\ps@headings{
 \let\@oddfoot\@empty\let\@evenfoot\@empty \def\@evenhead{\thepage\hfil\slshape\leftmark}
  \def\@oddhead{{\slshape\rightmark}\hfil\thepage}
   \def\chaptermark##1{
    \markboth{\MakeUppercase{
    \ifnum\c@secnumdepth > \m@ne \if@mainmatter \@chapapp\ \thechapter. \ 
     \fi \fi ##1}}{}}
      \def\sectionmark##1{
       \markright{\MakeUppercase{
        \ifnum\c@secnumdepth > \z@ \thesection. \ 
         \fi ##1}}}} 

\makeatletter
\newcommand*{\lodbib@citeorder}{}
\newcommand*{\lodbib@notcited}{}
\def\citation{%
  \forcsvlist{\citation@i}}
\def\citation@i#1{%
  \ifinlist{#1}{\lodbib@citeorder}
    {}
    {\listxadd{\lodbib@citeorder}{#1}}}
\let\ltxorig@lbibitem\@lbibitem
\let\ltxorig@bibitem\@bibitem
\def\@lbibitem[#1]#2#3{%
  \csdef{lodbib@savedlabel@#2}{#1}%
  \@bibitem{#2}{#3}}
\def\@bibitem#1#2{%
  \xifinlist{#1}{\lodbib@citeorder}
    {}
    {\listadd{\lodbib@notcited}{#1}}%
  \csdef{lodbib@savedentry@#1}{#2}}
\renewenvironment{thebibliography}[1]
     {\settowidth\labelwidth{\@biblabel{#1}}}
     {\def\@noitemerr
       {\@latex@warning{Empty `thebibliography' environment}}%
      \chapter{Bibliography}%
      \@mkboth{\MakeUppercase\refname}{\MakeUppercase\refname}%
      \list{\@biblabel{\@arabic\c@enumiv}}%
           {\leftmargin\labelwidth
            \advance\leftmargin\labelsep
            \@openbib@code
            \usecounter{enumiv}%
            \let\p@enumiv\@empty
            \renewcommand\theenumiv{\@arabic\c@enumiv}}%
      \sloppy
      \clubpenalty4000
      \@clubpenalty \clubpenalty
      \widowpenalty4000%
      \sfcode`\.\@m
      \lodbib@biblistloop
      \endlist}
\def\lodbib@biblistloop{%
  \forlistloop{\lodbib@bibitem}{\lodbib@citeorder}%
  \ifdefvoid{\lodbib@notcited}
    {}
    {\forlistloop{\lodbib@bibitem}{\lodbib@notcited}}}

\def\lodbib@bibitem#1{%
  \ifcsundef{lodbib@savedlabel@#1}
    {\ltxorig@bibitem{#1}}
    {\ltxorig@lbibitem[\csuse{lodbib@savedlabel@#1}]{#1}}%
  \csuse{lodbib@savedentry@#1}}
\makeatother

\emergencystretch=\maxdimen
\begin{document}

\frontmatter
\thispagestyle{empty}

\begin{center}
\includegraphics[width=5cm]{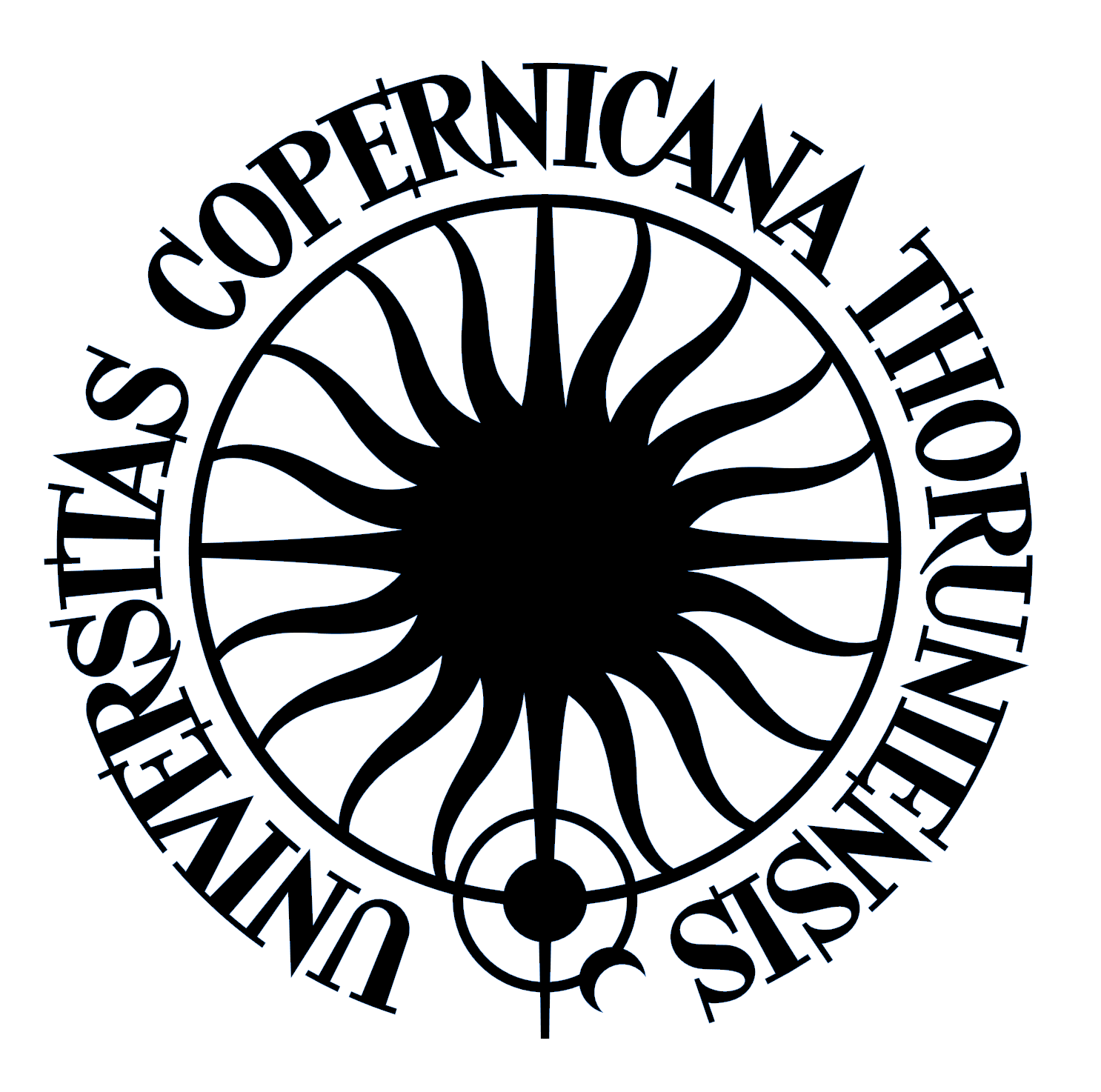} \\
{Nicolaus Copernicus University in Torun} \\
{ \itshape{} Faculty of Physics, Astronomy and Informatics \\
Institute of Physics \\
\vspace{1.5\baselineskip}}
\swelledrule
\end{center}

\vspace{5mm}
\noindent
\begin{center}
{\HUGE\scshape{}Applications} \\[1.5\baselineskip]
{\Huge\itshape{}of}\\[1.5\baselineskip]{\HUGE\scshape{}single-photon technology}
\end{center}

\begin{center}
\swelledrule \\
{\vspace{\baselineskip}{\itshape{}Doctoral Dissertation of} \\
 {\Large\scshape{}Marta Misiaszek -- Schreyner}}
\end{center}

\vspace{2cm}
\begin{flushleft}
{\itshape{}Supervisor \\ of the Doctoral Program: }\\
dr hab. Piotr Kolenderski \\
\vspace{1ex}
\end{flushleft}

\vspace{1.3cm}
\begin{center}
Toruń, Poland \\
2022
\end{center}

\cleartorecto

\thispagestyle{empty}

\mbox{}\vspace{50mm}
\noindent
\begin{quotation}
\hspace{2cm}
\begin{minipage}{6cm}

\begin{flushright}
{\LARGE Life is a great adventure.}\\
-- Jim Hawkins, \textit{Treasure Planet}, \\Walt Disney Pictures -- \\
\end{flushright}
\end{minipage}
\vspace{2cm}

\hspace{1cm}
\begin{minipage}{7cm}
\begin{flushleft}
\textit{\large \itshape{} So is physics!} \\
\vspace{5mm}
\noindent {\itshape{}Therefore, I dedicate this dissertation to those who inspired, supported, and accompanied me on a journey to become a scientist. \\
\vspace{3mm}
And to those whose journey has just begun.}
\end{flushleft}

\end{minipage}
\end{quotation}

\newpage

\cleartorecto
\noindent
\thispagestyle{empty}
\begin{center}
{\LARGE\itshape{}}\hspace{1.5em}{\Huge A few words of gratitude} \\[\baselineskip]
\end{center}

\vspace{\baselineskip}

\begin{quotation}
\noindent During my long studies, I have met many people who have supported me and patiently endured my whining. I am grateful to all of them, but I would especially like to thank a few people in particular.\\

\noindent I would like to thank dr hab. Michał Zawada for formal supervision during the studies, asking questions not always liked to answer, and pushing me to the right direction towards the finish line. \\

\noindent I am deeply grateful to dr hab. Piotr Kolenderski, who never pushed, but believed in me, my skills (even if I couldn't believed), and in that the work would be done on time. I have greatly benefited from observing how you set up SPA Lab in KL FAMO. It was a privilege to develop my skills under your guidance and supervision. \\

\noindent My research would have been impossible without the aid and support of Marysia Gieysztor. Thanks for spending restless days and nights above the laboratory table with me. Thanks for sharing the sorrows and joys of the experimental physicist. Also, thanks for all comments, suggestions, and warm encouragement while coming up with new (sometimes silly) ideas.\\

\noindent I owe a very important debt to Jakub Szlachetka and Artur Czerwiński. Discussions with you have been really deep and insightful. I believe that you contributed to a significant improvement in the quality of this thesis.\\

\noindent I received generous support from Karolina Słowik. Thanks for the certainty that in case of any problem, I could reach to you for help over a cup of coffee and a piece of cake. \\

\noindent My heartfelt gratitude goes to Mirek Bylicki, who kept his fingers crossed from the very beginning. I felt your supportive thoughtful comments and from them drew the strength to continue my work more than once. \\

\noindent Moreover, I have had the support and encouragement of Dominik Charczun. I am incredibly grateful that you tried to help every time I had any doubts on laboratory equipment, experiments I did, studies, or just life. Thanks for being in the right place at right time. Always. \\

\end{quotation}

\newpage
\thispagestyle{empty}

\mbox{}
\begin{center}
{\LARGE\itshape{}}\hspace{1.5em}{\Huge Acknowledgments} \\[\baselineskip]
\end{center}

The research presented within this thesis was supported and partially financed by:
\begin{itemize}
\item[$\rightarrow$] \textit{Single Photon Sources for Quantum Applications} project carried out within the \textit{First Team} Programme of the Foundation for Polish Science cofinanced by the European Union under the European Regional Development Fund (project ID First Team 2017-3/20);
\item[$\rightarrow$] \textit{Correlated Photon Triplets Observation} project carried out within the \textit{Preludium} Programme of National Science Centre in Poland (project ID 2018/31/N/ST2/02162);
\item[$\rightarrow$] special-purpose subsidy from the Ministry of Science and Higher Education, Republic of Poland, for \textit{System of time-resolved detection of single photons in the range of 500-2300 nm} (project ID IA/SP/0186/2016);
\item[$\rightarrow$] \textit{Optimization of  single photons coupling to the optical fiber}, IF UMK internal grant (project ID 1040-F);
\item[$\rightarrow$] UMK scholarship grant, \textit{Construction of the photon pair source
allowing to study  single photon -- single color center interaction}, WFAiS UMK resource for development of young scientists.
\end{itemize}

\begin{figure}[h!]
\centering
\includegraphics[width=0.9\linewidth]{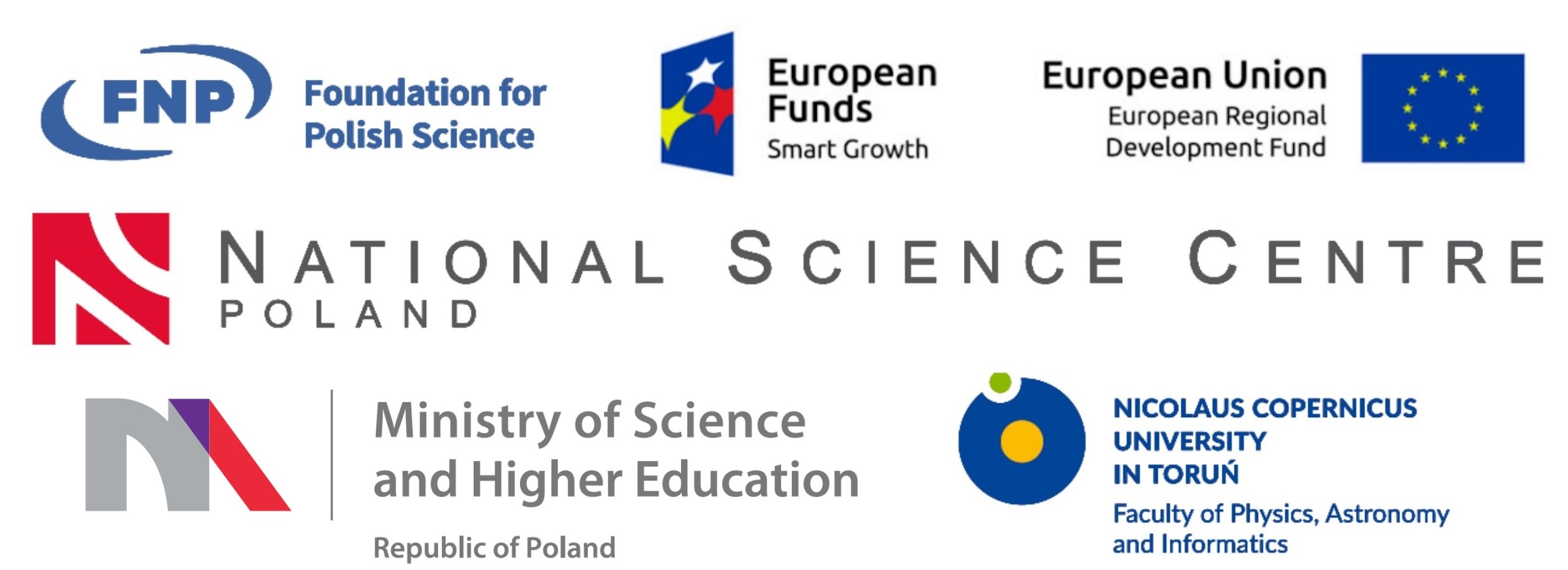}
\end{figure}

The research is also a part of the program of the National Laboratory FAMO in Toruń, Poland. \\
\begin{figure}[h!]
\centering
\includegraphics[width=0.3\linewidth]{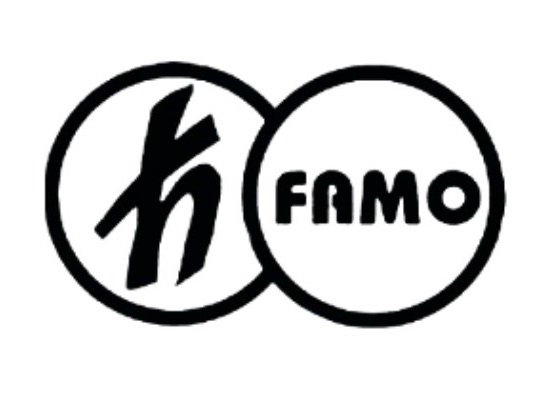}
\end{figure}

\newpage
\thispagestyle{empty}

\mbox{}\vspace{30mm}
\begin{center}
{\LARGE\itshape{}}\hspace{1.5em}{\Huge About this thesis} \\[\baselineskip]
\end{center}
\vspace{\baselineskip}

The main content of this doctoral dissertation is, as the title entails, the application of single-photon sources. The examples presented therein do not cover all possible implementations, they have been chosen to show the properties of specific single-photon sources. \\

This thesis was written with a special thought of people who would like to start their experimental work with single-photon sources or in a field of quantum optics in general. Therefore, some information may seem to be obvious for more advanced readers. The thesis consists of six chapters.\\

\textit{Chapter \ref{sec:ch1}} presents a motivation for research on single-photon sources, which is the quantum cryptography -- the single-photon technology, which is probably closer to commercial implementation than all others. It also shows a brief history of cryptography, as entitled, from letters encoded in fancy way, through modern cryptographic metod to quantum key distributed using single photons. \\
In \textit{Chapter \ref{sec:ch2}} the elements that build the quantum network are introduced, such as single-photon sources and single-photon detectors. There is also a quick glance on types of quantum channels with their pros and cons and methods for enhancing the quality of the transmission through them. \\
\textit{Chapter \ref{sec:ch3}} is the main part of this thesis. It consists of two parts, the first contains detailed description of two experimental setups, the second provides the descriptions of experimental methods for  characterization of single-photon sources.
In \textit{Chapter \ref{sec:ch4}}, three example applications of single-photon sources can be found. Two of them have been shown experimentally using the source which is described earlier. \\
\textit{Chapter \ref{sec:ch5}} is probably the most scientific part of this thesis. It contains definitions, calculations, detailed descriptions to which other chapters linked, thus, it is not recommended to read it separately. \\
And lastly, \textit{Chapter 6} includes full bibliography, without the abbreviations of scientific journals that may intimidate young adepts of quantum optics. \\

\newpage \thispagestyle{empty}

This dissertation includes two tables of contents, the short one with main chapters and sections and the long one that lists only the side information and notes, which are included in \textit{Chapter \ref{sec:ch5}}. Such a way of dividing content seemed to be more convenient and clearer for readers. \\

This thesis also contains a number of illustrations for the artistic value of which the reader should be apologized to. Hopefully, despite any inconveniences, they will help understand the ideas behind them better. Due to the complementarity of the text and figures, it should be noted that the list of figures has not been posted. \\

\vspace{1cm}
Formally, this doctoral dissertation is based on three scientific articles I coauthored:
\begin{itemize}
\item[$\rightarrow$] \textbf{M.~Misiaszek}, A.~Gajewski, P.~Kolenderski, \textit{Dispersion measurement method with down conversion process}, Journal of Physics Communications \textbf{2}, 065014 (2018) -- The results presented in this article are elaborated further in \textit{Section \ref{sec:ch3-visinfra}} and \textit{\ref{sec:ch3-qpm}}.
\item[$\rightarrow$]A.~Divochiy, \textbf{M.~Misiaszek}, Y.~Vakhtomin, P.~Morozov, K.~Smirnov, P.~Zolotov, P.~Kolenderski, \textit{Single photon detection system for visible and infrared spectrum range}, Optics Letters \textbf{43}, 6085 (2018) -- The results shown in this publication are discussed later in \textit{Section \ref{sec:ch4-detcalres}}.
\item[$\rightarrow$] M.~Gieysztor, \textbf{M.~Misiaszek}, J.~van der Veen, W.~Gawlik, F.~Jelezko, P.~Kolenderski, \textit{Absorption of a heralded single photon by a nitrogen-vacancy center in diamond}, Opics Express \textbf{29}, 564 (2021) -- The results demonstrated there are described in \textit{Section \ref{sec:ch5-nvres}}.
\end{itemize}

My contribution to the work presented therein is fully described in respective sections of this thesis. \\

\cleartorecto\tableofcontents*
\newpage
    \listofanswers*

\mainmatter
\chapter{From letters to photons}
\label{sec:ch1}

Throughout the history of our civilization, an information has been one of the most valuable matters. Future of empires depended on keeping secrets and passing messages smoothly between kings and their armies. It required inventing reliable methods both for securing the messages themselves and for sharing such secret messages between different, authorized people. Monarchs (or their advisers) were aware of threats and attempted to overcome them in numerous ways.

\section{Historical methods of encryption}
The oldest records of encrypted communication originate from \textit{The Histories}, written in 440 BC by Herodotus \cite{book:Singh}. The ancient Greek historian, who described long-lasting military conflict between Greek and Persian empire, mentioned two examples of sending a hidden message. One of them was sent as a wax tablet, but the message was hollowed directly onto the wooden backing before applying its beeswax surface. Since the wax tablets were widely used at that time, no one suspected a deceit. The other procedure was much more sophisticated, since the note carried by a servant was written onto his scalp. Although this way of communication provided relatively high level of security due to its unconventionality, it was also very time-consuming. It is because the sender of the message had to wait for the regrowth of the servant's hair, before sending him to the receiver. \\
The method of encryption described in the above examples is called \textbf{steganography}. It is based on hiding a message within the transmitted object using various methods, such as marking characters into a bigger part of a text, using of secret ink or photographically-produced microdots. It may be also applied to digital files. The example is shown in \textit{Section \ref{sec:notes-steganography}.}
\\
Despite being the oldest, steganography has one particular advantage over the other encryption methods, since the encoded message does not attract attention to itself. In consequence, the very existence of the message is hidden from third parties. Nevertheless, if a given person expects the message and has some clue about where to look for it, he/she can be able to easily read it. This can be seen as the biggest disadvantage of steganography and the reason why \textbf{cryptography} started to evolve, looking for better encryption methods. \\

The origin of the word \textit{cryptography} is Ancient Greek. It emerged by combining two words --  \foreignlanguage{greek}{κρυπτoς (kryptos)}, which means \textit{hiding}, and \foreignlanguage{greek}{γραφειν (graphein)}, which means \textit{writing} \cite{book:Bauer}. In contradiction to its meaning, the main goal of cryptography is to encode content of a message, instead of simply hiding its existence. In order to do this, encryption algorithms are used. The simplest and the oldest ones are based on substitution and transposition of single plaintext characters. \\

The basic idea of \textbf{transposition ciphers} is to rearrange the letters of the original message \cite{book:Buchmann,book:Stinson}. The same letters are used in both plaintext and ciphertext, but their order is changed in a specific way, known only to communicating parties. Due to large number of transposition ciphers, their working principle cannot be presented in common algorithm. The example of transposition ciphers can be found in \textit{Section \ref{sec:notes-transciphers}}. \\
Since transposition cipher merely rearranges characters, both the plaintext and encrypted message have the same length. On the contrary, in case of \textbf{substitution cipher}, the lengths of plaintext and ciphertext may be different. It is because the point of substitution ciphers is to replace each group of characters with another predefined group \cite{book:Buchmann,book:Stinson}. \\ 
Depending on a type of used algorithm, two main types of substitution ciphers can be distinguished -- \textbf{monoalphabetic}, where substitution over the entire message is fixed, and \textbf{polyalphabetic}, where substitution is changed cyclically and hinges on the character position in the plaintext. What is more, there is another classification, which relies on the number of substituted characters. It purports that there are \textbf{simple substitution ciphers}, where a single character is replaced by a single one; \textbf{homophonic} substitution ciphers, which replace single character with a larger group; and  \textbf{polygraphic ciphers} encrypting groups of characters with different ones. \\ Examples of each substitution cipher type can be found in \textit{Section \ref{sec:notes-subciphers}}.\\

It should be noted here that nowadays messages encrypted using any of the cipher types mentioned above may be easily decoded when intercepted. To perform such a task, a method called \textbf{frequency analysis} is typically used. It takes advantage of a fact that due to a specific structure of a language, certain letters and groups of letters have different frequency distribution than others \cite{book:Bauer,book:Buchmann}. This method may be found time-consuming, especially in case of polyalphabetic ciphers. Nevertheless, most of them had been broken by the end of XIX century \cite{book:Gomez}. \\

The continuous search for better cryptographic methods, prompted by constant insufficiency of the security of the existing ciphers, led to modern cryptography, which started to develop along with coding machines and later -- computers. \\

\section{Modern cryptography}
\label{sec:ch1-modcrypto}

\begin{figure}[t!]
\centering
\includegraphics[width=\linewidth]{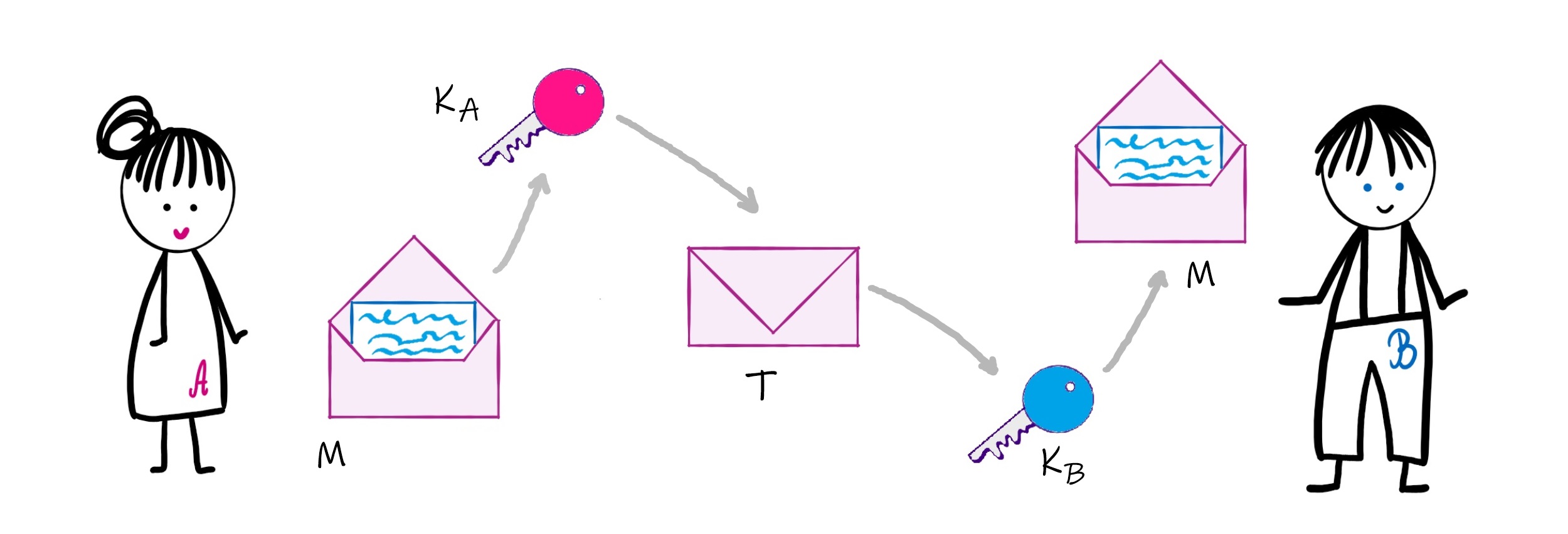}
\caption{A scheme of key encryption procedure. Symbols: M -- message, C -- ciphertext, $k_A$, $k_B$ -- secret keys. }
\label{pic:ABcryptoscheme}
\end{figure}

All ciphers described earlier reaped benefits from the fact that their algorithms were not publicly known. However, by the end of the XIX century, it became obvious that the replacement of characters is no longer strong enough to maintain secure communication. Due to rapid development of cryptographic machines, which were able to perform fast mathematical calculations, often simultaneously, the construction of ciphers' algorithms needed to be changed. \\

Currently utilized encryption algorithms are processed on computers or other electronic devices, such as mobile phones, tablets, etc. During encryption and decryption procedures, thousands of operations are performed, therefore the encryption process needs to be fast and efficient. Due to that fact, nowadays a concept of \textbf{secret key} is utilized. It means that the security of a cipher is provided by a string of characters (similarly to "keyword" or "phrase" in historical ciphers), which is not publicly known.  \\ 

The origins of a so-called \textbf{key cryptography} go back to the late XIX century, when Dutch linguist and cryptographer, {Auguste Kerckhoffs} (1835-1903), published his work on common but poor practices in French military cryptography. 
He expressed six rules of designing a military cryptosystem \cite{Kerckhoffs1883}. Most of them referred to telegraphic correspondence, so presently they are no longer in use. However, the second one, known as a \textbf{Kerckhoffs's principle}, may be translated as: 
\begin{quote}
\textit{A cryptographic system should be secure independently from the secrecy of any other part of the system except the key.}
\end{quote} 
A similar principle, \textit{"the enemy knows the system being used"}, has been formulated independently by Claude Shannon in 1949 \cite{Shannon1949}, and is known as \textbf{Shannon's maxim}.\\

Let us take a look at Fig \ref{pic:ABcryptoscheme}. It presents the basic scheme of key cryptography. Alice, a sender, wants to send message $M$ by using secret key $K_A$. Thus, the ciphertext $T$ is created, which is later forwarded to Bob. Bob, using key $K_B$ is decrypting ciphertext $T$ and gets the original message $M$. \\

Formally, defined over $(\mathscr{K},\;\mathscr{M}\;\mathscr{T})$, where $\mathscr{K}$, $\mathscr{M}$, $\mathscr{T}$ are sets of all possible keys, messages and ciphertexts, respectively, the cipher is a pair of algorithms $(\mathscr{E},\;\mathscr{D})$, that maps
\begin{equation}
\mathscr{E}:\mathscr{K}\times\mathscr{M}\rightarrow\mathscr{T} \qquad \mathscr{D}:\mathscr{K}\times\mathscr{T}\rightarrow\mathscr{M}
\end{equation}
and fulfills the correctness condition, given as:
\begin{equation}
\forall_{m\in\mathscr{M}, k\in\mathscr{K}} \; \mathscr{D}\big(k,\mathscr{E}(k,m)\big)=m \,.
\end{equation}

Currently, all kinds of modern cryptographic algorithms incorporate the use of a secret key. However, two main types of key cryptography may be distinguished -- symmetric and asymmetric. Their definitions are based on the number of required different keys and the way these keys are used. \\

\textbf{Symmetric encryption} procedures use the same key for encoding and decoding an information ($K_A=K_B$ in the Fig.~\ref{pic:ABcryptoscheme}). Depending on the length of a key, and hence, the method of encryption, one can differentiate stream and block ciphers. \\
\textbf{Stream ciphers} work on a continuous stream of plaintext data, which is encoded bit by bit using the secret key formed from an infinite stream of random numbers. In this case, the secret key is always about the size of the message that is to be encoded.
In case of \textbf{block ciphers}, the key is usually $128$ or $256$ bits long. It means that there are respectively $2^{128}$ or $2^{256}$ possibilities of encoding, so brute-force attack (checking all the possibilities), is not a realistic approach to break the key because of a very long time needed for its execution.\\
The use of block ciphers entails dividing plaintext data into smaller parts, called blocks, which are usually of a key size. Then, each of the blocks is processed using pseudorandom permutation functions \cite{book:Bauer,book:Buchmann}. In case of some algorithms, blocks may be processed independently, parallel to others. Also, such algorithms may provide additional security by combining data from different blocks.\\
Symmetric ciphers are fast and efficient. Although they require sophisticated methods to distribute the secret keys to both parties, they are commonly used. The most popular symmetric block cipher is \textbf{AES} \cite{Nechvatal2001}, developed by Vincent Rijmen and Joan Daemen in $1997$. Moreover it was approved as a federal encryption standard in the United States in $2002$ \cite{Nechvatal2001}. \\

\textbf{Asymmetric encryption} works in a different manner than symmetric one. First of all, there is a need to use a pair of keys, called \textbf{public key} and \textbf{private key}, which are related to each other. The public key is used for encoding a message and should be available to anyone who intends to communicate with its owner. Contrarily, the private key should remain secret and is used for decrypting messages. \\
The idea of asymmetric cryptography and example algorithm of such key distribution is more widely presented in \textit{Section \ref{sec:notes-pubkeycrypto}}. \\

There is also another use of assymetric encryption in the form of authentication. Since messages are signed with a private key, anyone who posses associated public key is able to verify the origin of the received data. \\

\begin{figure}[t!]
\centering
\includegraphics[width=\linewidth]{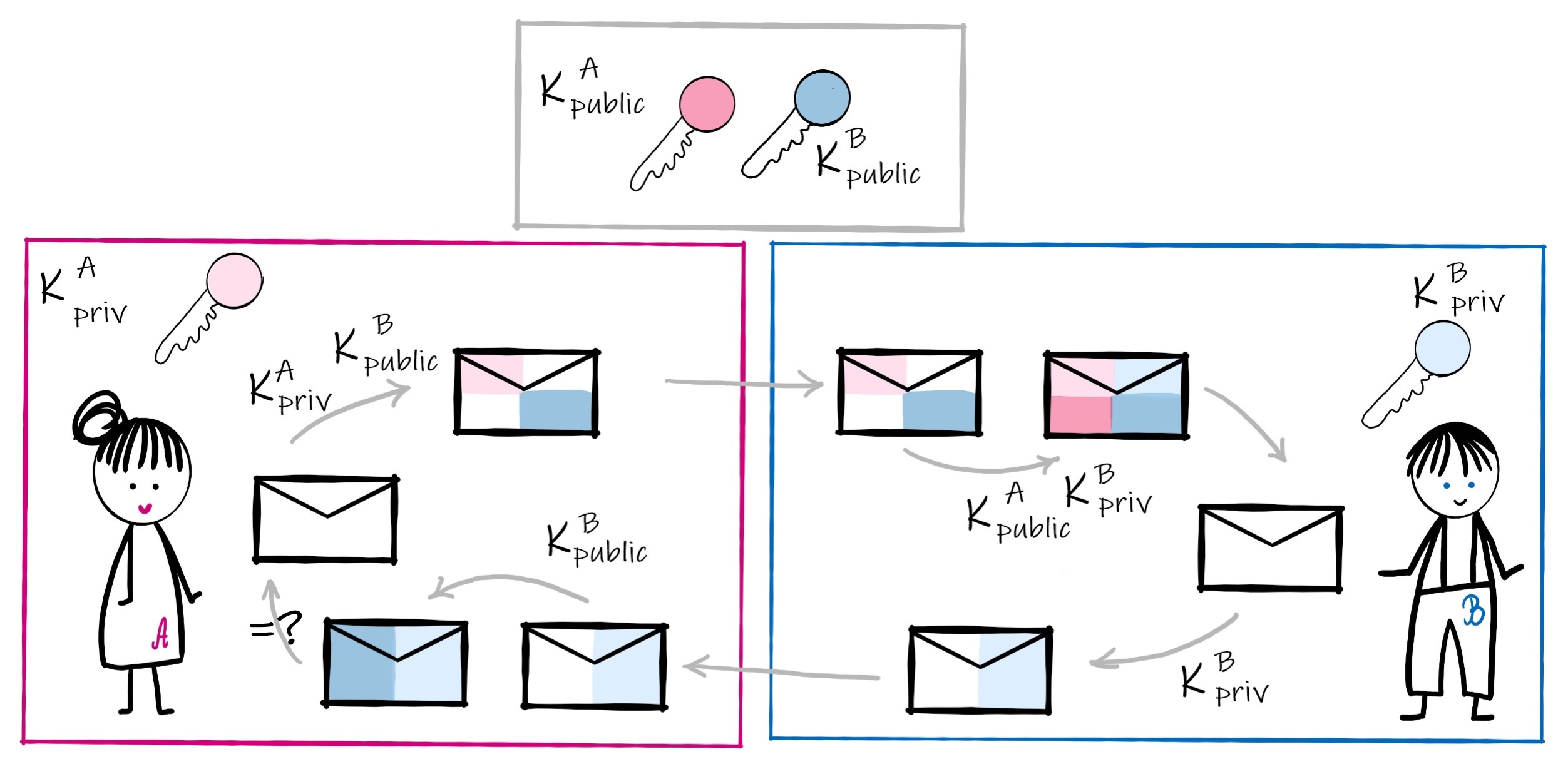}
\caption{A scheme of authentication with asymmetric encryption procedure. Symbols: $K_{public}^A, K_{public}^B$ -- Alice's and Bob's public keys, $K_{priv}^A, K_{priv}^B$ --  their private keys.}
\label{pic:ABaut}
\end{figure}

Sticking to Alice and Bob example, authentication process looks as depicted in Fig.~\ref{pic:ABaut}. Alice chooses a random message (or a hash function -- the function, which "shortcuts" the original message with known algorithm). Then, she encrypts it with Bob's public key and her own private key. Bob decrypts with his private key and Alice's public key. Considering the fact that the encryption and decryption procedures are based on multiplication, the order of operations is not important.\\
Subsequently, Bob reencrypts message with Alice's public key and sends it back. Upon receiving the original message from Bob, Alice knows that the sender must know Bob's private key, therefore it must be Bob. Furthermore, since the message was decrypted with Alice's public key, Bob knows that Alice is in possession of Alice's private key.  This is a part of identity system that allows to check which entity owns a particular private key. \\

Although asymmetric encryption solves the problem of key distribution, there are some disadvantages to using it. One of them is that, depending on the type of encryption and encrypted data, accidental disclosure of a private key may lead to widespread security compromise, when all sent messages could be read easily by an eavesdropper.    
Another factor is that public keys are not internally authenticated, which means that users need to prove the ownership. However, potential attacker does not have any benefits from impersonating the true owner of a key. Finally, due to the length of a key (typically $1024$ or $2048$ bits), asymmetric encryption algorithms are slow, therefore they are not suitable to encrypt a full communication. They are rather used for setting secure private channels, as in the case of \textbf{SSL handshake}. \\

SSL handshake is a part of \textbf{HTTPS protocol}, which enables encrypted communication between the web browser and the web server in both directions. To obtain this, the HTTPS protocol uses SSL (Secure Sockets Layer), a standard technology for estabilishing secured links. It is not an encryption algorithm, rather a mix of other known algorithms. \\
SSL handshake uses both asymmetric and symmetric coding. It may be described in 5 steps:
\begin{quote}
\begin{enumerate}
\item Browser sends a request for visiting specific website to a web server.
\item Web server sends back a digital certificate, which consists of public key $K_{public}$.
\item Browser generates a secret session key $K_{session}$, which will be used later for symmetric encryption. 
\item The secret key is encrypted using public key $E_{K_{public}}\big[K_{session}\big]$ and sent back to web server.
\item Server decrypts received secret key with his private one \\ $D_{K_{private}}\Big[E_{K_{public}}\big[K_{session}\big]\Big ]$.
\end{enumerate}
\end{quote} 
After that, both the web browser and web server, share a secret session key, which allows to set secure link between them. As the name suggests, the session key is used only once, during the ongoing connection. \\

The example above shows that symmetric and asymmetric cryptography is essential in everyday use to provide security for digital communication. Unfortunately all presently used ciphers, except \textbf{one-time pad} (see \textit{Section \ref{sec:notes-otp}}), rely on practical difficulties and computational complexity of algorithms. \\ 

In the near future, when the quantum computer allows for faster calculations, currently utilized ciphers (especially asymmetric ones) may be compromised and their use will not be secure any longer \cite{book:nielsenchuang}. As a consequence, novel type of encryption will have to be developed. \\

\section{Quantum means useful}
\label{sec:ch1-quantuse}

The threat for the security of the presently used encryption algorithms comes from the fact that quantum computers will be able to perform complicated mathematical calculations much more efficiently than the classical computers. Therefore, to counter this threat, it seems natural to look for some alternative encryption algorithms, security of which would have a more solid foundation. A promising idea is to exploit the fundamental laws of quantum mechanics to prevent a potential eavesdropper from obtaining classified information. \\ This novel type of encryption, based on \textbf{qubits} (see \textit{Section \ref{sec:ch1-qubit}}), is called \textbf{quantum cryptography}. \\

The best known example of quantum cryptography is \textbf{quantum key distribution} (\textbf{QKD}), which, involving features such as \textit{quantum entanglement} and \textit{quantum superposition} of a state, enables communicating parties to produce a shared random secret key \cite{Bennett1984}. \\

The first QKD protocol, nowadays called \textbf{BB84}, was proposed by Charles Bennett and  Gilles Brassard in 1984. Later, they said \cite{Bennett1992}
\begin{quote}
\textit{The main breakthrough came when we realized that photons were never meant to store information, but rather to transmit it.}
\end{quote}
On the base of that conviction, they utilized photons' property -- \textit{polarization} -- to encode information, and then transmit it (see \textit{Section \ref{sec:ch1-pol}}).

\hsubsection{BB84 protocol}
The proposed protocol, \textbf{BB84}, is for secure distribution of a symmetric private key between two parties for use in one-time pad encryption. If the sender, traditionally called Alice, wants to send a private key to the receiver, traditionally called Bob, it looks as follow:
\begin{quote}
\begin{enumerate}
\item Alice generates two strings $x$ an $y$ of random bits, \\ each of $n$ bits long.
\item Using these two strings, Alice creates a tensor product of $n$ qubits: \\
$|\psi\rangle= \stackbin[i=0]{n}{\otimes}|\psi_{x_i y_i}\rangle$ . \\
If $x_iy_i$ defines the bits value, this gives four qubit states: \\ $|\psi_{00}\rangle=|0\rangle$, \quad $|\psi_{10}\rangle=|1\rangle$, \\ $|\psi_{01}\rangle=\frac{1}{\sqrt{2}}|0\rangle+\frac{1}{\sqrt{2}}|1\rangle$ \\ 
$|\psi_{11}\rangle=\frac{1}{\sqrt{2}}|0\rangle-\frac{1}{\sqrt{2}}|1\rangle$.
\item Alice sends qubit state $|\psi\rangle$ to Bob over public quantum channel.
\item Bob measures received qubit using randomly chosen string  $y'_i$.
\item Bob announces publicly that he had received Alice's qubits. Alice announces $y_i$ string. They compare strings and discard qubits for which $y_i$ and $y'_i$ differ.
\item From remaining bits, the secret key is formed.
\end{enumerate}
\end{quote}
One can note that if Bob receives the qubit, due to the \textbf{no-cloning theorem}, any eavesdropper cannot have its copy (see \textit{Section \ref{sec:notes-noclo}}). The eavesdropper may perform measurements on intercepted qubits but risks disturbing each qubit with probability $1/2$. \\

The practical realization of BB84 exploits encoding qubits in single-photon polarization states. Using rectilinear and diagonal polarization bases, the ensemble of qubits used in this protocol takes a form:
\begin{quote}
$|\psi_{00}\rangle=|\lrarr\rangle$, \quad $|\psi_{10}\rangle=|\udarr\rangle$, \\ $|\psi_{01}\rangle=\frac{1}{\sqrt{2}}|\lrarr\rangle+\frac{1}{\sqrt{2}}|\udarr\rangle=|\ruarr\rangle$,  \\
$|\psi_{11}\rangle=\frac{1}{\sqrt{2}}|\lrarr\rangle-\frac{1}{\sqrt{2}}|\udarr\rangle=|\rdarr\rangle$.
\end{quote}

One can follow the procedure, shown in Tab. \ref{tab:ch1-bb84}. Symbols $\lrbase$ and $\udbase$ stand for the measurement's polarization bases.\\

\begin{table}[h!]
\begin{tabular}{cc|r|cccccccccc}
\hline
\parbox[t]{2mm}{\multirow{3}{*}{\rotatebox[origin=c]{90}{Alice's}}}&\parbox[t]{2mm}{\multirow{3}{*}{\rotatebox[origin=c]{90}{part}}}  & $x_i$ string &1&1&1&0&0&1&1&0&1&1\\
&& $y_i$ string &1&0&1&0&1&0&0&1&1&0\\
\cline{3-13}
&& qubit &$|\rdarr\rangle$&$|\udarr\rangle$&$|\rdarr\rangle$&$|\lrarr\rangle$&$|\ruarr\rangle$&$|\udarr\rangle$&$|\udarr\rangle$&$|\ruarr\rangle$&$|\rdarr\rangle$&$|\rdarr\rangle$\\
\hline
\hline
\parbox[t]{2mm}{\multirow{3}{*}{\rotatebox[origin=c]{90}{Bob's}}}&\parbox[t]{2mm}{\multirow{3}{*}{\rotatebox[origin=c|]{90}{part}}}  & $y'_i$ string&1&1&0&0&1&1&1&0&1&0\\
\cline{3-13}
&& bases &$\lrbase$&\textcolor{red}{$\times$}&\textcolor{red}{$\times$}&$\udbase$&$\lrbase$&\textcolor{red}{$\times$}&\textcolor{red}{$\times$}&\textcolor{red}{$\times$}&$\lrbase$&$\lrbase$\\
\cline{3-13}
&& result &$|\rdarr\rangle$&&&$|\lrarr\rangle$&$|\ruarr\rangle$&&&&$|\rdarr\rangle$&$|\rdarr\rangle$\\
\hline
\end{tabular}
\caption{Exemplary application of BB84 protocol with single photons.
}
\label{tab:ch1-bb84}
\end{table}

In practical implementation of BB84, the qubits are often transmitted between the legitimate parties using optical fiber. Its birefringence may influence qubit states, so at the receive end their appear changed. Hence, their states must be backtransformed into the original coordinate system. This can be done with polarization controller, hovewer, it may also introduce some noise. \\

Moreover, due to the randomness of genaration $y_i$ and $y'_i$ strings, after the comparison, statistically, half of sended qubits may be discarded. \\
Therefore this method of secret key distribution is not very efficient, which can be seen in the example. \\

\hsubsection{Single photons in the service}
So far, the main interest of quantum cryptography is the development of quantum key distribution protocols. Many of them use qubits in a form of single photons. Therefore the investigation of their sources is highly important the day before quantum computer era. 
\chapter[Quantum world]{Quantum world \\ {\LARGE of single-photon devices}}
\label{sec:ch2}

Quantum network is a facility which enables encoding and transferring information in a form of qubits as well as reading and operating onto them. Its main applications are quantum communication through quantum internet and distributed quantum computing (similar to clustering classical computers).\\
While the exact framework of a quantum network depends on its application, the basic structure is analogous to a classical network. As such, it consists of end nodes and transportation layer. \\

\section{Building blocks of quantum networks}
\hsubsection{Quantum processors}
\label{sec:ch1-quantnet}

Quantum processors are the end nodes, the crucial parts of quantum networks. In general, any system which is able to perform quantum logic gates (see \ref{sec:notes-qloggate}) on a certain number of qubits may be called a quantum processor, or in a  more catchy manner, a quantum computer. \\

Quantum processors can be made in many ways, employing various physical systems. The most evolved by now are \textit{superconducting quantum computers}, which are based on superconducting electronic circuits (Josephson junctions \cite{Josephson1962, Josephson1974}) \cite{supquantcomp}. This technology is developed by leading companies such as Google \cite{Castelvecchi2017}, IBM \cite{IBM}, or Intel \cite{Intel2020}. However, since superconducting quantum computers require using low temperatures, which is generally costly and complicates the setup, various other types of end nodes are currently under investigation. \\

Among many others, the following examples may be cited: 
\begin{itemize}
\item[--] \textit{quantum dot computer} -- qubit state defined by spatial position of an electron in double quantum dot \cite{Fedichkin2000} or by the spin states of trapped electrons \cite{Imamoglu1999};
\item[--] \textit{trapped ion quantum computer} -- qubit defined by internal states of trapped ions \cite{Monz2011};
\item[--] \textit{optical lattices} -- qubit states implemented by utilizing internal states of atoms trapped in an optical lattice \cite{Brennen1999};
\item[--] \textit{optical quantum computer} -- photon states processed through linear optical elements \cite{Knill2001};
\item[--] \textit{diamond-based quantum computer} -- qubit states realized by the electronic or nuclear spin of nitrogen-vacancy centers in diamond \cite{Nizovtsev2005, Dutt2007}.
\end{itemize}  
These are promising solutions, giving hope for implementation in the near future. \\

\hsubsection{Fiber and a free-space links}
\label{sec:ch1-ffquantnet}
As it is mentioned in \textit{Section \ref{sec:ch1-quantuse}}, transmission of information over long distances utilizes photon-based qubits and optical networks. According to this, the transportation layer may take a form of free-space networks or those based on existing telecommunication fiber links. \\

Since the biggest problems of quantum world are losses and noise, especially decoherence, the length of a communication link is crucial for network performance. Currently, the longest quantum fiber network, consisting of $2000$ km of fiber and $32$ quantum repeaters, is Beijing-Shanghai Trunk Line, finished in 2017 \cite{Zhang2018}. In the case of free-space communication, the exchange of single photons between ground and quantum satellite (see \textit{Section \ref{sec:ch4-sat}}), on a slant distance of $20000$ km, has been reported \cite{Villoresi2019}. Although distribution of entanglement has been achieved only over a distance of $1203$ km \cite{Yin2017}, these results look promising for further development of free-space quantum communication. \\

It is also important to notice that a type of a transmission layer often determines optimal protocols, which can be used for communication. With free-space link, qubits are usually encoded in photons' polarization states. On the other hand, fiber communication favors time-bin or phase encoding (see \textit{Section \ref{sec:notes-tbin-ph}}).

\hsubsection{Repeaters and error correction}
\label{sec:ch1-reperrcorr}

As it was mentioned earlier, due to the losses and decoherence, information transmission over long distances, especially through fiber, is one of the biggest challenges for quantum communication. A standard parameter to measure the quality of the transmitted data is the so-called signal-to-noise ratio. To improve it, two fundamentally different approaches are  used. One of them is to employ the quantum repeaters to the network, the other is to perform error correction with multi-qubit non-destructive measurements. \\

The idea of a quantum repeater came from traditional telecommunication networks, where the analogous device extends transmission to cover longer distances. Depending on the type of transmission, different kinds of repeaters are used. In the case of a telephone, it is a transistor-based amplifier in a telephone line. In the case of radio communication, the repeater is a combination of a receiver and a transmitter that retransmits a radio signal. In the case of fiber classical communication, it is an optoelectronic circuit that amplifies the light beam. \\

A quantum repeater extends quantum communication distance by distributing entanglement between the end nodes \cite{Kimble2008, Sangouard2011}. However, due to the quantum mechanical limitations such as \textit{no-cloning theorem} \cite{Wootters1982} it is not a trivial task. A possible solution solution comes from the same background as limitations -- \textit{quantum teleportation} \cite{Bennett1993} or \textit{entanglement swapping} \cite{Zukowski1993}. \\

\begin{figure}[hb!]
\centering
\includegraphics[width=0.8\linewidth]{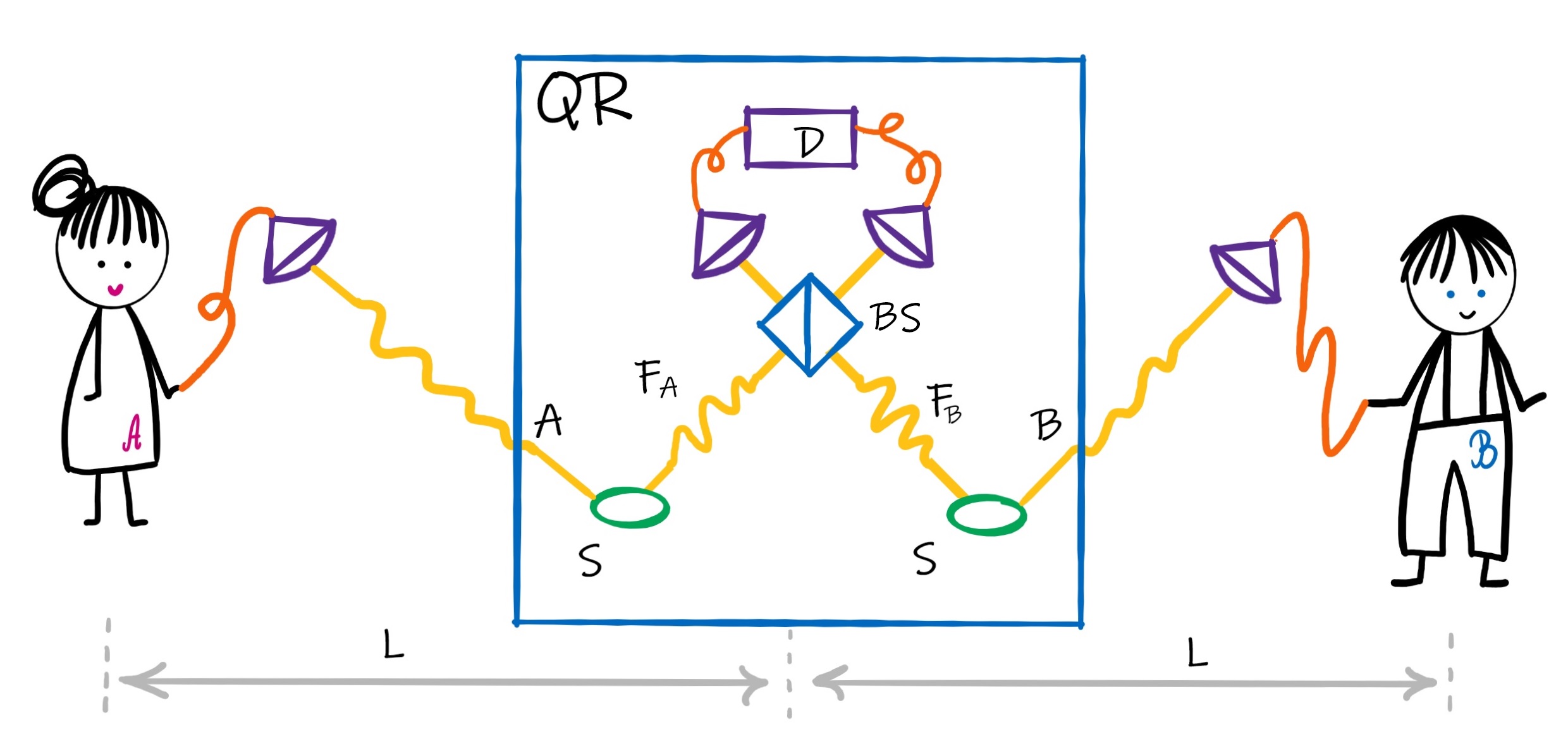}
\caption{Quantum repeater operational scheme. Symbols: QR -- quantum repeater, S -- photon-pair source, D -- BSM detection system, L -- distance between quantum repeater and the end node.}
\label{pic:rep}
\end{figure}

The operating mechanism of a quantum repeater is presented in Fig.~\ref{pic:rep}.

Let us assume that Alice and Bob want to share entangled qubits. The distance between them is  long; hence, they decide to use a quantum repeater. The device consists of two identical single-photon pair sources and optical detection system (see \textit{Section \ref{sec:ch2-bsm}}), which enables the Bell-state measurements (BSM) \cite{Lee2013}. Each photon source produces pair of entangled photons, marked as $(A,F_A)$ and $(B,F_B)$, creating a pair of qubits. Then one photon from each pair is send to communicating parties, $A$ and $B$, respectively. The other photons, $F_A$ and $F_B$, are send to the optical detection system, where the BSM is performed. As a result of this measurement, the maximally entangled quantum state of photons $A$ and $F_A$ is \textit{swapped} to photons $A$ and $B$. Hence, Alice and Bob now share the entangled pair of states $(A,B)$ on a distance twice as long as between the initially entangled pairs $(A,F_A)$ and $(B,F_B)$. When implemented with sufficiently high-quality setup elements, such a protocol can be successfully performed with higher probability than can be achieved in the case of the direct distribution of an entangled state to such a distance.\\

The other method of boosting the signal-to-noise ratio after the propagation is the use of quantum error correction algorithms. Similarly to classical error correction codes, the quantum ones utulize \textbf{syndrome measurements} to check the quality of transmitted data. The syndrome measurements retrieve the information about an error without getting any knowledge about encoded quantum state itself.\\ 

There are various types of such codes. Most of them use  \textbf{bit flip} or \textbf{phase flip} (see \textit{Section \ref{sec:notes-bferrcorr}} and \textit{\ref{sec:notes-phferrcorr}}) as a syndrome. One of commonly known correction codes is the \textit{Shore code} \cite{Shore1995, Devitt2013}, which combines both syndromes, enables for correction of an arbitrary single-qubit error at once. \\

In principle, there is also a completely different method that helps to deal with noise in the quantum channel. It is based on proper choice of type of encoding, in a way that the information carrier is not influenced by the decoherence. If the variables belong to the so-called \textbf{decoherence-free subspace}, their evolution is unitary, and thus, they are not affected by noise at all (see \textit{Section \ref{sec:ch2-decfree}}).\\

\section{End nodes}

From both, the classical and quantum point of view, the end nodes basically are the components of transmission network, that are located on both sides of the transportation channel. The most crucial parts, without which quantum communication would not be possible at all, are the single-photon sources and single-photon detectors. \\

\subsection{Single-photon sources}
\label{sec:ch1-sinphsource}
An ideal single-photon source is the one that emits  \textbf{indistinguishable single} photons (see \textit{Section \ref{sec:ch2-indistinguishability}} and \textit{Section \ref{sec:ch2-g2}}) at any arbitrary time specified by the communicating party. Therefore, the wavelength of emitted photons should be well defined, the \textit{repetition rate} of photon emission as high as possible, the probability of the single-photon emission should be equal to $1$, whereas the \textit{probability of multi-photon emission} is equal to $0$. \\
Since the ideal single-photon source does not exist, the realistic ones may be divided into two groups -- \textbf{deterministic} and \textbf{probabilistic} single-photon sources. It is worth to mention that every single-photon source has its advantages and disadvantages, thus it is not possible to choose the best one under all conditions and circumstances. Usually, the type of source is chosen carefully to fulfill requirements imposed by experimental setup. Nevertheless, the comparison of parameters of different types of sources may be found in Ref. \cite{Eisaman2010}, Ref. \cite{Chunnilall2014} and Ref. \cite{Urbasi2019}. \\

\begin{figure}[t!]
\centering
\includegraphics[width=0.6\linewidth]{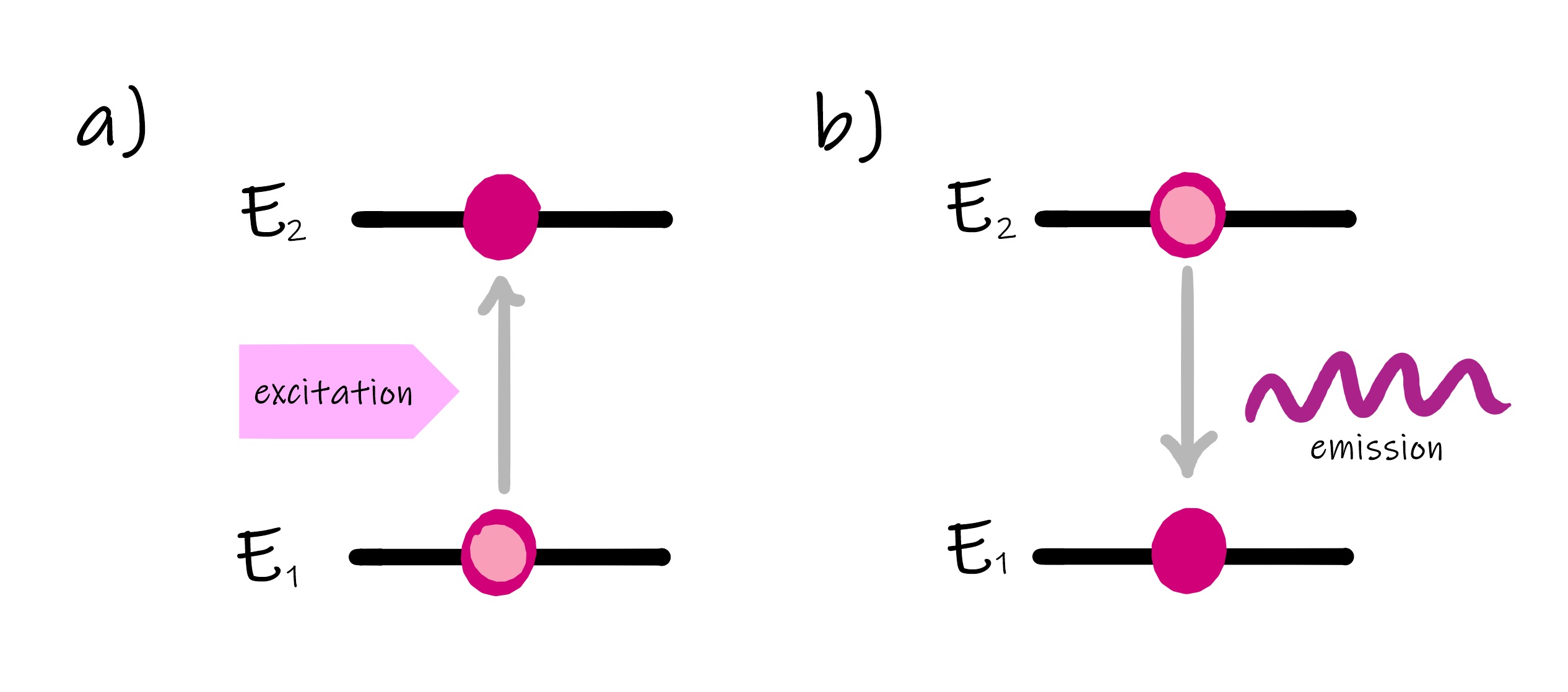}
\caption{Scheme of single-emitter quantum system, a) excitation, b) emission.}
\label{pic:sinemsys}
\end{figure}

\textbf{Deterministic sources} are based on single-emitter quantum systems. Such systems, schematically, consist of two internal levels (see \textit{Fig.~\ref{pic:sinemsys}}). It means that the system may be excited in some way and then emit a single photon.
There are many systems under investigation. Among them, worth to mention are for example:
\begin{itemize}
\item[--] \textit{single neutral atom} \cite{Hennrich2004,Higginbottom2016} or \textit{single ion} \cite{Keller2004,Dibos2018};
\item[--] \textit{single molecule} \cite{Moerner2004,Zhang2017};
\item[--] \textit{color center} \cite{Kurtsiefer2000, Khramtsov2018};
\item[--] \textit{quantum dot} \cite{Kako2006, Katsumi2019}.
\end{itemize} 

\begin{figure}[b!]
\centering
\includegraphics[width=0.4\linewidth]{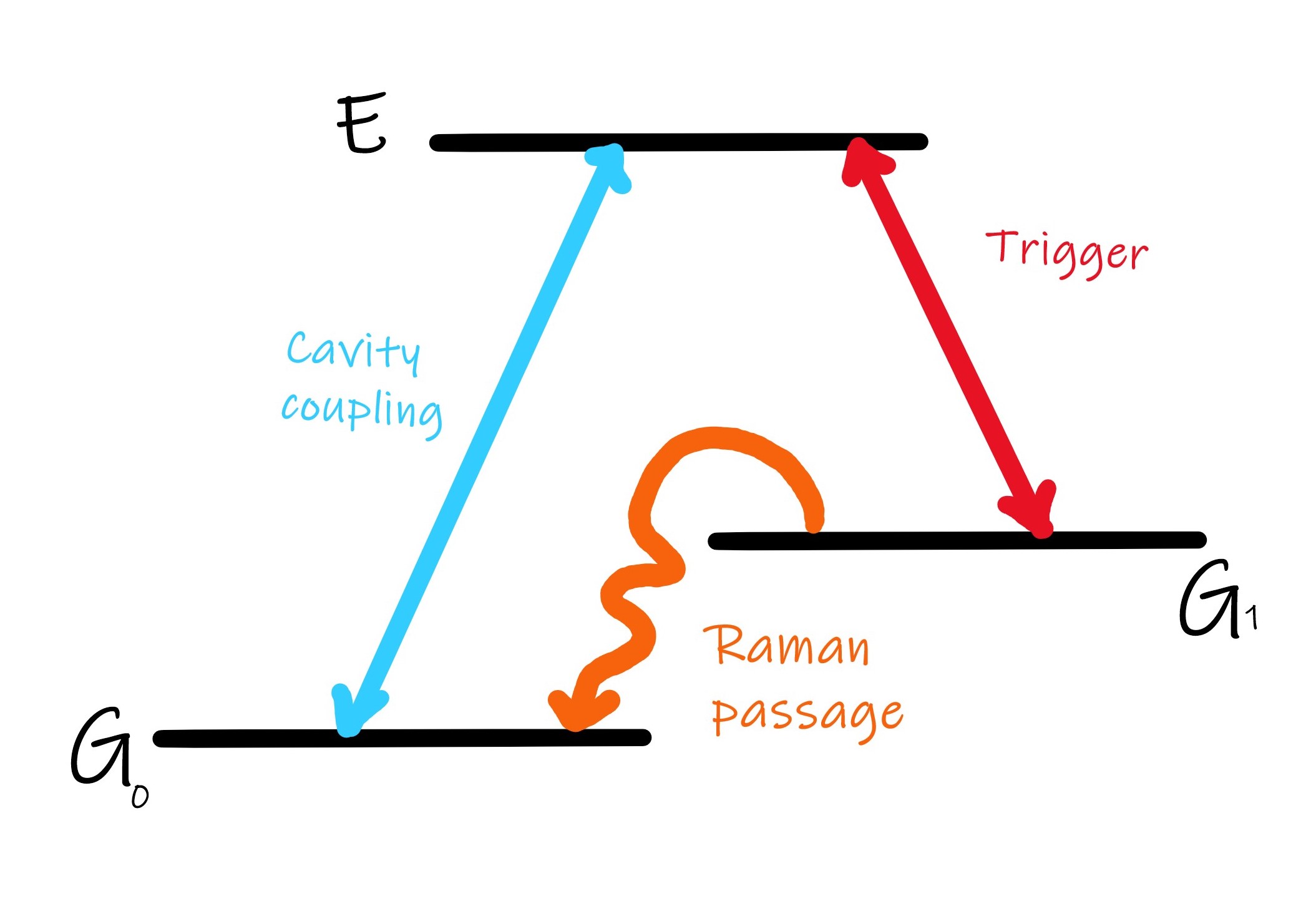}
\caption{Scheme of lambda-type single-photon source.}
\label{pic:lambdaemsys}
\end{figure}

The systems with \textbf{single neutral atom} or an \textbf{ion} are very similar. In both cases, atom is cooled and captured with a magneto-optical trap (MOT) inside the cavity. Such systems have the  $\Lambda$-type energetic structure with two metastable ground states and short-lifetime excited state (see Fig.~\ref{pic:lambdaemsys}). The transition between grounds and excited state is controlled by the pump laser pulse and atom-cavity coupling. Single photons are emitted via stimulated \textit{Raman adiabatic passage (STIRAP}) \cite{Bergmann2019} or \textit{Raman scattering} \cite{Raman1928, Singh2002} (only for ion source). Although the efficiency of the photon emission is close to unity, in such systems photons are emitted in all directions. Therefore, the problem concerns the efficiency of the photon collection. It may be solved by the use of strong cavity coupling, but it is not experimentally easy, especially with a charged particle. Also, to eliminate the multi-photon emission, only one atom should be captured inside a trap. This, due to the fact that the time spent by an atom in the cavity is finite and, thus, the atom needs to be replaced, influences the repetition rate of such a source and may require the proper engineering of the experimental setup. Despite that, this type of source has obvious advantage -- photons from successive pulses are indistinguishable, their wavelength is precisely defined by the energetic levels of and atom (ion). Therefore, to obtain desired wavelength with the atom (ion) single photon source, the atom with proper energetic structure needs to be chosen. It narrows down the possibilities for the use of such sources in quantum optical experiments. \\

Contrary to the sources based on single atoms, single photon sources based on \textbf{single molecules} can be used without the laser cooling systems, even in room temperatures \cite{Steiner2007}. However, till now, the indistinguishability of generated photons was demonstrated only for cryogenic temperatures \cite{Ahtee2009,Lombardi2021}.\\
The electronic structure of the molecule, embedded into some crystal structure or immersed in some chemical solution, may be modeled by 4-level system, with two pairs of vibrational states, one pair for ground and one for excited state. The non-radiation transition between vibrational states is fast and thanks to that there is no need for repumping the molecule to its initial state. Therefore, it is possible to get the high photon emission repetition rate. Another advantage of such a source is the possibility of controlling the polarization of the emitted photon \cite{Trebbia2010}. This is a feature, which is inevitable in some of the quantum key distribution protocols. Hovewer, due to the fact that the molecule source, similarly to single atom source, emits photons in any direction, without the use of a cavity, the coupling efficiency is low. This moves the single molecule photon source away from being used in any practical (commercial) not only experimental realization. \\

\begin{figure}[t!]
\centering
\includegraphics[width=0.5\linewidth]{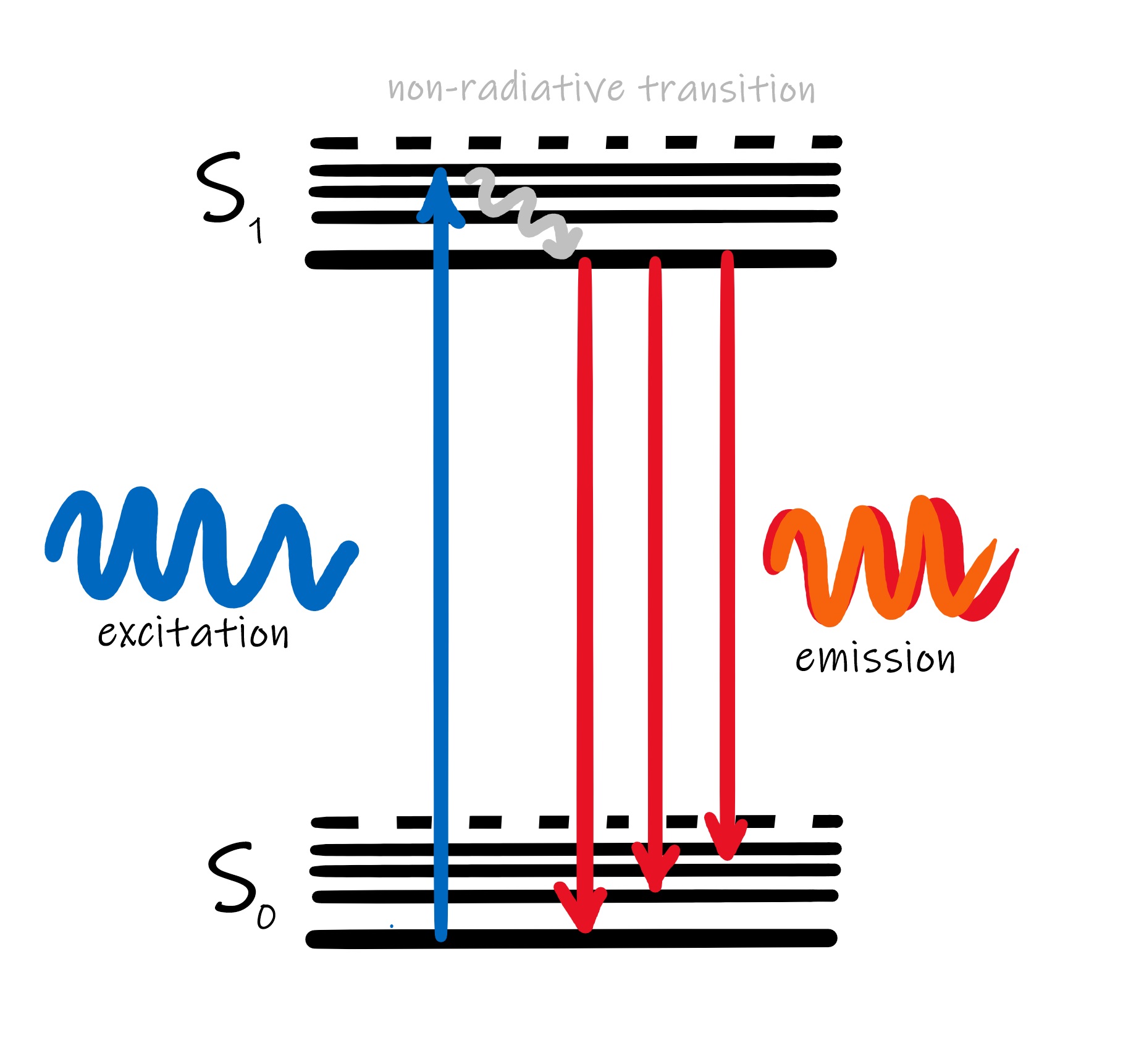}
\caption{Scheme of single molecule energy levels ($S_0$-ground state, $S_1$-excited state), excitation (blue lines) and emission (red lines) transitions.}
\label{pic:molemsys}
\end{figure}

\begin{figure}[t!]
\centering
\includegraphics[width=0.5\linewidth]{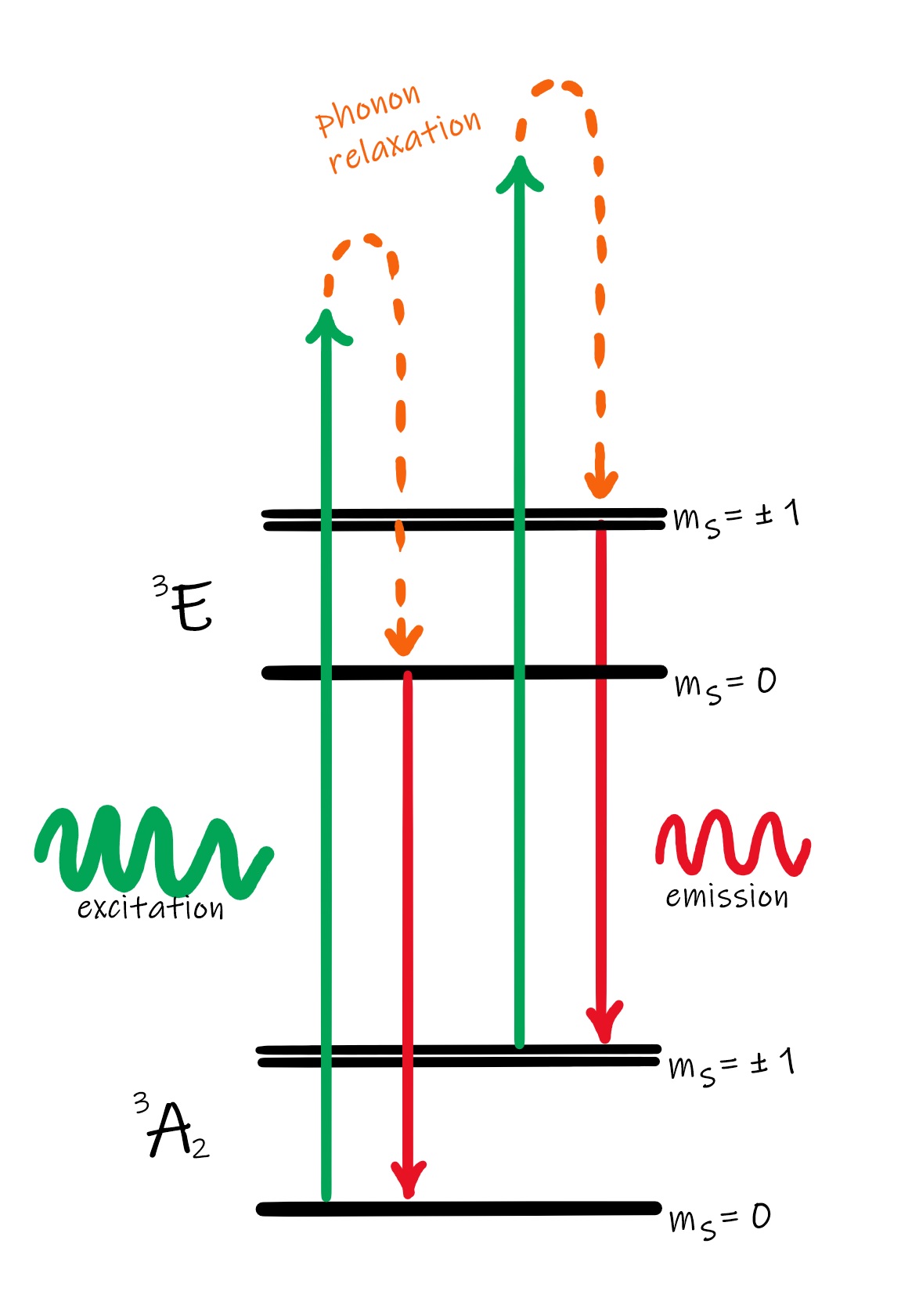}
\caption{Scheme of energy levels for NV center.}
\label{pic:nvemsys}
\end{figure}

Another candidate for single-photon source which is characterized by the polarization controlling of emitted photons are \textbf{color centers} \cite{Lutz2018, Brouri2000}. Based on defects in crystalline structure, primarily  nitrogen-vacancies (NV) in diamonds, a color center has broad absorption and emission spectrum, with the central lines at $575$ nm and $637$ nm, respectively \cite{Jelezko2001, Marysia2020}. Moreover, the radiative efficiency of NVs is close to unity, even in room temperatures \cite{Pezzagna2021}. Taking these into account, with broad emission spectrum and a proper spectral-line filters, color centers may be used as very convenient tunable photon sources. \\ 
The electronic structure of color centers may be modelled with two - ground and excited - energy levels with spin sublevels (for triplet states) and two intermediate levels (for singlets states) \cite{Auzinsh2019}, see Fig.~\ref{pic:nvemsys}. The liftime of the radiative transition influences the repetition rate for this type of source \cite{Lee2011}. Although the photons are emitted in any direction, the collection of photons may be improved by enclosing the color center in microcavity or nanowire. Unfortunately, the indistinguishability of photons has not been achieved yet, which is why such sources are excluded from usage in some quantum cryptography protocols, despite the fact that in general they are experiment friendly. \\

\begin{figure}[t!]
\includegraphics[width=\linewidth]{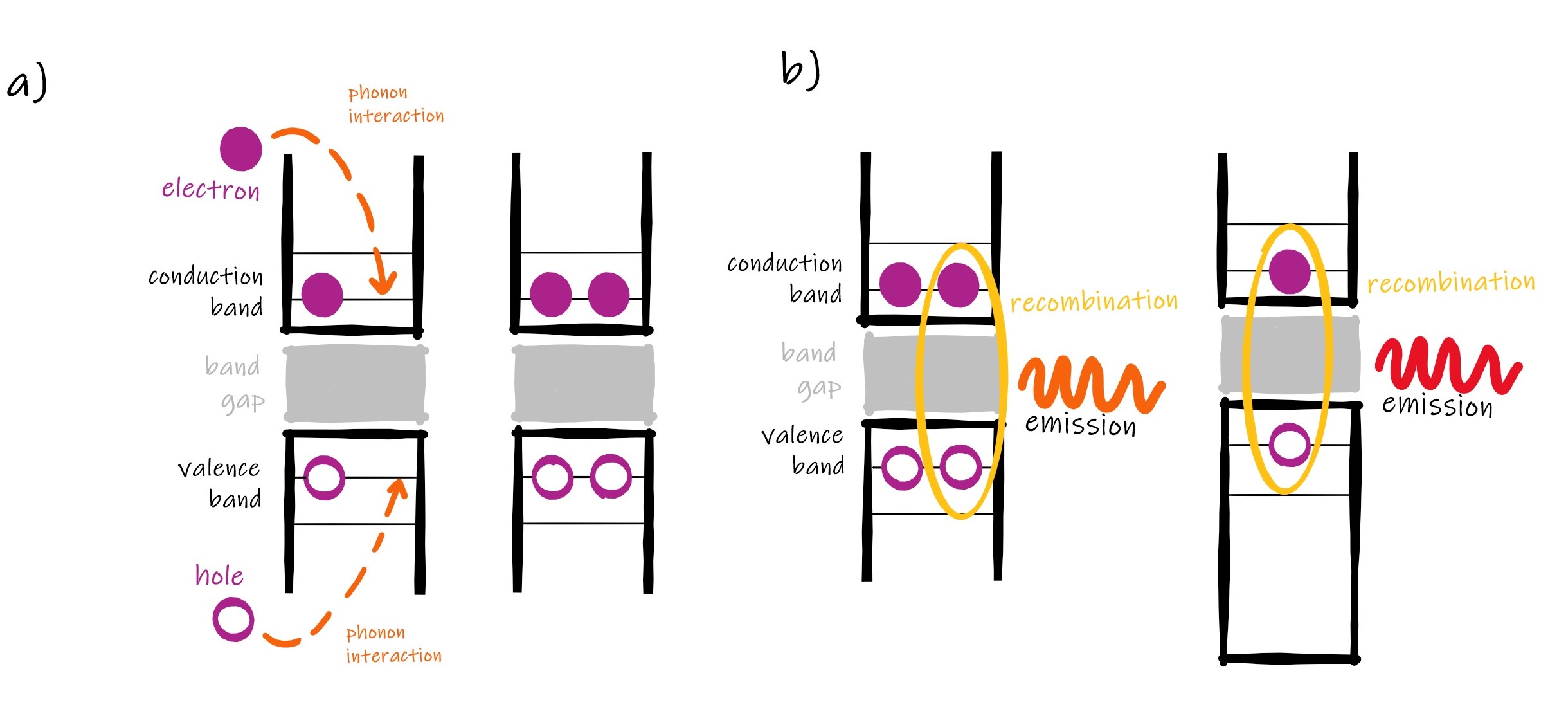}
\caption{Radiative cascade of carriers in a quantum dot. a) Electrons and holes created through carrier collisions or interaction with phonons (orange lines). b) Carriers radiatively recombine and emit photons in a cascade.}
\label{pic:qdemsys}
\end{figure}

Next promising single-photon source is the semiconductor \textbf{quantum dot}. Due to its quantum character, the dot has discrete energy structure (see Fig.~\ref{pic:qdemsys}). Therefore, when excited optically \cite{Brokmann2004} or electrically \cite{Zwiller2003}, in weak-excitation regime, when excitations are rare, electron-hole pair is created. The radiative recombination of this pair results in photon emission. Although the multi-photon emission may be induced with optical excitation, emitted photons have different wavelengths \cite{Oxborrow2005}. To eliminate this, the use of proper filters is needed. \\
Despite the fact that, due to the emission into the full solid angle, the collection of emitted photons from quantum dot is poor, it may be improved by integration with micro-cavity. Also, the fabrication of quantum dots, using epitaxial growth onto a prepared substrate, is very convenient and paves the way to miniaturization and integration of single-photon sources.  Unfortunately, to achieve the best results as a single-photon source, quantum dot needs to operate in cryogenic temperatures \cite{Senellart2017}.\\

All the photon sources described above are characterized by the fact that the emission of a photon is preceded by some kind of excitation (usually optical) of a system. Therefore, these types of sources are called "on-demand". Another approach is to produce a pair of photons, in such a way that one of them, when detected, heralds the existence of the other one. This is a so-called \textbf{probabilistic single-photon source}. It is based on production of a pair of photons in a probabilistic process, usually \textbf{spontaneous parametric down conversion (SPDC)} or \textbf{four-wave mixing (FWM)}, after the laser excitation of a nonlinear optical material. Since the laser pump beam may contain lots of photons and the nature of pair production process is statistical, to avoid the production of more than one pair, the average efficiency of pair-production levels must be held at low levels, much less than one. \\

\begin{figure}[t!]
\centering
\includegraphics[width=0.7\linewidth]{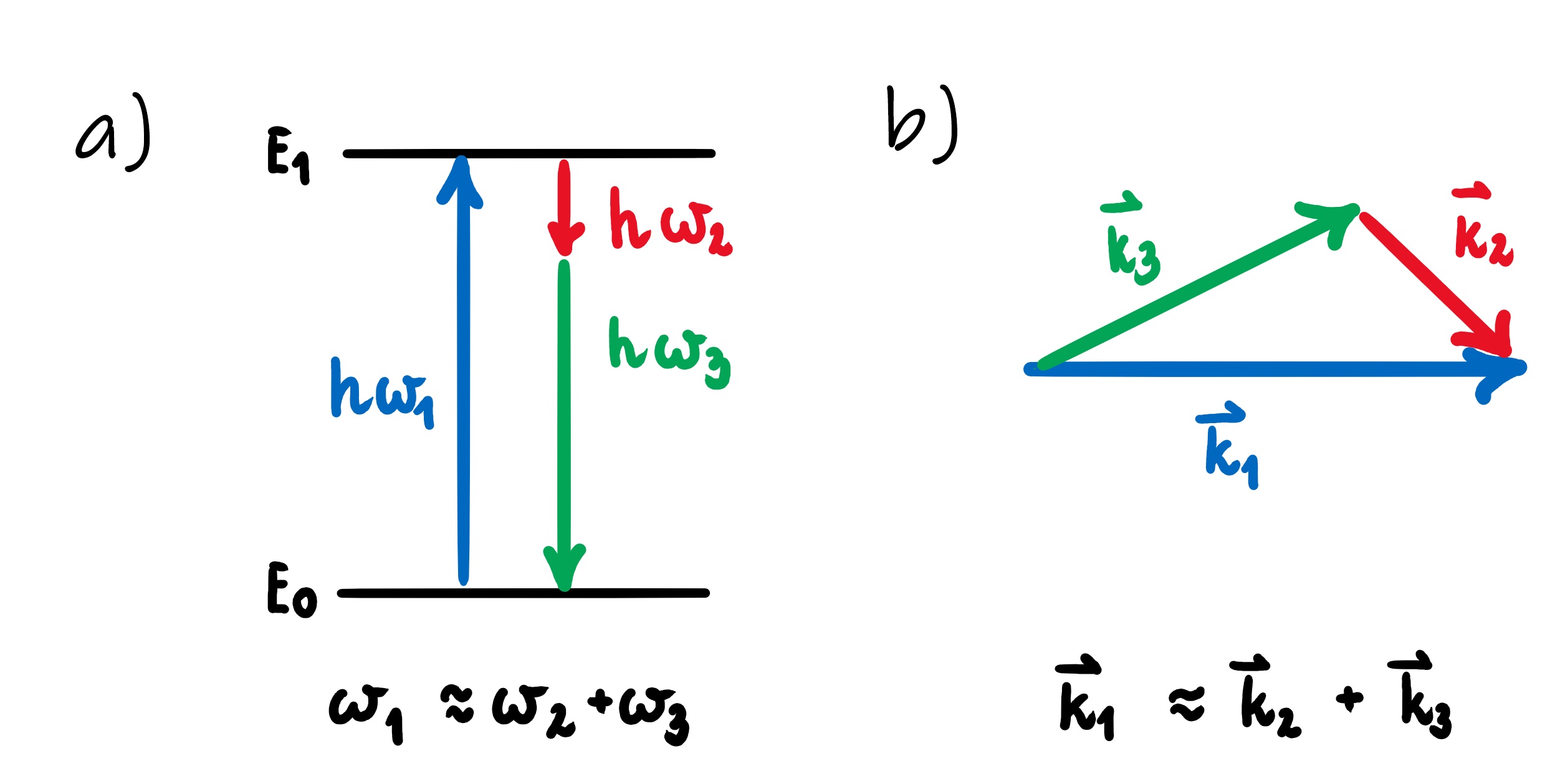}
\caption{a) The energy and b) momentum conservation for spontaneous parametric down-conversion process.  $E_0, E_1$ symbols stands for virtual energy states.}
\label{pic:corr-spdc}
\end{figure}

The \textbf{SPDC sources} exploit the nonlinear optical materials with non-zero elements of second order electric susceptibility tensor $\chi^{(2)}$. Such materials (crystals \cite{Burnham1970}, waveguides \cite{Chen2009}) enable the pump photon to be converted into a pair of photons, named signal and idler or heralded and heralding. This process, presented schematically in \textit{Fig.~\ref{pic:corr-spdc}}, was predicted theoretically in 1961 \cite{Louisell1961}, however it was not experimentally observed until 1969 \cite{Klyshko1970}.\\

\begin{figure}[h!]
\centering
\includegraphics[width=0.9\linewidth]{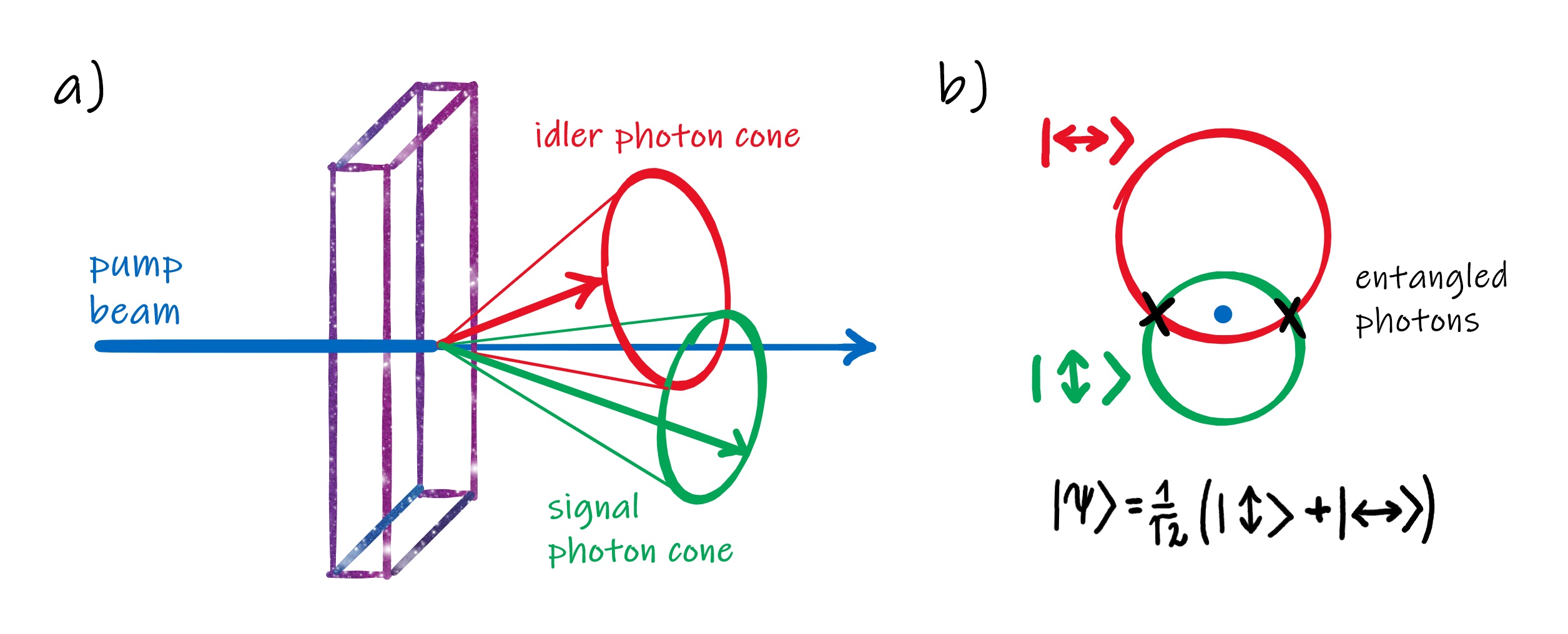}
\caption{The SPDC cones. a) Type-II process. b) Polarization-entangled photon pair. }
\label{pic:spdccones}
\end{figure}

In SPDC process, the pump photon is converted into a pair in accordance with the so-called phase-matching conditions, which consist of the law of conservation of energy and momentum \cite{book:Scully}. Due to that, photons are emitted into cones, as presented in Fig.~\ref{pic:spdccones}. Depending on the phase-matching relations, and thus, the polarization of all three photons, distinct types of SPDC processes can be obtained. \\
In \textit{type-0} all three photons have the same polarization, while the \textit{type-I} is characterized by the fact that both converted photons have the same polarization, which is perpendicular to the pump photon's polarization. On the other hand, in \textit{type-II}, polarizations of photons from the pair are perpendicular to each other, whereas the pump photon has the same polarization as the one from the pair. To clarify that, all SPDC types are collected in Tab. \ref{tab:SPDCtypes}. \\

\begin{table}[h!]
\centering
\begin{tabular}{r|c|c|c}
\hline
\multirow{2}{*}{\textbf{type of SPDC}} & \multicolumn{3}{c}{\textbf{polarization of a photon}}\\
&\textbf{pump}&\textbf{signal}&\textbf{idler}\\
\hline
\multirow{2}{*}{type-$0$}&$\udarr$&$\udarr$&$\udarr$\\
&$\lrarr$&$\lrarr$&$\lrarr$\\
\hline
\multirow{2}{*}{type-I}&$\udarr$&$\lrarr$&$\lrarr$\\
&$\lrarr$&$\udarr$&$\udarr$\\
\hline
\multirow{2}{*}{type-II}&$\lrarr$&$\lrarr$&$\udarr$\\
&$\udarr$&$\udarr$&$\lrarr$\\
\hline
\end{tabular}
\caption{The polarization of pump and converted photons for different SPDC process types.}
\label{tab:SPDCtypes}
\end{table}

It is noteworthy that in case of the type-II SPDC process, photons, which are at the intersection of cones (see Fig.~\ref{pic:spdccones}), are found in the polarization-entangled state. This property makes the type-II SPDC single-photon source widely used in quantum optical experiments \cite{Kwiat1995, Couteau2018, Kuo2020}. \\

\begin{figure}[b!]
\centering
\includegraphics[width=0.6\linewidth]{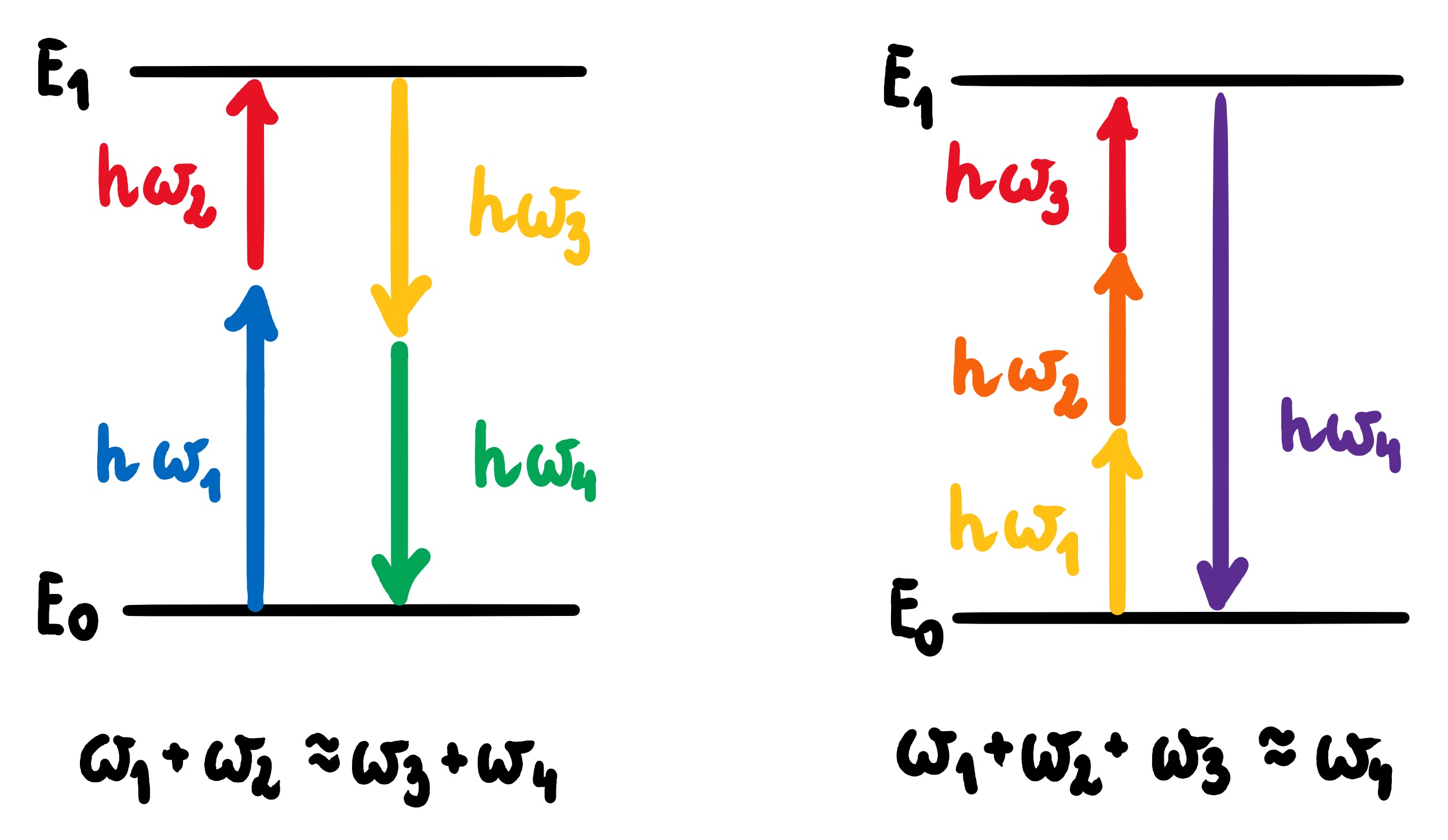}
\caption{The  scheme of FWM energy conservation relations. $E_0, E_1$ symbols stands for virtual energy states.}
\label{pic:corr-fwm}
\end{figure}

Similarly to the SPDC sources, \textbf{FWM sources} utilizes the nonlinear optical materials, but with non-zero elements of third order electric susceptibility tensor $\chi^{(3)}$. The medium allows for interaction between two or three photons, which results in the production of two or one photon respectively \cite{book:Scully, book:Fox}. The wavelength conversion relation is presented schematically in \textit{Fig.~\ref{pic:corr-fwm}}.\\

As it was  mentioned earlier, the probabilistic source is meant to be used as a heralded one, so the interest is to convert a pair of pump photons into other pair, signal and idler photon, as previously described. Such photons, due to the fact that the FWM sources are usually based on optical or crystal fibers \cite{Antunes2006,Alibart2006,Shukhin2020}, are fiber coupled, which makes them friendly to use. It makes them also a potential candidate for on-chip miniature of single-photon sources. \\

Although the efficiency of probabilistic sources is much lower than for deterministic ones and emission of a photon is not "on-demand", the heralding of photons is of much use in many quantum optical experiments. Adding to that, the experimental setups are much simpler and do not require the use of cryocoolers. Such sources are perfect candidates for miniaturization and commercial use in quantum cryptography applications. \\
\subsection{Single-photon detectors}
\label{sec:ch2-sinphdet}

For many quantum optical experiments, especially those that involve a research towards the security  of quantum communication protocols, a high-performance single photon detection is crucial. Usually, a single-photon detection is based on conversion of a photon into an electric signal, therefore, the detection efficiency comes from the design of a detector electronics (for example, in case of avalanche photodiode it is crucial when the carriers' avalanche is stopped). Also, it is noteworthy that to achieve the best possible results, the detector electronics are often as important as a detector itself. However, it is beyond the scope of this thesis.\\
The values of the characteristic parameters of the detector may differ significantly for individual models within one type. Because of that,  here, only the main aspect of detectors that are widely used in quantum optical laboratories, important in the experimental use, are described. The comparison of the detectors containing numerical data can be found in Ref. \cite{Eisaman2010}, Ref. \cite{Chunnilall2014} and Ref. \cite{You2020}. \\

An ideal single-photon detector is the one for which the \textit{detection efficiency} of an incident photon should reach $100$\%, but the \textit{dark count rate} should be $0$. It means that the probability of a successful detection of a single photon equals $1$, whereas the probability of a detection in the absence of photons equals $0$. Moreover, the \textit{dead time}, defined by the minimum time interval that the incidence of the two consecutive photon pulses must be separated, in order to detect them separately, should be equal to $0$. Furthermore, the time uncertainty  between the detection of a photon and the generation of an output signal, called \textit{timing jitter}, should be equal to $0$. Lastly, an ideal single-photon detector should have the ability to specify the number of photons in an incident pulse. \\

Since similarly to the case of  the ideal single-photon source, an ideal single-photon detector does not exist, among the detectors two general kinds may be distinguished -- \textbf{photon-number resolved} and, so called, \textbf{bucket detectors}. \\
The main difference between those two kinds of detectors is that, in contrast to the photon-number resolved detectors, which can detect the number of photons in the incident pulse, the bucket detector is not photon-number resolved. It means that such a detector can only distinguish between states with zero photons and more then zero photons.
Unfortunately, this is not a sharp division. Due to the losses and defects, the detection efficiency of photon-number resolved detectors is less than unity, which translates to the fact that such a detector never tells the true number of incident photons. What is more, the detectors considered as non-photon-number resolved, sometimes have a photon-number resolving capability \cite{ Eisaman2010, Nevet2011}. For example, it may happen that a photon's energy is too low to be detected, whereas a greater number ($N$) of photons in incident pulse may have the total energy sufficient to be detected. Then, such a detector distinguishes between less than $N$- and equal or more than $N$-photon-number states. \\

In the group of the \textbf{bucket detectors}, due to their prevalence and ease of purchase, worth to mention are:
\begin{itemize}
\item[--] \textit{photomultiplier tube (PMT)} \cite{Simon1968, Morton1968};
\item[--] \textit{single-photon avalanche photodiode (SPAD)} \cite{Lacaita1996, Zappa2007};
\item[--] \textit{supercondcting single-photon detector} (SSPD) \cite{Hadfield2005};
\item[--] \textit{up-conversion single-photon detector (UCSPD)} \cite{VanDevender2004}.
\end{itemize}

In the \textbf{PMT} \cite{Morton1968}, firstly used as a visible wavelength range detectors an incident photon knocks an electron out from a photocathode. Then this electron is accelerated towards the first dynode, where it knocks out more electrons. Since those electrons are also accelerated, this process is repeated. The signal created by the electrons is then detected by the electronics. Although the detection efficiency usually is high (in the range of $10$\% to $40$\%), the dark count rate is low (a few events per second \cite{Donati2014}), similarly to timing jitter and dead time, the performance of PMT relies on the vacuum technology used for fabrication. It limits the lifetime of such a detector and adversely affects their development \cite{Eisaman2010}.\\

The \textbf{SPAD}, another type of a bucket detector, works similarly to the PMT. An incident photon  creates an electron-hole pair, which starts the continous avalanche of creating charges. The avalanche is terminated by a specific electronic circuit in order to prepare SPAD for a subsequent photon pulse \cite{Zappa2007, Chen2020}. Despite that SPADs have higher detection efficiencies than PMTs, up to $85$\% for visible wavelength range (Si-based SPADs) \cite{Donati2014, Chen2020, Bronzi2016}, the efficiency for infra-red range (InGaS/InP-based SPADs), when cooled, is around $30$\% \cite{Donati2014, Bronzi2016, Itzler2011, Vines2019}. The timing jitter and dark counts are much higher than for PMTs. Also, due to the use of a semiconductor, the detection may suffer from \textit{afterpulsing effect} (the avalanche may be started by carriers released from a trap instead an incident photon). To prevent this, the dead time is prolonged, sometimes even to microsecond values. \\ 

Another type of a bucket detector, which, contrarily to SPAD, is designed to have high detection efficiency for infrared wavelength range (up to $90$\%) \cite{Marta2018}, is based on a superconducting NbN nanowire. In \textbf{SSPD}, the incident photon puts the wire into the state of normal resistance, thanks to which a voltage drop is detected. Due to that, such detectors have low timing jitter (around $20$ ps)\cite{Zolotov2018}. The dark count rate is also low, however, it is mostly a consequence of working in a cryogenic temperature (up to $4$ K). It is a main drawback of this kind of detector. On the other hand, it had been shown that when carrying the heralded type of measurement, even when the photon wavelengths are not optimal for the detection efficiency of a detector, the use of a pair of SSPDs leads to high total detection efficiency in the setup \cite{Marta2018}. \\

Similarly to SSPDs, the \textbf{UCSPD} is typically designed to detect the infrared photons with high efficiency. It is based on a nonlinear crystal, in which up-conversion of a single photon occurs, which results in generation of a photon from a visible wavelength range \cite{Ma2018}. Although the efficiency near the unity was reported for the up-conversion step \cite{VanDevender2007}, the overall detection efficiency is usually from $30$\% to $60$\% \cite{Eisaman2010}. Also, due to the up-conversion process, such detectors have the limited working wavelength range. \\

As it was mentioned earlier, bucket detectors sometimes may behave as if they were photon-number resolved. However, if during an experiment the photon-number resolution is needed, usually one of the following is used: 
\begin{itemize}
\item[--] \textit{photon-number resolved superconducting nanowire single photon detector (PNR SNSPD)} \cite{Divochiy2008};
\item[--] bolometers -- \textit{superconducting transition edge sensor (TES)} \cite{Cabrera1998} or \textit{hot electron bolometer (HEB)} \cite{Romestain2004};
\item[--] SPAD-based PNR detectors -- \textit{SPAD arrays} \cite{Jiang2007} or \textit{multipixel photon counters (MPPC)} \cite{Kalashnikov2011}.
\end{itemize}

In principle, the \textbf{PNR SNSPD} operates similarly to its bucket cousin, the only difference is that for PNR detectors multiple superconducting nanowires are put inside. Each nanowire can detect a single photon, therefore, by connecting them with proper electronics, such system is able to distinguish the number of photons in incident pulse. Recently, due to impressive detection efficiency as high as $86$\% for a telecom-wavelength range \cite{Moshkova2019}, low dark count rate and low jitter \cite{Zolotov2018, Zhang2019}, such detectors gain an attention. Nevertheless, they need to be cooled to cryogenic temperatures (from $2$ K to $4$ K).\\

The next kind of PNR detector, which will be looked at more closely, are those based on a device called bolometer, which measures the number oft photons via the thermal changes of a material it consists of. The energy of incident photons yields in a change of a resistance of such material, thus, the output voltage is proportional to the number of absorbed photons. There are two types of bolometers -- \textbf{TES} and \textbf{HEB}. Both detectors have sensors that are built from the thin layers of the superconducting film (for HEB it is only $5$ nm thick). Since they are temperature sensitive, they need to operate at very low temperatures (for TES it is mK scale, for HEB around $4$ K), however, the systems operating at or slightly above liquid nitrogen temperatures ($77$ K) had also been reported \cite{Romestain2004}. It is also noteworthy that unlike TES detectors, which deadtime is around $1$ $\mu$s, HEB detectors are quite fast (deadtime in nanosecond scale) \cite{Romestain2004}.  Another difference between these detectors is that TESes are typically designed for wavelengths in the range $1-2$ $\mu$m, whereas HEBs are appropriate for longer wavelengths, above $3$~$\mu$m.  Nevertheless, such detectors provide the detection efficiency up to $95$\% \cite{Zhang2019} (in case of TES it is even $99$\% at $850$ nm \cite{Lita2009}), which is very convenient in terms of photon-number-resolved capabilities and future applications. \\

Other types of PNR detectors that are worth to mention are those SPAD-based. There are two architectures of such detectors -- the \textbf{SPAD arrays}, which consisted of individual SPADs, with signal read out individually, and the \textbf{MPPC}, in which SPADs built a circuit to give single summed output signal with voltage proportional to the number if detected photons. Although the detection efficiencies of SPAD-based detectors is around $50$\% \cite{Kalashnikov2011, Cai2019}, they may feature \textit{the cross-talk effect}, in which the neighbouring pixels create additional dark counts due to the thermal instabilities \cite{Bronzi2016}. In site of this fact, the main advantage of such detectors is that they usually operate in room temperatures. They are also designed for visible-wavelength-range photons. These properties make them relatively cheap and affordable, which is a good sign in terms of their development. \\

By adding a comment to all the information above, although it may be obvious, the experimental purpose and design of the experimental setup are the most important factors when deciding what type of the detector should be used. Usually, the chosen detector has parameters that represent a compromise between all achievable possibilities, which range is much wider than of those listed subjectively above.\\

\chapter{SPDC-based single photon sources}
\label{sec:ch3}

In the beginning of this chapter, the two single-photon sources based on SPDC process are described. Later, the characterization parameters as well as the methods of their measurements are shown. \\


\section[On a lab table]{On a lab table \newline\small{Experimental realization}}
\subsection*{(S1) Source No 1 -- visible and infrared photons}
\label{sec:ch3-visinfra}
The S1 source, based on type-$0$ SPDC process, was designed to generate a pair of photons with $532$ nm and $1550$ nm central wavelengths when illuminated with $396.1$ nm pump beam. \\

\begin{figure}[t]
\centering
\includegraphics[width=0.8\linewidth]{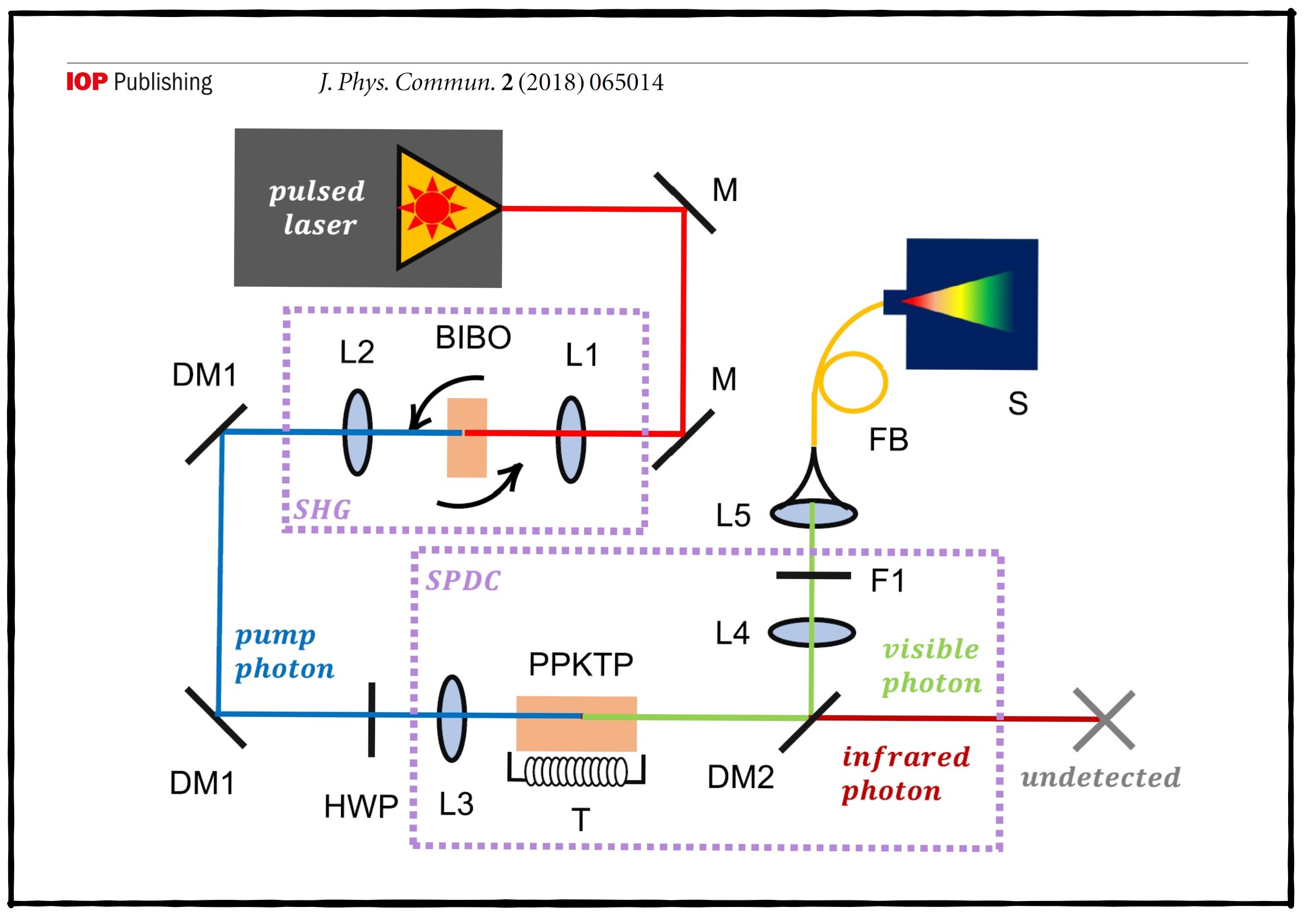}
\caption{Scheme of SPDC single-photon source based on PPKTP crystal and experimental setup that enables  characterization of its the phase-matching conditions. Symbols: pulsed laser -- Ti:Sapphire Chameleon Coherent Inc. laser, M -- mirror, L1,L2 -- lenses (focal lengh $f=7.5$ cm), BIBO -- bismuth triborate crystal, DM1 -- dichroic mirror (Semrock T425 LPXR), HWP -- half-wave plate, L3, L4 -- plano-convex lenses ($f=10$ cm, $12$ cm, respectively), PPKTP -- periodically poled potassium titanyl phosphate crystal, T -- temperature controller, DM2 -- dichroic mirror (Semrock 76-875 LP), F1 -- set of filters (Chroma ET500, Z532-rdc ), L5 -- aspheric lens ($f=1.51$ cm), FB -- fiber (Thorlabs SMF460B), S -- spectrometer (Ocean Optics USB2000+).}
\label{pic:s1}
\end{figure}

The scheme of the source along with experimental setup that enables the source characterization is presented in Fig.~\ref{pic:s1}. It consists of two parts: the first one, marked as SHG, is the frequency doubler for the pulsed laser, the second one is a photon pair generator based on nonlinear crystal. Let us follow the light beam inside the setup. \\

The femtosecond pulsed laser (pulse duration $120$ fs, period $12.5$ ns) emits light with $792$ nm central wavelength (marked with a red line in the figure), which reflects off the two silver mirrors (M) and is focused using plano-convex lens (L1) on the BiBO nonlinear crystal. The crystal doubles the frequency of incoming light, which results in generation of a light beam with $396$ nm central wavelength (blue line). The phase-matching conditions for doubling process (\textit{second harmonic generation (SHG)} \cite{book:Scully, book:Fox}) is satisfied by rotation and tilting of a BiBO crystal, which is marked by the arrows near the crystal in the experimental setup scheme. Both, the blue and the red beam, are then collimated using plano-convex UV lens (L2). However, whereas the SHG beam (blue line) is reflected off the two dichroic mirrors (DM1), the fundamental laser beam is filtered out (red line). \\
Next, the SHG beam is used to pump the SPDC source. It passes through the half-wave plate (HWP), which rotates its polarization and using another plano-convex UV lens (L3) is focused into the PPKTP crystal. The temperature of the crystal is stabilized by the temperature controller (T). The periodically polled crystal generates a pair of photons, which creates two overlapping  photon beams -- visible (green line) and infrared one (purple line). These beams are separated using the dichroic mirror (DM2). \\
For the purpose of experimental measurements, the infrared beam stays undetected, whereas the visible beam is collimated using plano-convex lens (L4). Then, the remainings of the fundamental beam (red), pump beam (blue) and generated infrared beam (purple) are filtered out using sets of spectral filters (F2). Next, the visible photon beam (green) passes through the filters and is collimated into the single-mode fiber (FB) using aspherical lens (L5). At last, the fiber is connected to the spectrometer (S).\\

The S1 source was firstly described in Journal of Physics Communications' article entitled \textit{Dispersion measurement method with down conversion process} \cite{Andrzej2018}. My contribution to the work presented there included building the experimental setup, conducting measurements, and collecting the spectra and proper coincidence histograms for characterization of the source. I also prepared all schemes and figures for the publication. The obtained results are discussed further in this chapter. \\

\FloatBarrier
\subsection*{(S2) Source No 2 -- infrared photons}
\label{sec:ch3-redinfrared}

\begin{figure}[h!]
\centering
\includegraphics[width=0.8\linewidth]{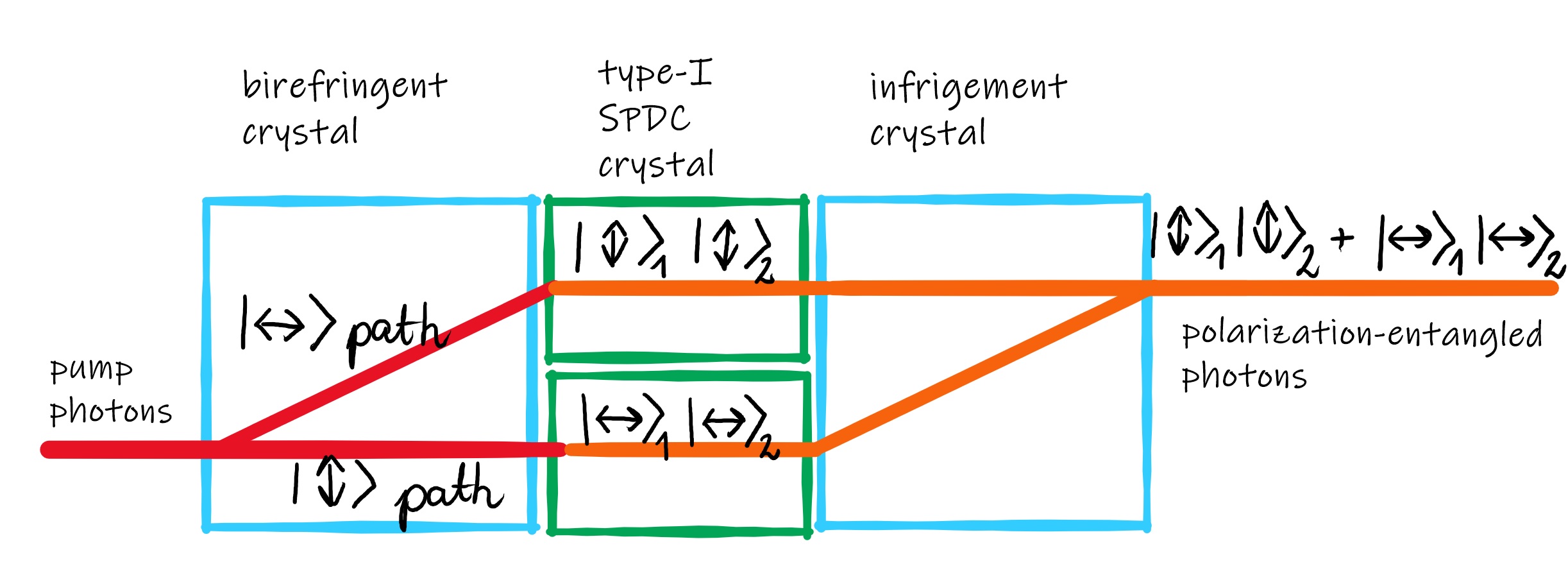}
\caption{Entangled-photon source scheme.}
\label{pic:s2ent}
\end{figure}

The S2 source, based on type-I SPDC process, was designed to produce infrared entangled photon pairs (central wavelength $1550$ nm) in polarization degree of freedom.\\
The main element of such a source is the set of nonlinear crystals, two SPDC crystals sandwiched between a pair of birefringent ones. As presented in Fig.~\ref{pic:s2ent}, the incident polarized beam is split onto first birefringent crystal. Then, each SPDC crystal, type-I phase-matched, generates pair of infrared photons in specified polarization state, orthogonal to each other. One crystal produces a pair with vertical $|\udarr_1\udarr_2\rangle$, the other in horizontal $|\lrarr_1\lrarr_2\rangle$ polarization. The last crystal is the infrigement one, where both paths are overlapped. That allows to obtain polarization-entangled state $|\udarr_1\udarr_2\rangle+|\lrarr_1\lrarr_2\rangle$ \cite{Fiorentino2008}. \\

In our case, commercially available system is used, patented by Rolf. T. Horn \cite{Hornpatent}. It consists of six crystals, two SPDC crystals and four birefringement ones \cite{Horn2019}. Such configuration enables obtaining better indistinguashability of generated photons, which gives better quality of experimental results. Nonetheless, the results may be further improved by boosting the coupling efficiency by optimization of the focal lengths of both, input and output, lenses.\\

\begin{figure}[t]
\centering
\includegraphics[width=0.8\linewidth]{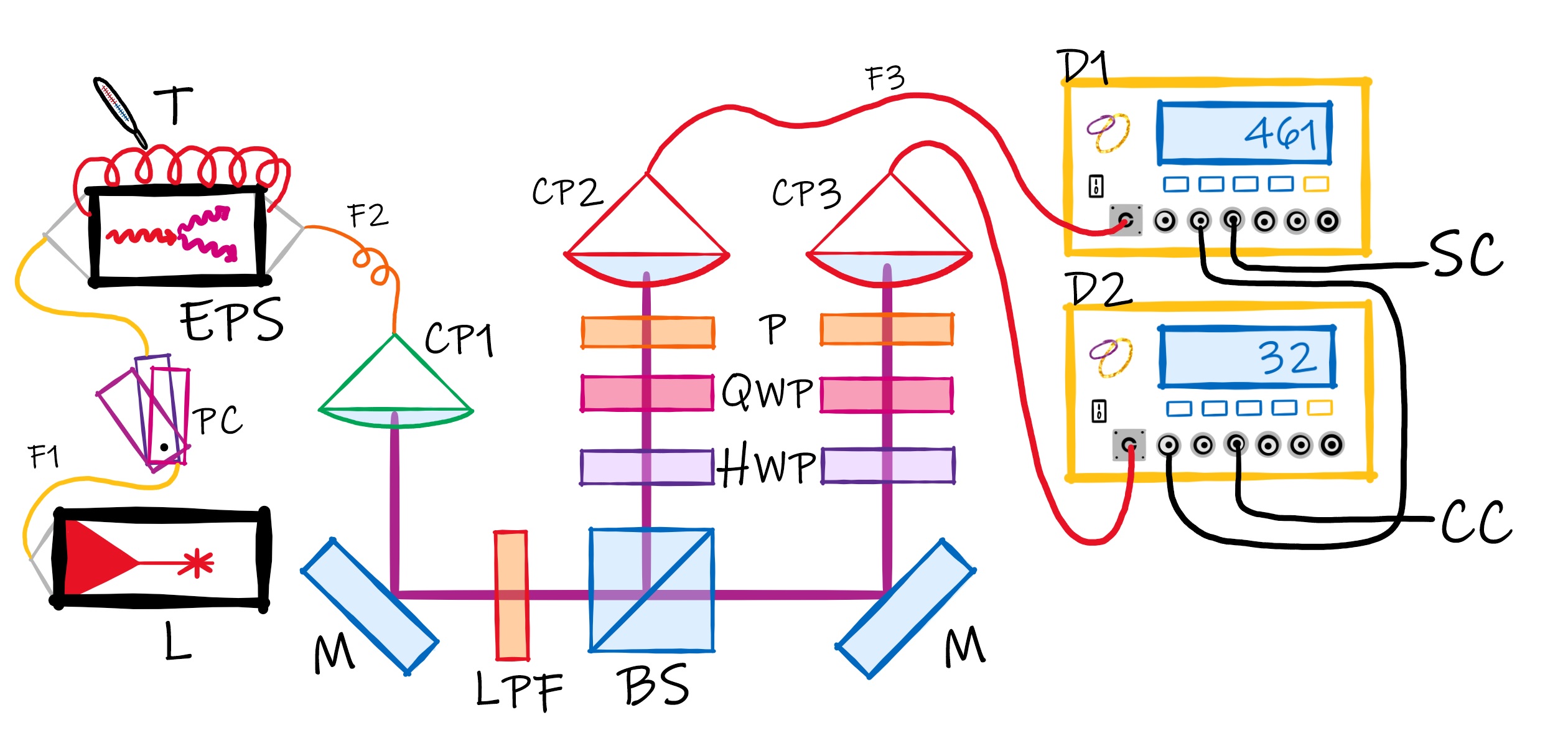}
\caption{Scheme of infrared entangled photon source and experimental setup that enables quantum-state tomography measurement. Symbols:
L -- pump laser (QPhotonics QFLD-775-10SB), F1 -- single-mode fiber (Thorlabsa SMF630A), PC -- polarization controller, EPS -- entangled-photon source, T -- temperature controller, F2 -- single-mode fiber (Fibrain SMX-28), CP1 -- collimation package (Thorlabs F260FC-1550), M -- silver mirror (Thorlabs PF10-03-P01), LPF -- longpass filter (AHF F76-1321, BLP01-1319R-25), BS -- beams-splitter cube (Thorlabs BS006), HWP -- half-wave plate (Altechna 2APW-L4-012D), QWP -- quater-wave plate (Altechna 2-APW-L4-012D), P~-- polarizer (Thorlabs LPNIR100-MP2), CP2, CP3 -- collimation aspherical lenses (Thorlabs A260TM-C), F3 -- single-mode fiber (Fibrain SMX-28), D1 , D2 -- single-photon InGaAs detectors (IdQuantique Id201), SC -- single counts' port, CC -- coincidence counts' port.}
\label{pic:s2}
\end{figure}

The scheme of the source along with experimental setup that enables the quantum-state tomography measurement of generated photons is presented in Fig.~\ref{pic:s2}. Let us follow the light path through the setup. \\

The $780$ nm laser, either continuous wave or pulsed, is fiber coupled (F1). Then, the fiber is wound on polarization controller (PC) and connected to the thermally controlled (T) entangled photon source (EPS), which allows to pump sets of crystals and therefore, to generate polarization-entangled photons. Infrared photons with $1550$ nm central wavelength are then coupled into the fiber (F2) and collimated with collimation package (CP1) for further measurements. The rest of the experimental setup is used to conduct the quantum-state tomography measurements. \\
The infrared photon beam (marked with the purple line) reflects off the mirror (M) and passes through the long-pass filter, which blocks the remainings of pump beam. Then, the beam is divided into the beam-splitter, which is followed by sets of half-wave plate (HWP), quater-wave plate (QWP), and polarizer (P), identically in both arms of the setup. Then, the photons are coupled into the fiber with collimation packages (CP2) and detected using the InGaAs detectors (D1, D2). \\
The quantum-state tomography measurement procedure and its result is elaborated further in \textit{Section~\ref{sec:ch3-qst}}. \\

The S2 source was built and characterized in the FamoLab for testing an \textit{electronic quantum entanglement generator controller} designed by \textbf{Syderal Polska} company as a part of their R\&D project financed by NCBiR. \\

\section[Mathematics in practice]{Mathematics in practice\newline\small{Theoretical description}}

Most quantum optical experiments are based on SPDC single-photon source. When designing such an experiment, one of the most important things is the knowledge about the performance of such a source. Therefore, the first step is its characterization, both theoretical and experimental. 
Parameters of the single-photon source, which are tested during the characterization, include spectra of generated photons, their polarization, and photon-number statistics. \\

Typically, theoretical consideration of SPDC process starts from single-mode electromagnetic field in the nonlinear optical materials with non-zero elements of second-order electric susceptibility tensor $\chi^{(2)}$. After quantization of electromagnetic field and other mathematical transformations, which can be found in Ref. \cite{book:Scully}, these considerations result in the second order interaction hamiltonian that describes the SPDC process \cite{book:Scully, book:Loudon}. From this, using the interaction picture, the probability amplitude function of photon pair generation can be calculated. If the photon pair with a specific correlated wavelength is generated using continuous-wave laser, it can be written as:
\begin{equation}
\Psi (\omega_s,\omega_i)=C\,d_{eff}\, L\cdot sinc \Big( \frac{\Delta k_z (\omega_s, \omega_i)}{2}\Big)\; .
\label{eq:spdcampl}
\end{equation}
This function consistes of two parts, the first one characterizes properties of the medium ($d_{eff}$ is an effective nonlinear coefficient of a crystal, $L$ is its length), the second describes photons which propagates inside ($\Delta k_z (\omega_s,\omega_i)$ is a so-called phase-mismatch). There is also constant $C$, correlated with electromagnetic field amplitudes. It is important to mention that since the function presented in Eq.~\ref{eq:spdcampl} originates from the second order of interaction hamiltonian, it is not normalized in the sense that $\langle\Psi|\Psi \rangle \neq 1$. Thanks to this, it may be used in further calculations. \\
The full function, which includes all orders of emission events (not presented here), is normalized in a standard way \cite{book:Scully, book:Loudon}.\\

\subsection{Conversion efficiency}
\label{sec:ch3-eff}
The conversion efficiency $\eta$ of SPDC process operationally can be defined as:
\begin{equation}
\eta=\frac{N_p}{N}\,,
\end{equation} 
where $N_p$ is the number of generated photon pairs and $N$ is the number of pump photons. These two values, $N_p$ and $N$, can be also measured and thus, neglecting the losses inside crystal and setup, give the estimation of SPDC conversion efficiency. Maximal conversion efficiency values reported in literature are in the order of $10^{-9}$ for bulk crystals, $10^{-8}$ for periodically-poled crystals, and $10^{-6}$ for waveguides \cite{UrbasiTAB}. Typically, bulk and periodically poled crystals, commercially easily available and widely used in laboratories, have conversion efficiency in the order of $10^{-10}$ and $10^{-9}$, respectively.   \\

In an ideal case, for single-mode pump photon beam, the number of generated pairs is equal to $N_p=|\Psi|^2\cdot N$. The more photons are generated, the greater the conversion efficiency. Since the number of photons generated depends directly on the effective nonlinear crystal coefficient (see Eq.~\ref{eq:spdcampl}), the choice of the right material is crucial for this process. Let us take a closer look at the origin of the effective nonlinear crystal coefficient $d_{eff}$. \\

The propagation of a light beam inside any medium is described by Maxwell equations. The interaction between light and dielectric medium changes its properties and reveals as the \textit{polarizability} $\alpha_e$ of the medium, described by Clausius-Mossotti equation \cite{Rysselberghe1932}. In case of homogeneous and isotropic dielectric material with density $d$, the equation takes the form:
\begin{equation}
\frac{M}{d}\cdot\frac{\epsilon_r-1}{\epsilon_r+2}=\frac{N_A\alpha_e}{3\epsilon_0}\, ,
\end{equation}
where $M$ is the molar mass, $\epsilon_0$ is the permittivity of a vacuum, $\epsilon_r$ is the relative permittivity of a material, $N_A$ is an Avogadro's number. Also a polarization $\vec{P}$ of a material, defined as a total dipole moment per unit volume, is related to polarizability:
\begin{equation}
\vec{P}=N\cdot\alpha_e\vec{E}=\epsilon_0\chi\vec{E}\, ,
\end{equation} 
where $N$ is the number of molecules per unit volume and $\chi$ is the electric susceptibility, $\chi=\epsilon_r-1$. \\

In the general case of a medium with complicated structure, which is not homogeneous or isotropic, the polarization $\vec{P}$ is defined using following expansion:
\begin{equation}
\begin{split}
&\vec{P}=\sum_{\alpha=1}^3 \vec{P}(\vec{r},t)\cdot \vec{e}_{\alpha}\,=\,\\
=\, \epsilon_0 \sum_{\alpha=1}^3\Big(\sum_{\beta=1}^3\chi^{(1)}_{\alpha\beta}E_{\beta}\,+\, \sum_{\beta,\gamma=1}^3&\chi^{(2)}_{\alpha\beta\gamma}E_{\beta}E_{\gamma}\,+\, \sum_{\beta,\gamma,\delta=1}^3\chi^{(3)}_{\alpha\beta\gamma\delta}E_{\beta}E_{\gamma}E_{\delta}\,+\,...\Big)\vec{e}_{\alpha}\, ,
\end{split}
\label{eq:polgen}
\end{equation}  
where $\vec{e}_{\alpha}$ is a basis versor and  $\chi^{(n)}$ is the n-th order electric susceptibility tensor that characterizes the n-th order of light-matter interaction. For example, the second order tensor corresponds to nonlinear processes such as second-harmonic generation and spontaneous parametric down-conversion, thus, description below is focused on this term.\\

Eq.~\ref{eq:polgen} shows that the second order of electric susceptibility tensor has 27 elements. Fortunately, due to the use of Kleinman symmetry conditions for permutations of elements' indices \cite{book:Boyd,book:He,Kleinman1962}:\\
\begin{center}
\begin{tabular}{c|cccccc}
$\beta\gamma$&11&22&33&23 or 32&13 or 31&12 or 21\\
$\zeta$ &1&2&3&4&5&6 \\
\end{tabular}
\end{center}
tensor reduces to 18 elements, therefore, after the substitution $\chi^{(2)}_{\alpha\beta\gamma}\;\rightarrow \;d_{\alpha\zeta}$, the tensor takes a form:
\begin{equation}
d_{\alpha\zeta}=\Bigg(\;\begin{matrix}
d_{11}&d{12}&\cdot\cdot\cdot&d_{16}\\
d_{21}&d{22}&\cdot\cdot\cdot&d_{26}\\
d_{31}&d{32}&\cdot\cdot\cdot&d_{36}\\
\end{matrix}\;\Bigg)\; .
\end{equation}
Later, after introducing the net symmetry of the medium, the tensor is further simplified to eight or ten nonzero elements for crystals with \textit{monoclinic class 2} and \textit{m} structure, respectively. The exact forms of tensor $d_{\alpha\zeta}$ can be found in Ref. \cite{book:Dmitriev} and Ref. \cite{Tzankov2005}. \\

Next step in obtaining a single-value effective nonlinearity $d_{eff}$, is to find the direction of unit polarization vectors inside a crystal for propagating \textit{slow} and \textit{fast beam} \cite{book:Boyd,book:He}. Using them, known as \textit{direction cosines}, the effective nonlinearity can be represented as a sum of products of proper $d_{\alpha\zeta}$ elements and unit polarization vectors'. General expressions that allow to calculate effective nonlinearity for diferent types of crystals can be found in Ref. \cite{book:Dmitriev}. Parameters, including effective nonlinearity values, for nearly all commercially available crystals can be found in Ref. \cite{book:Nikogosyan}. \\

Operationally, for experimental use the most preferable option is to choose the crystal with the highest effective nonlinearity value for which desired phase-matching conditions are satisfied. \\


\subsection{(Quasi)phase matching}
\label{sec:ch3-qpm}
The spectral and spatial properties of generated photons in SPDC process are described by \textbf{phase-matching conditions} $\Delta k_z\,=\,0$, which, in principle, are the energy and momentum conservation relations that in case of collinear beams can be written as $k_p=k_s+k_i$, as described in \textit{Section \ref{sec:ch1-sinphsource}} and presented in Fig. \ref{pic:corr-spdc}. \\

Theoretically, phase-matching conditions can be derived in many ways. One approach is to further consider  the probability of generating photon pair, described by Eq. \ref{eq:spdcampl} that results in maximization of $sinc(\cdot)$ expressions, therefore:
\begin{equation}
sinc\Big(\frac{\Delta k_z (\omega_s,\omega_i) L}{2}\Big)=1 \;\rightarrow\; \Delta k_z (\omega_s,\omega_i) \,=\,0 \;.
\label{eq:qpmbk}
\end{equation}

In practice, it is important to check the transmission and absorption of a nonlinear crystal at desired wavelength range.  \\

\begin{figure}[t]
\centering
\includegraphics[width=0.5\linewidth]{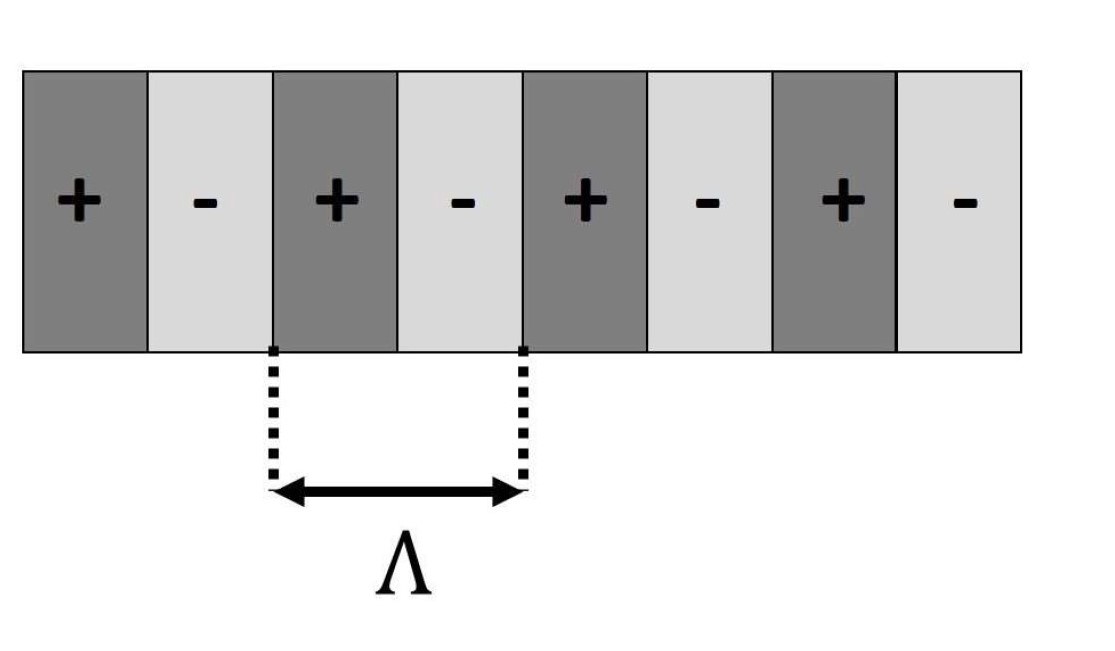}
\caption{Perodically poled crystal model. Symbols '+' and '--' stand for the sign of effective nonlinear coefficient $d_{eff}$, $\Lambda$ is crystal's poling period.}
\label{pic:ppcryst}
\end{figure}

Currently, the most popular are the periodically poled crystals, which spatial nonlinear properties are modulated during the fabrication process. The structure of such a crystal consists of small domains, for which the sign of effective nonlinear coefficient is periodically reversed, as presented in Fig. \ref{pic:ppcryst}. Mathematically, this requirement can be written as \cite{Houe1995, Arie2007}:
\begin{equation}
d_{eff}(z)=d_0\sum_{m=-\infty}^\infty G_m e^{i\,\frac{2\pi m}{\Lambda}\,z} \; ,
\label{eq:deff}
\end{equation}
where $d_0$ is the magnitude of interaction, $\Lambda$ is the crystal poling period and $G_m$ is Fourier coefficient, which for rectangular grating with duty cycle $D$ is equal to:
\begin{equation}
G_m=\frac{2}{\pi\, m} sin\big(\pi \, m \,D\big) \; .
\end{equation}
More examples of Fourier coefficients for various types of crystals can be found in Ref. \cite{Arie2007}. \\

\begin{figure}[b!]
\centering
\includegraphics[width=0.8\linewidth]{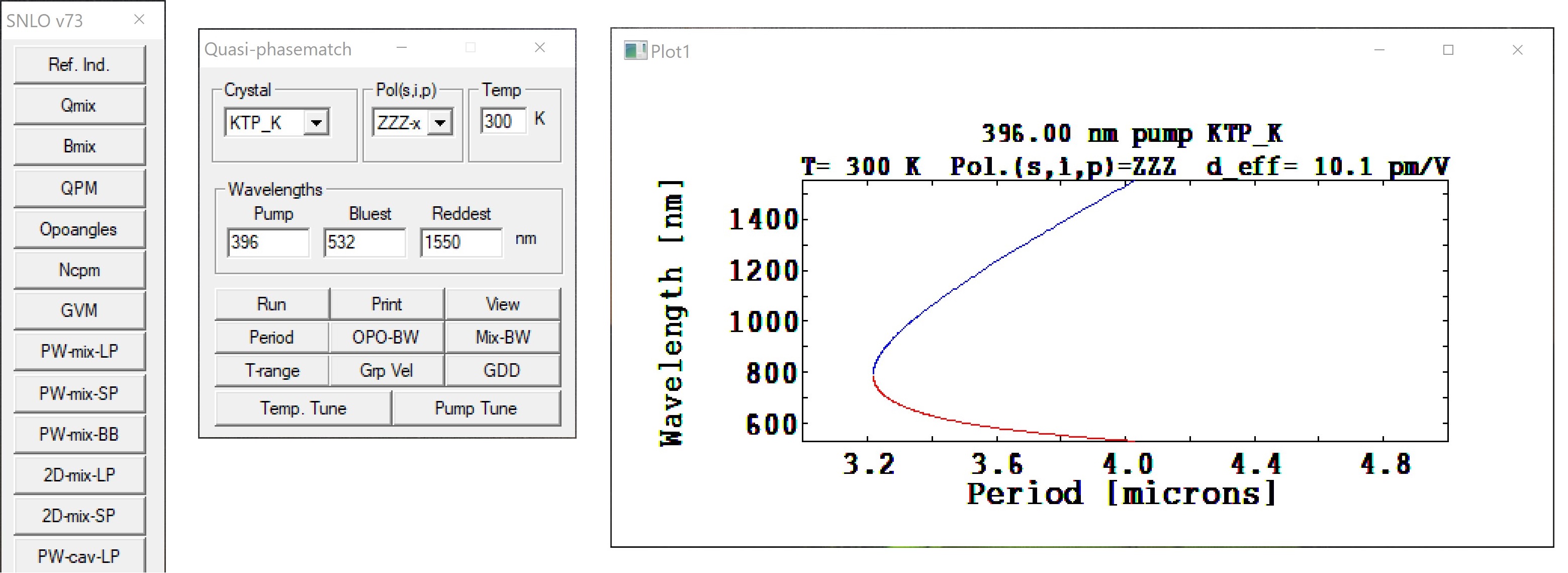}
\caption{Quasi-phase-matching conditions in a function of crystal poling period. Data for \textbf{PPKTP} crystal, obtained using SNLO software \cite{snlo}, presented as an example (see \textit{Section \ref{sec:ch5-snlo}}).}
\label{pic:snlo-qpm}
\end{figure}

Due to the exponential form of effective nonlinear coefficient (Eq. \ref{eq:deff}), for periodically poled crystals, only small amendment in phase-matching conditions (Eq. \ref{eq:qpmbk}) is needed \cite{book:Nikogosyan}, thus:
\begin{equation}
\Delta k_z-\frac{2 \pi m}{\Lambda}=0 \; ,
\label{eq:qpmpp}
\end{equation}
where $m$ is the order of interactions (usually for crystals designed for single wavelength $m=1$). Eq. \ref{eq:qpmpp} presents so-called  \textbf{quasi-phase-matching conditions}. Thanks to this, by adjusting the parameter $\Lambda$ of a crystal, nonlinear interactions even with big phase mismatch ($\Delta k_z \,\neq\, 0$) are allowed. The dependence between quasi-phase matching conditions and poling period of a crystal is presented in Fig. \ref{pic:snlo-qpm}.\\ 

Unfortunately, a large mismatch may lead to short poling period, which may be hard to achieve in fabrication process. Currently, the smallest commercially available poling period for PPKTP crystal is around $2.7$ $\mu$m. Such crystal is suitable to work under very specific circumstances -- generation of a photon pair with big difference in central wavelengths, while maintaining collinear propagation of these beams. Typically, in PPKTP crystals that generate photons of the same wavelength through type-II SPDC process, the poling period is around $40$ $\mu$m, which is not a challenge for manufacturing companies.  \\

Experimentally, (quasi-)phase-matching conditions can be checked by spectrum measurement for two selected photons. For visible wavelength range sources, single photon spectra can be measured using standard spectrometer that is based on diffraction gratings after a long exposure time of photosensitive matrix. For infrared-wavelength source the fiber-based spectrograph, which utilizes chromatic group velocity dispersion in a single-mode fiber \cite{Avenhaus2009}, may be used. \\

\begin{figure}[t]
\centering
\includegraphics[width=\linewidth]{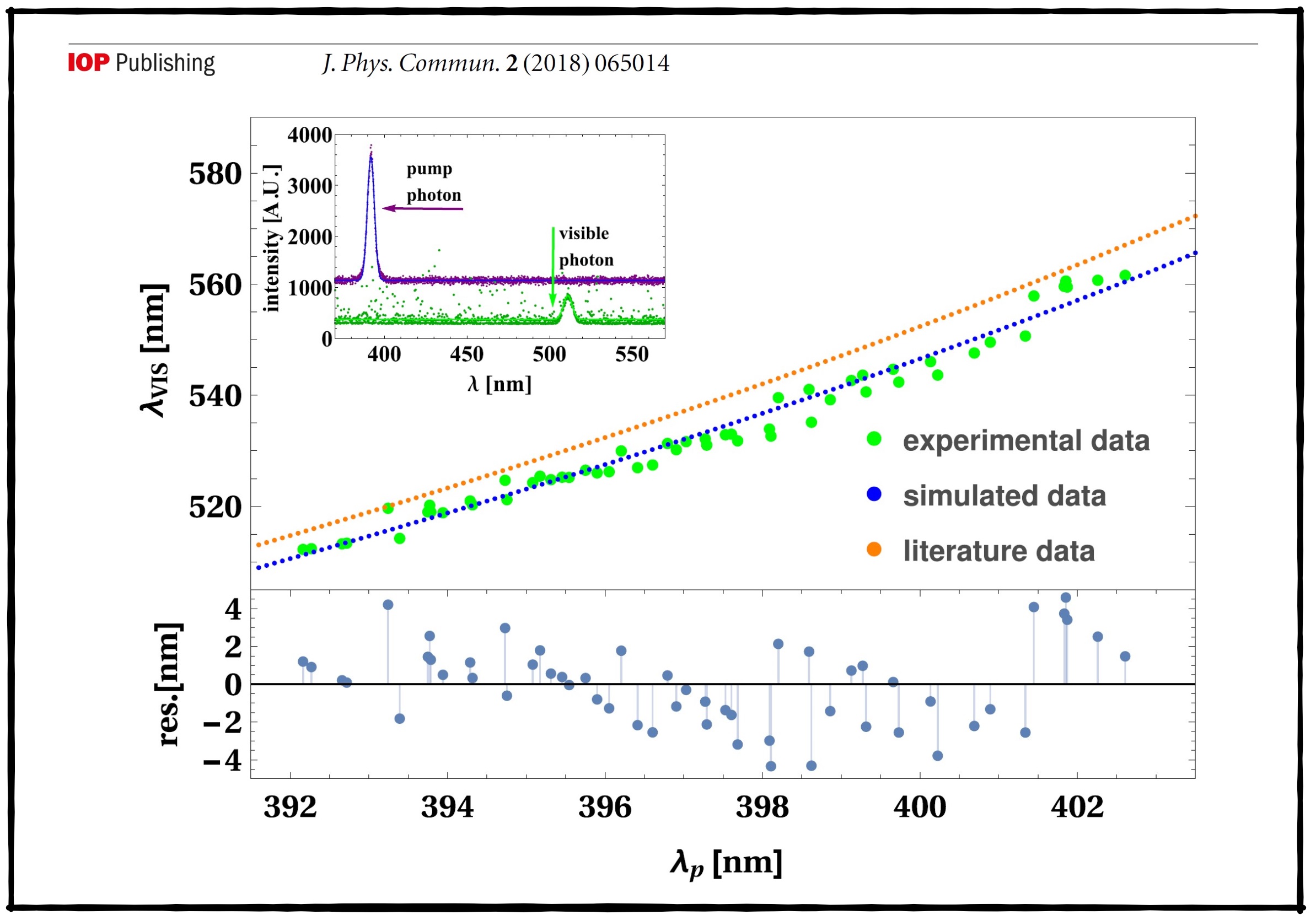}
\caption{Characterization of S1 source. Up: Dependence of the central wavelength of the visible and pumping photons. Comparison with theoretical calculations. Down: differences between literature data and theoretical results. Inset: an exemplary measurement of a visible and pump photon spectra.}
\label{pic:qphppktp}
\end{figure}

\begin{figure}[t]
\centering
\includegraphics[width=0.8\linewidth]{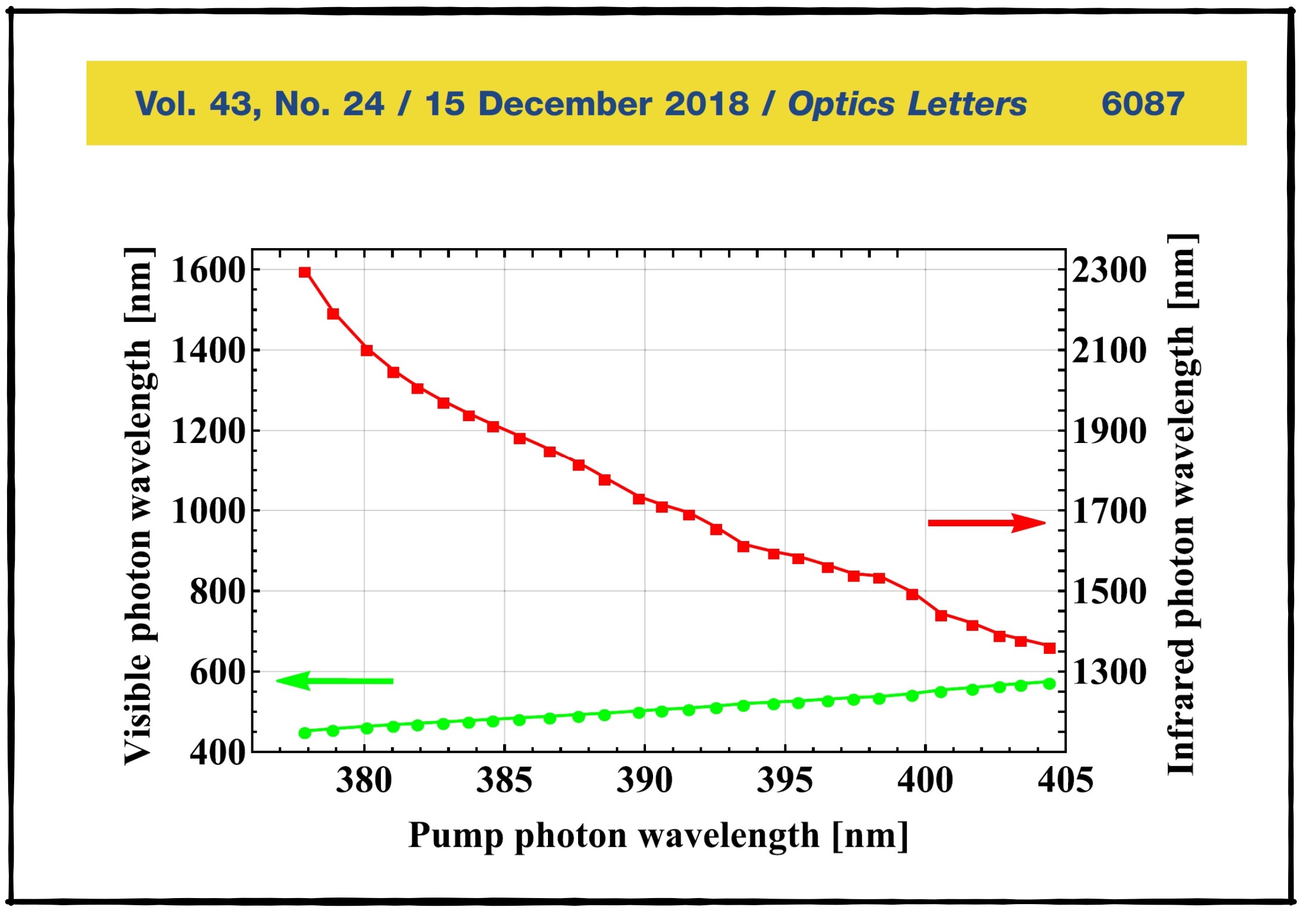}
\caption{Characterization of S1 source. Phase matched wavelengths for generated photons in PPKTP crystal. Green dots show measured wavelengths of the visible photon, and red ones are calculated from the energy conservation relation.}
\label{pic:detresphmatch1}
\end{figure}

Such measurements are presented in Fig. \ref{pic:qphppktp} (see Ref. \cite{Andrzej2018}), where the quasi-phase matching for PPKTP crystal is tested and compared to theoretical calculations, and in Fig. \ref{pic:detresphmatch1}, where the long-range phase-matching conditions for the same crystal is presented (see  Ref. \cite{Marta2018}). \\

The spectral characterization of SPDC source usually includes spectra measurement of two photons, pump and one from generated pair. The central wavelength of the third photon is calculated using energy conservation relation, as presented in Fig. \ref{pic:detresphmatch1}. \\

For each measurement point that can be denoted as $(\lambda_p,\lambda_s)$, the central wavelength $\lambda_i$ of idler photon was computed as:
\begin{equation}
\lambda_i=\frac{\lambda_p\lambda_s}{\lambda_s-\lambda_p} \; .
\end{equation}

Using the measurement uncertainties $\Delta\lambda_p$, $\Delta\lambda_s$ the uncertainty  $\Delta\lambda_i$ can be calculated:
\begin{equation}
\begin{split}
\Delta\lambda_i\;&=\;\Big|\frac{\partial}{\partial\lambda_p}\Big(\frac{\lambda_p\lambda_s}{\lambda_s-\lambda_p} \Big)\Big|\Delta\lambda_p + \Big|\frac{\partial}{\partial\lambda_s}\Big(\frac{\lambda_p\lambda_s}{\lambda_s-\lambda_p} \Big)\Big|\Delta\lambda_s \;=\\
&=\;\frac{\lambda_s^2}{\big(\lambda_p-\lambda_s\big)^2} \Delta\lambda_p\,+\,\frac{\lambda_p^2}{\big(\lambda_p-\lambda_s\big)^2} \Delta\lambda_s\; .
\end{split}
\end{equation}
The calculated uncertainties are smaller than markers plotted in Fig. \ref{pic:detresphmatch1}. \\
\FloatBarrier

\subsection{Singleness of the source}
As it is mentioned in \textit{Section~\ref{sec:ch1-sinphsource}}, the use of a true single-photon source is one of the conditions for the security of the transmission. The method which is used in checking the quality of singleness of the source is the second-order correlation function $g^2(\tau)$ measurement, described in more detail in \textit{Section~\ref{sec:ch2-g2}}. This method is based on coincidence measurement.\\

\begin{figure}[t!]
\centering
\includegraphics[width=0.6\linewidth]{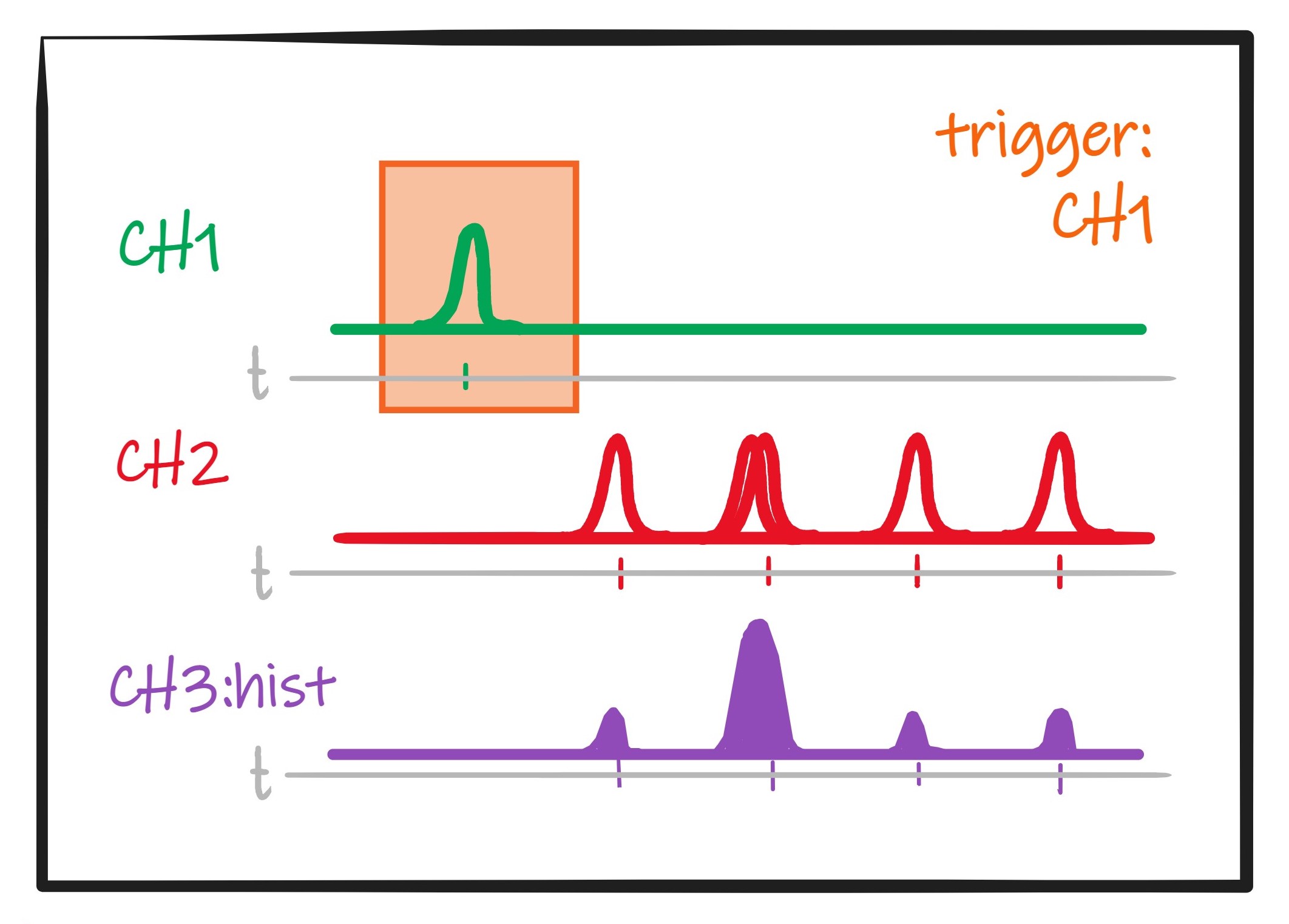}
\caption{Example result of a time-arrival coincidence measurement between two channels, registered on a oscilloscope with a long waveform persistence. Waveforms on \textit{channel 2} (red) are triggered with signal on \textit{channel 1} (green) that is included in the orange trigger box. \textit{Channel 3} presents delta-time measurement histogram. Full description of the measurement is in the text below.}
\label{pic:coincosc}
\end{figure}

An example result of a time-arrival coincidence measurement recorded on an oscilloscope with a long waveform persistence is presented in Fig.~\ref{pic:coincosc}. It shows signals from single-photon detectors, connected to two separate channels among which one is the trigger for the other. Here, it means that the arrival of a waveform registered on \textit{channel 1}, included in orange trigger box, causes start of signal recording on \textit{channel 2} of the oscilloscope. As it can be seen, the signal recorded on \textit{channel 2} is repeated at regular intervals, corresponding to time intervals between consecutive laser pulses. The oscilloscope performs a so-called \textit{delta-time measurement}, i.e., it measures time differences between signals on \textit{channel 1} and \textit{2}, which, as a result, gives the histogram presented on \textit{channel 3}. \\

It is important to note, that as presented on \textit{channel 3} when recording histograms with a long persistance, one peak becomes highest and in the long run may cover the others. It indicates that this particular peak corresponds to the true time-arrival coincidence, which means that both signals were detected in the appropriate detection windows. \\
In case of photons generated through the SPDC process, separated and detected with different detectors, the highest peak corresponds to the detection of photons from the same pair (produced from same pump photon). Smaller peaks are obtained from accidental coincidences, which are observed when one photon from a pair is not detected due to its loss. If the SPDC single-photon source is well optimized, the small peaks are almost invisible. On the other hand, if it is not optimized at all, or simply detected photons are not from the same pair, the peaks have equal or nearly equal height. \\

Depending on the single-photon source use case, the measurement of $g^2(\tau)$ may differ. For example, an SPDC source may be used as a single-photon pump for an other sample, such as quantum dot or color center. Thus, it is useful to check the contributions from various orders of SPDC (meaning generation of pairs, quadruplets, octuplets, etc.). If the generated beams are split and this consisting of only one photon from a pair is used, it may be done in two different ways. \\
First one is to put the 50:50 beam splitter in the path of the characterized photon beam and to collect the time-arrival coincidence measurement histograms when changing the path lengths between the beam splitter and detectors. \\
The other way is very similar, the only difference is that the detectors are triggered by the other photon from a pair, which existence was not taken into account in the previous way. This method is faster, especially when measuring $g^2(0)$  (the path difference is equal to 0). \\

Although the first method allows for observing the anti-bunching effect for SPDC sources (see \textit{Section~\ref{sec:ch2-g2}}), the data collection is time consuming and burdened with a high noise detection. Also, due to the electronic limitations, collection of proper statistics of detection events may be difficult to achieve experimentally. On the other hand, the second method gives $g^2(0)$, which is the most important value when characterizing the signleness of single-photon source.  \\

The results of second method, i.e., $g^2(0)$ measurements for S1 source are presented in Ref.~\cite{Marysia2020}. The measurement was done in two power regimes, for low ($1$ mW) and high ($5$ mW) pump power, and gives $g^2_l(0) = 0.0011(2)$ and $g^2_h(0) = 0.0111(5)$ respectively. It is worth to mention that the first value is the lowest reported in literature for that type of SPDC sources. \\


\subsection{Quantum-state tomography}
\label{sec:ch3-qst}

As it was mentioned earlier, to fully describe photon beams generated by single-photon source is to determine their statistics, spectra, and polarization states. Measurements for the first two parameters were presented in previous subsections, here the measurement that allows to determine the polarization state is presented. This measurement is called a \textbf{quantum-state tomography} \cite{PST2005, Jakub2021, Artur2022}. \\

Since the measurement itself causes the collapse of a quantum state that is under investigation, the idea of quantum-state tomography is based on multiple measurements, performed on many samples of a given state. If the measurement is chosen correctly, its set gives the full picture and provides all information for the reconstruction of the quantum state. \\

The mathematics that stands behind the algorithms which are used for reconstruction of quantum state density matrix is described in detail in \textit{Section \ref{sec:ch5-qst}}. Here, only an experimental point of view is presented. \\

The S2 source, which is used for presentation of state reconstruction, generates two polarization-entangled qubits (see \textit{Section \ref{sec:ch3-redinfrared}}). Therefore, as showed in the Fig. \ref{pic:s2}, part of the setup that allows for determining the polarization state consists of a beamsplitter, where photons are separated due to the reverse Hong-Ou-Mandel effect (see \textit{Section \ref{sec:notes-HOM}}). Then, the beamsplitter is followed by a quarter-wave plate, half-wave plate, polarizer, and detector for each experimental arm. \\

\begin{figure}[h!]
\centering
\includegraphics[width=0.9\linewidth]{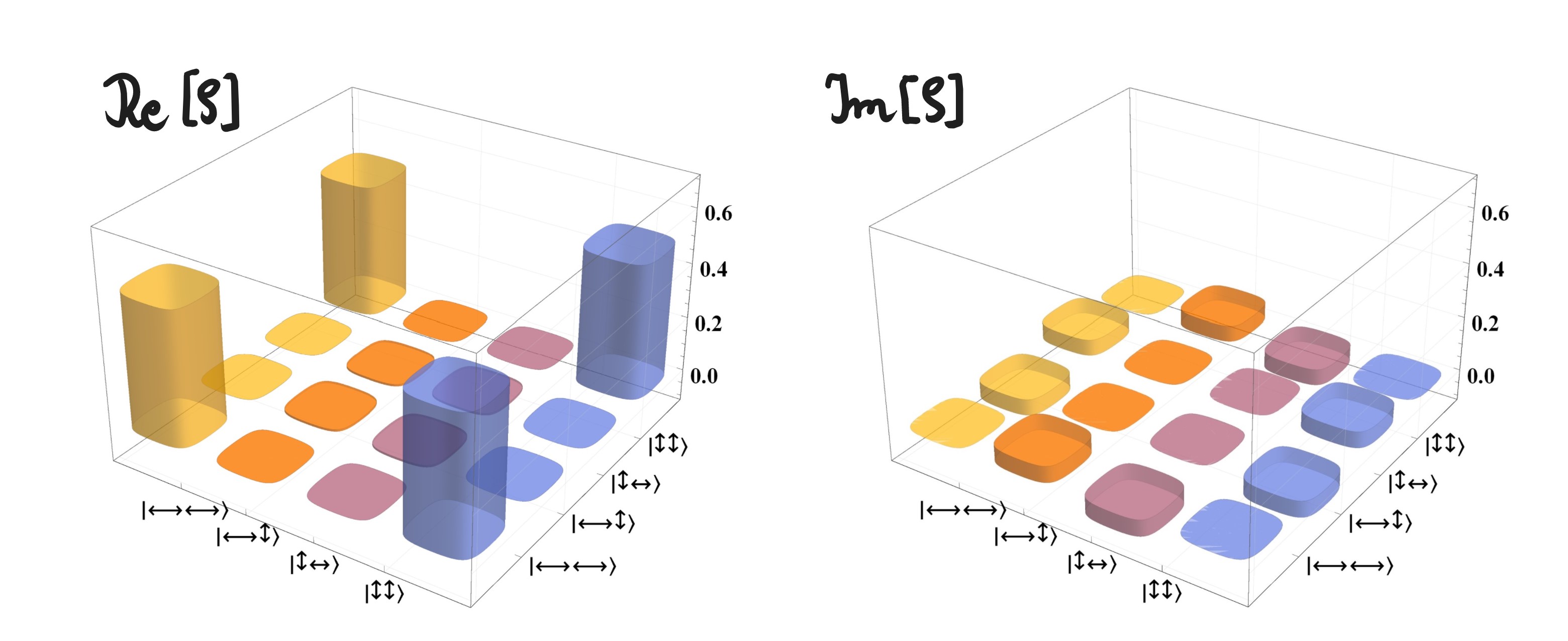}
\caption{The results of quantum-state tomography measurement. Real (a) and  imaginary (b) part of the density matrix representing the entangled state produced by the S2 source. Details for this particular density matrix reconstruction are provided in \textit{Section \ref{sec:ch5-expexamp}}. Reconstruction algorithm was performed by dr Artur Czerwiński.}
\label{pic:s2-qst}
\end{figure}

The performed measurement is based on coincidence counts between detectors on both arms for each setting angle of the waveplates (16 positions to cover all polarization states, for HWP -- $\{0^o, 22.5^o\}$, for QWP -- $\{0^o, 45^o\}$).  \\
These measurements allow for using the iterative algorithm that finds the closest elements of the density matrix that would produce the state measured. The key element of this algorithm is \textit{an estimator} that associates measured coincidence counts with expected ones (see \textit{Section \ref{sec:ch5-est}} and \textit{\ref{sec:ch5-alg}}).\\
Obtained results are presented in Fig. \ref{pic:s2-qst}. \\

\vspace{1cm}

For the reconstructed density matrix, parameters that characterize the amount of entanglement were calculated. The fidelity was equal to $F(\rho)=0.972(2)$ and concurrence was equal to $C(\rho)=0.985(2)$ (see \textit{Section \ref{sec:ch5-fidelity}} and \textit{\ref{sec:ch5-conc}}, respectively). Furthermore, the Bell parameter was calculated according the Ref. \cite{Hu2013} and was equal to $B_{max}(\rho)=2.78735$. \\
These calculations as well as the reconstructed density matrix proved that the S2 source generates the polarization-entangled qubits. \\
\chapter[Future is now]{Future is now. \\ {\LARGE Applications of single-photon sources}}
\label{sec:ch4}

\section[Secure communication over long distances]{Secure communication over long distances \newline \small{Long-distance quantum communication with satellite}}
\label{sec:ch4-sat}

The ability of secure quantum key distribution (QKD) over long distances is a crucial feature of worldwide quantum network. As it was described in \textit{Sections \ref{sec:ch1-ffquantnet}} and \textit{\ref{sec:ch1-reperrcorr}}, due to the decoherence and losses, the transmission through a fiber link is limited to several hundreds of kilometers \cite{Tsai2021}, depending on a type of a photon source and protocol used. Utilizing a satellite as a node, which transfers the key between two communicating parties, gives a hope to increase this distance. \\

\begin{figure}[b!]
\centering
\includegraphics[width=0.8\linewidth]{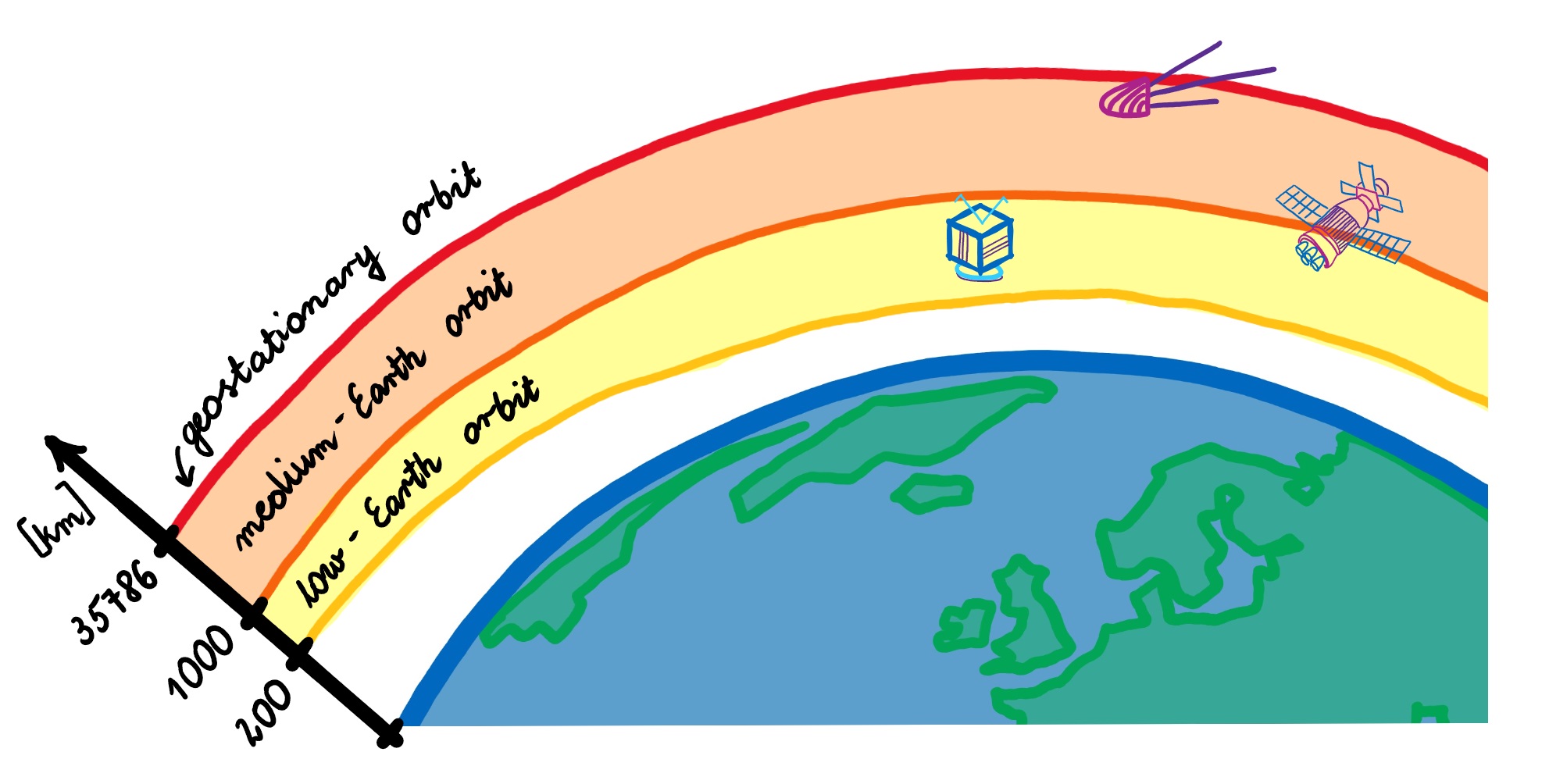}
\caption{Earth orbits on which a quantum sattelite may be placed. The individual elements are not drawn to scale.}
\label{pic:satqkd}
\end{figure}

A QKD satellite can be located at three typical orbits -- low-Earth orbit (LEO) \cite{Liao2017}, medium-Earth orbit (MEO) \cite{Vallone2016}, and geostationary orbit (GEO), that ranges between $200$-$1000$ km, $1000$-$35786$ km and  $35 786$ km respectively, from Earth ground \cite{ESA}. The chosen orbit determines the time that is needed to lap the Earth. For example, for LEO it is approximately 90 minutes, which gives around 10 minutes visibility from the ground station (in practice it is rather $3$-$5$ minutes \cite{Piotr2021}), whereas for GEO it is exactly 23 hours 56 minutes and 4 seconds, which makes a satellite appear to be fixed above some point on the Earth ground (see \textit{Section \ref{sec:ch4-sattime}}). Due to that fact, different types of photon sources need to be used. For LEO QKD  applications, it is usually faint laser with coherent pulses, while the GEO orbit enables use of SPDC single-photon sources \cite{Lee2019}. \\ 

Schemes of QKD witha  satellite can be divided into two general types, a \textit{downlink} and an \textit{uplink} scenario \cite{Lee2019}. In the downlink scenario, a single-photon source is placed on the satellite, while in the uplink scenario is the other way round -- the source is on the ground. Regardless of the scenario, the QKD protocols may utilize both, encoding in polarization \cite{book:Hughes1999, Ecker2021} and time-bin \cite{Vallone2016}. Nonetheless the first solution is much more popular \cite{Bedington2017}. As a qubit source, especially a polarization one, sources based on SPDC process are typically used. The operating wavelength depends on the scenario 
and the transmission windows for which the overall atmospheric transmission is highest. Hovewer, in downlink type, windows around $665$–$685$ nm, $775$–$785$ nm, $1000$–$1070$ nm, and $1540$–$1680$ nm were found \cite{Bourgoin2013}. Usually, due to the relatively high quantum efficiency of a single-photon detectors, wavelengths from $775$–$785$ nm range are used \cite{Ecker2021}.  \\ 

There are many projects underway to develop and study quantum communications using satellites, some of them are at an advanced stage, including:
\begin{itemize}
\item[--] \textbf{QUESS}, Quantum Experiments at Space Scale --  China-led research project that demonstrated secure QKD connection between Beijing and Vienna in 2016 using the Micius satellite;
\item[--] \textbf{QEYSSat}, Quantum Encryption and Science Satellite -- demonstration of QKD in uplink scenario using micro-satellite, carried out by CSA and Institute for Quantum Computing (IQC) at University of Waterloo, satellite launch scheduled for the end of 2022;
\item[--] \textbf{CubeSat} platform, \textbf{SPEQTRE} The Space Photon Entanglement Quantum Technology Readiness Experiment -- demonstration of QKD in downlink scenario, mission in international collaboration between RAL Space, part of the UK Science and Technology Facilities Council (STFC), and the Centre for Quantum Technologies (CQT), National University of Singapore (NUS), satellite launch scheduled for the middle of 2023;
\item[--] \textbf{HydRON}, High thRoughput Optical Network, hosted by ESA’s \textbf{ScyLight} (Secure and Laser communication technology) programme -- the aim of a project is to develop the world first all-optical network at terabit capacity, and extension of terrestrial fiber networks utilizing satellites and \textit{fiber in the sky} technology, demonstration phase of the project is scheduled for 2024;
\item[--] \textbf{SAGA}, Security And cryptoGrAphic mission -- Europe-wide quantum communications satellite system in downlink scenario, located at GEO, under development by ESA, launch of the system is scheduled for 2027.
\end{itemize}

There are various methods that enable safe QKD with satellite \cite{Liao2017, Bedington2017, Bacsardi2013}. One of them is presented in Fig.~\ref{pic:satqkd}. There are two ground stations, $A$ and $B$, each of them exchanges a secret key, $k_A$ and $k_B$, respectively, with the satellite. Later, one of the shared keys, let us assume that it is $k_A$, is encoded using the other one, so $k_{AB}=E_{k_B}\big[ k_A\big]$. The result of encoding, namely $k_{AB}$ is sent to station $B$, where the decoding procedure  takes place. Since the satellite uses the $k_B$ to encode the key, the decoding procedure is as following: $D_{k_B}\big[k_{AB}\big]=D_{k_B}\Big[ E_{k_B}\big[ k_A\big]\Big]=k_A$, and therefore, both ground stations now share the same secret key. \\

\begin{figure}[b!]
\centering
\includegraphics[width=0.8\linewidth]{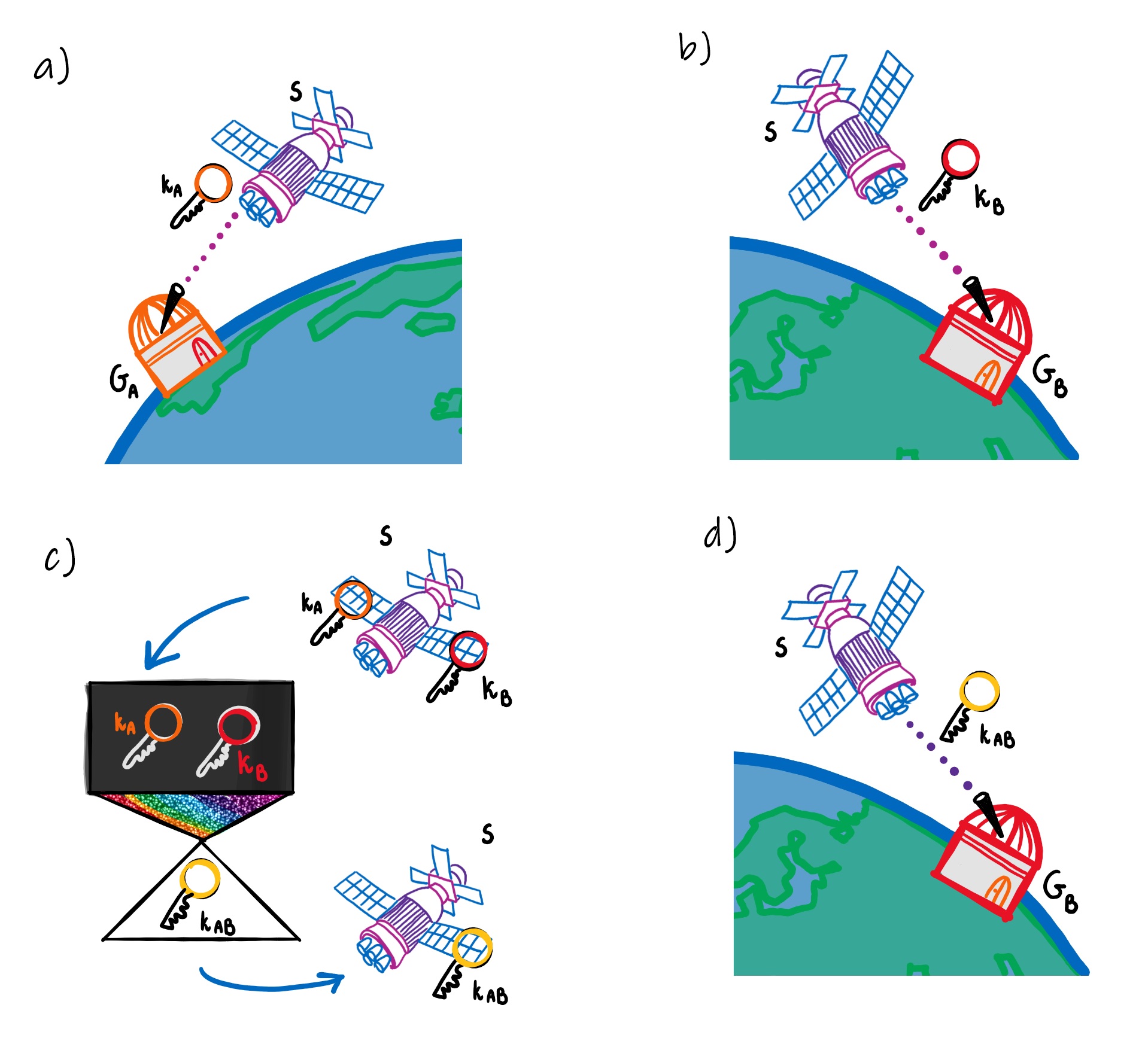}
\caption{Scheme of quantum key distribution utilizing a satellite.}
\label{pic:satqkd}
\end{figure}

\begin{figure}[t!]
\centering
\includegraphics[width=0.8\linewidth]{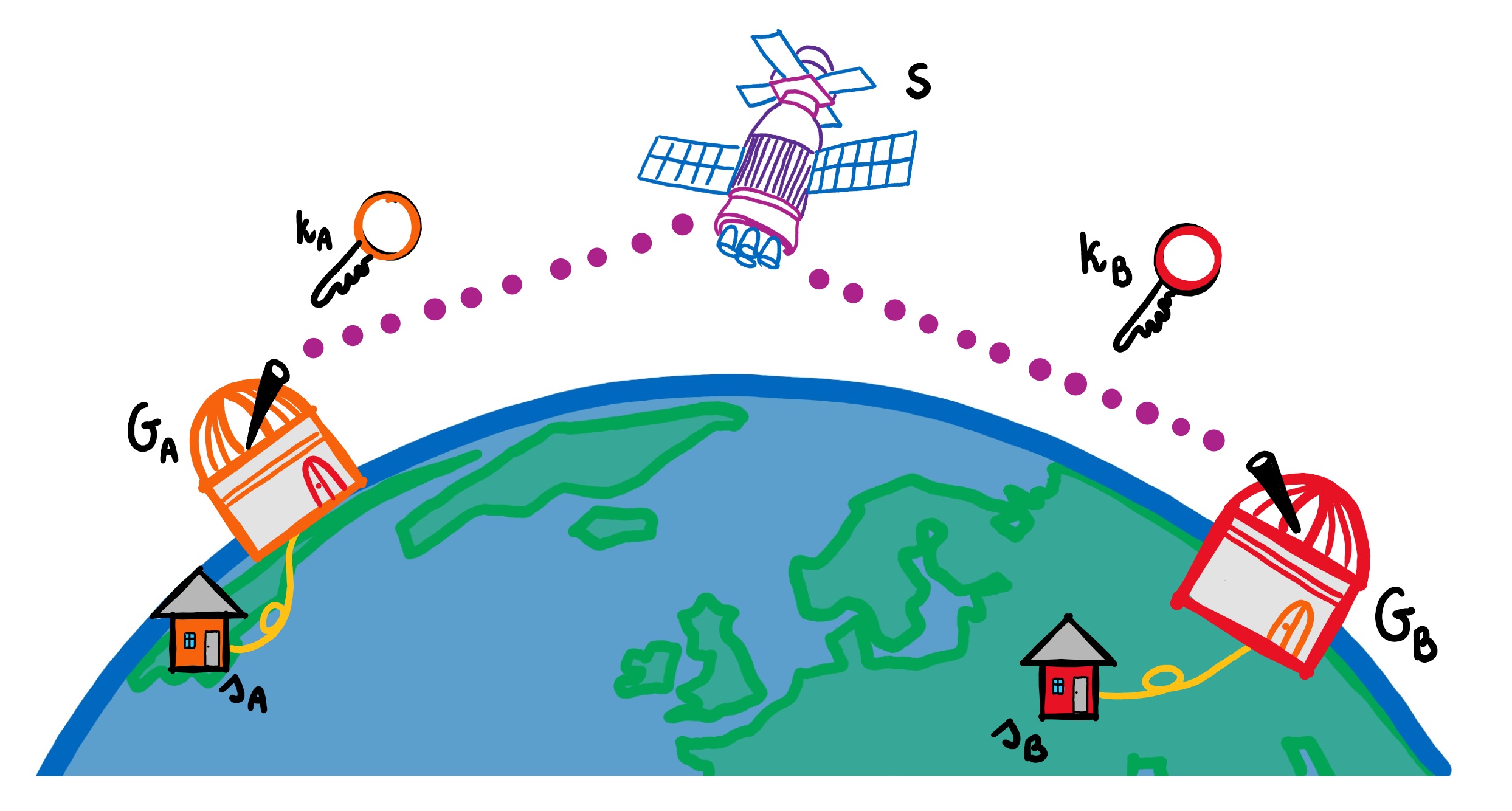}
\caption{The idea of integration of free-space and fiber channels for QKD applications. Symbols: S -- moving satellite, $G_A$, $G_B$ -- ground stations, where single-photons sources are located, $s_A$, $s_B$ -- nearby node ground stations, $k_A$, $k_B$ -- secret keys, that are exchanged between ground stations and a satellite (one at a time).}
\label{pic:satfiberqkd}
\end{figure}

Utilization of satellites for QKD applications, opens a possibility for integration of both free-space and fiber transmission channels.  A single-photon source described in \textit{Section \ref{sec:ch3-visinfra}} was designed for this application. Each photon from a pair is sent through a different channel. \\
Fig.~\ref{pic:satfiberqkd} presents the idea of free-space and fiber QKD channel integration. Single-photon source, that emits a pair of correlated photons, with visible and infrared central wavelengths, is located in ground station $G_A$. Visible photon is sent towards a satellite $S$, whereas the infrared goes through a fiber to the nearby station $s_A$. Both keys, distributed between ground station $G_A$ and satellite and stations $G_A$ and $s_A$, are correlated, therefore, both can be denoted as $k_A$. Similar situation can be observed for the distant ground station $G_B$ and another nearby station $s_B$, where the key $k_B$ is exchanged with the satellite. Finally, an approach presented previously in Fig.~\ref{pic:satqkd} results in a secure QKD among the four users of this network. \\
Even though the chosen visible wavelength is far from being perfect for uplink free-space transmission, quantum efficiency of single-photon detectors, such as SPADs, reaches maximum around this wavelength. They are also moderately inexpensive, comparing to other detector types, which is important for further commercialization. It should also be noted that the advantage of such a network is its scalability. \\

\section[Setting standards for quantum metrology]{Setting standards for quantum metrology \newline \small{Single-photon detector calibration}}
\label{sec:ch4-detcalres}

Currently, one of the most rapidly developing fields of quantum applications is the quantum metrology, which purpose is high-sensitive measurements of physical parameters. By developing novel measurement methods, utilizing entangled and squeezed states, precision higher than in the classical measurements is obtained.  \\

As an example of an application, detection of gravitational waves with the use of Mach-Zehnder interferometer may be mentioned, however, despite remarkable proposals \cite{Kimble2000, Agasi2013}, this has not yet been demonstrated on a quantum level. On the other hand, quantum metrology is successfully applicable in biology  \cite{Taylor2014}, through applications such as positron emission tomography with entangled photons \cite{Kumar2018}, quantum optical coherence tomography \cite{Sylwia2020, Jakub2021} or fluorescence microscopy \cite{Langhorst2009, Szymanski2013}. \\

Without a doubt, all the applications mentioned above are important, however, the core of quantum metrology is the use of single-photons, which defines units of measurement for other high-precision research. Therefore, what is the most essential in this field is setting standards for single-photon sources, and even more importantly, detectors. \\

\begin{figure}[t]
\centering
\includegraphics[width=0.7\linewidth]{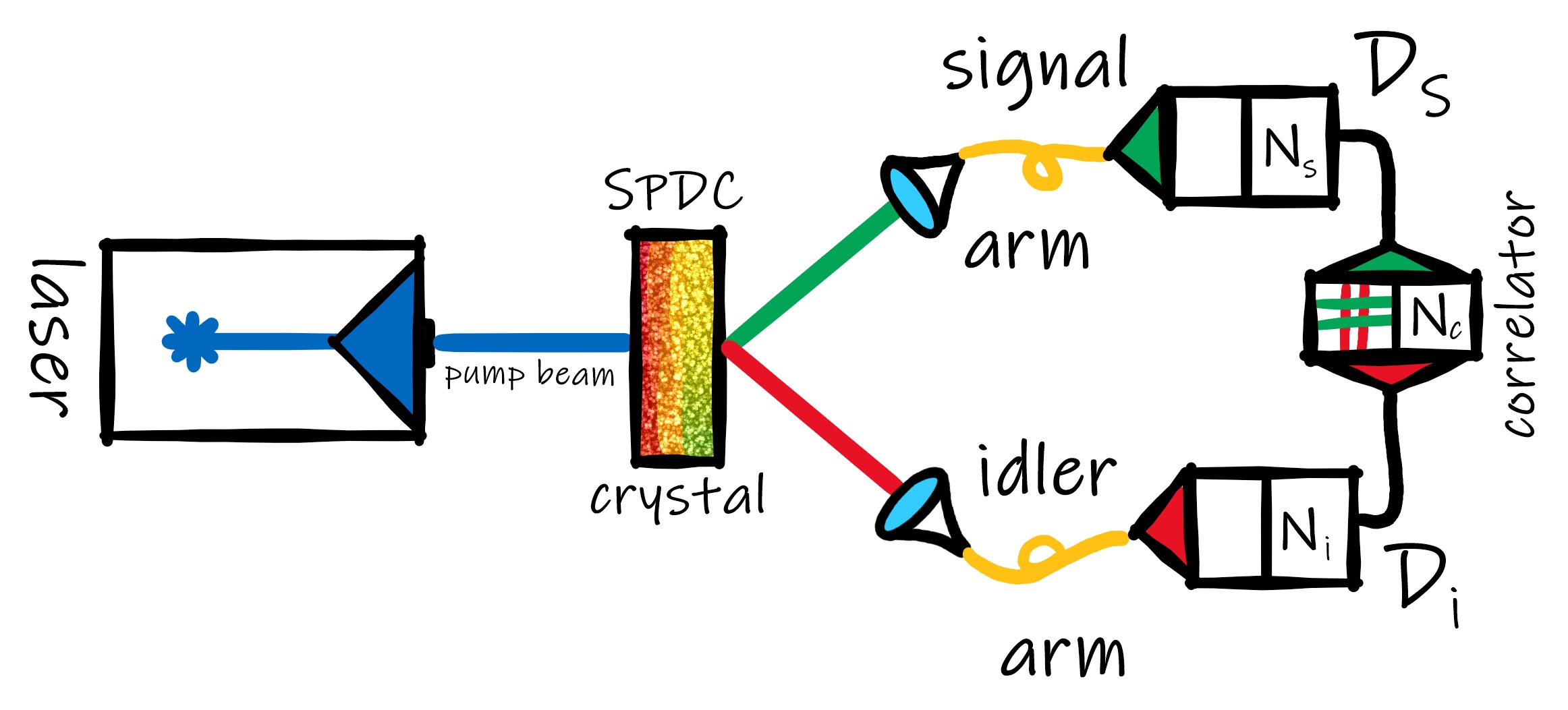}
\caption{Scheme of experimental setup that enables system detection efficiency measurement.}
\label{pic:detcal}
\end{figure}

Methods for single-photon detector calibration are well described in the literature \cite{Brida2005, Brida2007, Cohen2018}. All of them utilize the SPDC single-photon source and the correlation between photons from a pair. 
The idea that stands behind detector calibration, and thus, detection efficiency measurement, is presented in Fig.~\ref{pic:detcal}. If the single photon source emits $N$ photon pairs in some time interval $T$, then integrating over the same time interval, detectors should  be recorded $N_s$, $N_i$ and $N_c$ for signal photons, idler photons, and coincidence events, respectively. Assuming that $\eta_s$ and $\eta_i$ are system detection efficiencies containing the losses in each path the measured values should be equal to:
\begin{equation}
\begin{split}
N_s=& \;\eta_s \cdot N \; ,\; N_i=\eta_i \cdot N \; , \\
&N_c=\eta_s\eta_i\cdot N \; ,
\end{split}
\end{equation}
thus the system detection efficiencies are equal to:
\begin{equation}
\eta_s =\frac{N_c}{N_i} \;, \; \eta_i =\frac{N_c}{N_s} \;.
\end{equation}

It should be noticed here that the knowledge of system detection efficiency of one path is not important to indicate efficiency of the other. \\
Also, similar method can be used for absolute callibration of free-space single-photon detector \cite{Brida2005}, in low-power regime, using the irises that select only first spatial mode and free-space transmission, in order not to introduce additional losses related to coupling efficiency and optical fiber imperfections. \\
This method can be also adopted for use in multi-photon detector callibration \cite{Cohen2018, Avella2011}. \\

\hsubsection{Results}
The aim of our work, published in Optics Letters' article entitled \textit{Single photon detection system for visible and infrared spectrum range} \cite{Marta2018} was to characterize superconducting single-photon detectors (see \textit{Section \ref{sec:ch2-sinphdet}}). We demonstrated that detectors are sensitive in $452$-$2300$ nm range and the system photon detection efficiency of our setup is as high as $64$\% at $1550$ nm and $15$\% at $2300$ nm. Furthermore, we showed that in heralding scheme measurement, the use of two SSPDs for photon detection gives much better results  than any other configuration of detectors. \\ 

My contribution to this work consisted of building the experimental setup, testing detectors by collecting coincidence histograms and calculating signal-to-noise ratio for different kinds of detectors. I also prepared all schemes and figures for the publication. \\

\begin{figure}[t!]
\centering
\includegraphics[width=0.8\linewidth]{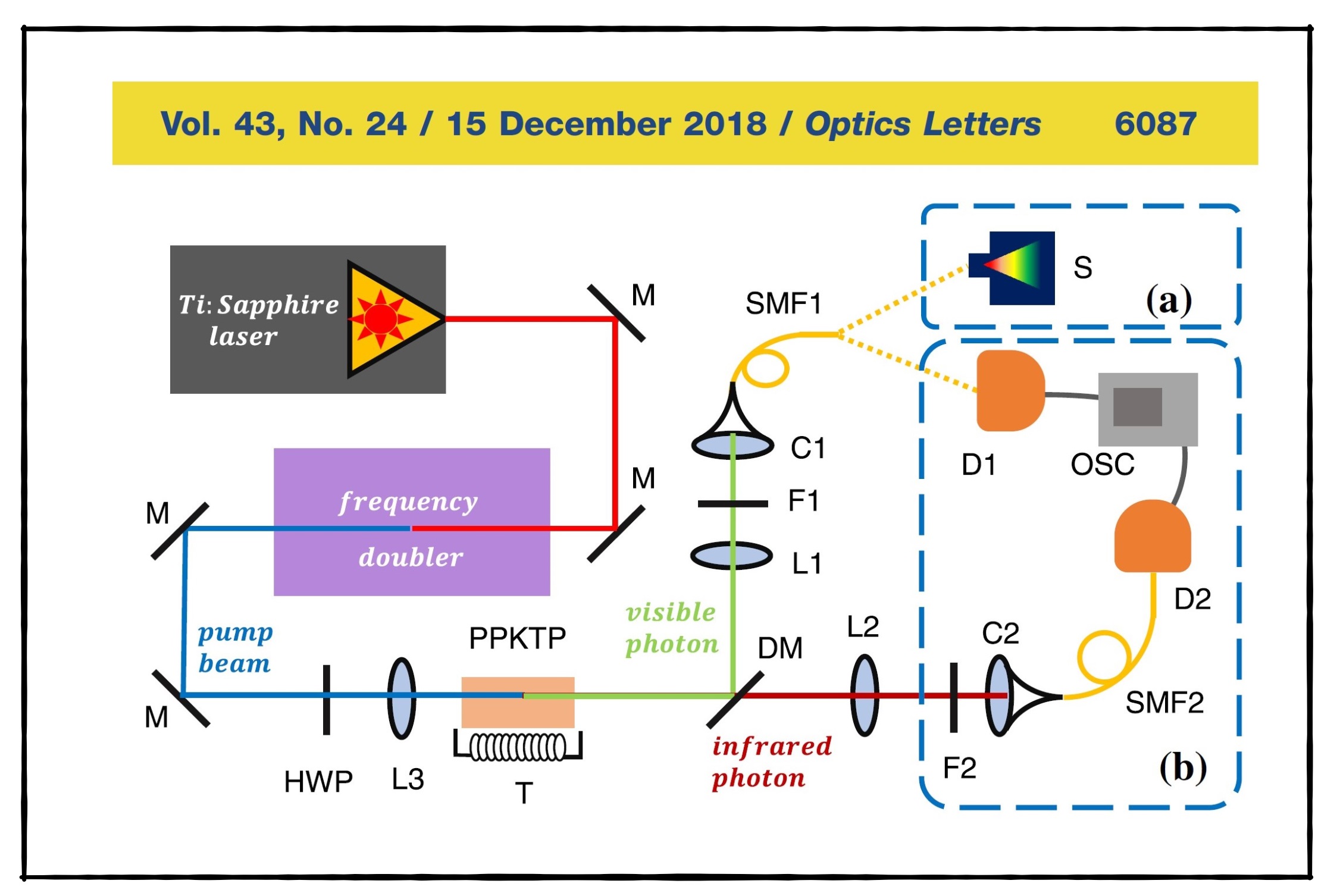}
\caption{Scheme of an experimental setup used for SSPD callibration. Symbols: Pulsed Ti:sapphire laser, M—mirror, L3—lens (focal lengh $f=10$ cm), PPKTP—periodically poled potassium
titanyl phosphate crystal, T—temperature controller, DM—dichroic mirror (Semrock 76-875 LP), L1, L2—plano–convex lens ($f=12$ cm, $15$ cm), F1—set of filters [three settings: (1) Chroma ET500 and Z532-rdc, (2) 2 pcs of AHF 442 LP, (3) 2pcs of AHF442 LP and Thorlabs FESH0700], C1, C2—single-mode fiber coupler ($f=0.8$ cm, $1.51$ cm), F2—long-pass filter (Semrock BLP01-1319R), SMF1, SMF2—single-mode fiber [two settings: (1) Thorlabs
SMF460B and SMF1550, (2) SMF780 and SMF2000]. Detection setups: (a) S—spectrometer (Ocean Optics USB2000+); (b) D1,D2—detectors, OSC—oscilloscope.}
\label{pic:detresset}
\end{figure}

\begin{figure}[t]
\centering
\includegraphics[width=0.7\linewidth]{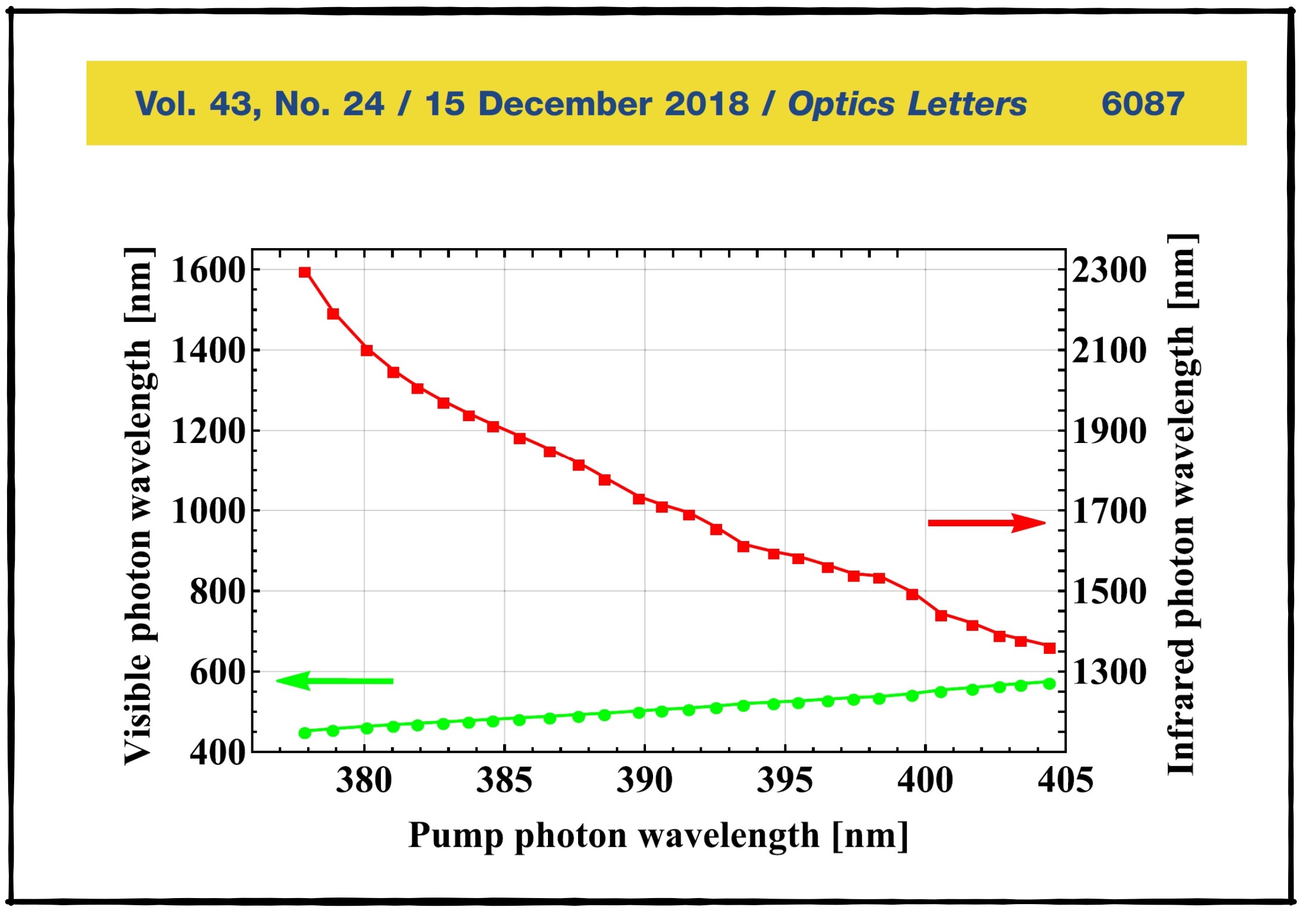}
\caption{Phase matched wavelengths for generated photons. Green dots show measured wavelengths of the visible photon, and red ones are calculated from the energy conservation relation. }
\label{pic:detresphmatch}
\end{figure}

The experimental setup is presented in Fig.~\ref{pic:detresset}. The first step was to build the single-photon source, that consists of pulsed laser, frequency doubler, and PPKTP crystal, where pairs of visible and infrared photons via the SPDC process were generated. Then, the spectrum of pump and visible photons was measured. Its central wavelengths were used for calculation of the infrared photon's wavelength. The result is presented as tuning curves in Fig.~\ref{pic:detresphmatch}.\\

\begin{figure}[t]
\centering
\includegraphics[width=\linewidth]{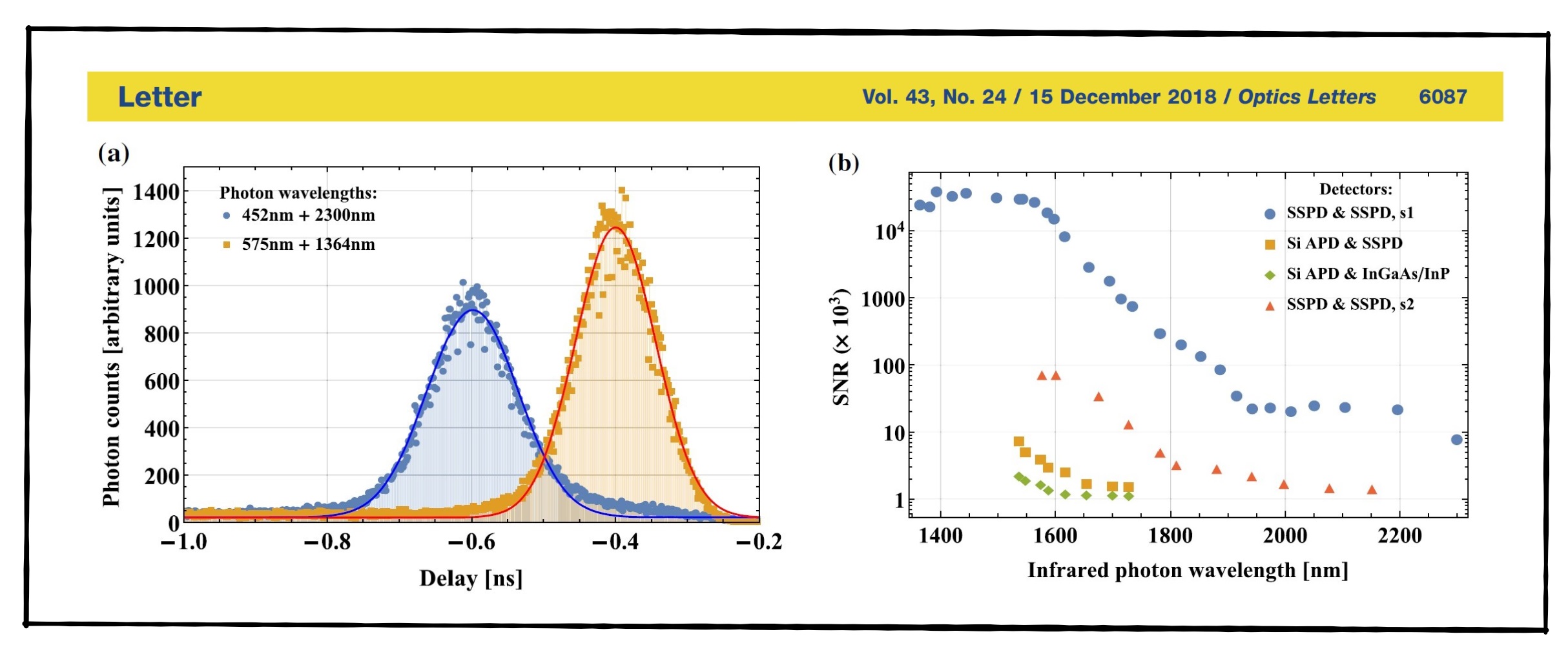}
\caption{(a) Arrival time coincidence histograms for two example settings with different phase-matching conditions. The measured setup timing jitter is approximately 86 ps (FWHM). (b) SNR for four settings with different pairs of detectors. Plot markers are bigger than error bars, excluding SNR values close to 1.}
\label{pic:detreseff}
\end{figure}

Later, for each pump setting and detectors configuration (SiAPD, InGaAs, and SSPD), the arrival time coincidence histograms were collected and the Gaussian functions were fitted to the data. Two example histograms, for SSPDs in both arms of experimental setup and two distinct pump settings, are presented in Fig.~\ref{pic:detreseff} (a). \\
It is noteworthy that the maximum value of coincidence peaks is different for each case, which comes from the difference of total detection efficiency. What is more, there is time shift between histograms. The reason for the time shift is that in each case photon had different group velocities inside fibers into which they were coupled.\\

It should also be noticed that total timing jitter of experimental setup is equal to $86$ ps, which gives roughly $60$ ps for each detector, however, the measured wavelengths for which the detectors were used, were far from optimized. 

Parameters from Gaussian function fittings were used to calculate the signal-to-noise ratio for each setting. The fitting function took a form: 
\begin{equation}
f_G(t)=b+a exp\Big(\frac{-4ln2(t-t_0)}{\sigma^2}\Big),
\end{equation} 
where $b$ stands for background noise, $a$ is the signal amplitude, $\sigma$ is the timing jitter, $t_0$ is the time for weach the peak reach maximum value, and the SNR value was computed as:
\begin{equation}
SNR=\frac{a}{b}\sqrt{\frac{\pi}{16}log(2)}erf\big(2\sqrt{log(2)}\big)\approx 0.52\frac{a}{b} \; .
\end{equation}
The comparison of SNR values for different settings is presented in Fig.~\ref{pic:detreseff} (b). 

Additionally, detector configuration with calculated SNR value for comparable setup settings are listed in Tab. \ref{tab:detconf} below. \\

\begin{table}[!h]
\centering
\begin{tabular}{c|c|c}
\hline
\multicolumn{2}{c|}{\textbf{type of a detector}}                                     & \multirow{2}{*}{\textbf{SNR value}} \\
\multicolumn{1}{c|}{\textbf{visible photon}} & \multicolumn{1}{c|}{\textbf{infrared photon}} &                            \\ \hline
SiAPD& InGaAs & $8.0(6)$\\
\hline
SiAPD& SSPD & $44.1(7)$\\
\hline
SSPD & SSPD & $1027.3(8)$ \\
\hline
\end{tabular}
\caption{The configuration of detectors used in the experiment and highest SNR ratio for comparable filter setting (denoted as S2 in Fig.~\ref{pic:detresset}).}
\label{tab:detconf}
\end{table}

It should be emphasized here that even if the detectors are not optimized for the wavelengths of detected photons (i.e., fiber input of a detector is SMF2000, which results in high dark count rate), the use of SSPDs in terms of obtained SNR value is the best possible choice. What is more, as showed in Fig.~\ref{pic:detreseff} (b), the SNR values may be even more incredible if the proper filters and fibers are chosen. The highest calculated SNR value ($>20000$) were obtained for wavelengths up to $1600$ nm. For wavelengths in the range $1800$-$2300$ nm, SNR values were $>8$, which is equal to best result of $SiAPD\&InGaAs$ configuration. \\

\FloatBarrier
\section[Towards a quantum computer]{Towards a quantum computer \newline \small{Single photon -- single color center interaction}}
\label{sec:ch5-nvres}

As it was mentioned in \textit{Section \ref{sec:ch1-quantnet}}, currently, the most developed quantum computers are those based on superconducting qubits. However, the major inconvenience of their use is that, due to their design, they need cryocooler circuits to work properly, which makes them more challenging to miniaturize and commercialize. Therefore, the quantum computer that operates at room temperature is desirable. One of such computers is \textit{solid-state quantum computer} based on point defects in diamond structure, such as commonly studied candidate -- NV color center.  \\

Even though the NV color centers had been studied from 1970s \cite{Davies1976}, the first remarkable results came in 1997 \cite{Gruber1997} when the fluorescence microscopy of single NV$^-$ centers and \textit{optically-detected magnetic resonance (ODMR)} \cite{Gruber1997, Jelezko2006, Dolde2014} at room temperature were demonstrated. That proved the photostability of this type of  color centers and raised hope for the rapid development in this field. \\

\begin{figure}[t!]
\centering
\includegraphics[width=0.8\linewidth]{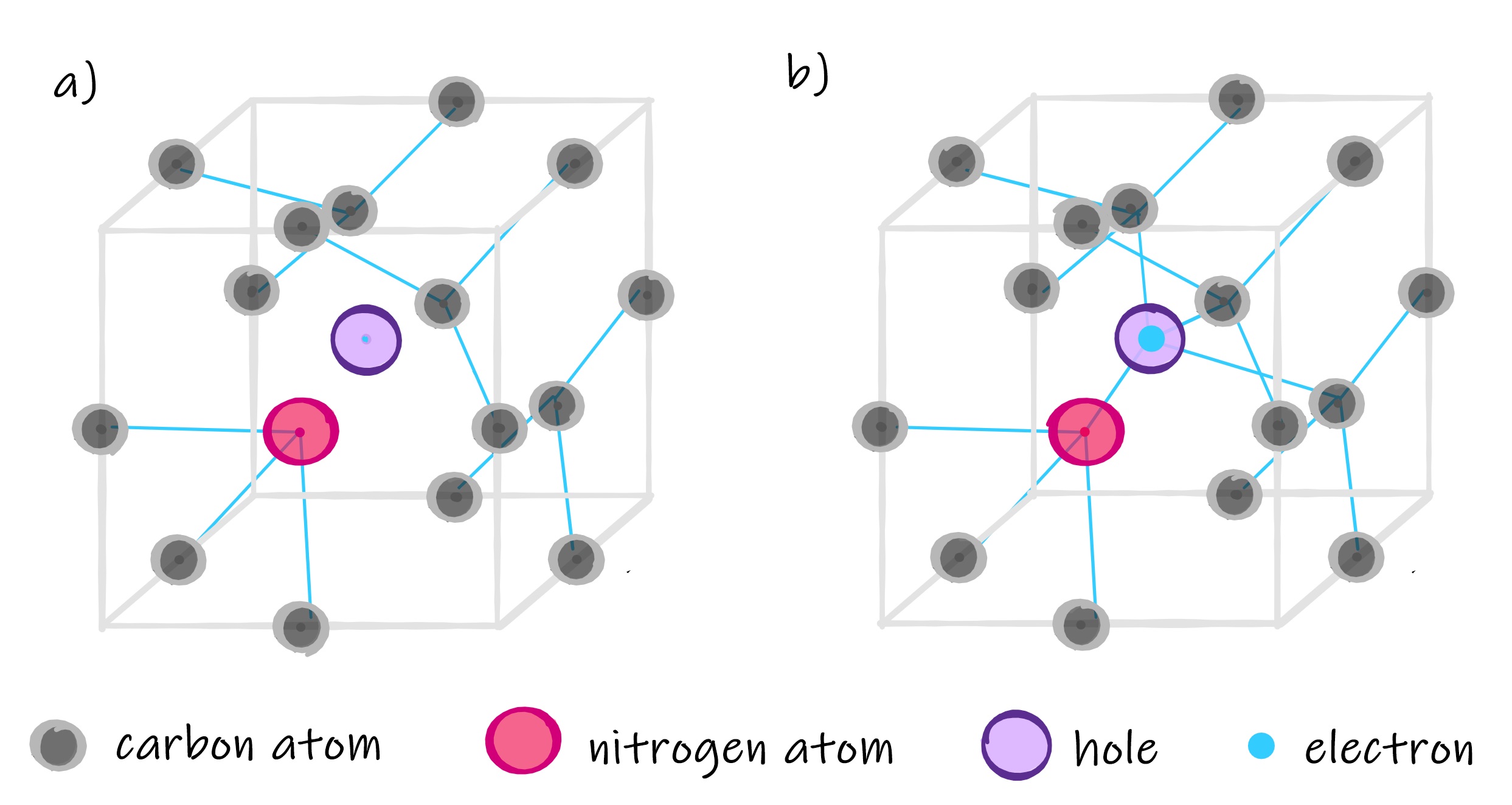}
\caption{The lattice of of a) NV$^0$, b) NV$^-$ color center in diamond.}
\label{pic:nvlatt}
\end{figure}

\begin{figure}[b!]
\centering
\includegraphics[width=0.6\linewidth]{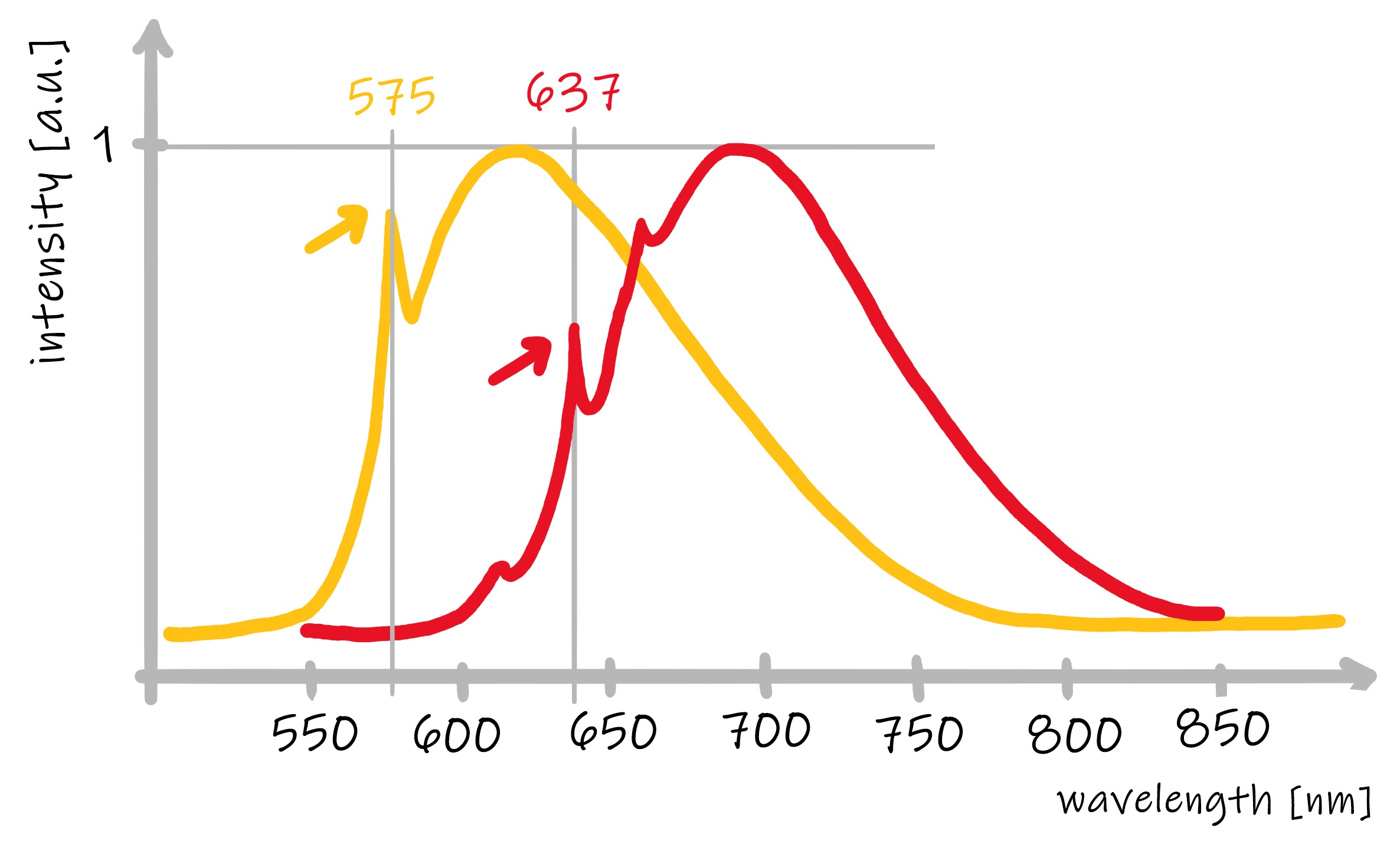}
\caption{Fluorescence spectrum for NV$^0$ (yellow line) and NV$^-$ (red line) color centers in diamond. Peaks marked with arrow stand for zero-phonon line in each case.}
\label{pic:nvemspect}
\end{figure}

Due to the nature of defects in diamond lattice (see Fig.~\ref{pic:nvlatt}), there are different kinds of NV color center -- NV$^0$, when two carbon atoms are substituted, one by nitrogen atom (N), the other one by a vacancy (V), and NV$^-$, when instead of the vacancy, borrowed electron is trapped \cite{Haque2017}. Also, the NV$^+$ color center can be produced by interaction with additional holes in the crystalline structure; however, due to the fact that it is not optically active \cite{Pezzagna2021}, it is not in our interest. The emission spectra of the other two, collected at room temperature, are presented in Fig.~\ref{pic:nvemspect}. As it can be seen, in both cases fluorescence spectrum is broad, however zero-phonon line peak is easily visible. If the fluorescence emission was tested at lower temperature, width of the spectrum would be norrower, resulting in zero-phonon line when no phonons are excited (at lowest possible temperature -- absolute zero). \\

The fabrication of NV centers is relatively easy. Typically, they are formed by a diamond lattice irradiation, which generates vacancies, followed by ion implantation and annealing at high temperatures (above 700$^o$C). They may be created both in a single crystal diamond or in nano-diamonds; however, it is easier to initialize the \textit{graphitization} process in nano-diamonds than in diamond thin films \cite{Haque2017}. \\

\begin{figure}[b!]
\centering
\includegraphics[width=\linewidth]{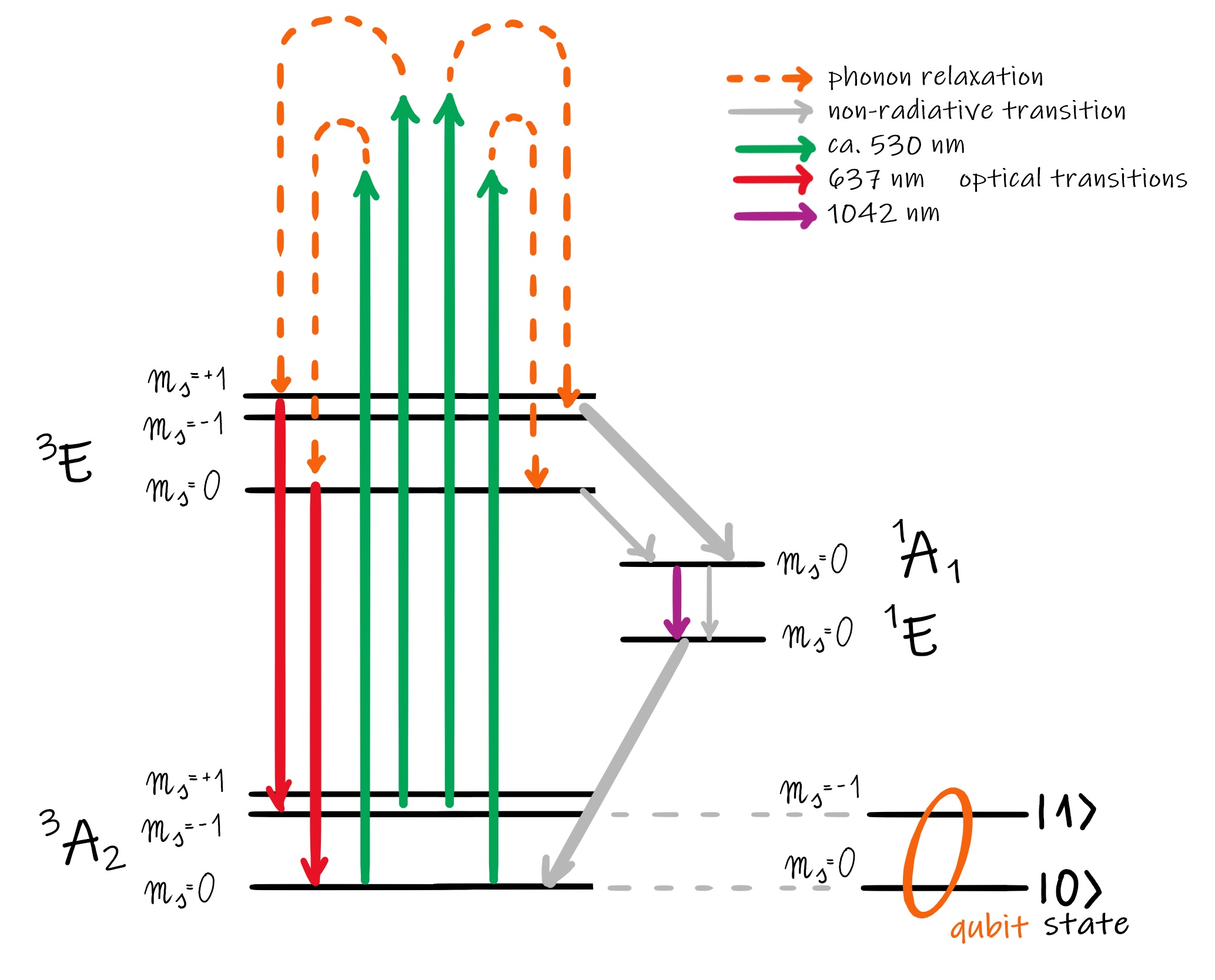}
\caption{The electronic structure of NV$^-$ center (left-hand side) with electron spin qubit states $|0\rangle$, $|1\rangle$ marked (right-hand side).}
\label{pic:nvqubit}
\end{figure}

The NV centers generally have a tendency to trap an extra electron, forming a negatively charged NV$^- $ center. Therefore, the following description focuses specifically on them. Also, the optically detected properties of NV$^-$ center, such as fluorescence spectrum splitting in  \textit{ODMR} experiment \cite{Jelezko2006, Dolde2014}, that shows the transition between different spin states, were the one that brings attention to them and suggested the use as a qubit host. \\
Although the electronic structure has already been described in \textit{Section \ref{sec:ch1-sinphsource}, Fig.~\ref{pic:nvemsys}}, let us take a closer look at energy levels. Unlike the most defects in diamond \cite{Pezzagna2021}, both NV states -- ground ($^3A_2$) and exited ($^3E$) -- are triplet states, and the intermediate states -- $^1E$ and $^1A_1$ -- are both singlets. The triplet state contains three spin states, for $m_s=0$ and $m_s=\pm1$. Thanks to this, the states $|0\rangle$ and $|1\rangle$ may be respectively defined as states with $m=0$ and $m=-1$, as presented in Fig.~\ref{pic:nvqubit}. \\

\begin{sloppypar}
What is noteworthy here is that there are both radiative and non-radiative transitions, but the important property of this system is that the non-radiative transition ${\{^3A, m_s=\pm1\}\rightarrow \{^1A_1, m_s=\pm1\} }$ is much more probable that ${ \{^3A, m_s=0\}\rightarrow \{^1A_1, m_s=\pm1\} }$ transition. This allows for a so-called \textit{spin-state initialization} or \textit{optical spin-polarization} procedure, resulting in transferring the population to the $|0\rangle$ state after few excitation cycles \cite{Tetienne2012}.
\end{sloppypar}
Furthermore, the spin-state of NV center can be controlled by using the microwave $\pi$ pulses, allowing for conversion between $|0\rangle$ and $|1\rangle$ states. This can be realized in approximately $7$ ns \cite{Fuchs2010} and observed as Rabi oscillations \cite{Jelezko2006}. \\

While controlling spin of a single NV center is enough to create a qubit, it is definitely insufficient to perform useful operations, such as error correction or even the use of two-qubit quantum gates, such as \textit{controlled-NOT} or \textit{SWAP} gate. On the one hand, the use of these requires longer coherence times than  NV$^-$ spin state provides (which is several milliseconds at room temperature); on the other, the coupling between few qubits is desirable. \\

\begin{figure}[b!]
\centering
\includegraphics[width=0.65\linewidth]{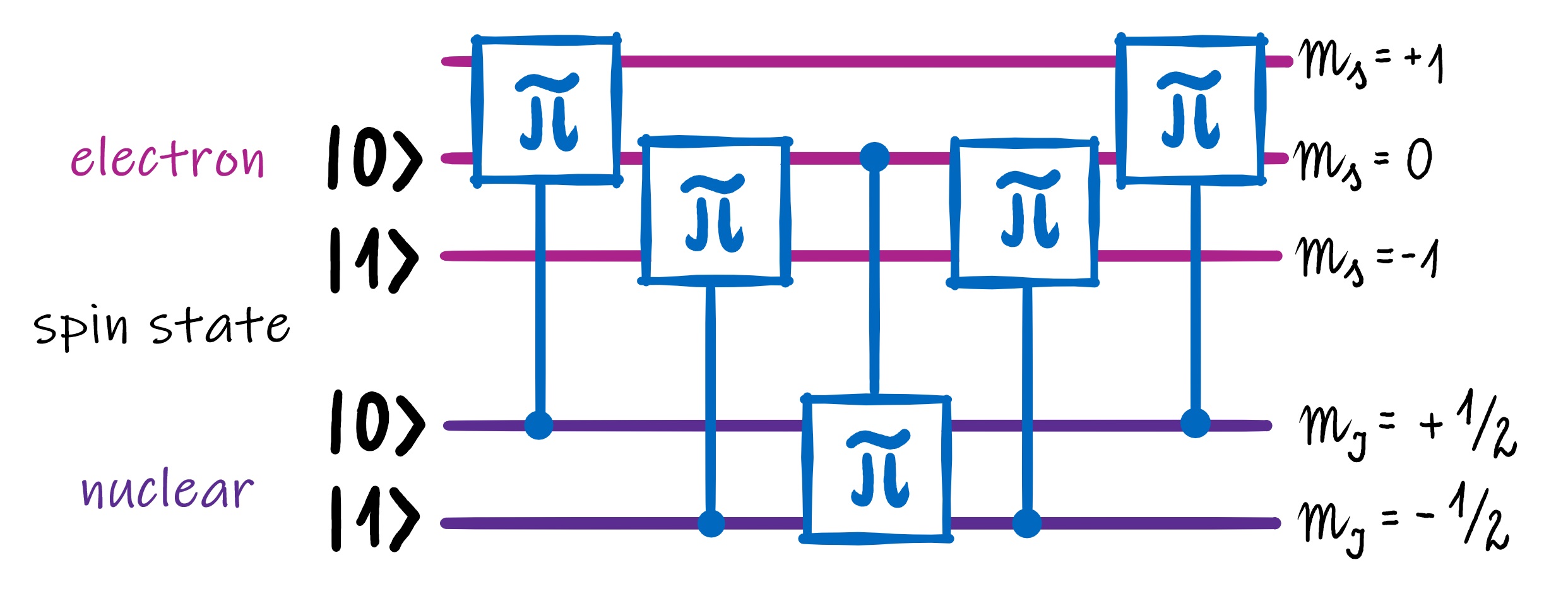}
\caption{Scheme of algorithm that allows for transferring the electron spin state to nuclear  spin state of the NV$^-$ center. $\pi$ gates represents the microwave $\pi$-pulses. Detailed description can be found in Ref. \cite{Dolde2014}.}
\label{pic:nvspinswap}
\end{figure}

In order to prolong the coherence time, up to seconds \cite{Maurer2012} or even minutes \cite{Taminiau2019}, the hyperfine coupling between spins states of trapped electron and $^{14}N$ nucleus can be used. The method allowing to swap between the electron and nuclear spin state is presented in Fig.~\ref{pic:nvspinswap}. It is based on microwave $\pi$-pulses, that change the electron spin state. Since the electron spin state is correlated with a nuclear one, this procedure allows for manipulation with nuclear spin state.\\

To couple qubits, the magnetic dipole-dipole interaction is used. This allows for interaction between them, as well as performance of multiqubit operations. It is noteworthy that too weak coupling strength may lead to long switching times for two-qubit gates, while too strong one reduces the coherence time. Since this interaction is dependent on the separation $r$ of the parties, and due to that changes quickly, proportionally to $r^{-3}$, obtaining the optimal structure is very challenging. It is also important that the proper structure can be obtained only by the adjustment of the distance between NV color centers, determined during the fabrication process. \\

Although it seems that recently the field of solid-state quantum computers based on NV centers is drenched by different proposals and results, the efficient large-scale system, containing several qubits and allowing readout as well as manipulation, has not been built yet. Therefore, the further study is needed, especially one involving the interaction between NV color centers and single photons, which are, in principle, the messengers of quantum information. \\

\hsubsection{Results}

\begin{figure}[b!]
\centering
\includegraphics[width=0.7\linewidth]{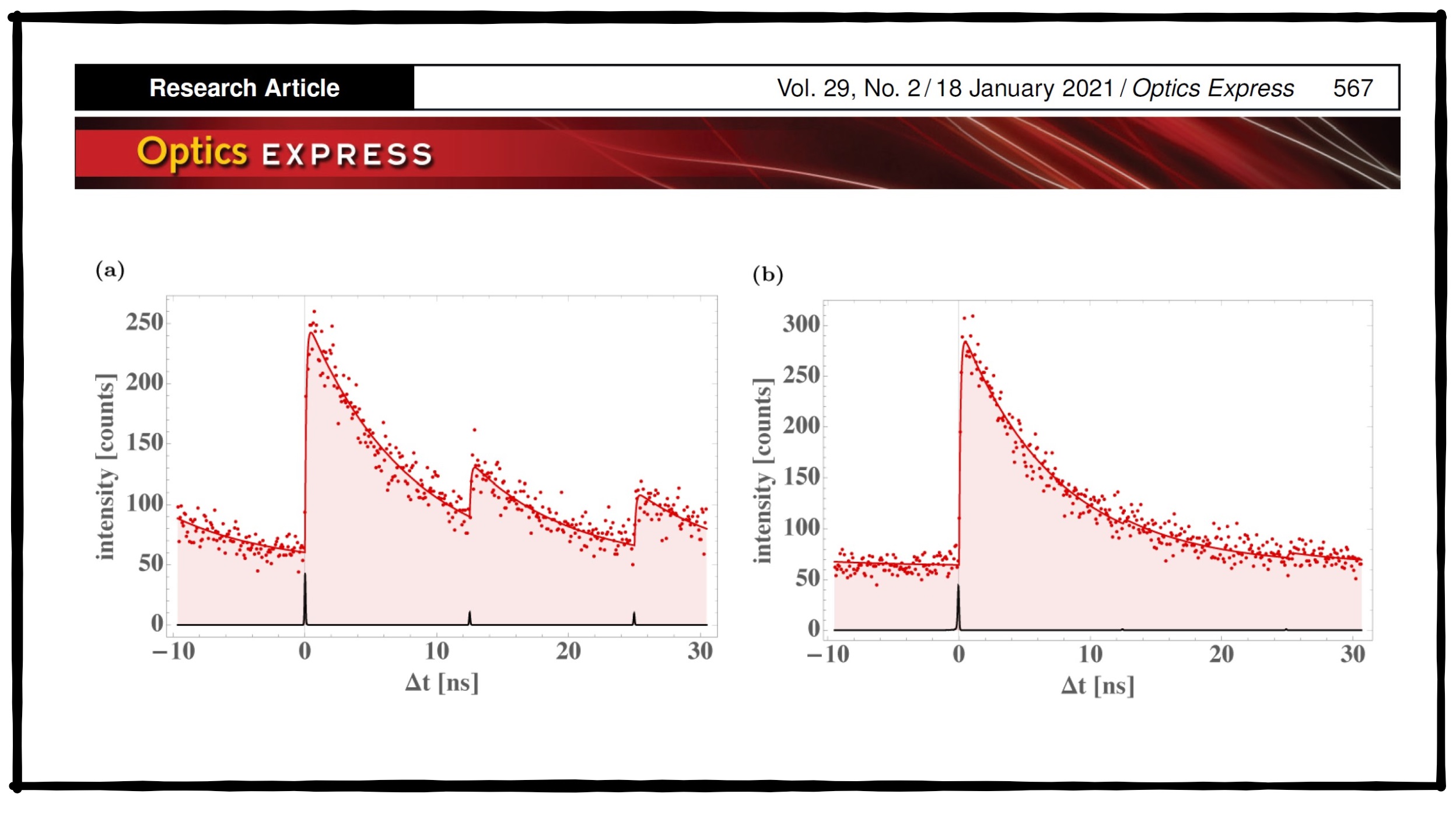}
\caption{NV$^-$ fluorescence decay measurement pumped with single photons. Single photons with wavelength of $532$ nm are generated in the SPDC process in (a) higher and (b) lower pumping power setting \cite{Marta2018}. Red dots show experimental data and red curve represents fitted model. Black curve shows a histogram of photon coincidences measured from single-photon source. Large and small peaks correspond to proper and accidental coincidences, respectively.}
\label{pic:nvresdectime}
\end{figure}

The aim of our study, published in Optics Express' article entitled \textit{Absorption of a heralded single photon by a nitrogen-vacancy center in diamond} \cite{Marysia2020}, was the demonstration of interaction between single NV center and single photon. The NV$^-$ center after the absorption of a photon emits the florescence photon. After many measurements, time histograms of coincidence measurement between the infrared and fluorescent photon were created, as presented in Fig.~\ref{pic:nvresdectime}. Followed by fitting the theoretical model, these measurements lead to obtaining the decay time for NV$^-$ centers. \\

\begin{figure}[t!]
\centering
\includegraphics[width=0.8\linewidth]{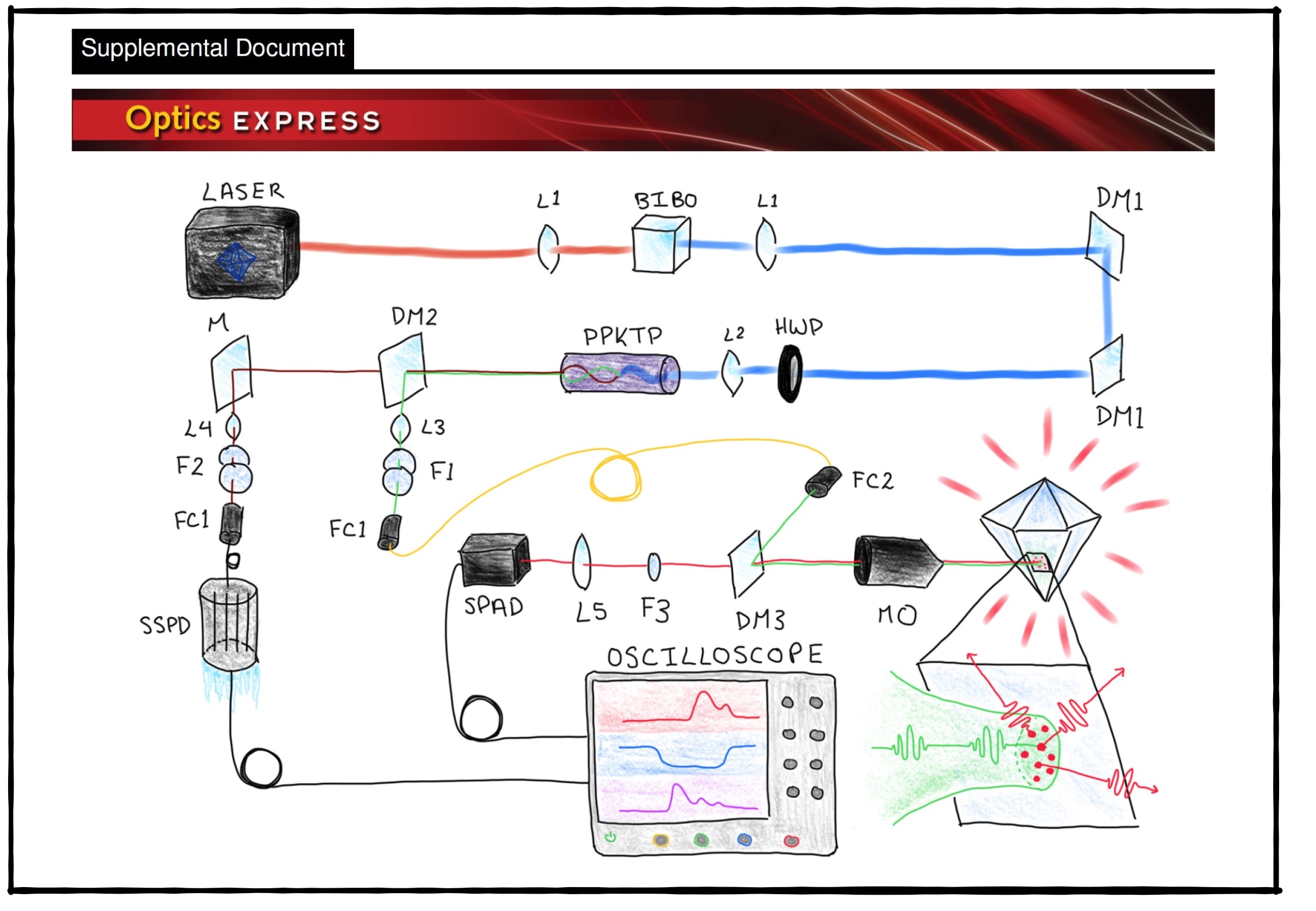}
\caption{Experimental setup for fluorescence decay time measurements. Symbols: LASER -- Ti:Sapphire pulsed laser, L1 -- lens (focal length $f = 7.5$ cm), BIBO -- nonlinear SHG crystal, DM1 --
dichroic mirror (Chroma AHF T425LPXR), HWP -- half-wave plate, L2 (focal length $f
= 10.0$ cm), PPKTP -- type-II SPDC nonlinear crystal,  
DM2 -- dichroic mirror (Chroma AHF 76-875-LP), L3 -- lens (focal length $f = 15.0$  cm), F1 -- filter (ET500
and FF550/88), FC1 -- fiber collimation package with focal length  $f=1.1$ cm and SMF780 single mode fiber or $f=1.51$ cm and F1-2000-FC-1 single mode fiber, L4 -- lens (focal length $f = 10.0$ cm), F2  -- filter (AHF 1319 LP), and coupled
M -- mirror, SSPD -- superconducting nanowire single photon detector, 
FC2 -- collimation package (F671FC-405), DM3 -- dichroic mirror (Semrock FF573-DI01), MO -- microscope objective, F3 -- spectral filter (Thorlabs FELH550/FELH700/BP650-40), L5 -- lens (focal length $f = 3.0$ cm), SPAD -- single-photon avalanche photodiode, Oscilloscope. }
\label{pic:nvresset}
\end{figure}

My contribution to this work consisted of building single-photon source that emits photons, which were then absorbed by the NV centers, its characterization and, since the source was the least stable element of whole setup, collecting the time arrival histograms. \\

The experimental setup is shown in Fig.~\ref{pic:nvresset}. It can be divided into two parts -- one is the single-photon source, the other is confocal microscope that allows for observation of fluorescence from the sample. The single-photon source generates a pair of photons, visible and infrared one, with $532$ nm and $1550$ nm central wavelengths, respectively, via SPDC process, which occured in PPKTP crystal. Then the visible photon is used to excite the NV center. This is followed by the emission of a red fluorescence photon. At the end, the coincidence measurement between infrared and red fluorescence photon is performed. The arrival times are recorded by the oscilloscope. \\

\begin{figure}[t!]
\centering
\includegraphics[width=0.7\linewidth]{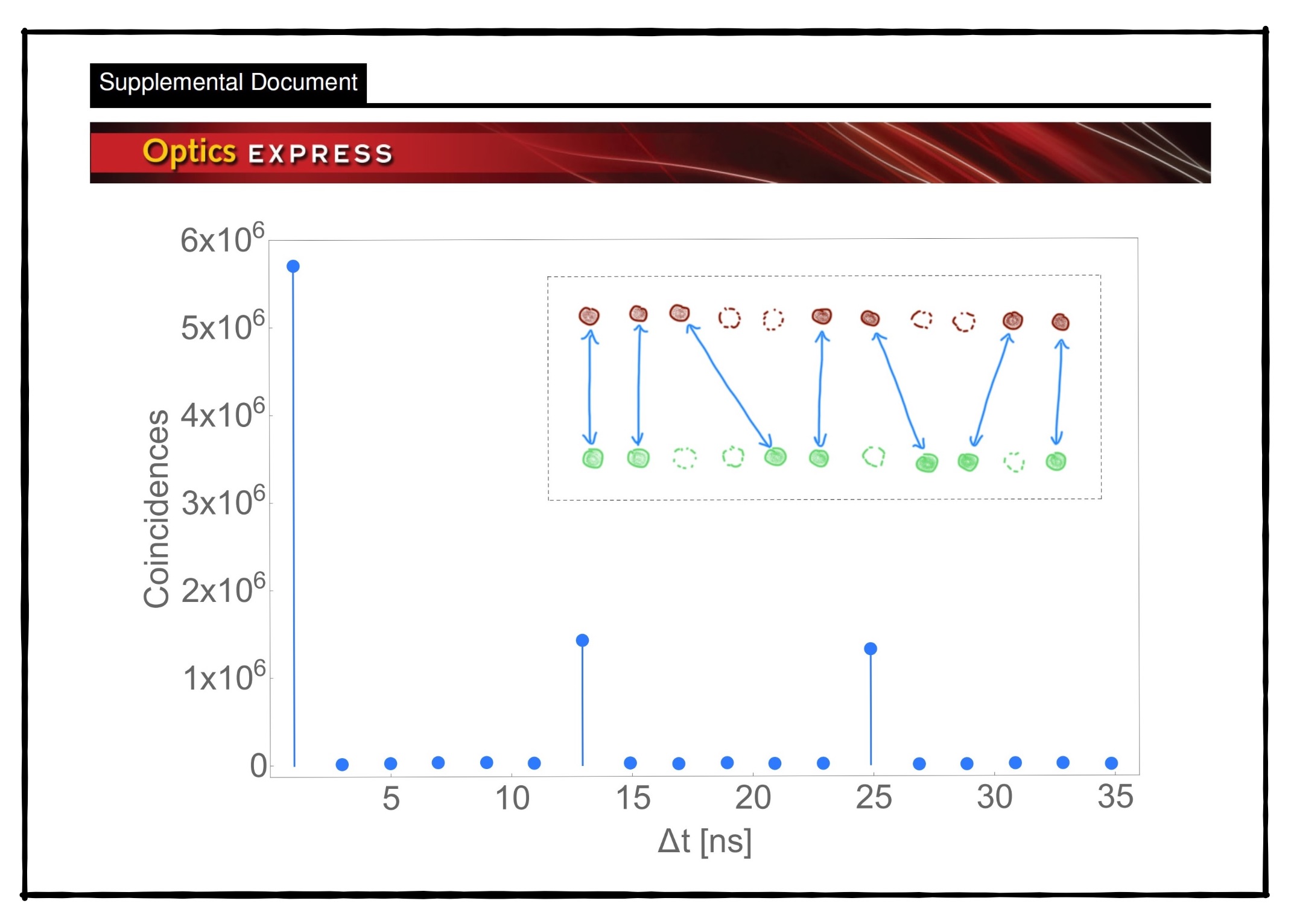}
\caption{Modeled coincidence histogram. The big blue peak corresponds to proper coincidence counts, whereas small ones to accidental coincidence counts. The temporal difference between peaks depends on the pulsed laser repetition rate. The inset presents the concept of accidental coincidence counts. If one photon from SPDC pair (solid, colored circle) is vanished (dashed, empty circle), the coincidence count is detected between photons from neighbouring pairs.}
\label{pic:nvacccoinc}
\end{figure}

The histograms were collected for two setting of the source, labeled as \textit{high power} and \textit{low power}. They differ in photon number in pulse, which is revealed in different second-order coherence function value. For \textit{high power} setting, the $396$ nm pump power setting was $5$ mW and second-order coherence function was equal to $g^2(0)=0.111(5)$, whereas for \textit{low power}, the $396$ nm pump power setting was $1$ mW and second-order coherence function was equal to $g^2(0)=0.0011(2)$. Also, for \textit{high power} setting, peaks corresponding to accidental coincidences, that comes from the photon from neighbouring pairs, appear in coincidence histogram (Fig.~\ref{pic:nvresdectime} a)). The reasoning behind the cause of accidental coincidence counts appearing in the measurement is presented in Fig.~\ref{pic:nvacccoinc}. \\

\begin{sloppypar}
It should be stressed that the result of second order coherence function $g^2(0)= 0.0011(2)$ obtained for the low power setting ($1$ mW average power in pulsed pump beam) is the lowest reported result in the literature for SPDC single-photon sources to date \cite{Urbasi2019}. \\
\end{sloppypar}
\chapter{Side information and notes}
\label{sec:ch5}

\hsection{Digital steganography}
\hnote{Digital steganography}
\label{sec:notes-steganography}

Modern steganography came to life in $1985$ thanks to Datotek Inc.,  manufacturer of cryptographic solutions, based in Dallas (Texas, USA). Two engineers, Barrie Morgan and Mike Barney, created and fielded steganographic systems (called \textit{c0™ / See Zero™ / Sea Zero™}) for personal use \cite{web:m2b2}. Although the first tools worked slowly, their development occurred rapidly. \\

Similarly to its ancient predecessor, digital steganography is based on hiding message into existing data, such as image, audio, or text files \cite{Adak2013}. Its principle is to add or replace bits \cite{book:Stallings}. \\
In case of adding bits, secret information may appear in two ways: as a \textit{file header} (which, for example in .tex file, is a preamble) or after \textit{the end of file mark}. Unfortunately, this simple file modification is easily noticeable. Therefore, it is better to replace bits. The method which helps with that is called \textbf{Least Significant Bit (LSB)}. \\
 
Due to the architecture of computer systems, data is saved as a string of bits, represented by zeroes \textbf{0} and ones \textbf{1}. Every $8$ bits compose $1$ byte, the smallest addressable unit of digital memory, which is needed to encode a single character of a text (see ASCII code \cite{book:Buchmann}) or an image pixel. In case of pictures, bits needed to define color of a single pixel have uneven contributions. The last two bits are the least important, which makes them ideal candidates for replacement without a significant change of color. \\

To illustrate\textbf{ LSB }method better, let us encrypt a word into a picture. Utilizing \textbf{RGB color model} \cite{book:Stallings}, every pixel is defined by three additive primary colors -- red, green, and blue -- with $2^8=256$ levels of saturation. These three colors need to be encoded separately, so each pixel offers the opportunity to encode $2\cdot3=6$ bits of secret information. As it is mentioned above, $8$ bits are needed to encode one letter. As a consequence, one pixel in not enough to encode one character. But three characters ($3\cdot8=24$) may be encoded into four pixels ($4\cdot6=24$). \\

\begin{figure}[t!]
\centering
\includegraphics[scale=0.47]{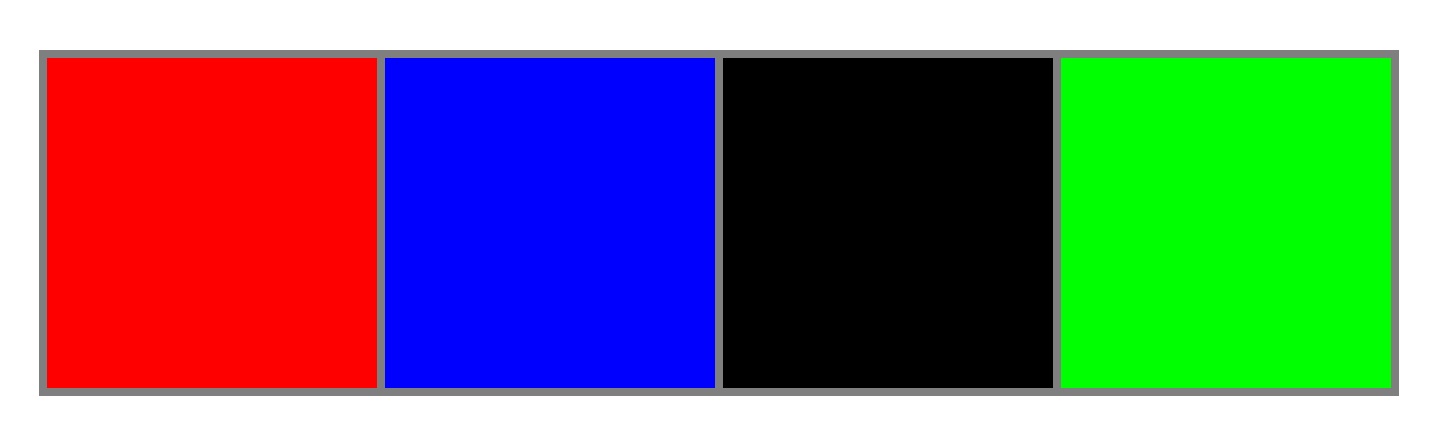}
\caption{Original file.}
\label{pic:stegorig}
\end{figure}

Let us use Fig.~\ref{pic:stegorig} as an example. The color of each pixel is introduced as a vector $(R,G,B)$, where $R$, $G$ and $B$ may take values from $0$ to $255$. Later, these values must be represented in a binary system to get a string of bits. We obtain:
\begin{quote}
1st pixel: $(255,0,0)\rightarrow (\color{red}11111111\color{black},\color{green}00000000\color{black},\color{blue}00000000\color{black})$, \\
2nd pixel: $(0,0,255)\rightarrow (\color{red}00000000\color{black},\color{green}00000000\color{black},\color{blue}11111111\color{black})$, \\
3rd pixel: $(0,0,0)\rightarrow (\color{red}00000000\color{black},\color{green}00000000\color{black},\color{blue}00000000\color{black})$, \\
4th pixel: $(0,255,0)\rightarrow (\color{red}00000000\color{black},\color{green}11111111\color{black},\color{blue}00000000\color{black})$.
\end{quote}

Let us encode the word $cat$ into the \ref{pic:stegorig}. The ASCII representation for each letter is:
\begin{quote}
$c \rightarrow 01100010$, $a \rightarrow 01100001 $ and $t \rightarrow 01110100 $,
\end{quote}
so full $cat$ word representation is $011000100110000101110100$. \\

\begin{figure}[b!]
\centering
\includegraphics[scale=0.47]{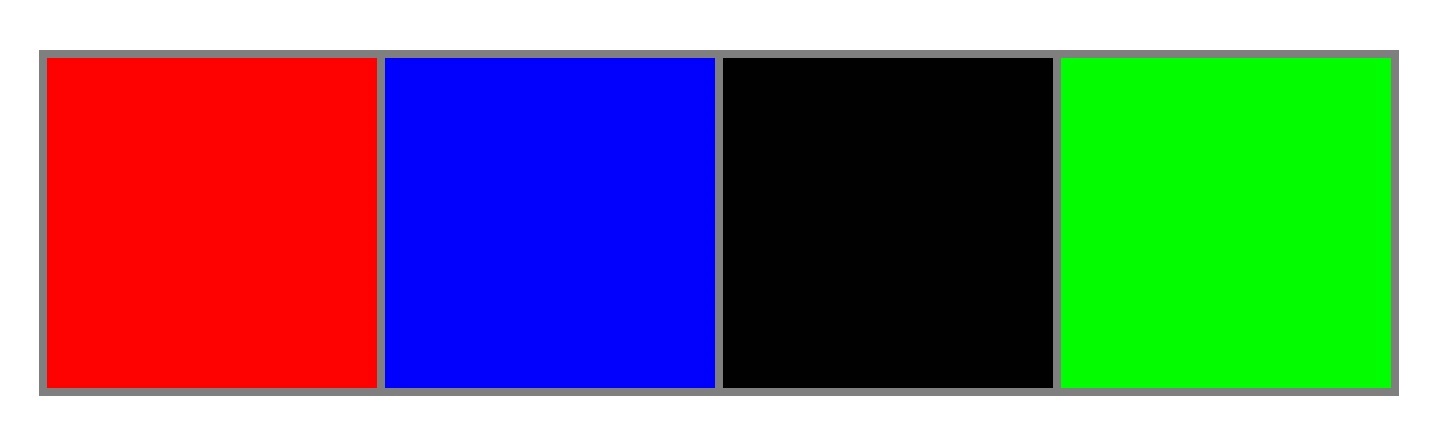}
\caption{File with encrypted message.}
\label{pic:stegenc}
\end{figure}

Next, the encoding procedure is to replace the last two bits of each color with two bits from the word we want to encode. In case of the first pixel we have:
\begin{quote}
$(\color{red}111111\color{black}01,\color{green}000000\color{black}10,\color{blue}000000\color{black}00)$,
\end{quote}
which gives $(253,2,0)$. In the end, entire encryption procedure results in:
\begin{quote}
1st pixel: $(11111101,00000010,00000000)\rightarrow(253,2,0)$,\\
2nd pixel: $(00000010,00000001,11111110)\rightarrow(2,1,254)$,\\
3rd pixel: $(00000000,00000001,00000001)\rightarrow(0,1,1)$,\\
4th pixel: $(00000011,11111101,00000000)\rightarrow(3,253,0)$.
\end{quote}
Fig.~\ref{pic:stegenc} shows an image with the hidden word. It is worth noting that due to small changes of saturation, both figures look similar and differences between them are imperceptible by human eye. Regardless of that fact, they can be easily spotted in numeric representation of colors.

\hsection{Transposition ciphers}
\hnote{Transposition ciphers}
\label{sec:notes-transciphers}

Since transposition ciphers are based on changing the position of letters in the plain text, their most important feature is the method of ordering. 
One of the approaches, for instance, is called \textbf{railway fence}, where all characters of the message are written in a zigzag pattern, downwards and upwards on successive \textit{rails} of an imaginary fence and then read off in rows. The number of \textit{rails} $R$ defines the cycle $C$ of a cipher, as $C=2R-2$, which allows receiver to decipher the message. \\

Let us follow the encryption procedure using a three-row fence, in which $R=3$, and an exemplary sentence $PHYSICS\; IS\; FUN!$. As mentioned earlier, the first step is to form the zigzag pattern:
\begin{quote}
\begin{center}
\begin{tabular}{ccccccccccccccc}
\hline
P&&&&I&&&&I&&&&U&& \\
\hline
&H&&S&&C&&&&S&&F&&N& \\
\hline
&&Y&&&&S&&&&&&&&! \\
\hline
\end{tabular}
\end{center}
\end{quote}
Next, the encrypted message is read off in rows, so we obtain \\$PIIUHSC\;SFNYS\;!$. \\

The decryption procedure starts with counting number $N$ of characters including the spaces (in the example $N=15$) and finding the number of "full" and "partial" cycles, which is needed to recreate zigzag pattern. \\
Here $R=3$, so $C=2\cdot3-2=4$, and $N:C=15:4=3.75$. Therefore, we have $3$ "full" cycles and $3$ characters in the "partial" one. It means that in the first row, there is $4$ characters ($3$ from "full" cycles, $1$ from the "partial"), in the second one, $7$ characters ($2\cdot3=6$ from "full" cycles, $1$ from the "partial") and in the third one, again $4$ characters. Dividing the ciphertext into rows, we have $PIIU$, $HSC\;SFN$, $YS\; !$, respectively. This way the "rail fence" may be reproduced. \\

\hsection{Substitutions ciphers}
\hnote{Substitutions ciphers}
\label{sec:notes-subciphers}

\hsubsection{Simple substitution cipher}
\label{sec:notes-subsimple}
There are many examples of simple substitution ciphers. One of the widest known is \textbf{Caesar cipher} \cite{book:Bauer}, named after Roman emperor Gaius Julius Caesar ($100$-$44$ BC), who used it for communication with friends and allies. He encrypted messages by replacing each letter with a letter shifted three positions down the alphabet. For instance, using this method, a word $cat$ is written as $zxq$. \\
Another example of a simple substitution cipher is \textbf{ROT13}, where letters are shifted by 13 positions (which, in case of Latin alphabet makes pairs of letters mapping each other). \\

In general, if we assign successive integers (from $0$ to $n-1$) to individual letters, the encoding of letter $x$ may be written mathematically as:
\begin{equation}
E(x)=(x+s) \mod{n},
\end{equation}
and decoded as:  
\begin{equation}
D(x)=(x-s) \mod{n},
\end{equation}
where $s$ represents the magnitude of a shift. \\

Unfortunately, due to the fact that mathematical description of encoding procedure is uncomplicated, it is easy to break. \\

\hsubsection{Homophonic substitution cipher}
\label{sec:notes-subhomo}
Homophonic cipher was invented as an improvement of simple substitution cipher. The purest example of such an algorithm is the one that is based on replacing each letter with a pair of adjacent ones. In this case, a word $cat$ is encoded as $bdzbsu$. Note that replacing every single letter by a larger group of characters, instead of another single letter, changes frequency distribution of characters in ciphertext, which makes a homophonic cipher harder to break. \\
A popular type of homophonic cipher is a \textbf{book cipher} \cite{book:Bauer,book:Stinson}. As the name suggests, it requires to use a book, the same one for both, encoding and decoding procedures. In this case, each letter of a plain text is coded with three numbers -- page, verse, and the letter number. \\

\hsubsection{Polygraphic cipher}
\label{sec:notes-subpolygra}
In polygraphic ciphers, a group of letters is replaced by different groups. To apply such an encoding method, the plaintext needs to be divided into smaller parts. Let us assume that a pair of letters is mapped into a different pair. To find the encoded pair, one can use an sample table presented below. The table consists of $25$ letters. 
\begin{quote}
\centering
\begin{tabular}{|c|c|c|c|c|}
\hline
Q&R&F&I&O \\
\hline
A&E&V&K&L \\
\hline
Z&D&B&M&P \\
\hline
W&C&N&U&G \\
\hline
C&X&H&Y&T \\
\hline
\end{tabular}
\end{quote}
Hovewer, Latin alphabet comprises $26$ letters, so one of the rare letters (for example \textit{x} or \textit{q}) should be skipped or letters \textit{i} and \textit{j} should be treated as one letter. \\
Although the filling of a table may be done randomly, its form must be known for both communitating parties. \\
The encryption procedure relies on dividing the table into rectangles in such a way that every  pair of letters from the plaintext is in opposite corners or sides of a single rectangle. Then, the encrypted pair lies in the other two sides or corners. \\

To see it better, we utilize the phrase $Physics\; is\; fun$. In the tables below, each pair of plaintext is marked with blue color, whereas the corresponding pair from ciphertext is marked with red (if the pair is mapped into itself, letters are bicolor). \\
\textbf{
\begin{center}
\begin{minipage}{0.3\linewidth}
\begin{tabular}{|c|c|c|c|c|}
\hline
Q&R&F&I&O \\
\hline
A&E&V&K&L \\
\hline
Z&D&\color{red}B\color{black}&M&{\color{blue} P}\color{black} \\
\hline
W&C&N&U&G \\
\hline
S&X&{\color{blue} H\color{black}}&Y&\color{red}T\color{black} \\
\hline
\end{tabular}
\end{minipage}
\vspace{0.5cm}
\begin{minipage}{0.3\linewidth}
\begin{tabular}{|c|c|c|c|c|}
\hline
Q&R&F&I&O \\
\hline
A&E&V&K&L \\
\hline
Z&D&B&M&P \\
\hline
W&C&N&U&G \\
\hline
{\twocollet[S]{red}{blue}}&X&H&{\twocollet[Y]{red}{blue}}&T \\
\hline
\end{tabular}
\end{minipage}
\vspace{0.5cm}
\begin{minipage}{0.3\linewidth}
\begin{tabular}{|c|c|c|c|c|}
\hline
Q&{\color{red}R}\color{black}&F&{\color{blue} I\color{black}}&O \\
\hline
A&E&V&K&L \\
\hline
Z&D&B&M&P \\
\hline
W&{\color{blue} C\color{black}}&N&\color{red}U\color{black}&G \\
\hline
S&X&H&Y&T \\
\hline
\end{tabular}
\end{minipage}
\begin{minipage}{0.3\linewidth}
\begin{tabular}{|c|c|c|c|c|}
\hline
\color{red}Q\color{black}&R&F&{\color{blue} I\color{black}}&O \\
\hline
A&E&V&K&L \\
\hline
Z&D&B&M&P \\
\hline
W&C&N&U&G \\
\hline
{\color{blue} S\color{black}}&X&H&\color{red}Y\color{black}&T \\
\hline
\end{tabular}
\end{minipage}
\vspace{0.5cm}
\begin{minipage}{0.3\linewidth}
\begin{tabular}{|c|c|c|c|c|}
\hline
\color{red}Q\color{black}&R&{\color{blue} F\color{black}}&I&O \\
\hline
A&E&V&K&L \\
\hline
Z&D&B&M&P \\
\hline
W&C&N&U&G \\
\hline
{\color{blue} S\color{black}}&X&\color{red}H\color{black}&Y&T \\
\hline
\end{tabular}
\end{minipage}
\vspace{0.5cm}
\begin{minipage}{0.3\linewidth}
\begin{tabular}{|c|c|c|c|c|}
\hline
Q&R&F&I&O \\
\hline
A&E&V&K&L \\
\hline
Z&D&B&M&P \\
\hline
W&C&\twocollet[N]{red}{blue}&\twocollet[U]{red}{blue}&\color{black}G \\
\hline
S&X&H&Y&T \\
\hline
\end{tabular}
\end{minipage}
\end{center}
}

The result looks as follows:
\begin{quote}
\begin{tabular}{rcccccc}
plaintext:& PH & YS & IC & SI & SF & UN \\
ciphertext: &TB & SY & UR & QY & HQ & NU \\
\end{tabular}
\end{quote}

The cipher of this type is known as \textbf{Playfair cipher}, named after Scottish politician and scientist, Lyon Playfair, who popularized its use \cite{book:Singh,book:Gomez}. In his version, when filling table cells, one should use a secret phrase from which duplicated letters should be skipped. All the remaining letters should be entered into the table, without changing their original order. Before doing that, the parties should decide in which order the table ought to be filled (for example, row by row from left to right and from top to bottom). The rest of the table cells should be filled with the remaining letters in the ordinary alphabetical order. \\

Modifications of the Playfair ciphers, \textbf{two-square cipher} and \textbf{four-square cipher}, which respectively utilize two or four alphabetic tables also exist. \\ 
In the case of a two-square cipher, the tables should be placed in such a way that the rows or columns are aligned, giving 2x1 or 1x2 pattern. Next, the first letter from each pair should be found in the left (or upper) table and the second one in the right (bottom) table. Encrypted letters are those from corners of the defined square, each from a different table. Similarly, in case of four-square cipher, tables should be aligned vertically and horizontally, giving 4x4 pattern. The first letter from each pair should be found in the upper left table and the second one in the bottom right table. Encrypted letters should be read from other tables.\\
Alphabet tables may be filled out randomly or started from a given phrase. \\

\hsubsection{Monoalphabetic cipher}
\label{sec:notes-submono}
Among all types of ciphers, \textbf{monoalphabetic ciphers} are the simplest and the oldest \cite{book:Singh}. They retain the substitution of letters along the encryption procedure, which means that the mapping from the plaintext to the ciphertext alphabet is fixed. \\ 
Examples of this kind of ciphers are presented in \textit{Sections \ref{sec:notes-subsimple}, \ref{sec:notes-subhomo}} and \textit{\ref{sec:notes-subpolygra}}. \\

\hsubsection{Polyalphabetic cipher}
\label{sec:notes-subpoly}
\textbf{Polyalphabetic ciphers} were invented to disguise the plaintext letter frequency and make the application of frequency analysis harder for the potential eavesdropper. In principle, these are any ciphers that use multiple alphabets for subsequent letter substitution. The mapping between alphabets changes repeatedly and is dependent on the position of a letter in the plaintext.  \\
The best-known example of such a cipher is \textbf{Vigenère cipher}. It is named after a French diplomat, cryptologist, and alchemist Blaise de Vigenère (1523--1596). It is noteworthy that this type of cipher was first described in 1553 by an Italian cryptologist Giovan Battista Bellaso \cite{book:Singh}. \\

In this case, an alphabet table called \textit{tabula recta} or \textit{Vignère square} is used in the encryption process \cite{book:Buchmann, book:Stinson}. The table is presented on the next page. It consists of cyclically shifted alphabets. Therefore, Vigenère cipher is a mix of Caesar ciphers with different shifts. To encode a message, communicating parties first have to determine the way of changing the alphabet, which is usually based on a keyword that is repeated during the procedure. Individual letters from a keyword sets current alphabet -- the row from tabula recta which should be used to encrypt particular letter. \\

Let us follow the encryption algorithm for the $Physics\; is \;fun$ phrase with a keyword $cat$. As mentioned above, a keyword must be repeated during the procedure. Each letter defines a row, so the first letter $P$ is encrypted using row $C$, second $H$ with row $A$, third $Y$ with row $T$ and so on. The result is the following:
\begin{quote}
\begin{tabular}{rcccccccccccccc}
plaintext:&P&H&Y&S&I&C&S&&I&S&&F&U&N\\
keyword:&C&A&T&C&A&T&C&&A&T&&C&A&T\\
ciphertext:&R&H&R&U&I&V&U&&I&L&&H&U&G\\
\end{tabular}
\end{quote} 

As one can notice, the keyword defines the number of alphabets utilized during the encryption procedure. Therefore, the keyword should generally be as long and complicated as possible to ensure safe communication. \\ Moreover, the main weakness of the Vigenère cipher is the cyclic nature of the substitution. Because of it after frequency analysis through long sample of a ciphertext, length of a keyword may be guessed. After that, ciphertext may be divided into separated samples and treated as a Caesar cipher with unknown alphabet shift. \\

\begin{tikzpicture}
\foreach \i in {0,...,25} {
  \foreach \j in {0} {
    \edef\k{\ifnum\numexpr\i+\j\relax>25
        \the\numexpr\i+\j-26\relax
      \else
        \the\numexpr\i+\j\relax
      \fi}
  \node[draw,fill=black!10,minimum size=0.5cm,inner sep=0pt]
    at (\i*0.5,-\j*0.5) {\strut\symbol{\numexpr`A+\k\relax}};
  }}
\foreach \i in {0} {
  \foreach \j in {0,...,25} {
    \edef\k{\ifnum\numexpr\i+\j\relax>25
        \the\numexpr\i+\j-26\relax
      \else
        \the\numexpr\i+\j\relax
      \fi}
  \node[draw,fill=black!10,minimum size=0.5cm,inner sep=0pt]
    at (\i*0.5,-\j*0.5) {\strut\symbol{\numexpr`A+\k\relax}};
  }}
\foreach \i in {1,...,25} {
  \foreach \j in {1,...,25} {
    \edef\k{\ifnum\numexpr\i+\j\relax>25
        \the\numexpr\i+\j-26\relax
      \else
        \the\numexpr\i+\j\relax
      \fi}
  \node[draw,minimum size=0.5cm,inner sep=0pt]
    at (\i*0.5,-\j*0.5) {\strut\symbol{\numexpr`A+\k\relax}};
  }
}
\end{tikzpicture}

\vspace{0.8cm}

Although it is not commonly known, the most popular encryption machine -- \textbf{Enigma} -- also utilized the polyalphabetic substitution cipher \cite{book:Gomez}. However, in that particular case, position of mechanical rotors, and then electric circuits, assigned letter substitution. \\
The strength of the Enigma cipher lied in the number of substitution possibilities, which, considering rotors only, was equal to $16900$. Real number of possibilities with unknown starting position of other parts of the Enigma mechanism was around $158 962 555 217 826 360 000$  (nearly $159$ quintillion, or $2^{67}$, which gives $67$ bits). For this reason Enigma users were convinced that the number of substitution would be enough to deter any brute-force attack. \\
\hsection{Public key cryptography}
\hnote{Public key cryptography}
\label{sec:notes-pubkeycrypto}

As it is described in \textit{Section \ref{sec:ch1-modcrypto}}, in asymmetric encryption, a pair of keys associated with each other is used. Since one of them is publicly known, the issue with creating algorithm that allows encryption with one key and decryption with the other boils down to mathematical problem of producing one-way function. These functions, the values of which are easy to compute, but which inverses are not easy to find, are often called \textbf{trapdoor functions}. \\

First consideration of this problem was done in the XIX century, when British logican and economist William Stanley Jevons (1835-1882) wrote in his book entitled \textit{The Principles of Science} \cite{book:Jevons}:
\begin{quote}
\textit{Can the reader say what two numbers multiplied together will produce the number 8616460799? I think it unlikely that anyone but myself will ever know.}
\end{quote}
Presently, it is known that $89681\cdot 96079$ is the answer. However, the factorization of the product of large prime numbers is still a computational issue. Therefore, this idea may be used for generating asymmetric keys. \\

There are two widely used asymmetric algorithms -- Diffie-Hellman key exchange protocol and RSA. 
They are applied in slightly different conditions. The first one is a method used to share a secret symmetric key through insecure channel, whereas the second one is a cryptosystem for digitally signed messages and data transmissions. To illustrate them better, one can use a color analogy and the codes from RGB color model. In this adaptation, it is assumed that the encryption is a color mixing procedure (mathematically -- adding RGB codes modulo 256). Additionally, separating colors from its mixture is possible only if at least one of them is known. \\

\begin{figure}[t!]
\centering
\includegraphics[width=\linewidth]{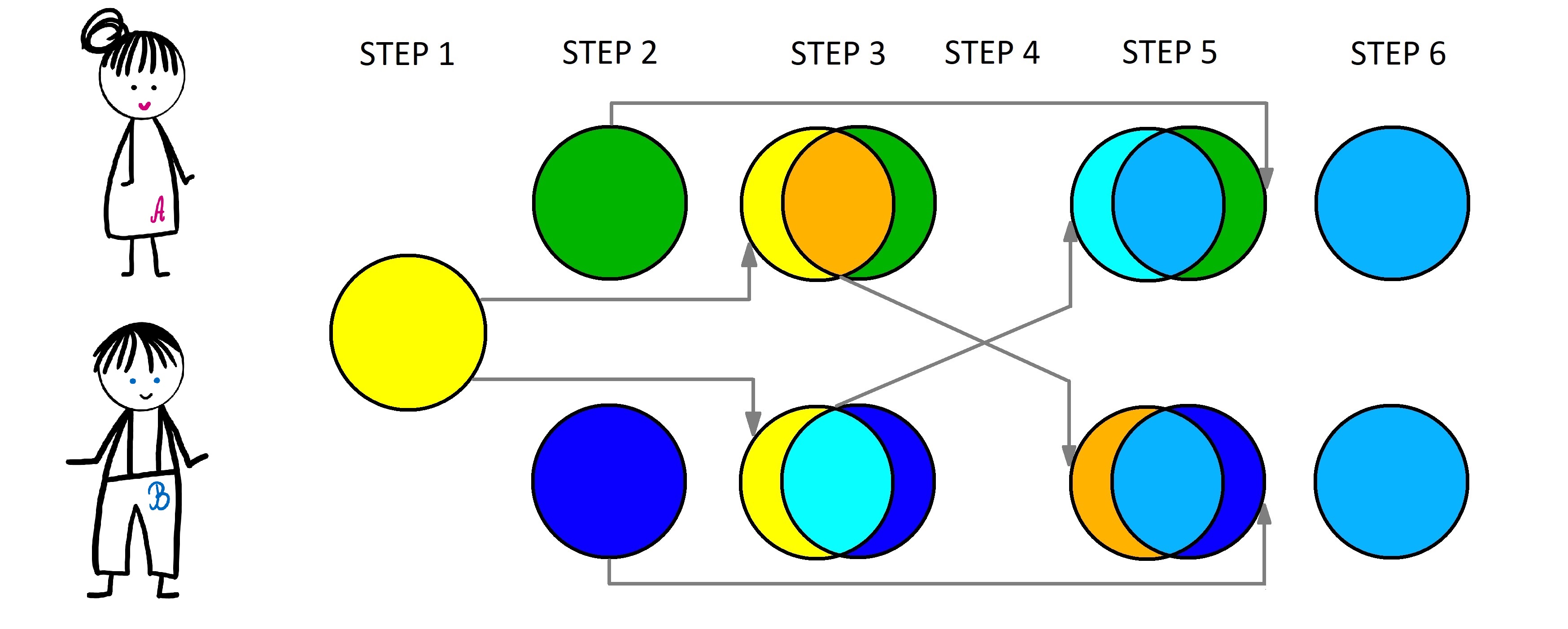}
\caption{Colour analogy scheme for Diffie-Hellman exchange key protocol.}
\label{pic:diff-hell}
\end{figure}

The \textbf{Diffie-Hellman algorithm} was first published by Whitfield Diffie and Martin Hellman in $1976$, but allegedly mathematicians from Government Communications Headquarters (GCHQ), the British signals intelligence agency, showed how this concept could be achieved as early as $1969$. \\
Nowadays, Diffie-Hellman key-agreement protocol is widely used, i.e., it is a part of the SSL protocol, which sets the secure connection between web browser and web server. \\
The algorithm is presented in Fig.~\ref{pic:diff-hell} and it looks as follows:

\begin{quote}
\begin{enumerate}
\item Alice and Bob agree on their mutual paint, in our case it is \colorbox{dhc1}{$(255,255,0)$}.
\item Each of them chooses the secret color, Alice chooses \colorbox{dhc2}{$(0,180,0)$} and Bob \colorbox{dhc3}{$(10,0,250)$}. The secret colors are their \textbf{private keys}.
\item They mix secret and common colors (encrypting one with another) and get \colorbox{dhc4}{$(255,179,0)$} and \colorbox{dhc5}{$(9,255,255)$}, for Alice and Bob, respectively. This mixture constitutes \textbf{public keys}. 
\item Alice and Bob share their mixtures and ...
\item ... add their secret colors to them.
\item Both Alice and Bob get \colorbox{dhc6}{$(9,179,255)$}, which is their joint \textbf{symmetric key} for further communication.
\end{enumerate}
\end{quote}

Mathematically, operations in Diffie-Hellman protocol use \textit{the multiplicative group of integers modulo p}, where \textit{p} is prime, and \textit{g} is a primitive root modulo\textit{ p}.
Therefore, the steps above may be written in this manner:
\begin{quote}
\begin{enumerate}
\item Alice and Bob agree to use a pair $(p,g)$.
\item Each of them chooses secret integer, Alice chooses $a$ and Bob $b$. The secret numbers are their \textbf{private keys}.
\item They compute $A=g^a \mod p$, $B=g^b \mod p$. These results are their \textbf{public keys}. 
\item Alice and Bob share their public keys.
\item Alice calculates $s=B^a \mod p$, Bob  $s=A^b \mod p$ .
\item Alice and Bob now share a \textbf{secret number} $s$.
\end{enumerate}
\end{quote}
One can notice that Alice and Bob share the same secret number $s$, because \\ $A^b \mod p = (g^a \mod p)^b \mod p = g^{ab} \mod p = g^{ba}\mod p = B^a \mod p$. \\
Moreover, due to the specific property of exponentiation operation, Diffie -- Hellman key exchange algorithm is not limited to negotiating a key by only two participants. It is a huge advantage of this method. \\

Another algorithm which is worth mentioning, \textbf{RSA}, was first publicly described by Ron Rivest, Adi Shamir, and Leonard Adleman in $1977$. However, the British GCHQ had already developed its equivalent in $1973$. It is a public-key cryptosystem, which allows for secure data transmission without sharing the secret key. Therefore, it is more convenient than other key-exchange protocols. \\

\begin{figure}[t!]
\centering
\includegraphics[width=0.5\linewidth]{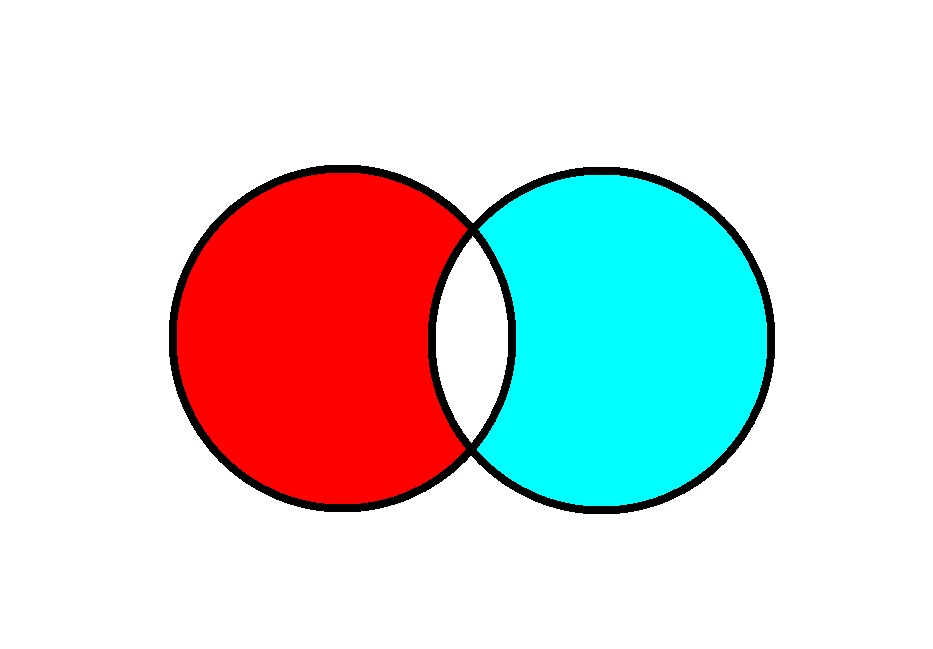}
\caption{Complementary colours \textcolor{rsak1}{$(255,0,0)$} and \textcolor{rsak2}{$(0,255,255)$} as a private and public key in RSA.}
\label{pic:rsakeys}
\end{figure}

In the RSA protocol, the receiver creates a pair of keys and publishes the public one. The keys need to be related to each other. In the color analogy, we can associate them with complementary colors (which when added give white -- in RGB system -- \colorbox{black}{\textcolor{white}{$(255,255,255)$}}\, , as shown in Fig.~\ref{pic:rsakeys}.\\

Let us follow the procedure of encoding a message with RSA. Alice wants to send a message to Bob, so:
\begin{quote}
\begin{enumerate}
\item Bob creates a pair of keys. He decides that \colorbox{rsa1}{$(200,55,200)$} is the public and \colorbox{rsa2}{$(55,200,55)$} the private one. 
\item The public key is shared with other RSA users.
\item Alice encrypts the plaintext using Bob's public key, which is color coded with \colorbox{rsa3}{$(255,200,55)$}. 
\item The result \colorbox{rsa4}{$(199,255,255)$} is sent to Bob.
\item Bob decrypts the ciphertext using his private key and gets \colorbox{rsa5}{$(255,200,55)$}.
\end{enumerate}
\end{quote}

The mathematical version of this protocol, using prime numbers, is as follows:
\begin{quote}
\begin{enumerate}
\item Creating a pair of keys:
\begin{enumerate}
\item Bob chooses two large prime numbers $p$ and $q$.
\item He calculates $n=p \cdot q$ and Euler's totient function \\ $\phi(n)=(p-1)(q-1)$.
\item Bob chooses an integer $e$, which is coprime with $\phi(n)$.
\item Bob computes $d=e^{-1} \mod \phi(n)$.
\item The private key is formed by a pair $(n,d)$, whereas public key takes the form $(n,e)$. 
\end{enumerate}
\item The public key is shared with other RSA users.
\item Alice encrypts the message $m$ with the public key $(n,e)$ using formula \\ $c = m^e \mod n$.
\item The result is sent to Bob.
\item Bob decrypts ciphertext using his private key $(n,d)$ and gets \\ $c^d \mod n = (m^e)^d \mod n\equiv m $.
\end{enumerate}
\end{quote}

\vspace{3ex}
It should be noted that the protocol RSA is able to generate relatively strong keys which are very hard to break. It is because its security is based on both inability of classical computers to efficiently factorize big numbers and inability to extract $e$ from given $d$ (known as the \textbf{RSA problem} \cite{book:Stallings}). \\
These are strong points of RSA cryptosystem, however, with the use of quantum computers, both mentioned operations could be done in acceptable time making it obsolete \cite{book:nielsenchuang}.\\

\begin{figure}[h!]
\centering
\includegraphics[width=\linewidth]{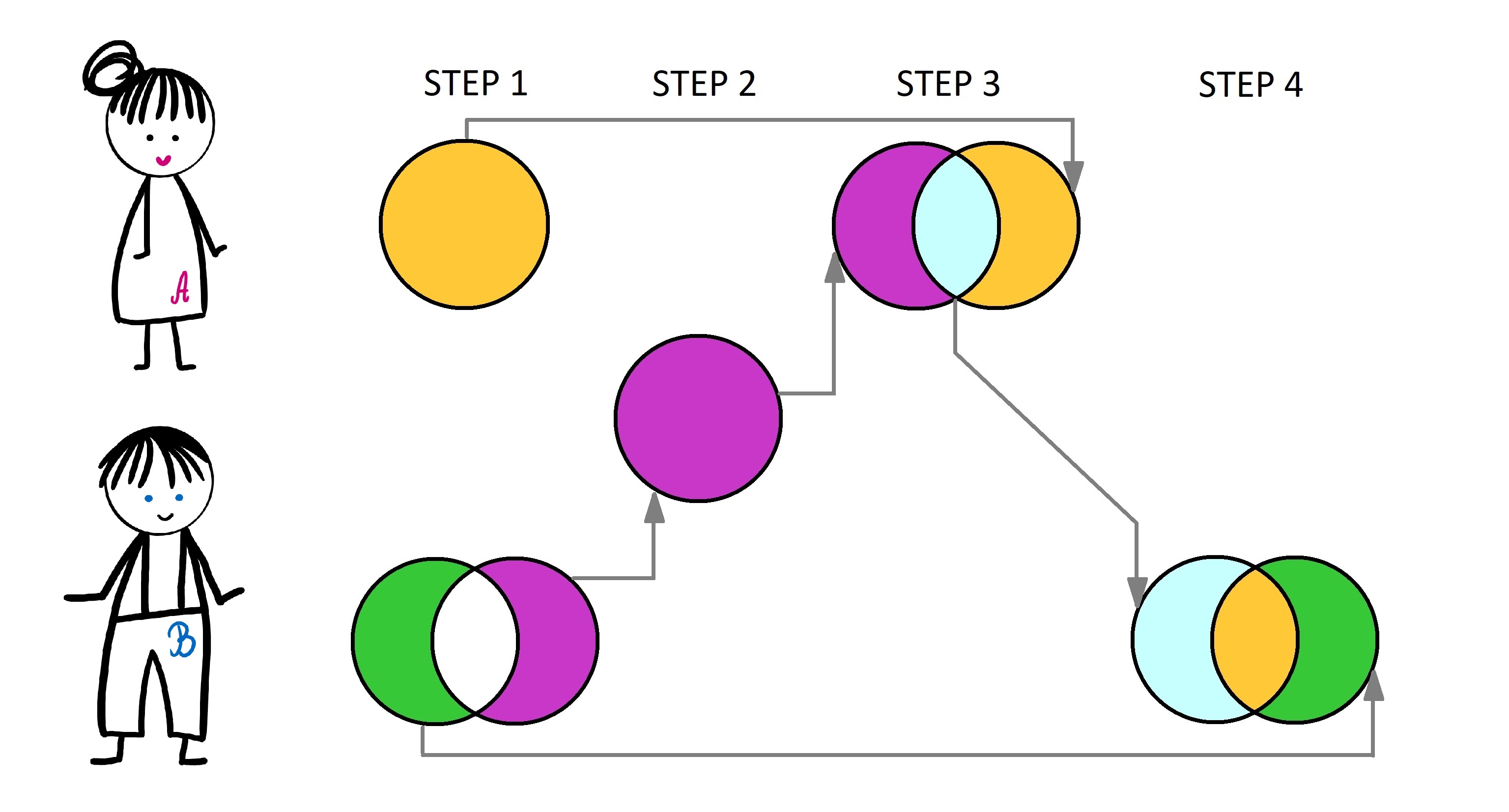}
\caption{Colour analogy scheme for RSA.}
\label{pic:rsa}
\end{figure}

\hsection{One-time pad}
\hnote{One-time pad}
\label{sec:notes-otp}

\textbf{One-time pad} is an encryption method, which is considered to provide the perfect secrecy when used properly. It was first proposed in 1882 by Frank Miller \cite{book:Gomez}, American banker and cryptologist, in telegraphic transmission context. Later, in 1917, Gilbert Vernam re-invented it, adding the so-called $XOR$ operation ( denoted as $\oplus$) to the encryption process (logical operation that outputs "true" only when inputs differ) \cite{Vernam1926}. \\

A recipe for OTP is simple:
\begin{enumerate}
\item Fix an integer $l$. Then, the cipher is defined over $(\mathscr{K},\mathscr{M},\mathscr{T})$ space of keys, messages, and ciphertext, respectively, where each of them is equal to $\{0,1\}^l$.
\item The key $k \in \mathscr{K} $ is formed as a string of zeros and ones, according to the uniform distribution. This means that each bit should be chosen randomly. 
\item Encryption $\mathscr{E}$ is given as $t:=k\oplus m$, for $t\in\mathscr{T}$, $m\in \mathscr{M}$.
\item Decryption $\mathscr{D}$ is given as $m:=k\oplus t$.\\
\end{enumerate}

The example of coding and decoding message with one-time pad is presented in Tab. \ref{tab:notes-otp} below. \\

\begin{table}[h!]
\begin{tabular}{r|ccccccc}
\multicolumn{8}{c}{\textbf{Encryption procedure}} \\
\hline \hline
plaintext&1&0&0&1&1&0&0\\
\hline
key &0&0&0&1&1&0&1\\
\hline
ciphertext &1&0&0&0&0&0&1\\
\hline
\end{tabular}
\quad
\begin{tabular}{r|ccccccc}
\multicolumn{8}{c}{\textbf{Decryption procedure}} \\
\hline \hline
ciphertext &1&0&0&0&0&0&1\\
\hline
key &0&0&0&1&1&0&1\\
\hline
plaintext&1&0&0&1&1&0&0\\
\hline
\end{tabular}
\caption{Exemplary use of one-time pad with XOR operation.}
\label{tab:notes-otp}
\end{table}

One-time pads are considered to be perfectly secure. From Shannon's definition, in general, a cipher $(\mathscr{E},\mathscr{D})$ has \textit{perfect secrecy} if $\forall_{m_1, m_2\in\mathscr{M}}$ with the same length and $\forall_{t\in\mathscr{T}}$ the probabilities of encoding $m_1$ and $m_2$ into $t$ with a key $k$ are equal
\begin{equation}
Pr\big[\mathscr{E}(k,m_1)=c\big]=Pr\big[\mathscr{E}(k,m_2)=c\big].
\label{eq:perfsec}
\end{equation}
It simply means that the ciphertext provides no information about the original message except its length. \\
In case of the one-time pad, $Pr\big[t|m\big]=Pr\big[t=m\oplus k|m\big]=Pr\big[k=m\oplus c|\big]=\frac{1}{2^l}$. \\
Since it holds for all distributions, the Eq.~ \ref{eq:perfsec} is fulfilled. \\

Although one-time pad is perfectly secure, it is based on the use of a one-time key of at least the same size as the plaintext. The key should be generated in a truly random way (the condition of a uniform distribution) and shared safely between communicating parties. These requirements make it challenging for every-day use. \\

\hsection{Qubits}
\hnote{Qubits}
\label{sec:ch1-qubit}
As it was mentioned in \textit{Section \ref{sec:ch1-quantuse}}, one of the tools of quantum cryptography are qubits. A \textbf{qubit}, similarly to its classical equivalent, \textbf{bit}, is the unit of quantum information. However, instead of representing the classical states "0" and "1", due to its quantum nature, qubit represents the coherent superposition of both states. In other words, a qubit is simply a quantum two-level (or two-state) system. \\

\begin{figure}[b!]
\centering
\includegraphics[width=0.6\linewidth]{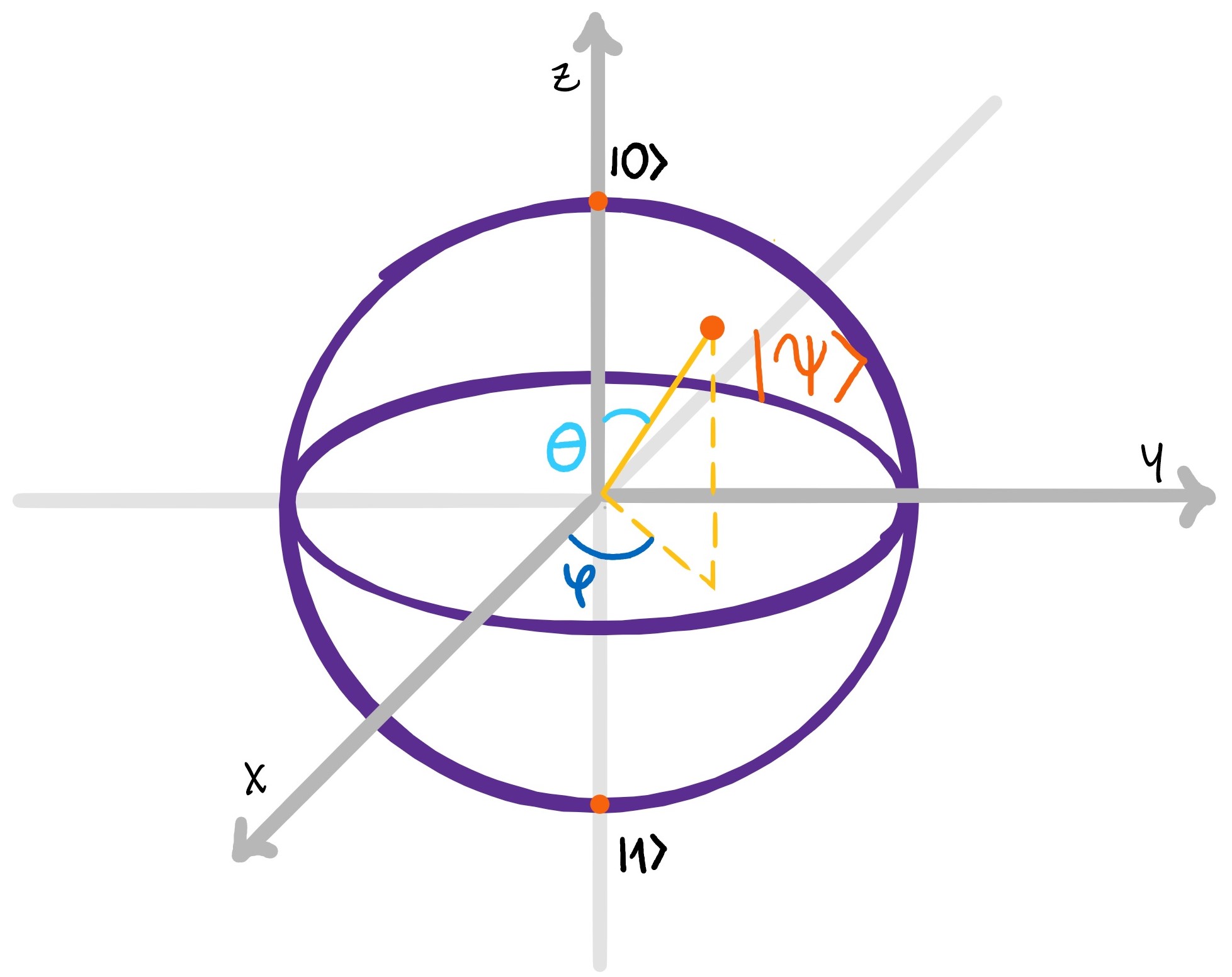}
\caption{Qubit state on a Bloch sphere.}
\label{pic:qubloch}
\end{figure}

Mathematically, a single qubit \textit{pure state} is described by linear combination of the basis states, in the example of $\{|0\rangle,|1\rangle\}$:
\begin{equation}
|\psi\rangle=\alpha|0\rangle+\beta|1\rangle \;,
\end{equation}
where $\alpha$, $\beta$ are the probability amplitudes of finding the system, after the measurement, in either state. Thus the relation $|\alpha|^2+|\beta|^2=1$ must be satisfied. In general, $\alpha$ and $\beta$ may be complex numbers. \\

It is worth to mention here that higher-level quantum systems, called \textit{qu\textbf{d}its} (\textit{d} stands for the dimension), can be described similarly. Although such states are considered to be the feature of quantum computing \cite{Wang2020}, their experimental realization is complicated. \\

\hsubsection{Bloch sphere}
Qubit state may be presented visually as a point on \textbf{Bloch sphere} (also known as \textbf{Poincar\'{e} sphere}). Bloch sphere, named after Swiss physicist Felix Bloch (1905-1983), is  a geometrical representation of the pure state space of a two-level quantum mechanical system \cite{book:nielsenchuang, book:vlatko}. \\
The orthogonal state vectors are at antipodal points of a sphere. Points on the surface of a sphere correspond to the \textit{pure states}, the interior points represent the \textit{mixed states} of the system. The Bloch sphere may be generalized to an n-level quantum system \cite{Dietz2006, Kolenderski2010}, but then its geometrical visualisation is challenging. \\

The coordinates of a point, which represents the $|\psi\rangle$ state on the Bloch sphere, may be found using the \textit{Hopf substitution} \cite{Treisman2009}:
\begin{equation}
\bigg\{
\begin{matrix}
\alpha=e^{i\phi}cos\frac{\theta}{2}\\
\beta=e^{i(\phi+\psi)}sin\frac{\theta}{2}
\end{matrix} \, ,
\end{equation}
where $i$ is an imaginary number. Since in experiments only a relative phase is measured, we can arbitrarily choose that $\phi=0$. Then, we get the spherical coordinates in a form:
\begin{equation}
\bigg\{
\begin{matrix}
\alpha=cos\frac{\theta}{2}\\
\beta=e^{i\psi}sin\frac{\theta}{2}
\end{matrix} \, ,
\end{equation}
where $\psi$ and $\theta$ are the angles marked in Fig.~\ref{pic:qubloch}. \\

\hsubsection{Quantum entanglement}
\label{sec:ch1-ent}
One of the most valuable features of qubits is that they can interact with each other in such a way that their quantum state cannot be described separately. This feature is called \textbf{quantum entanglement}. \\ 

It was first described by Albert Einstein, Boris Podolsky, and Nathan Rosen in 1935 \cite{EPR1935}.
The researchers found that for entangled particles the measurements of physical properties are perfectly correlated. For example, when positronium, the elementary particle with spin 0, decays into two particles with spin $1/2$ but with opposite momentum, the sign measurement on the first particle determines the sign on the second one. What is important, the result of the measurements does not depend on the distance between those two newly created particles. That property caused further considerations, including the \textit{superluminal} exchange of information, which is, as they all agreed, forbidden by the \textit{theory of relativity}. The authors claimed that, due to the fact that given a specific experiment, in which the outcome of a measurement is known before the measurement takes place, there must exist some property of reality, which determines the measurement results. Furthermore, their work ended with a sentence:
\begin{quote}
\textit{We are thus forced to conclude that the quantum-mechanical description of physical reality given by wave functions is not complete.}
\end{quote}
This statement induced instant response by Niels Bohr \cite{EPR-resp}, where he stated that the reasoning presented by scientists was wrong. He stressed that since, for example, measurements of position and of momentum are complementary, making the choice to measure one excludes the possibility of measuring the other, therefore:
\begin{quote}
\textit{arguments do not justify their conclusion that the quantum description turns out to be essentially incomplete.}
\end{quote}
Presently, these considerations, which lead to further work on \textit{local realism} \cite{Wiseman2015}, or rather \textit{non-locality} \cite{Brunner2014,Popescu2014, Khrennikov2017} and \textit{hidden-variable theory} \cite{Bohm1966, Dzhafarov2019, Acuna2021}, are known as \textit{EPR paradox} \cite{Sch1935}.\\

Moving back from philosophical to physical considerations, it should be emphasized that the entanglement may be introduced not only using the spin, but also other physical properties. Currently, the main source of entangled states are polarization-entangled photons. \\

Let us take a look at the mathematical description of two systems, denoted as $A$ and $B$. Its quantum states, they are either \textit{pure} $|\psi\rangle_A$ and $|\phi\rangle_B$ or \textit{mixed} $\rho_A$ and $\rho_B$, belong to respective \textit{Hilbert space} $\mathcal{H}_A$ and $\mathcal{H}_B$. Then, the Hilbert space of whole system is the tensor product $\mathcal{H}_A\otimes\mathcal{H}_B$ and, similarly, the state is either:
\begin{equation}
\begin{split}
|\Psi\rangle_{AB}&=|\psi\rangle_A\otimes|\phi\rangle_B\, , \\
\rho_{AB}&=\rho_A\otimes\rho_B \; .
\end{split}
\label{eq:sep}
\end{equation} 
States, which take such form are called \textit{separable states}. \\
However, not all states are separable. The most general form of a state that belongs to $\mathcal{H}_A\otimes\mathcal{H}_B$ is either:
\begin{equation}
|\Psi\rangle_{AB} =\sum_{i,j}c_{ij}|i\rangle_A\otimes|j\rangle_B \; ,
\label{eq:inseppure}
\end{equation}
or
\begin{equation}
\rho_{AB}=\sum_j q_j \rho_j^A \otimes \rho_j^B \; ,
\label{eq:insepmix}
\end{equation}
where ${|i\rangle_A}$ and ${|j\rangle_B}$, $\rho_j^A$ and $\rho_j^B$, are the basis of spaces $\mathcal{H}_A$ and $\mathcal{H}_B$, respectively. Note that only if $c_{ij}=c_i^A\cdot c_j^B$, the state from Eq.~\ref{eq:inseppure} takes a form from Eq.~\ref{eq:sep}. All other states, which do not fulfill that condition, are \textit{inseparable}. If the state of a composite system is inseparable, it is called an \textbf{entangled} \cite{4H2007}. \\

There are many ways to determine the amount of entanglement in the system \cite{Vedral1997, Plenio2007, Regula2016} (see also \textit{Sections \ref{sec:ch5-fidelity}} and \textit{\ref{sec:ch5-conc}}). One of them is the introduction of \textbf{entanglement witness} \cite{4H2007, Brandao2005}. It is a mathematical operator, which expectation values are positive in the class of separable states. If such operator, during the measurement, gives negative expectation value, it is a direct indication of an entanglement. The theory of entanglement witness came from Hahn-Banach theorem \cite{Narici1997}, but since it still under investigation \cite{Chruscinski2014, Frerot2021, Zhang2021}, it will not be further elaborated on in this work. Nonetheless, it is worth to mention that in photonic experiments one of the most useful forms of an entanglement witness is functional $f=2-S$, where $S$ is a Bell parameter (see \textit{Section \ref{sec:notes-BellIneq}}) \cite{Franson1989}. \\

Another way is the measurement of \textbf{entanglement entropy} $E\big( |\psi\rangle\langle\psi|\big)$ \cite{Witten2020}. For a pure state it is defined by:
\begin{equation}
E\big( |\psi\rangle\langle\psi| \big):= S\Big(Tr_A\big(|\psi\rangle\langle\psi|\big)\Big) = S\Big(Tr_B\big(|\psi\rangle\langle\psi|\big)\Big),
\end{equation}
where $S(\cdot)$ is von Neumann entropy \cite{book:Stenholm} and $Tr_A$, $Tr_B$ stands for partial trace operation over one of the subsystems, $A$ or $B$ respectively. The von Neumann entropy, reported in 1935 by John von Neumann, Hungarian-American mathematician and physicist, is defined for a quantum system described using the \textit{density matrix} $\rho$ and takes a form:
\begin{equation}
S(\rho)=-Tr\big(\rho log (\rho)\big)\,.
\end{equation}
It is worth to notice that $S(\rho)$ is invariant under a unitary transformation. Thus, after the diagonalization of the density matrix $\rho=\sum_i p_i|\phi_i\rangle\langle\phi_i|$, where ${|\phi_i\rangle}$ forms an orthonormal basis and $\sum_i(p_i)^2=1$, the von Neumann entropy can be written as:
\begin{equation}
S(\rho)=-\sum_i p_i log (p_i) \,.
\end{equation}

\hsubsection{Bloch basis}
\label{sec:ch5-blochbasis}
Let us now return to pure state of entangled qubits. The bases are given by $\{|0\rangle,|1\rangle\}_A$ and $\{|0\rangle,|1\rangle\}_B$ for qubit $A$ and $B$, respectively. The four qubit entangled states may be written as:
\begin{equation}
\begin{split}
|\Phi^{\pm}\rangle=\frac{1}{\sqrt{2}}\big(|0\rangle_A\otimes|0\rangle_B \pm |1\rangle_A\otimes|1\rangle_B \big)=\frac{1}{\sqrt{2}}\big( |00\rangle\pm|11\rangle\big)\, ,\\
|\Psi^{\pm}\rangle=\frac{1}{\sqrt{2}}\big(|0\rangle_A\otimes|1\rangle_B \pm |1\rangle_A\otimes|0\rangle_B \big)=\frac{1}{\sqrt{2}}\big( |01\rangle\pm|10\rangle\big)\, .
\end{split}
\label{eq:bellbasis}
\end{equation} 
These states, commonly known as \textbf{Bell states}, named after Northern-Irish physicist John S. Bell, form a maximally entangled basis of the four-dimensional Hilbert space \cite{Bell1964}. \\

To check if, for example, $|\Phi^{+}\rangle$ Bell state is entangled, the entanglement witness in the earlier mentioned form $f=2-S$ is chosen. As it is shown in  \textit{Section \ref{sec:notes-BellIneq}}, Bell parameter for this state is equal to $S=2\sqrt{2}$, so computing the $f$ value, we get the negative value $f=2-2\sqrt{2}$. Thus, the state $\Phi^{+}$ is entangled. \\
Now, in order to check if it is maximally-entangled, we can calculate entanglement entropy:
\begin{equation}
E\big( |\Phi^+\rangle\langle\Phi^+|\big)=S\Big( Tr_B(|\Phi^+\rangle\langle\Phi^+|)\Big) \; .
\end{equation}
Since:
\begin{equation}
\begin{split}
Tr_B(|\Phi^+\rangle\langle\Phi^+|)\Big)=Tr_B\Big( \frac{1}{2}\big(|00\rangle+|11\rangle\big)\cdot\big(\langle 00|+\langle 11|\big) \Big)=\\
=Tr_B\Big(\frac{1}{2}\big( |00\rangle\langle00|+|00\rangle\langle11|+|11\rangle\langle00|+|11\rangle\langle11|\big)\Big)=\frac{1}{2}\big( |0\rangle\langle0|+|1\rangle\langle1|\big) \; ,
\end{split}
\end{equation}
the entanglement entropy is given as:
\begin{equation}
E\big( |\Phi^+\rangle\langle\Phi^+|\big)=-\frac{1}{2}log\big(\frac{1}{2}\big) -\frac{1}{2}log\big(\frac{1}{2}\big)=-log\big(\frac{1}{2}\big)=log (2) =1\; .
\end{equation}

\begin{figure}[t!]
\centering
\includegraphics[width=0.4\linewidth]{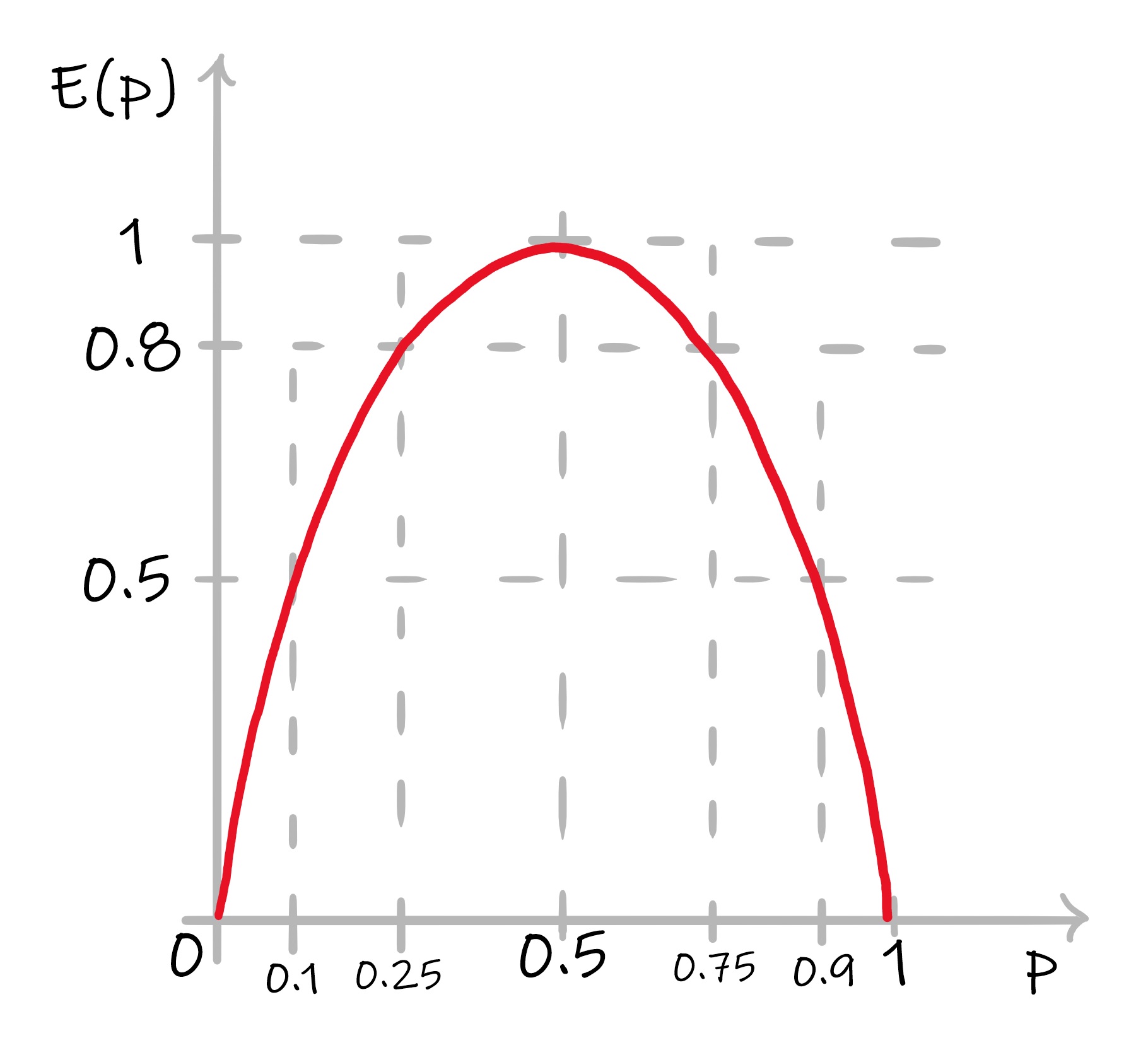}
\caption{The entanglement entropy for a qubit state $|\phi\rangle = \sqrt{p}|00\rangle+\sqrt{1-p}|11\rangle$ as a function of the probability $p$.}
\label{pic:qubent}
\end{figure}

For generalization of the calculations above, we can find the entanglement entropy for qubit state with the similar form:
\begin{equation}
|\phi\rangle= \sqrt{p}|00\rangle+\sqrt{1-p}|11\ \; ,
\end{equation} 
where $p$ is the probability that the qubit is found in the $|00\rangle$ state. Then, the trace over $B$ system is equal to:
\begin{equation}
Tr_B\big(|\phi\rangle\langle\phi|\big)=p|0\rangle\langle0|+(1-p)|1\rangle\langle1| \; .
\end{equation}
Thus, the entanglement entropy is given as:
\begin{equation}
E\big( |\Phi^+\rangle\langle\Phi^+|\big)=-p log (p) - (1-p)log(1-p) \;.
\end{equation}
As it can be seen, the entanglement entropy is the function of the probability amplitude $p$. Fig.~\ref{pic:qubent}. What is important, for $p=1/2$ the entropy achieves its maximum. It proves that the state $|\Phi^+\rangle$ is maximally entangled \cite{4H2007}. Similar calculations can be done for other Bell states. \\

It is noteworthy to mention here another useful but indirect method of measuring the strength of an entanglement -- \textbf{Bell state measurement}. Its description can be found in \textit{Section \ref{sec:ch2-bsm}}. \\

\hsection{Bell inequalities}
\hnote{Bell inequalities}
\label{sec:notes-BellIneq}

As a result of the inconsistency in the view on measurement in quantum mechanics among the scientific community, known as \textit{EPR paradox} \cite{EPR1935, EPR-resp, Sch1935},  Bell's theorem on this subject appeared \cite{Bell1964}. A further analysis of entanglement states led to the formulation of inequalities, presently known as \textbf{Bell inequalities}. Although these inequalities belong to a quite general class of inequalities of probability theory, genuinely described by George Boole in 1862, were reinvented almost century later and reformulated several times. The most famous examples of Bell-like inequalities were derived by John Clauser, Michael Horne, Abner Shimony, and Richard Holt (\textit{CHSH inequality}) \cite{Clauser1978, Hess2016}, and  Anthony J. Leggett and Anupam Garg  (\textit{Leggett–Garg inequality}) \cite{Leggett1985, Hess2016}. \\

In order to trace the reasoning leading to the formulation of Bell-like inequalities, let us make an assumption that two particles $A$ and $B$ coming from the same source can be characterized by the same "hidden parameter" $\lambda$, where $\lambda \in \Lambda$. Then, the density distribution $\rho(\lambda)$ fully defines the values of the measurements that are done on the state of these particles denoted as $A_a(\lambda)$ and $B_b(\lambda)$ for particles $A$ and $B$, respectively. To simplify the calculations, the measurement results for either particle suppose to take a $\pm1$ value. Then, the correlation function between the measurements is defined by:
\begin{equation}
C(a,b)=\int_{\Lambda}A_a(\lambda)B_b(\lambda)\rho(\lambda)d\lambda \;.
\end{equation}
Now, let us take a closer look at the expression $C(a,b)-C(a,b')$ (to shorten formulas, the dependency on the parameter $\lambda$ is omitted in the notation):
\begin{equation}
\begin{split}
C(a,b)-C(a,b')&= \int_{\Lambda}\big(A_a B_b-A_a B_{b'}\big)\rho(\lambda)d\lambda=\\
&= \int_{\Lambda}\big(A_a B_b - A_a B_{b'}\pm A_aB_bA_{a'}B_{b'}\mp A_aB_bA_{a'}B_{b'}\big)\rho(\lambda)d\lambda=\\
&=\int_{\Lambda}\Big(A_aB_b\big(1\pm A_{a'}B_{b'}\big)-A_aB_{b'}\big(1\pm A_{a'}B_{b}\big)\Big)\rho(\lambda)d\lambda=\\
&=\int_{\Lambda}A_a \Big(B_b\big(1\pm A_{a'}B_{b'}\big)-B_{b'}\big(1\pm A_{a'}B_{b}\big)\Big)\rho(\lambda)d\lambda \;.
\end{split}
\end{equation} 

Then, using the inequality for the integral of $f(x)$ function, $|\int f(x)dx|\leq\int|f(x)|dx$ and the fact that $|A_a|=|B_b|=1$:
\begin{equation}
\begin{split}
\big|C(a,b)-C(a,b')\big| &\leq \int_{\Lambda}| A_a| \Big|B_b\big(1\pm A_{a'}B_{b'}\big)-B_{b'}\big(1\pm A_{a'}B_{b}\big)\Big|\rho(\lambda)d\lambda \\
&\leq  \int_{\Lambda} \Big||B_b|\big(1\pm A_{a'}B_{b'}\big)+|B_{b'}|\big(1\pm A_{a'}B_{b}\big)\Big|\rho(\lambda)d\lambda =\\
&= \int_{\Lambda}\big|2\pm\big(A_{a'}B_{b'}+A_{a'}B_b\big)\big|\rho(\lambda)d\lambda= \\
&=2\pm\big( C(a',b')+C(a',b)\big)
\end{split}
\end{equation} 
Thus, after proper signs reconciliation, we get the inequality:
\begin{equation}
-2\leq S\leq 2 \;,
\end{equation}
where $S= C(a,b)+C(a,b')+C(a',b)-C(a',b')$ is Bell parameter \cite{Franson1989}. This particular form, $|S|\leq 2$, is called \textbf{CHSH inequality}. \\

\begin{figure}[ht!]
\centering
\includegraphics[width=0.9\linewidth]{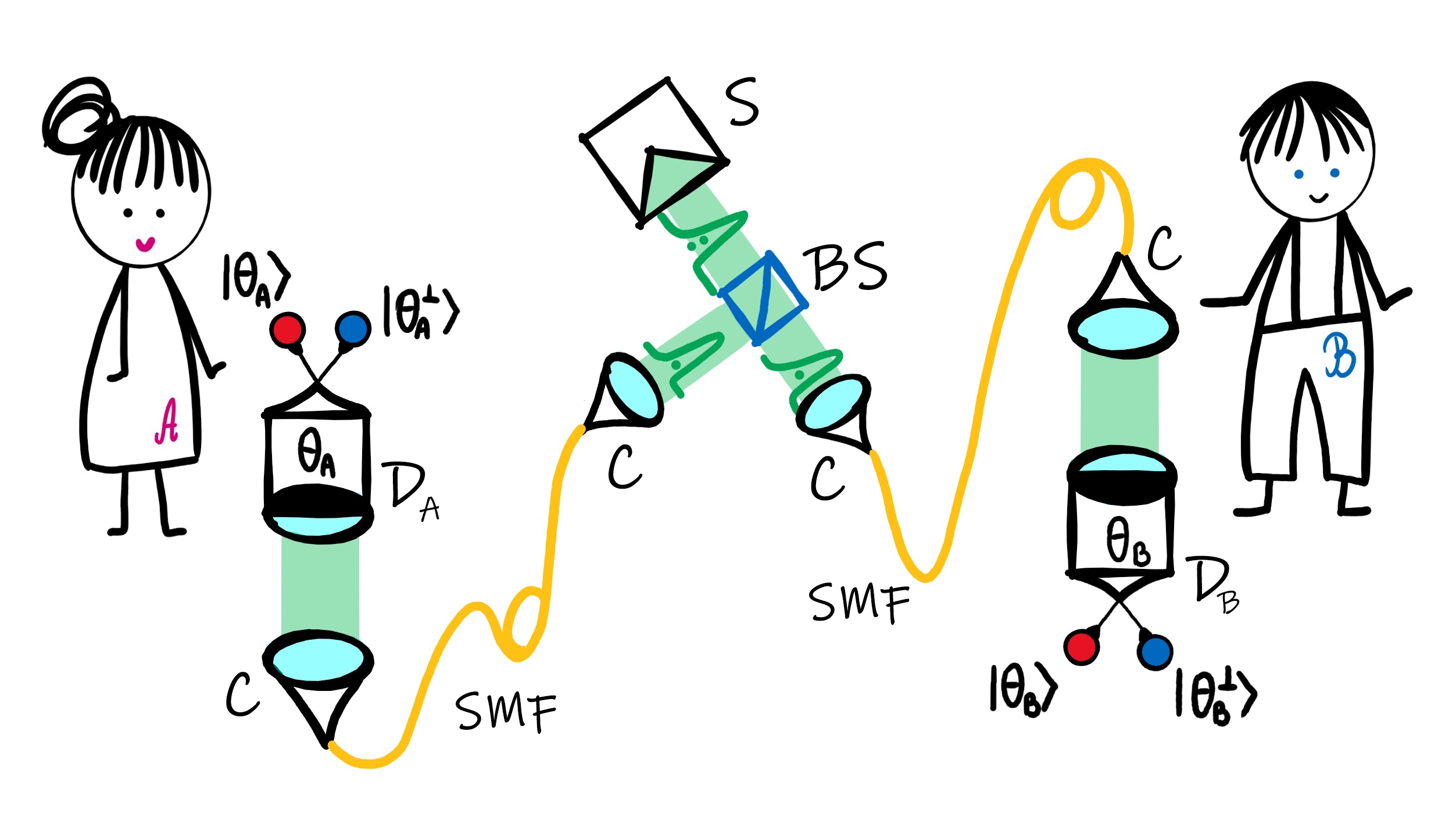}
\caption{Experimental setup scheme for Bell parameter measurement. A single-photon source generates entangled qubit, that are sent to Alice and Bob. Each of them has the detector, which detection depends on arbitrary parameter $\theta_A$ or $\theta_B$ for Alice and Bob respectively. Thus the projection onto orthogonal states ${|\theta_A\rangle,|\theta_A^{\perp}\rangle}$ and ${|\theta_B\rangle,|\theta_B^{\perp}\rangle}$ is measured. Symbols: S -- entangled qubits source, BS -- beam splitter, C -- coupler, SMF -- single mode fiber, D$_A$, D$_B$ -- detectors for Alice and Bob.}
\label{pic:CHSHset}
\end{figure}

The CHSH and others Bell-like inequalities were widely tested experimentally \cite{Brunner2014, Kupczynski2020} through the Bell-state measurement experiments (see \textit{Section \ref{sec:ch2-bsm}}). 
Typically, when pohotonic qubits are investigated, the experimental setup is similar to those presented schematically in Fig.~\ref{pic:CHSHset}. The correlation $C(a,b)$ is estimated by collecting proper coincidence counts for each selected value of $a$ and $b$. \\

Let us assume that the source produces entangled-qubits in  $|\Psi^{-}\rangle$ Bell state. Then, the measurement operator is defined as $|\phi\rangle\langle\phi|$, where $|\phi\rangle=|\theta_A\rangle|\theta_B\rangle$ and, for $i=\{A,B\}$, $\; |\theta_i\rangle=sin\frac{\theta_i}{2}|0\rangle+cos\frac{\theta_i}{2}|1\rangle$ . \\
The correlation of a measurements is determined rather instinctively by probability of coincidence counts $p(\theta_A,\theta_B)$ from specific detection ports:
\begin{equation}
C(\theta_A,\theta_B)=p(\theta_A,\theta_B)+ p(\theta_A^{\perp},\theta_B^{\perp}) - p(\theta_A^{\perp},\theta_B) -p(\theta_A,\theta_B^{\perp}) \; .
\end{equation}
Such probabilities may be calculated as:
\begin{equation}
\begin{split}
p(\theta_A,\theta_B)&=\langle\Psi^{-}|\phi\rangle\langle\phi|\Psi^{-}\rangle= |\langle \phi|\Psi^{-}\rangle|^2=|\langle\theta_A\theta_B|\Psi^{-}\rangle|^2= \\
&=\big|\big(sin\frac{\theta_A}{2} sin\frac{\theta_B}{2}\langle 00|+cos\frac{\theta_A}{2} sin\frac{\theta_B}{2}\langle 01|+sin\frac{\theta_A}{2} cos\frac{\theta_B}{2}\langle 10|+\\
&+cos\frac{\theta_A}{2} cos\frac{\theta_B}{2}\langle 11|\big)|\cdot\frac{1}{2}\big( |01\rangle-|10\rangle \big)\big|^2= \\
&=\big|\frac{1}{2} \big( cos \frac{\theta_A}{2} sin \frac{\theta_B}{2}- sin \frac{\theta_A}{2}  cos\frac{\theta_B}{2}\big)\big|^2= \\
&=\frac{1}{2}sin^2\big( \frac{\theta_A-\theta_B}{2} \big) \; .
\end{split}
\end{equation}
Then, the probabilities for different sets of specific detection angles can be computed, for example $\theta_A=0, \theta_{A'}=\pi/2, \theta_B=-\frac{3\pi}{4}, \theta_{B'}=\frac{3\pi}{4}$, minding that $\theta^{\perp} =\theta-\pi$, and later, proper correlations of measurements. We get $C(0,-\frac{3\pi}{4})=\frac{1}{\sqrt{2}}$, $C(0,\frac{3\pi}{4})=\frac{1}{\sqrt{2}}$, $C(\frac{\pi}{2},-\frac{3\pi}{4})=\frac{1}{\sqrt{2}}$, and $C(\frac{\pi}{2},\frac{3\pi}{4})=-\frac{1}{\sqrt{2}}$; thus, the Bell parameter $S$ is equal to $S=2\sqrt{2}$. \\

It should be emphasized here, that since the Bell parameter value for Bell state is equal to $S=2\sqrt{2}$ \cite{Brunner2014, Kupczynski2020}, therefore the violation of CHSH inequality is predicted by the quantum mechanics itself. Since at the beginning of whole derivation, the assumption of independence of the two measurements were taken, resulting in multiplying the separate probabilities to obtain the joint probabilities of pair as an outcome for any "hidden variable" $\lambda$, the violation of CHSH inequality proves that quantum mechanics and hidden variables theory does not get along each other. \\

\hsection{Polarization of a photon}
\hnote{Polarization of a photon}
\label{sec:ch1-pol}

\hsubsection{Monochromatic plane wave}

In classical optics, propagation of electromagnetic light in vacuum is described by \textit{Maxwell's equations}, which, in their simplest form, for the linear isotropic medium without any free electric charges, may be presented as:
\begin{equation}
\bigg\{
\begin{matrix}
\bigtriangledown^2\vec{E}=\mu_0\epsilon_0\frac{\partial^2\vec{E}}{\partial t^2},\\
\bigtriangledown^2\vec{B}=\mu_0\epsilon_0\frac{\partial^2\vec{B}}{\partial t^2},
\end{matrix}
\label{eq:maxeq}
\end{equation}
where $\mu_0$ and $\epsilon_0$ are the magnetic and electric constants \cite{book:Hecht}. As it is taught at schools and universities, one of the solutions is a monochromatic plane wave propagating in $z$ direction:
\begin{equation}
\bigg\{
\begin{matrix}
\vec{E}(z,t)=\hat{x}E_0 exp\big(i(kz-\omega t) \big),\\
\vec{B}(z,t)=\hat{y}B_0 exp\big(i(kz-\omega t) \big),
\end{matrix}
\label{eq:horpol}
\end{equation}
where $\omega$ is the angular frequency and $k=2\pi/\lambda$ is the absolute value of the wavevector. \\
This solution of Eq.~\ref{eq:maxeq}, a monochromatic wave with the elctromagnetic field oscillating in $x-z$ plane and magnetic field oscillating in $y-z$ plane, is \textbf{horizontally polarized}. Similarly, a mirror solution of  Eq.~\ref{eq:maxeq}, with the electromagnetic field oscillating in $y-z$ plane and magnetic field oscillating in $x-z$ plane, is \textbf{vertically polarized}.  \\


\begin{figure}[t!]
\centering
\includegraphics[width=0.65\linewidth]{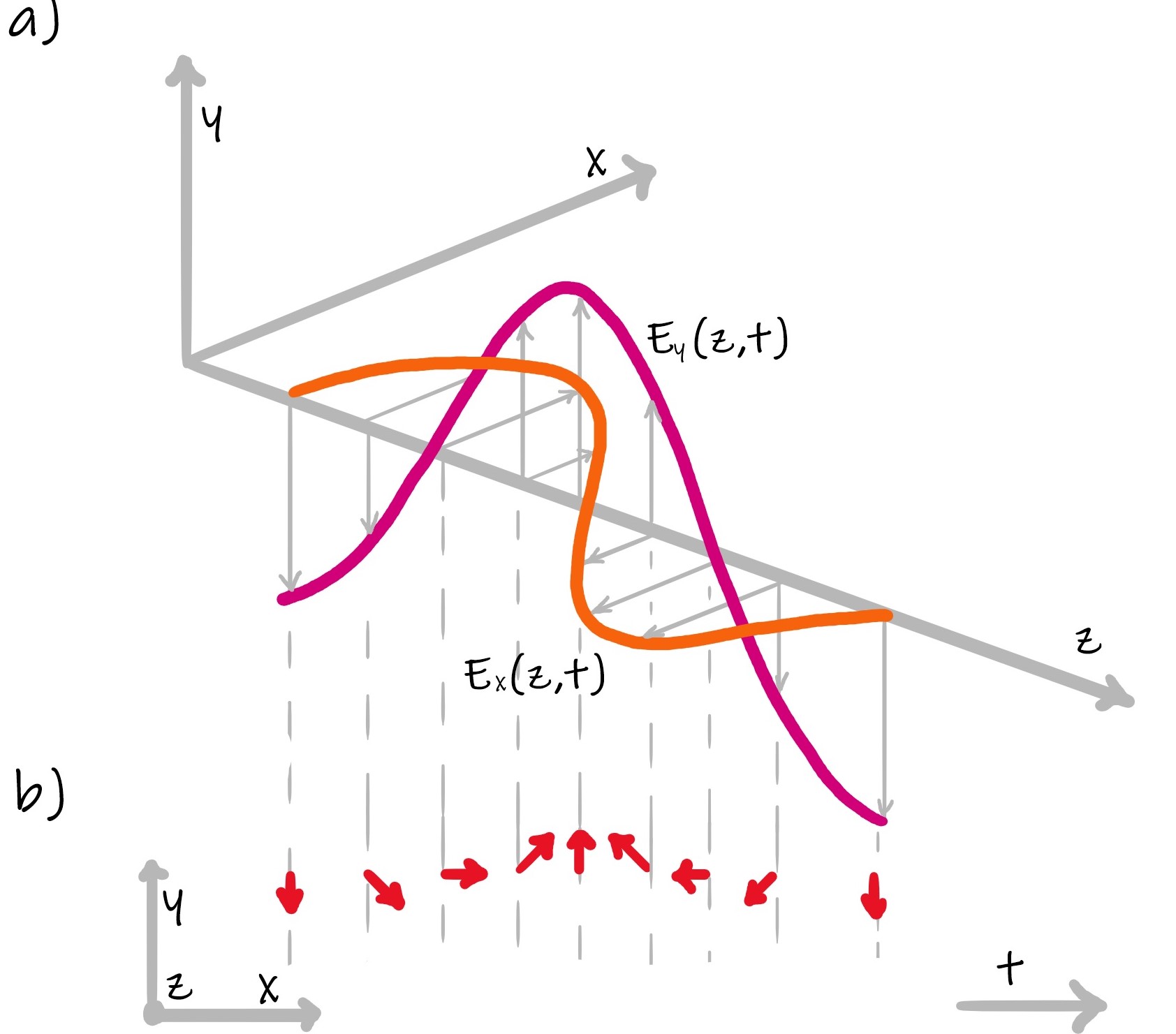}
\caption{Polarization of monochromatic plane wave for phase shift parameter $\theta=\pi/2$. \newline a) Electric fields $E_x(z,t)$, $E_y(z,t)$. b) The resultant electric field vector.}
\label{pic:polgen}
\end{figure}

In general, any superposition of these solutions, when properly normalized, is also a valid solution of Eq.~\ref{eq:maxeq}. Therefore, the most complex form, describing the electric field with a phase shift $\theta$ between vertical and horizontal component and parameter $\phi$, which relates to their relative amplitudes, takes the following form:
\begin{equation}
\vec{E}(z,t)=(cos\phi \, \hat{x}+sin\phi \, e^{i\theta}\hat{y})\,E_0exp\big(i(kz-\omega t)\big).\\
\label{eq:phshift}
\end{equation}
It should be stressed here that the $\phi$ parameter specifies the inclination angle of the polarization axis on the $x-y$ plane.\\

Fig.~\ref{pic:polgen} a) presents the monochromatic plane wave with phase shift $\theta=\pi/2$ and $\phi=\pi/4$. Let us follow the resultant vector of electric field over time, which is Fig.~\ref{pic:polgen} b). \\

\begin{figure}[b!]
\centering
\includegraphics[width=\linewidth]{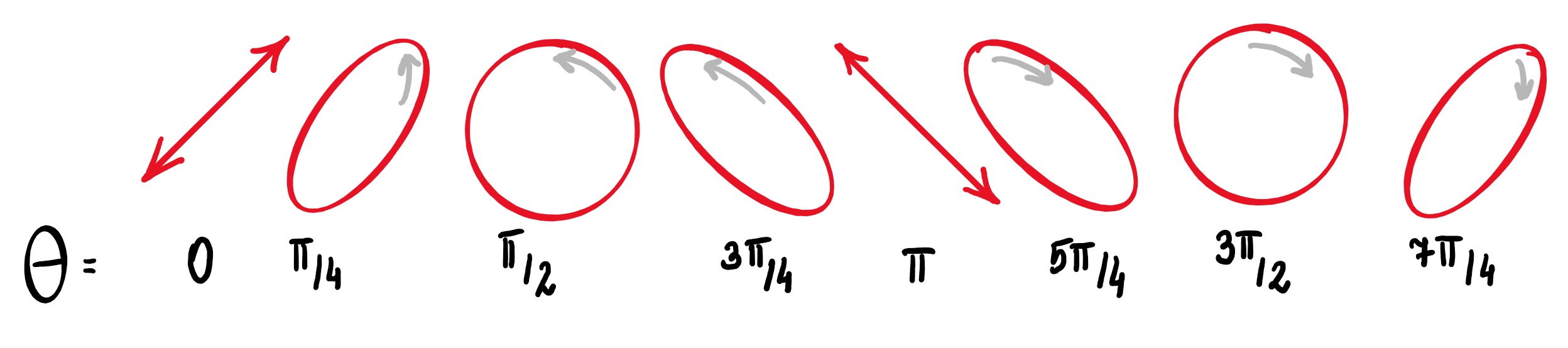}
\caption{Types of elliptical polarization states determined by phase shift parameter $\theta$. \newline The amplitudes of $E_x$ and $E_y$ are equal ($\phi=\pi/4$).}
\label{pic:polel}
\end{figure}

As it can be seen, the electric field vector traces out a circle. Thus, this type of a wave is referred to as \textbf{circularly polarized}. This is a special case of elliptical polarization, which is determined by phase shift parameter $\theta$ in Eq.~\ref{eq:phshift}. Other types of elliptical polarizations together with their corresponding $\theta$ parameters are presented in Fig.~\ref{pic:polel}. \\

In general, polarization of the plane wave and its change during the propagation through given physical system may be mathematically described as a column vectors and $2x2$ complex matrices, called \textbf{Jones vectors} and \textbf{Jones matrices} \cite{Jones1941}. This method, which is extensively used in classical optics, makes it easier to consider the change of polarization states. \\

The most general form of a \textbf{Jones vector}, related to the polarization state, which arises directly from Eq.~\ref{eq:phshift}, takes the form:
\begin{equation}
\hat{\epsilon}_{\theta} =\Big( \begin{matrix}cos\phi\\sin\phi \, e^{i\theta}\end{matrix} \Big)\:. 
\end{equation}

From this equation, specific Jones vectors can be computed. Vectors for the typical polarization states, in the case of equal amplitude components of electromagnetic field ($sin\phi=cos\phi=1$) are collected in Tab. \ref{tab:polvec}. \\

\begin{table}[t!]
\centering
\begin{tabular}{c|c|c}
\textbf{polarization state} & \textbf{Jones vector} & \textbf{symbol }\\
\hline \hline
linear polarization along x-axis & $\Big(\begin{matrix}1\\ 0\end{matrix}\Big)$&$|\lrarr\rangle$\\
linear polarization along y-axis & $\Big(\begin{matrix}0\\ 1\end{matrix}\Big)$&$|\udarr\rangle$\\
$+45^o$ linear polarization & $\frac{1}{\sqrt{2}}\Big(\begin{matrix}1\\ 1\end{matrix}\Big)$&$|\ruarr\rangle$\\
$-45^o$ linear polarization & $\frac{1}{\sqrt{2}}\Big(\begin{matrix}1\\ -1\end{matrix}\Big)$&$|\rdarr\rangle$\\
right-hand circular polarization & $\frac{1}{\sqrt{2}}\Big(\begin{matrix}1\\ i\end{matrix}\Big)$&$|\circleft\rangle$\\
left-hand circular polarization & $\frac{1}{\sqrt{2}}\Big(\begin{matrix}1\\ -i\end{matrix}\Big)$&$|\circright\rangle$\\
\hline
\end{tabular}
\caption{Jones vector for polarization states and its symbols.}
\label{tab:polvec}
\end{table}

Next, in order to illustrate the use of the \textbf{Jones matrix}, an example of a \textbf{wave plate} may be studied. It is an optical element that has different refraction index along orthogonal polarization axes. This property, called \textbf{birefringence}, allows one to retard the phase of one component (on the \textit{slow axis} of the incident electric field by exactly the $\theta$), relatively to the field on the orthogonal axis (called \textit{fast axis}). For instance, a \textbf{half-wave plate}, with $\theta=\pi$, changes the polarization states to $\frac{1}{\sqrt{2}}(\hat{x}+(e^{i\pi}\hat{y})=\frac{1}{\sqrt{2}}(\hat{x}-\hat{y})$, if the initial polarization state is $\frac{1}{\sqrt{2}}(\hat{x}+\hat{y})$. Thus, the properly orientated half-wave plate rotates a linear polarization by $90^o$.\\

Using \textbf{Jones calculus}, the polarization states before and after the half-wave plate may be easily written as:
\begin{equation}
\begin{split}
\text{before wave-plate:}\quad \hat{\epsilon}_{+45^o} =\frac{1}{\sqrt{2}}\Big(\begin{matrix}1\\1\end{matrix} \Big)\:, \\
\text{after wave plate:}\quad \hat{\epsilon}_{-45^o} = \frac{1}{\sqrt{2}}\Big(\begin{matrix}1\\-1\end{matrix} \Big)\:, 
\end{split}
\label{eq:wavevecpol}
\end{equation}
if the orthonormal basis with linearly polarized waves along \\ \textit{x-axis} $\hat{\epsilon}_{x} =\Big(\begin{matrix}1\\0\end{matrix} \Big)$ and \textit{y-axis} $\hat{\epsilon}_{y} =\Big(\begin{matrix}0\\1\end{matrix} \Big)$ is used. \\

Now, on the basis of the vectors from Eq.~\ref{eq:wavevecpol}, it can be seen that the matrix that transforms $\hat{\epsilon}_{+45^o}$ polarization state into $\hat{\epsilon}_{-45^o}$ is given by:
\begin{equation}
J_{HWP}=\Big(\begin{matrix}
1&0\\
0&-1\\
\end{matrix}\Big) \; .
\end{equation}
All polarization changes caused by any optical elements may be described with Jones matrices. Some of these matrices are listed in Tab. \ref{tab:polmat}.\\

\begin{table}[h!]
\begin{tabular}{c|c}
\textbf{optical element} & \textbf{Jones matrix} \\
\hline \hline
half-wave plate ($\theta=\pi$)& $\Big(\begin{matrix}
1&0\\
0&-1\\
\end{matrix}\Big)$\\
quarter-wave plate ($\theta=\frac{\pi}{2}$) &$
\Big(\begin{matrix}
1&0\\
0&-i\\
\end{matrix}\Big)$\\
linear polarizer aligned to x-axis& $\Big(\begin{matrix}
1&0\\
0&0\\
\end{matrix}\Big)$\\
half-wave plate at arbitrary angle $\phi$ & $\Big(\begin{matrix}
cos2\phi&sin2\phi\\
sin2\phi&-cos2\phi\\
\end{matrix}\Big)$\\
quarter-wave plate at arbitrary angle $\phi$ &$
\bigg(\begin{matrix}
cos^2\phi-i\cdot sin^2\phi&\frac{1}{2}\big(sin2\phi+isin2\phi\big)\\
\frac{1}{2}\big(sin2\phi+isin2\phi\big)&sin^2\phi-i\cdot cos^2\phi\\
\end{matrix}\bigg)$\\
linear polarizer at arbitrary angle $\phi$&$\Big(\begin{matrix}
cos^2\phi&cos\phi sin\phi\\
cos\phi sin\phi&sin^2\phi\\
\end{matrix}\Big)$\\
\hline
\end{tabular}
\caption{Jones matrices for polarization optics elements.}
\label{tab:polmat}
\end{table}

It should be mentioned here that the matrices listed in Tab. \ref{tab:polmat} relate to optical elements, which are aligned to the $x-y$ plane. When tilted, matrices should be transformed using rotation matrix $R(\Phi)= \Big(\begin{matrix}
cos\phi&-sin\phi\\
sin\phi&cos\phi\\
\end{matrix}\Big)$:
\begin{equation}
J'(\Phi)=R(\Phi)JR^{-1}(\Phi)\; .
\end{equation}

\hsubsection{Single photon}
\label{sec:polstatesing}

Let us take a look at the $+45^o$ polarized wave propagating through the vertical polarizer. The electric field before and after polarizer are given by:
\begin{equation}
\begin{split}
\text{before polarizer:}\quad \vec{E}_{in}(z,t)=\frac{1}{\sqrt{2}}(\hat{x}+\hat{y})E_0exp\big(i(kz=\omega t)\big)\:, \\
\text{after polarizer:}\quad\vec{E}_{out}(z,t)=\frac{1}{\sqrt{2}}\hat{y}E_0exp\big(i(kz=\omega t)\big) \:.
\end{split}
\label{eq:polvecpol}
\end{equation}
The change of intensity, related to the square of field amplitude may be calculated as a $I_{out}/I_{in}$ ratio, which gives $\frac{I_{out}}{I_{in}}=\frac{1}{2}$. Thus, the output intensity is 2 times smaller than the input one. \\

Now, suppose that the input intensity is so small that it contains a single photon. This simple calculation leads to the conclusion that on the output only the fraction of a photon should be endured. This, of course, due to the \textit{quantization of a field}, is not possible. \\

However, the classical-to-quantum-optics correspondence allows to use similar equations, remembering that instead of intensity of light, operations on probability amplitude must be done.\\ Thus, Jones calculus may be applied at the quantum mechanical level, in the sense that now Jones vectors describe the polarization states of single photons. In fact, Tab. \ref{tab:polvec} lists these states in the last column. \\

It is worth to notice that using  $|\leftrightarrow\rangle$ and $|\updownarrow\rangle$ as a basis states, all other polarization states may be written:
\begin{equation}
\begin{split}
|\ruarr\rangle&=\frac{1}{\sqrt{2}}\big(|\leftrightarrow\,\rangle+|\updownarrow\,\rangle \big) \; ,\\
|\rdarr\rangle&=\frac{1}{\sqrt{2}}\big(|\leftrightarrow\,\rangle-|\updownarrow\,\rangle \big) \; ,\\
|\circright\rangle&=\frac{1}{\sqrt{2}}\big(|\leftrightarrow\,\rangle+i|\updownarrow\,\rangle \big) \; ,\\
|\circleft\rangle&=\frac{1}{\sqrt{2}}\big(|\leftrightarrow\,\rangle-i|\updownarrow\,\rangle \big)\; .
\end{split}
\end{equation}

Using basis set listed above, the quantum mechanical representations of optical elements may also be examined. It may be easily shown that matrices obtained will be identical to those presented in Tab. \ref{tab:polmat}. \\

\hsubsection{Polarization states on Bloch sphere}
\label{sec:ch5-pauli}
Another useful representation of polarization states of single photons is an assignment of vectors on Bloch sphere to them. 

\begin{figure}[t!]
\centering
\includegraphics[width=0.7\linewidth]{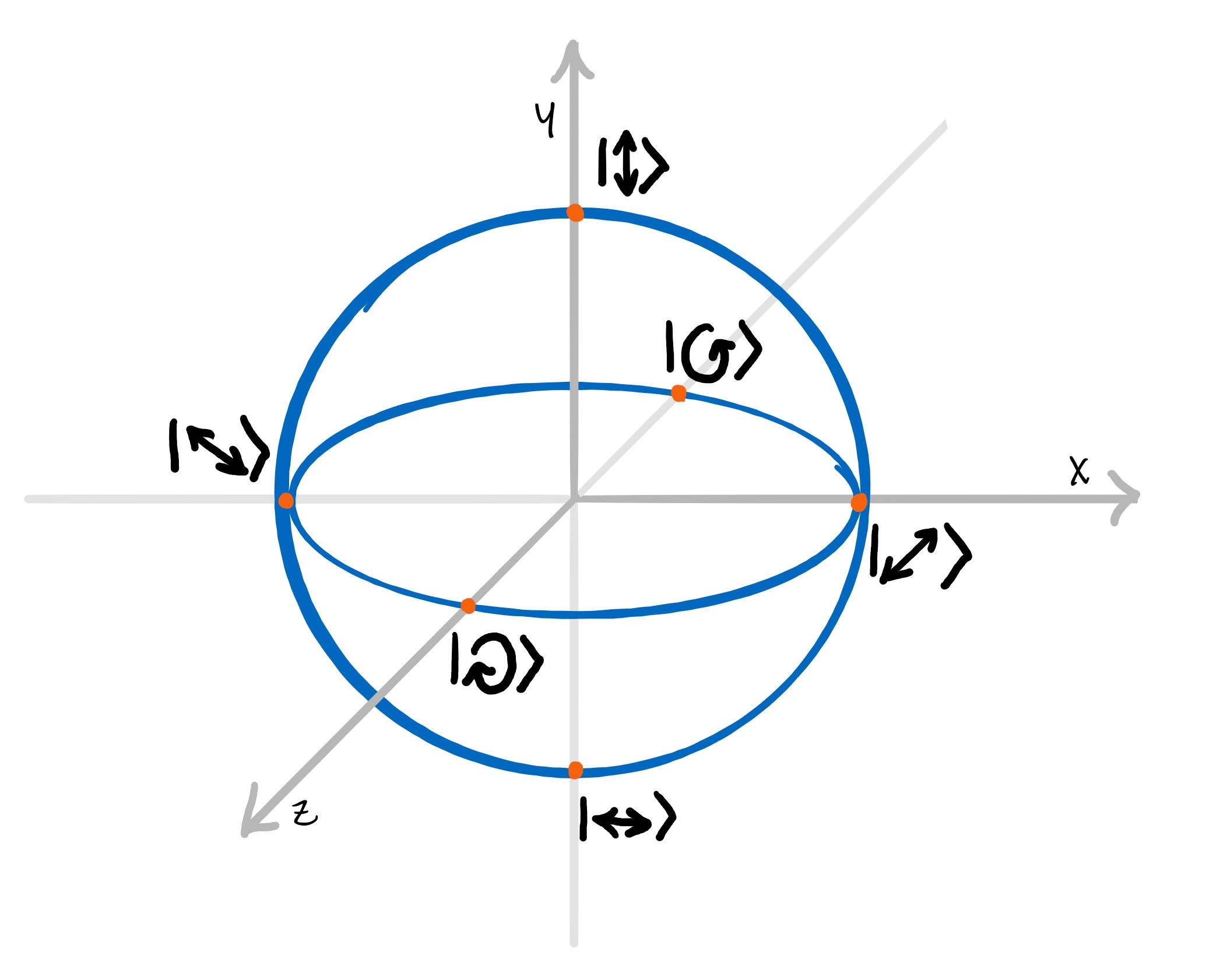}
\caption{Polarization states of single photons on Bloch sphere.}
\label{pic:Blochsphere}
\end{figure}

Mapping between quantum states and Bloch vectors may be found using the density operator of a state and its expansion using \textbf{Pauli matrices}:
\begin{equation}
\sigma_x=\Big(\begin{matrix}
0 & 1\\
1 & 0 \\
\end{matrix} \Big), \;
\sigma_y=\Big(\begin{matrix}
0 & -i\\
i & 0 \\
\end{matrix} \Big),\;
\sigma_z=\Big(\begin{matrix}
1 & 0\\
0 & -1 \\
\end{matrix} \Big).\\
\label{eq:paulimat}
\end{equation}
Then, this expansion follows:
\begin{equation}
\hat{\rho}=\frac{1}{2}\big(\mathds{1}+\vec{b}\cdot\vec{\sigma}\big) \; ,
\label{eq:blochvec}
\end{equation}
where $\vec{b}$ stands for Bloch vector and $\vec{\sigma}$ vector is defined as $\vec{\sigma}=(\sigma_x,\sigma_y,\sigma_z)$. \\

In the case described in \textit{Section \ref{sec:polstatesing}}, the orthonormal basis states $|\leftrightarrow\rangle$ and $|\updownarrow\rangle$ are at antipodal points of the Bloch sphere. The representation of the rest is calculated using Eq.~\ref{eq:blochvec}. All are shown in Fig.~\ref{pic:Blochsphere}. \\

It should be marked here, that elliptical polarizations shown in Fig.~\ref{pic:polel} are located on the equator of the Bloch sphere.\\

As it can be seen, due to the geometrical location of polarization states, the transformations between vectors may be associated with their rotations on the Bloch sphere. This property makes the Bloch representation incredibly useful. \\

\hsection{No-cloning theorem}
\hnote{No-cloning theorem}
\label{sec:notes-noclo}
As it was mentioned in \textit{Section \ref{sec:ch1-quantuse}}, a quantum operator, which could copy a state, would be very useful. Not only for the eavesdroppers. For example, the problem of distinguishing two unknown states $|\phi\rangle$ and $|\xi\rangle$ may be considered. If these states are orthogonal, they are perfectly distinguishable. If they are not ortogonal, states may be distinguished with some constant probability of error. However, if copying an unknown state was possible, the optimal measurement could be performed many times, and thus, the probability of error would be reduced to an arbitrarily small value. \\

The \textbf{no-cloning theorem} says that copying an unknown state is not physically possible. With a single unitary operator, only sets of orthogonal states can be copied. \\

Let us look closer at the "mathematics of copying". Firstly, we have two states -- an unknown state $|\psi\rangle$ and an "empty state register" $|0\rangle$ (register of the "cloning machine"). The first state should be copied into the "empty" one. Thus, the cloning operator $U_{CL}$ should work like this: 

\begin{equation}
U_{CL}\Big(|\psi\rangle \otimes |0\rangle \Big) \rightarrow |\psi\rangle \otimes |\psi\rangle = |\psi\rangle|\psi\rangle =|\psi\psi\rangle\; .
\end{equation}
Now, taking an unknown state $|\phi\rangle=\alpha|0\rangle+\beta|1\rangle$, we should get:
\begin{equation}
\begin{split}
U_{CL}\big(|\psi\rangle|0\rangle \big)\,\rightarrow|\,\psi\rangle|\psi\rangle = \big(\alpha|0\rangle+\beta|1\rangle \big) \big(\alpha|0\rangle+\beta|1\rangle \big)=\\
= \alpha^2|00\rangle+\alpha\beta|01\rangle+\beta\alpha|10\rangle+\beta^2|11\rangle \, .
\label{eq:non1}
\end{split}
\end{equation}
On the other side, due to the linearity of quantum-mechanical operators, we get:
\begin{equation}
U_{CL}\Big(\big(\alpha|0\rangle+\beta|1\rangle \big)|0\rangle\Big)\,\rightarrow\, \alpha|00\rangle+\beta|11\rangle \,.
\label{eq:non2}
\end{equation}

These two copying results, Eq. \ref{eq:non1} and Eq. \ref{eq:non2}, should be equal, but in the second one, we do not have the cross-terms. Thus, we can deduce that the only possibility, when they are identical is either $\alpha=0,\, \beta=1$ or $\alpha=1,\, \beta=0$. It means that the states must be orthogonal. Otherwise, such a unitary operator $U_{CL}$ cannot exist. \\

\hsection{Time-bin and phase encoding}
\hnote{Time-bin and phase encoding}
\label{sec:notes-tbin-ph}
\textbf{Time-bin encoding} is a method used for encryption of information in quantum communication. Due to its nature, this method is very robust against decoherence \cite{Franson1989}. Therefore it is a perfect method for transmitting information through fiber links, resulting in higher fidelities in quantum communication protocols. \\

\begin{figure}[b!]
\centering
\includegraphics[width=0.65\linewidth]{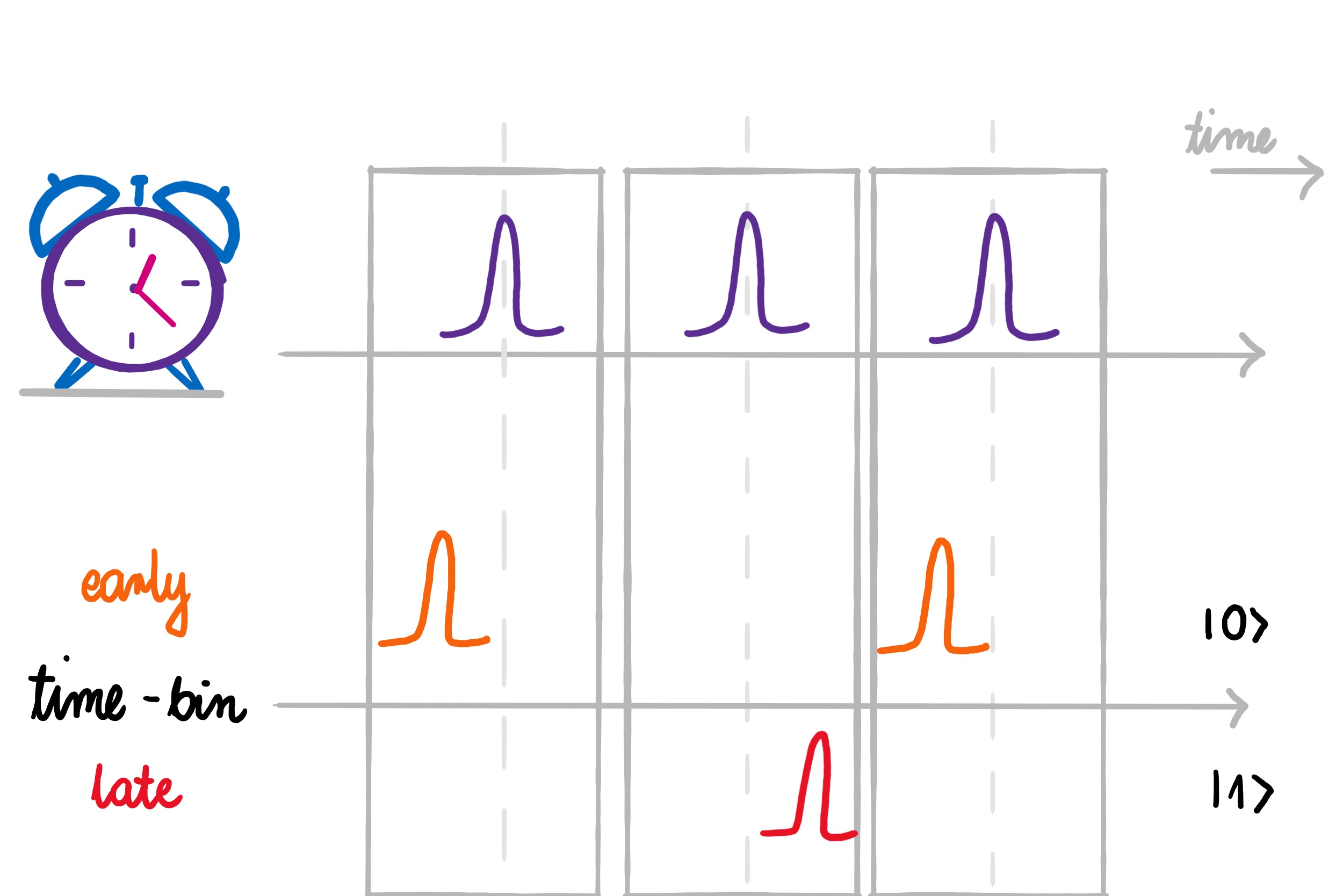}
\caption{Scheme of time-bin encoding. Early- and late- time-bin in relation to the external clock.}
\label{pic:tbe}
\end{figure}

Time-bin-encoded qubits may be simply created by single photons traveling paths of different lengths. As a result, photons' arrival time, compared to external clock, differ and may be assigned as the early- and late-time-bin (see Figure \ref{pic:tbe}). \\

\begin{figure}[t!]
\centering
\includegraphics[width=0.7\linewidth]{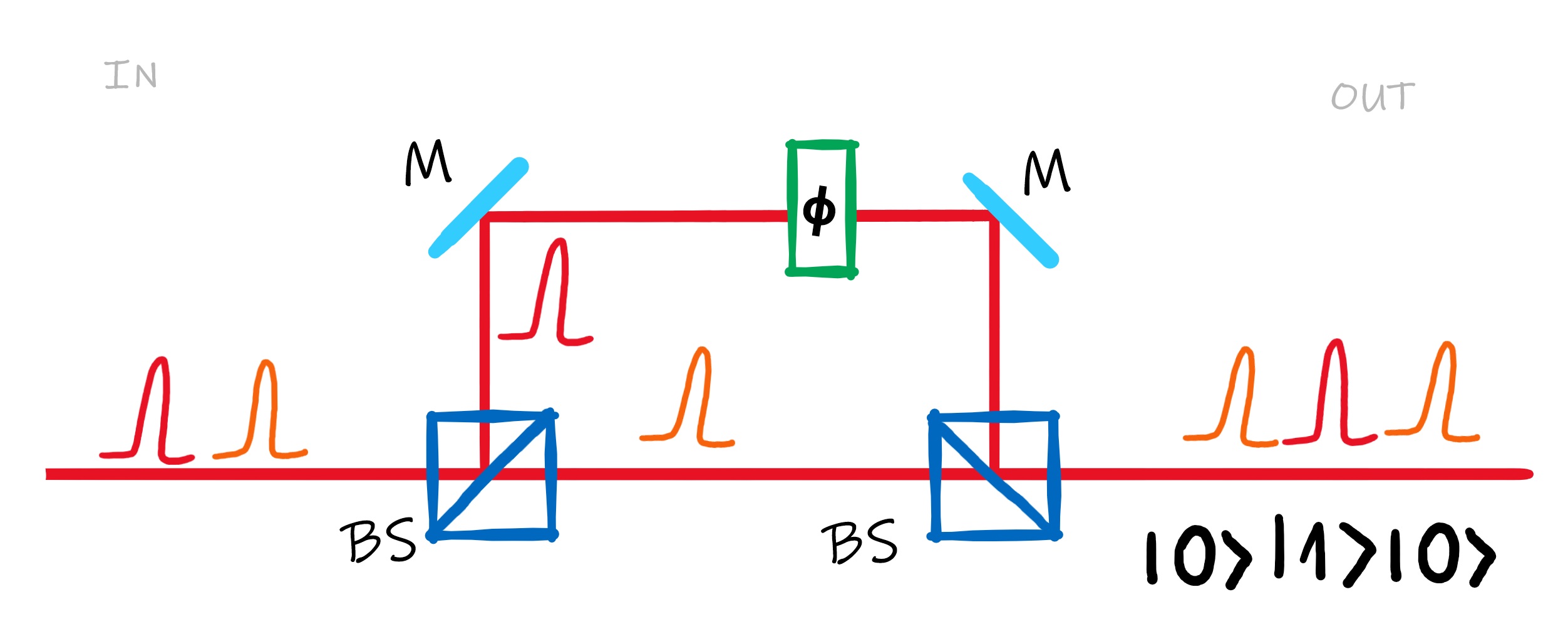}
\caption{The idea of time-bin encoding, utilizing single photons. Optical paths are marked with red line. Optical elements: BS  --  beam splitter, M -- mirror, $\phi$ -- total phase change on longer path.}
\label{pic:tbin}
\end{figure}

An example of a typical experimental realization, shown in Fig.~\ref{pic:tbin}, involves unbalanced Mach-Zehnder interferometer. The path difference, along with the additional setup element inserted in the longer path, introduces an additional phase $\phi$, which usually can be controlled in an experiment. Assuming that state of a photon on a short path may be written as $|0\rangle$ and on a long path as $|1\rangle$, its superposition takes the form:
$\frac{1}{\sqrt{2}}(|0\rangle+e^{i\phi}|1\rangle)$. \\

Measurement in the $\{|0\rangle,|1\rangle\}$ basis is done by measuring photons' arrival time. Therefore, the difference in path length must be longer than the coherence length of the photon. Otherwise, photons may not be distinguished correctly. \\
Sometimes, this type of encoding is called \textbf{phase encoding}. It is often used in quantum-key distribution protocols \cite{Inoue2002,Stucki2005}.\\

\hsection{Bell-state measurement}
\hnote{Bell-state measurement}
\label{sec:ch2-bsm}

\begin{figure}[b!]
\centering
\includegraphics[width=0.9\linewidth]{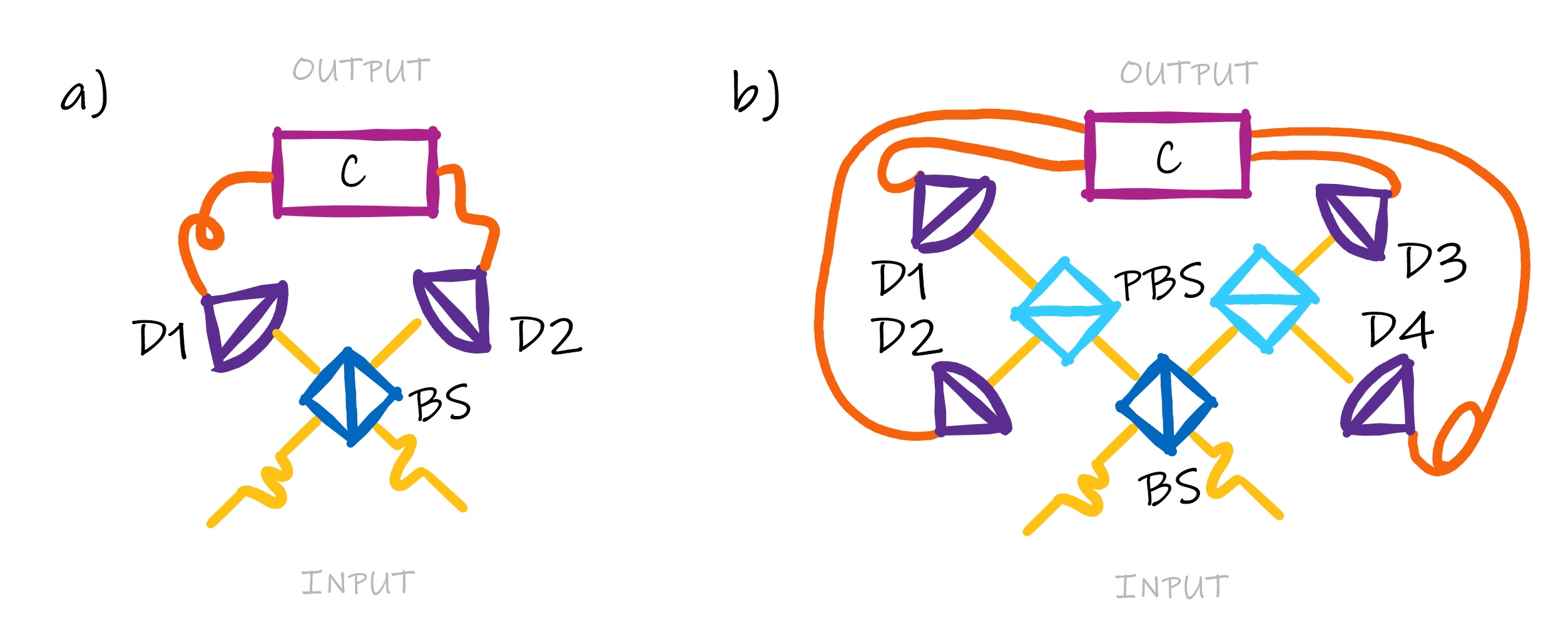}
\caption{The Bell-state measurement setup for photonic qubits, in case a) time-bin qubits, b) polarization qubits. Symbols: BS -- beam splitter, PBS -- polarizing beam splitter, D1, D2, D3, D4 -- detectors, C -- coincidence correlator.}
\label{pic:BSMset}
\end{figure}

The Bell-state measurement (BSM) \cite{Lutkenhaus1998} is an important tool for many quantum communication protocols, such as \textit{quantum teleportation} \cite{Kim2000} or \textit{superdense coding} \cite{Williams2016}.
Its purpose is the projection of the two qubit states onto one of the maximally entangled Bell states (see \textit{Section \ref{sec:ch1-qubit}}, Eq.~\ref{eq:bellbasis}). The working principle is based on \textbf{Hong-Ou-Mandel interference effect} \cite{HOM1987} (see \textit{Section \ref{sec:notes-HOM}}). Thus, in case of photonic qubits, it is usually implemented using a beam splitter \cite{Walborn2002}.\\
The BSM setups are presented in Fig.~ \ref{pic:BSMset}.\\

\begin{figure}[t!]
\centering
\includegraphics[width=0.7\linewidth]{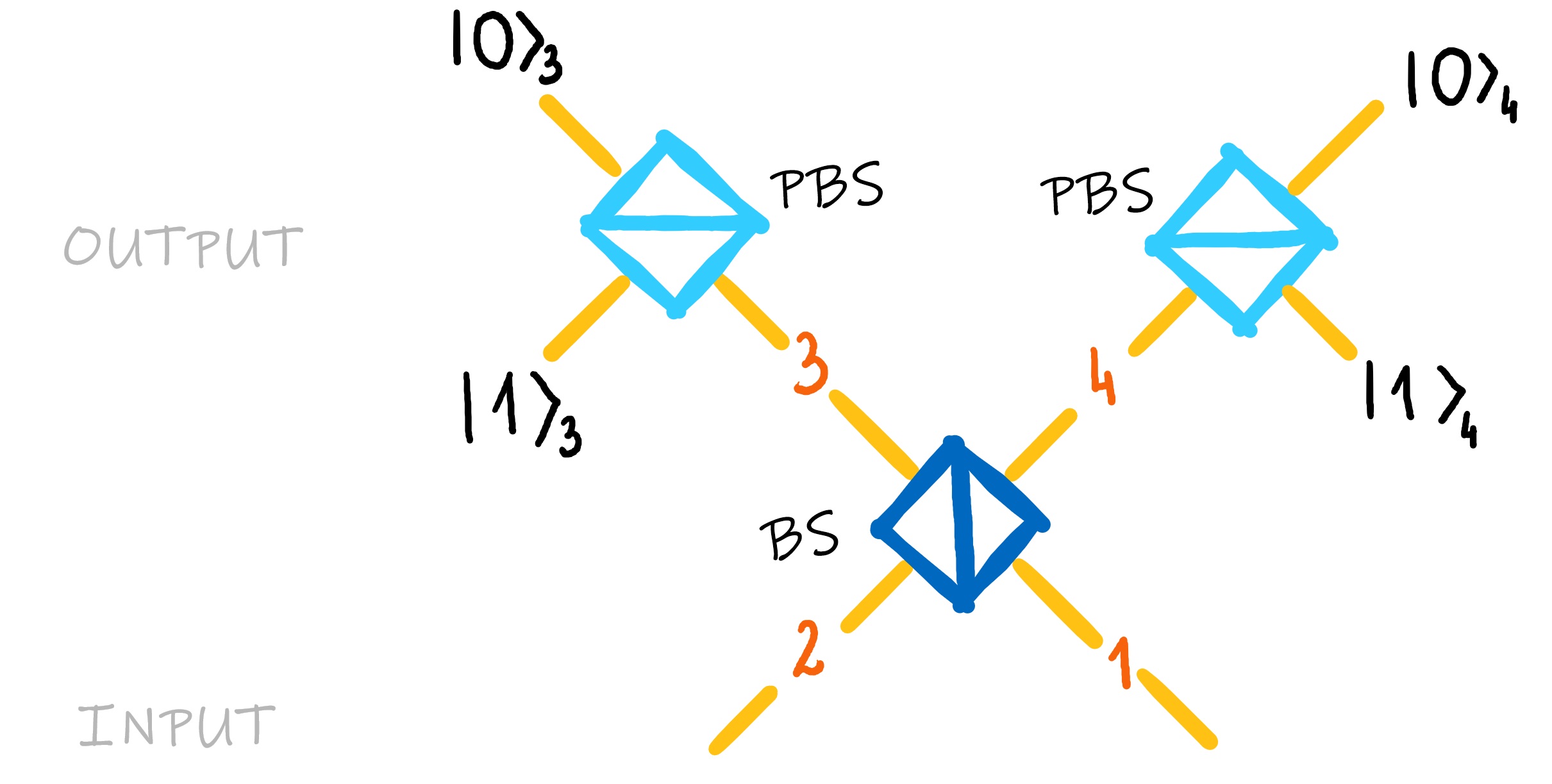}
\caption{Beam splitter input (1,2) and output (3, 4) ports. Photon states for specific outputs are marked.}
\label{pic:BSM-phset}
\end{figure}

Let us take a closer look at the BSM measurement for photonic qubits. The input and output photon spatial modes are labelled as $1$, $2$ and $3$, $4$, respectively. For each output the photon states are marked, as presented in Fig.~ \ref{pic:BSM-phset}.

A beam splitter transforms the Bell states (Eq.~\ref{eq:bellbasis}) in the following way:
\begin{equation}
\begin{split}
|\Phi^{+}\rangle_{12}&=\frac{1}{\sqrt{2}}\big( |00\rangle_{12}+|11\rangle_{12}\big)\\
\rightarrow \,& \frac{i}{2}\big( |00\rangle_{33}+|00\rangle_{44}+ |11\rangle_{33}+|11\rangle_{44}\big) =|\Phi^{+}\rangle_{34}
\, ,\\
|\Phi^{-}\rangle_{12}&=\frac{1}{\sqrt{2}}\big( |00\rangle_{12}+|11\rangle_{12}\big)\\
\rightarrow \,& \frac{i}{2}\big( |00\rangle_{33}+|00\rangle_{44} - |11\rangle_{33}-|11\rangle_{44}\big) =|\Phi^{-}\rangle_{34}
\, ,\\
|\Psi^{+}\rangle_{12}&=\frac{1}{\sqrt{2}}\big( |01\rangle_{12}+|10\rangle_{12}\big)\\
\rightarrow \,& \frac{i}{\sqrt{2}}\big(|01\rangle_{34}+|01\rangle_{44} \big) = |\Psi^{+}\rangle_{34}
\, ,\\
|\Psi^{-}\rangle_{12}&=\frac{1}{\sqrt{2}}\big( |01\rangle_{12}-|10\rangle_{12}\big)\\
\rightarrow \,& \frac{1}{\sqrt{2}}\big(|10\rangle_{33}-|01\rangle_{44} \big) = |\Psi^{-}\rangle_{34}
\, .
\end{split}
\label{eq:bellstateBSM}
\end{equation} 

From these equations, it can be seen that the photon coincidences in the same beam splitter output port manifest the projection onto $|\Psi^+\rangle$ state, whereas the coincidences in different output port evince the projection onto $|\Psi^-\rangle$. It should be also noted that the photon coincidences in the same output do not distinguish between $|\Phi^+\rangle$ and $\Phi^-\rangle$ states. \\

Summing up, setups presented in Fig.~ \ref{pic:BSMset} allow to distinguish two out of four Bell states. Although it had been proven that with linear optical elements the efficiency of the BSM is limited to $50$\% \cite{Lutkenhaus1998}, the nonlinear interactions can be used for distinguishing between all four Bell states \cite{Kim2000, Williams2016}.\\

\begin{figure}[b!]
\centering
\includegraphics[width=0.75\linewidth]{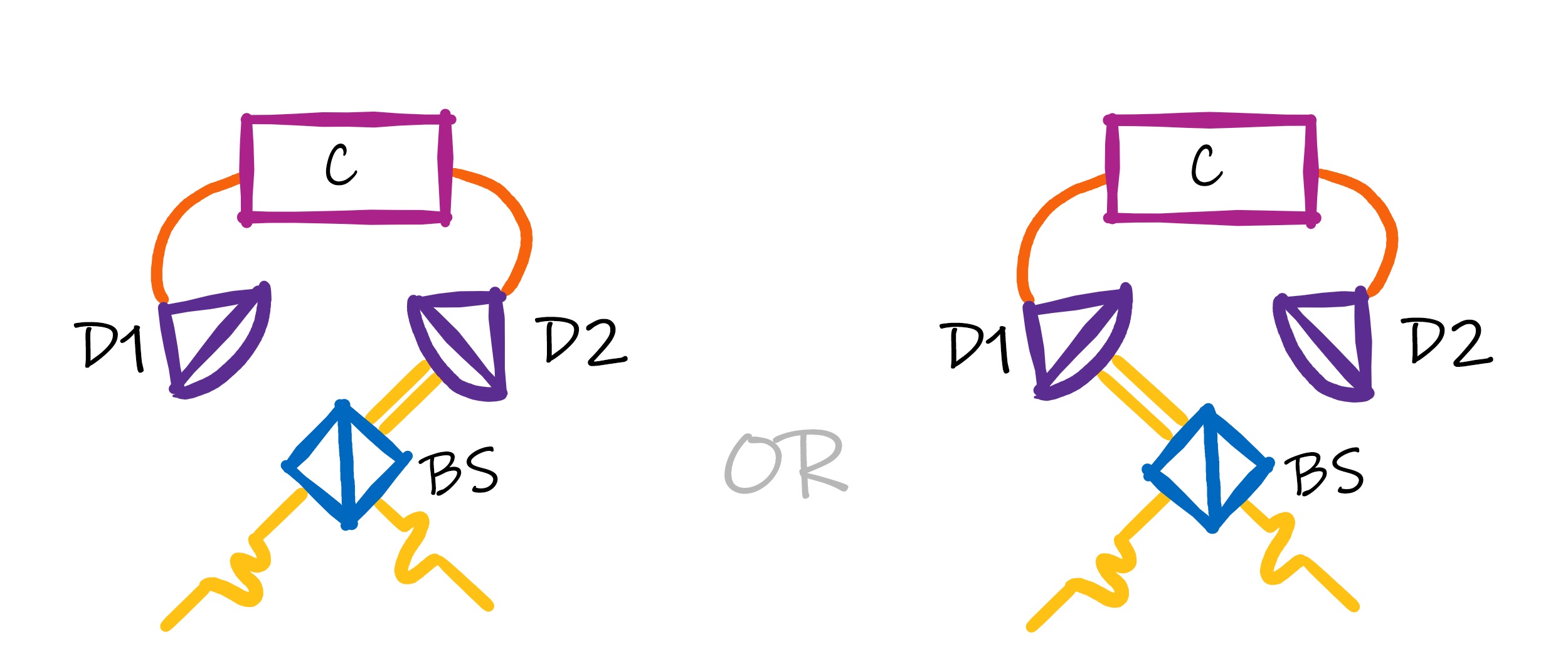}
\caption{The Hong-Ou-Mandel interference effect. Symbols: BS -- beamsplitter, D1,D2 -- detectors, C -- correlator.}
\label{pic:HOMbs}
\end{figure}

\hsection{Hong-Ou-Mandel interference}
\hnote{Hong-Ou-Mandel interference}
\label{sec:notes-HOM}
The \textbf{Hong-Ou-Mandel interference} effect \cite{HOM1987} was firstly demonstrated in $1987$ by three physicists Chung Ki Hong, Zhe Yu and Leonard Mandel. It is a two-photon interference effect, which occurs when two identical photons enter a beamsplitter, one in each port (see Fig.~ \ref{pic:HOMbs}).\\

If the photons are perfectly overlapping each other inside the beamsplitter, they always exit the beamsplitter through the same output port. Since it is a $50$:$50$ beamsplitter, photons have equal chance of exiting in both output modes. \\

\begin{figure}[t!]
\centering
\includegraphics[width=0.25\linewidth]{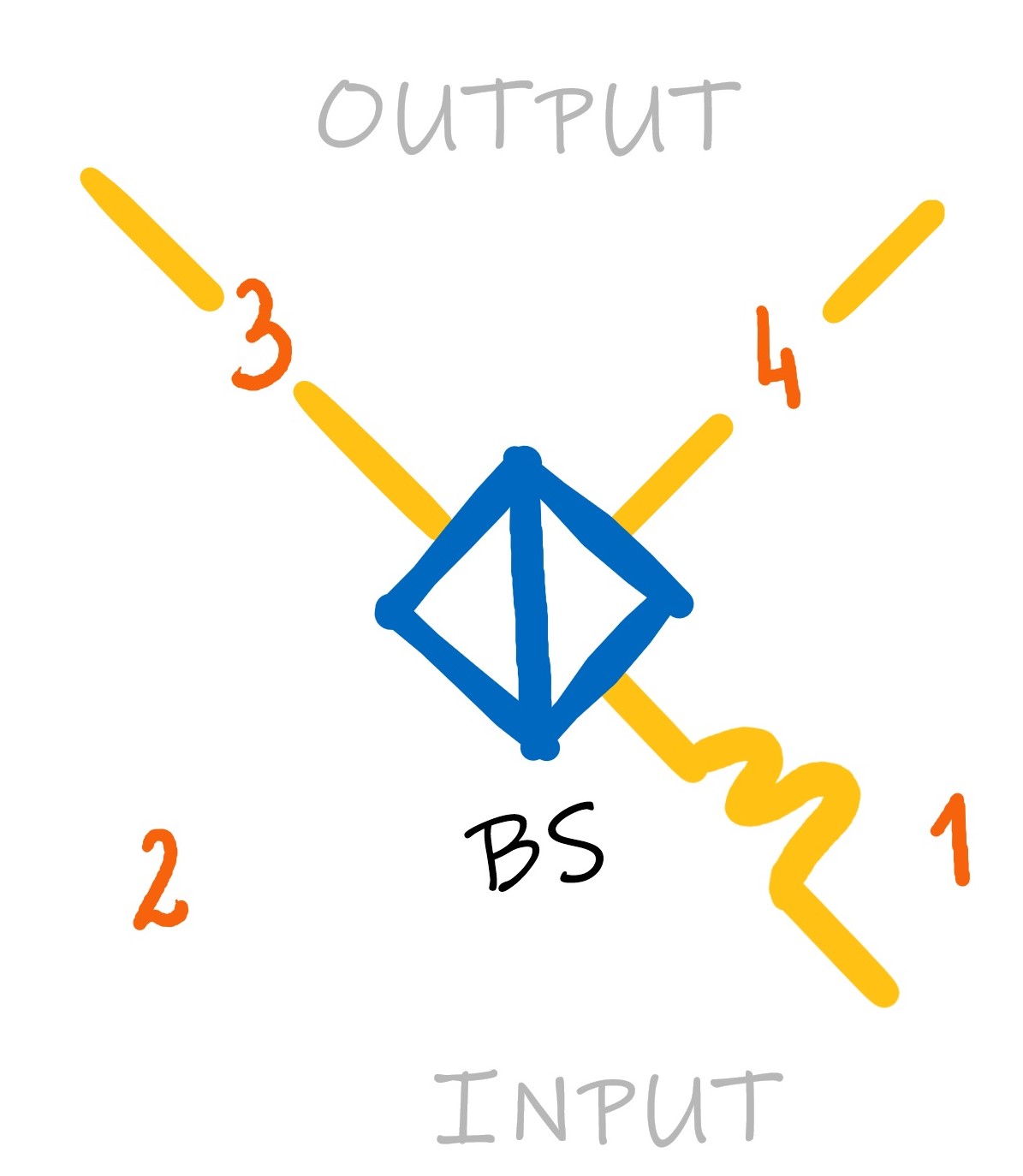}
\caption{Single photon incidents the 50:50 beamsplitter. All symbols are described in the text.}
\label{pic:BS}
\end{figure}

Let us first consider the transformation of the single photon state by the symmetrical $50$:$50$ beamsplitter (see Fig.~ \ref{pic:BS}). The photon enters the beamsplitter in port $1$, and leaves it through ports $3$ or $4$ with equal probability. Then, the Fock state can be written as:
\begin{equation}
|1\rangle_1|0\rangle_2\,\rightarrow \,\frac{1}{\sqrt{2}}\big(|1\rangle_3|0\rangle_3+i|0\rangle_4|1\rangle_4 \big),
\end{equation} 
where $i$ is the imaginary number that appears as a consequence of the phase shift of $\pi/2$ upon a reflection. \\

Let us now assume that we have two photons enter the beamsplitter in different inputs, then:
\begin{equation}
\begin{split}
|1\rangle_1|1\rangle_2\,\rightarrow& \,\frac{1}{2}\big(|1\rangle_3|0\rangle_3+i|0\rangle_4|1\rangle_4 \big)\big(i|1\rangle_3|0\rangle_3+|0\rangle_4|1\rangle_4 \big)=\\
&=\frac{i}{2}\big(|2\rangle_3|0\rangle_4+i|0\rangle_3|2\rangle_4 \big) \, .
\end{split}
\end{equation}

\begin{figure}[b!]
\centering
\includegraphics[width=0.68\linewidth]{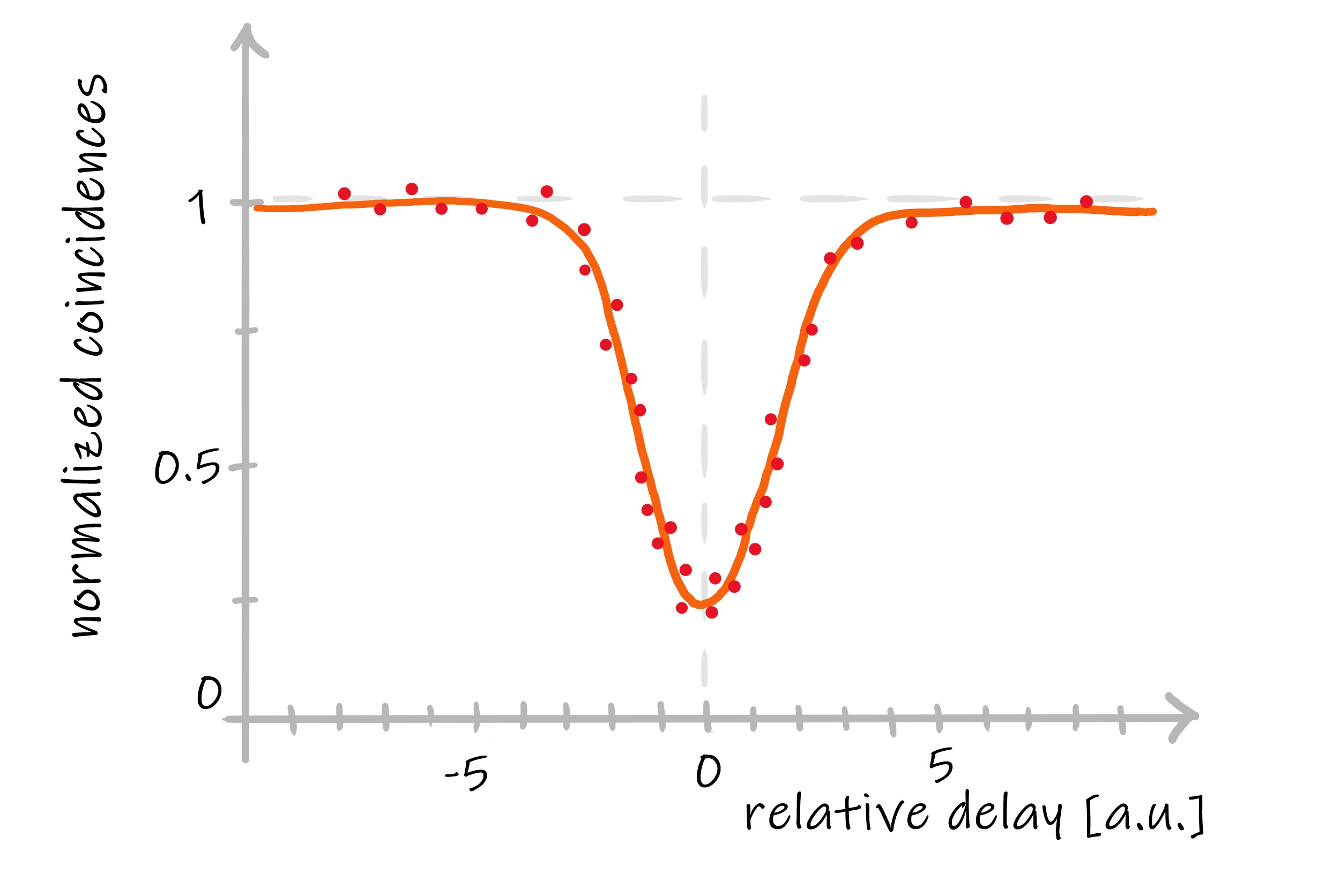}
\caption{The HOM interference dip. Normalized coincidence counts versus relative delay between photons.}
\label{pic:HOM-res}
\end{figure}

As it can be noticed, the mathematics of interference prohibits photons from taking separate paths, but only if photons are indistinguishable.  Depending on the distinguishability between two photons (derived from arrival time, photon polarization or energy), the number of photon coincidences detected at the output of the beamsplitter changes from maximal $C_{max}$ to minimal $C_{min}$ value. Since these parameters may slightly differ between successive photons, some of them may overlap their partners better than others. Therefore, the number of photon coincidences detected between them will be in the range $[C_{min}, C_{max}]$. This creates a dip in coincidence histogram, which is showed in Fig.~ \ref{pic:HOM-res}. The width of the dip corresponds to the temporal width of photon pulses.\\

The visibility $V$ of a HOM dip is given by:
\begin{equation}
V=\frac{C_{max}-C_{min}}{C_{max}}\,.
\end{equation}

Assuming conditions such as perfect source of single indistinguishable photons and the lack of noise in the experimental setup, the visibility should reach $V=1$. Despite that, in more realistic, coherent sources, which create a Poissonian photon number distribution (i.e., faint laser), the theoretical maximum of the visibility is $V=1/2$ \cite{Chen2016}, whereas in thermal sources it is equal to $V=1/3$ \cite{Jin2015}. \\
Therefore, the visibility can be used to quantify the indistinguashability of photons produced by sources with known photon number distributions. \\

It should be mentioned here that also the \textbf{reverse Hong-Ou-Mandel effect} occurs \cite{Chen2007, Chen2018}. Then, a pair of identical perfectly overlapped photons that incidents a beamsplitter (in the same input port) is separated and leaves the beamsplitter on different output ports. This effect may be used for designing quantum deterministic beamsplitter \cite{Chen2007}. \\

\hsection{Quantum logic gate}
\hnote{Quantum logic gate}
\label{sec:notes-qloggate}

\begin{figure}[b!]
\centering
\includegraphics[width=0.55\linewidth]{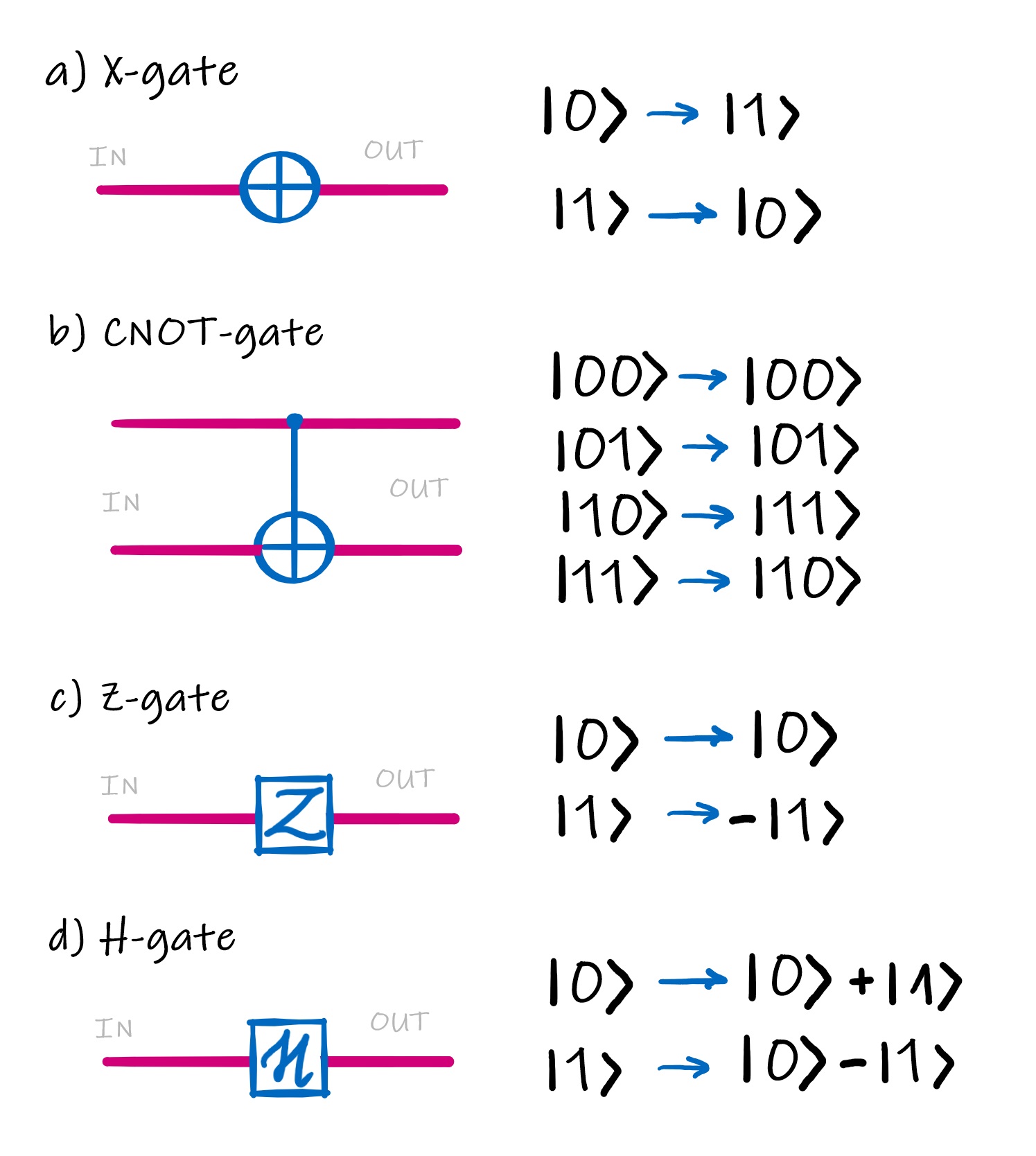}
\caption{Symbols and transformations of simple quantum-logic gates: a) X-gate, b) CNOT gate.}
\label{pic:quantloggate}
\end{figure}

A quantum logic gate or, more simply, a \textbf{quantum gate}, is any circuit that allows to change the qubit state in a desired way. Unlike their classical analogs, transformations introduced by quantum gates are reversible. \\
There are several different types of quantum gates \cite{book:nielsenchuang}. The simplest ones, which operate on qubits, are represented by $2x2$ matrices (in general $d^nxd^n$ for \textit{n-qudit state}).\\

In Bloch-sphere state representation, an operation of a quantum gate may be associated with a vector rotation of an arbitrary angle. The generators of such a transformation are Pauli matrices, which makes the mathematical calculations relatively easy. \\
As an example that will be used later, the \textit{X-gate} is presented here. It is the single-qubit equivalent of the \textit{NOT gate}. Since it maps $|0\rangle \rightarrow |1\rangle$ and $|1\rangle \rightarrow |0\rangle$, it corresponds to a rotation around the X-axis of the \textit{Bloch sphere} by $\pi$  radians and is represented by Pauli matrix $\sigma_x=\Big(\begin{matrix}
0 & 1\\
1 & 0 \\
\end{matrix} \Big)$. \\

Unfortunately, the physical realization of higher dimensional ($>1$) quantum gates is complicated \cite{Babazadeh2017}. Since qubits need to interact with each other, high nonlinearity in the optical system should be provided \cite{Azuma2008}. Theoretically, it is possible to use sets of nonlinear crystals as a quantum gate, however it was not proven experimentally \cite{Salmanpour2015}. \\

In the case of two-qubit \textbf{CNOT} gate (the quantum analog of classical \textbf{XOR}, mention earlier in \textit{Section \ref{sec:notes-otp}}), optical design of such a gate usually includes five beam splitters \cite{KLM,OBrien2004}. Despite the implementational challenges, quantum gate circuits are continuously miniaturized \cite{Zeuner2018}. This paves the way for quantum computers, which may be more convenient for future users.  \\

\hsection{Quantum error correction}
\hnote{Quantum error correction}

Error correction codes base on \textit{repetition codes} \cite{Devitt2013}. Clasically, to protect against the noisy symmetric channel, every bit of information is sent in three copies (or more -- it depends on the code). Assuming that the error probability is low, it is reasonable to posit that the correct result of the transmission is the bit which is repeated most. This is a so-called \textit{majority voting}. \\

The idea presented above is simple, but unfortunately its straightforward quantum adaptation, creation of multiple copies of unknown state $|\psi\rangle= \alpha|0\rangle+\beta|1\rangle$, is not allowed due to the \textbf{no-cloning theorem} (see \textit{Section \ref{sec:notes-noclo}}). Also, since the measurement process destroys encoded information, the desired qubits cannot be measured. \\

\begin{figure}[t!]
\centering
\includegraphics[width=0.6\linewidth]{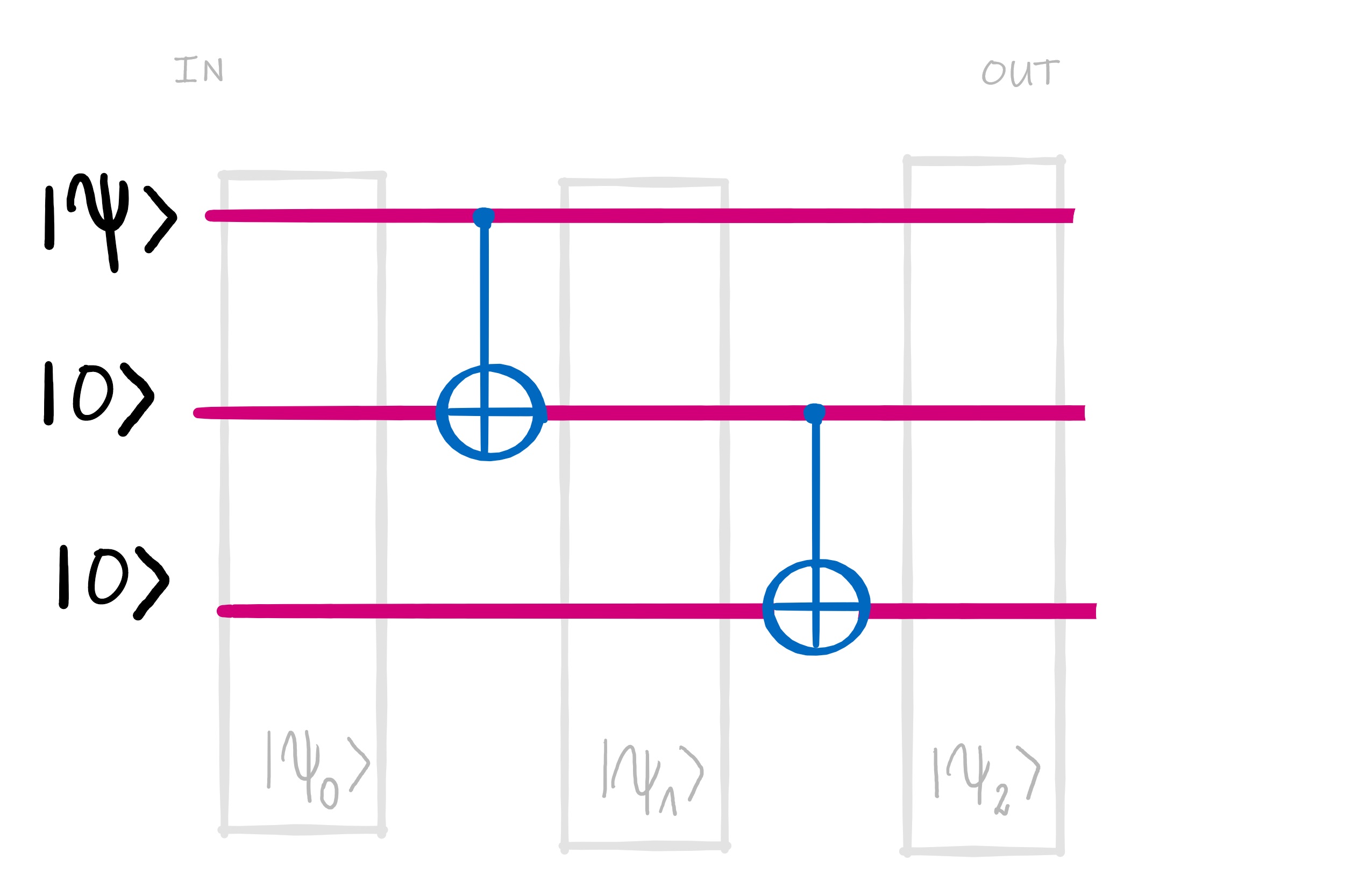}
\caption{Coding circuit into logical bits. For easier explanation, which may be found on the previous page, light grey boxes and $|\psi_0\rangle$, $|\psi_1\rangle$, $|\psi_2\rangle$ are marked. }
\label{pic:3bf}
\end{figure}

To overcome these obstacles, firstly, the logical qubits $|0\rangle_L=|000\rangle$ and $|1\rangle_L=|111\rangle$ are introduced \cite{book:nielsenchuang}. So a single qubit may be encoded in logical qubits as follows: \\
\begin{equation}
|\psi\rangle= \alpha|0\rangle+\beta|1\rangle \rightarrow \alpha|000\rangle+\beta|111\rangle = \alpha|0\rangle_L+\beta|1\rangle_L = |\psi\rangle_L. \\
\end{equation}
An example of a logic circuit, which implements the logical qubit encoding is presented in Fig.~ \ref{pic:3bf}. It consists of two CNOT gates. 

Let us follow the transformation of input qubit through this circuit:
\begin{equation}
\begin{split}
|\psi_0\rangle=\alpha|000\rangle+\beta|100\rangle \:, \\
|\psi_1\rangle=\alpha|000\rangle+\beta|110\rangle \:, \\
|\psi_2\rangle=\alpha|000\rangle+\beta|111\rangle \:. \\
\end{split}
\end{equation}
As it can be seen, thanks to the encoding circuit, a quantum version of the repetition codes may be used. \\

\begin{figure}[t!]
\centering
\includegraphics[width=0.6\linewidth]{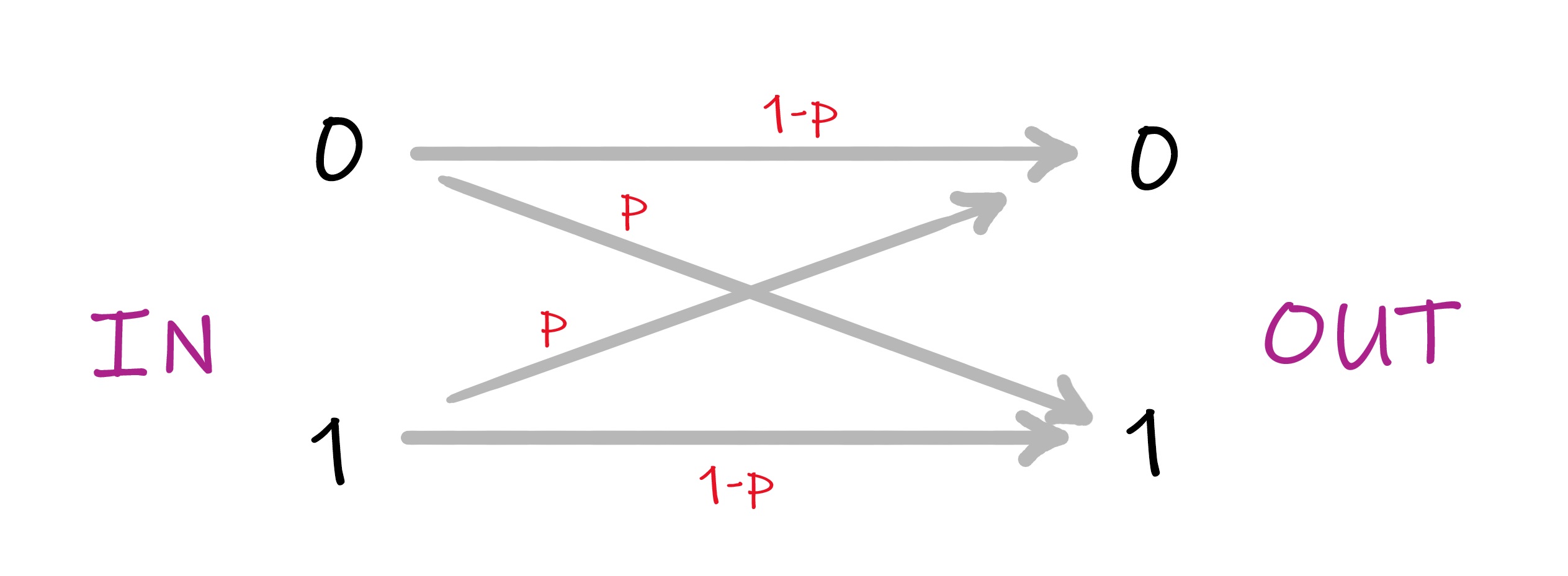}
\caption{Scheme of bit-flip error in a symmetric channel. Symbol $p$ stands for the probability of a flip.}
\label{pic:bf}
\end{figure}

\hsubsection{Bit-flip error}
\label{sec:notes-bferrcorr}
There are several ways that noisy channel may affect qubit. One of them is \textbf{bit-flip error}, which scheme is presented in Fig.~ \ref{pic:bf}. It simply reverses value of bits send through the channel. Thus, for a single qubit, it may be associated with the Pauli X-gate (see Fig.~ \ref{pic:quantloggate}). \\

It is worth noticing, that if the bit is flipped with probability \textit{p}, total probability error is equal to $p_e=p^3+3p^2(1-p)$.
It means, that in the case of 3-bit repetition codes  information sent can be recovered correctly if no more than one bit is flipped. Also, this method is efficient only if $p_e < p$. \\
 
After an imperfect transmission process which introduces errors, error syndrome diagnosis is performed. The error syndrome, a result of proper measurement, tells which bit is flipped (if any). \\ 
For the bit-flip channel and 3-bit repetition code, there are four error syndromes:
\begin{equation}
\begin{split}
\text{no error:}\quad \hat{S}_0 \equiv |000\rangle\langle 000|+ |111\rangle\langle 111| \:, \\
\text{error on 1st bit:}\quad \hat{S}_1 \equiv |100\rangle\langle 100|+ |011\rangle\langle 011| \:, \\
\text{error on 2nd bit:}\quad \hat{S}_2 \equiv |010\rangle\langle 010|+ |101\rangle\langle 101| \:, \\
\text{error on 3rd bit:}\quad \hat{S}_3 \equiv |001\rangle\langle 001|+ |110\rangle\langle 110| \:. \\
\end{split}
\label{eq:errsym}
\end{equation}
It should be noted that the state of a system is not changed after the measurement. It is equal to $\frac{\hat{S}_i}{\sqrt{s(i)}}|\psi\rangle$, where $s(i)=\langle\psi|\hat{S}_i|\psi\rangle$. Therefore, the syndrome measurement gives the information of the type of error, without revealing any information about the state itself ($\alpha$ or $\beta$ value). \\

As it may be expected, the recovery procedures depend on the error syndrome. Basically, they are flipping procedures on proper bits. \\

\begin{figure}[b!]
\centering
\includegraphics[width=0.5\linewidth]{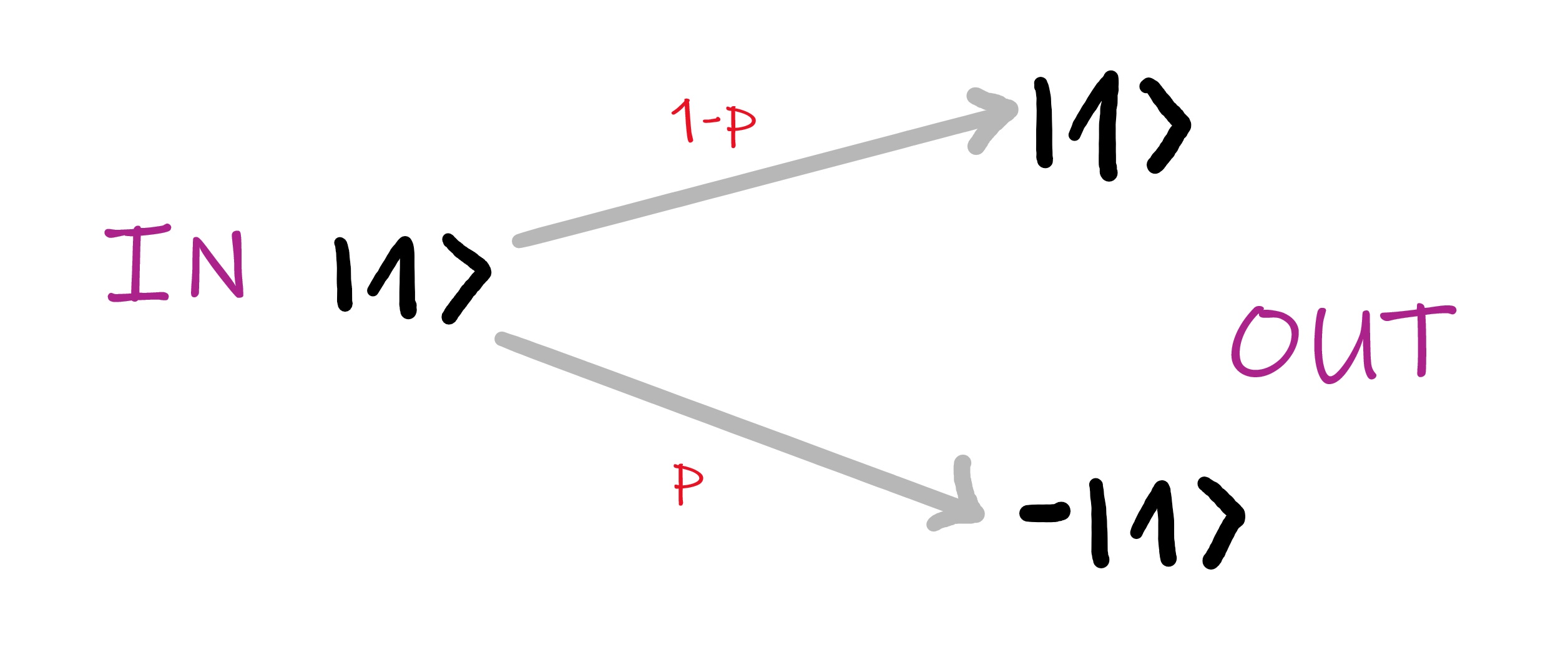}
\caption{Scheme of a sign-flip error in a symmetric channel. Symbol $p$ stands for the probability of a flip.}
\label{pic:phf}
\end{figure}

\hsubsection{Phase-flip error}
\label{sec:notes-phferrcorr}
The other way the noise may influence the qubit is the \textbf{phase-flip} (called also \textbf{sign-flip}) \textbf{error} \cite{Babar2018}. Its scheme is presented in Fig.~ \ref{pic:phf}. \\

As it can be seen, with some probability \textit{p}, putting in the input $|\psi_{IN}\rangle=\alpha|0\rangle+\beta|1\rangle$, we may get $|\psi_{OUT}\rangle=\alpha|0\rangle-\beta|1\rangle$ at the output. This type of error may be associated with transformation given by Pauli Z-gate (see Fig.~ \ref{pic:quantloggate}), which in matrix form may be written as $\hat{\sigma}_z=\Big(\begin{matrix}
1 & 0\\
0 & -1 \\
\end{matrix} \Big)$. \\ 

Unfortunately, when measuring in $\{|0\rangle,|1\rangle\}$ basis, detection of this type of error is not possible. Therefore, the basis must be  changed and here, the quantum gates come to the aid. 

It is worth to notice, that Pauli Z-gate may be transformed into X-gate, with a help of a matrix given as $\hat{\sigma}_H=\frac{1}{\sqrt{2}}\Big(\begin{matrix}
1 & 1\\
1 & -1 \\
\end{matrix} \Big)$ \, in the following way: 
\begin{equation}\hat{\sigma}_x=\hat{\sigma}_H\cdot\hat{\sigma}_Z\cdot\hat{\sigma}_H^{\dagger}.
\end{equation}

The $\hat{\sigma}_H$ matrix describes the Hadamard- or H-gate. \\

Thanks to that, the basis is changed from $\{|0\rangle,|1\rangle\}$ to $\{|+\rangle,|-\rangle\}$, where:
\begin{equation}
\begin{split}
|+\rangle=\frac{1}{\sqrt{2}}(|0\rangle+|1\rangle) \, ,\\
|-\rangle=\frac{1}{\sqrt{2}}(|0\rangle-|1\rangle) \, .\\
\end{split}
\end{equation}
In this basis, the Pauli Z-gate $\hat{\sigma}_Z$ switches between vectors:  
\begin{equation}
\begin{split}
\hat{\sigma}_Z|+\rangle=|-\rangle \, ,\\
\hat{\sigma}_Z|-\rangle=|+\rangle \, .\\
\end{split}
\end{equation}
Following that, Z-gate transformation in $\{|+\rangle,|-\rangle\}$ basis looks like the X-gate in $\{|0\rangle,|1\rangle\}$ and thus, phase-flip error becomes a bit-flip error, which is much easier to detect. \\

\begin{figure}[t!]
\centering
\includegraphics[width=0.8\linewidth]{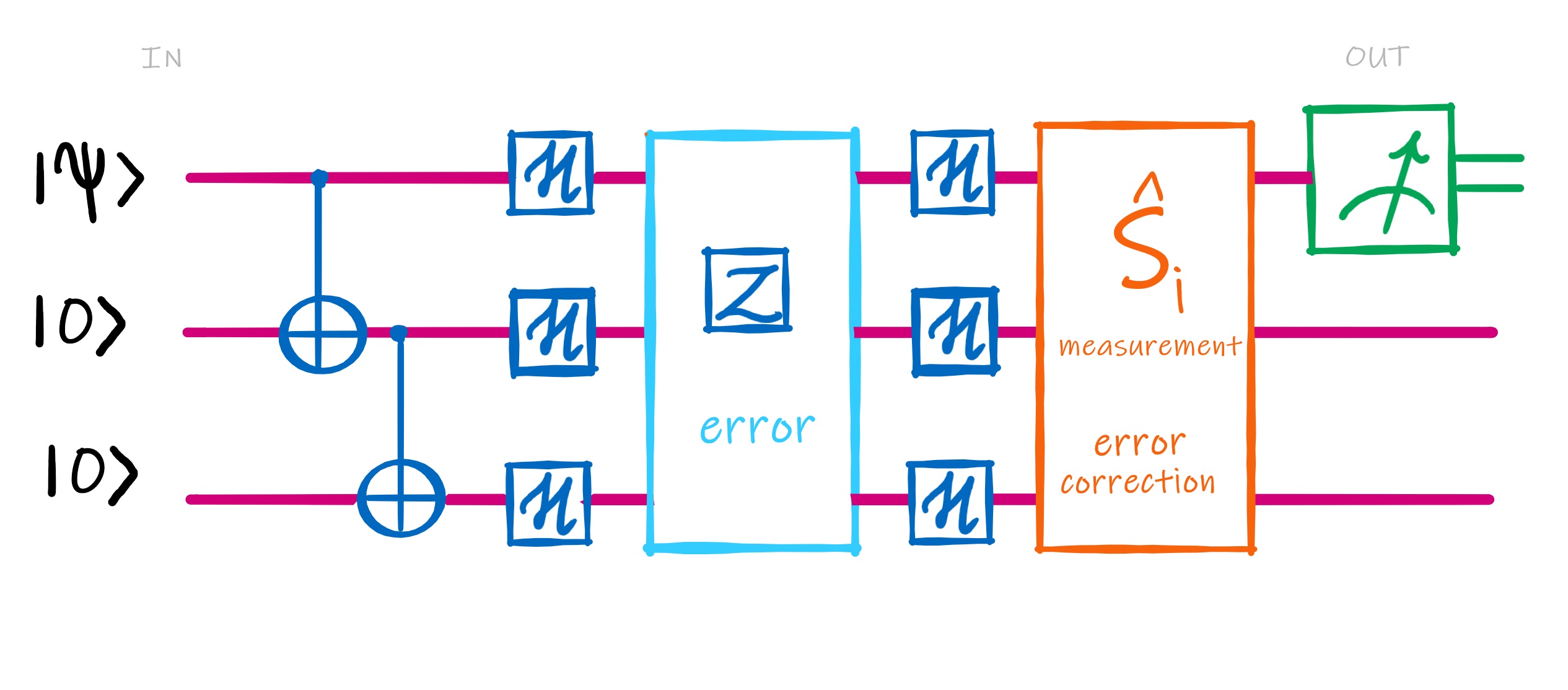}
\caption{Scheme of a 3-phase-flip error correction circuit. Green box at the right hand-side stands for qubit measurement.}
\label{pic:pherrcorr}
\end{figure}

In this case, 3-phase-flip repetition code uses logical qubits in a form: \\$|0\rangle_L=|000\rangle=|+++\rangle$ and $|1\rangle_L=|111\rangle=|---\rangle$. \\Then, the input state is given as:
\begin{equation}
|\psi\rangle= \alpha|0\rangle+\beta|1\rangle \rightarrow \alpha|+++\rangle+\beta|---\rangle = |\psi\rangle_L.
\end{equation}

As it may be predicted, error syndromes are similar to those presented earlier (Eq.~\ref{eq:errsym}):
\begin{equation}
\begin{split}
\text{no error:}\quad \hat{S}_0 \equiv |+++\rangle\langle +++|+ |---\rangle\langle ---| \:, \\
\text{error on 1st bit:}\quad \hat{S}_1 \equiv |-++\rangle\langle -++|+ |+--\rangle\langle +--| \:, \\
\text{error on 2nd bit:}\quad \hat{S}_2 \equiv |+-+\rangle\langle +-+|+ |-+-\rangle\langle -+-| \:, \\
\text{error on 3rd bit:}\quad \hat{S}_3 \equiv |++-\rangle\langle ++-|+ |--+\rangle\langle --+| \:. \\
\end{split}
\end{equation}

Before the error detection, with the use of another H-gate, measurement basis should be reversed again to the standard $\{|0\rangle,|1\rangle\}$ basis. Then, recovery operations on particular bits must be done. The whole procedure is presented schematically in Fig.~ \ref{pic:pherrcorr}. \\

\hsection{Decoherence-free subspace}
\hnote{Decoherence-free subspace}
\label{sec:ch2-decfree}

Encoding in parameters of a system that belong to \textbf{decoherence-free subspace} of a Hilbert space is an alternative method of dealing with noise in the quantum channel. \\
The system (which is predominantly coupled to the environment) behaves as an isolated one in the part that belongs to decoherence-free subspace.
In other words, this subspace, if carefully chosen, makes it possible to describe the evolution of these parameters with unitary operations, which may be reversed \cite{book:Lidar}. \\

Evolution of such a system may be described in the following way \cite{Lidar1998}:
\begin{equation}
\hat{H}=\hat{H}_S\otimes\mathds{1}_E+\mathds{1}_S\otimes\hat{H}_E+\hat{H}_I \; ,
\label{eq:dfs-h}
\end{equation}
where $\hat{H}_S$, $\hat{H}_E$ are the system and the environment operators, defined on the Hilbert subspaces $\mathcal{H}_S$ and $\mathcal{H}_E$, respectively. The interaction operator $\hat{H}_I$ takes the form:
\begin{equation}
\hat{H}_I=\sum_{i}\hat{S}_i\otimes\hat{E}_i \;.
\label{eq:dfs-hin}
\end{equation}
The $\hat{S}_i$ and $\hat{E}_i$ operators for system and environment respectively are the only parts of the hamiltonian $\hat{H}$ that represent their interaction with each other.\\

The evolution of a system may be described using the density matrix $\hat{\rho}_S$ and the $\hat{K}_{\alpha}$ Kraus operator sum representation, thus:
\begin{equation}
\hat{\rho}_S(0)\;\rightarrow\; \hat{\rho}_S(t)=\sum_{\alpha} \hat{K}_{\alpha}(t)\hat{\rho}_S(0)\hat{K}_{\alpha}^{\dagger}(t) \; ,
\end{equation}
where the Kraus operators satisfy the relation $\; \forall_t \; \sum_{\alpha}\hat{K}_{\alpha}^{\dagger}(t)\hat{K}_{\alpha}(t)=\mathds{1}_S$. \\

Kraus operators usually describe non-unitary evolution. However, if the partition of a system subspace $\mathcal{H}_S$ into $\mathcal{H}_S=\mathcal{H}_U\oplus\mathcal{H}_N$, where $U$ stands for unitary evolution and $N$ for the noise, is possible, the Kraus operator may be also partitioned to:
\begin{equation}
\hat{K}_{\alpha}(t)=g_{\alpha}\hat{U}\oplus \hat{N}_\alpha=\Big(\begin{matrix}
g_{\alpha}\hat{U}&0\\
0&\hat{N}_{\alpha}
\end{matrix}\Big).
\end{equation}
The operators $\hat{U}$ and $\hat{N}_{\alpha}$ act in $\mathcal{H}_U$ and $\mathcal{H}_N$ respectively, $g_{\alpha}\in \mathds{C}$ satisfies $\sum_{\alpha}g_{\alpha}g^*_{\alpha}=1$ and $\sum_{\alpha}\hat{N}_{\alpha}^{\dagger}(t)\hat{N}_{\alpha}(t)=\mathds{1}_N$. \\

If the initial state is partitioned in the same manner, where $\hat{\rho}_S(0)=\hat{\rho}_U(0)\oplus\hat{\rho}_N(0)$, then the evolution is described as:
\begin{equation}
\begin{split}
\hat{\rho}_S(0)\;\rightarrow\; \hat{\rho}_S(t)=\sum_{\alpha}(g_{\alpha}\hat{U}\oplus \hat{N}_{\alpha})\big(\hat{\rho}_U(0)\oplus\hat{\rho}_N(0)\big)(g^*_{\alpha}\hat{U}^{\dagger}\oplus \hat{N}_{\alpha}^{\dagger})= \\
= \Bigg(\begin{matrix}
\hat{U}\hat{\rho}_U(0)\hat{U}^{\dagger}&0\\
0&\sum_{\alpha}\hat{N}_{\alpha}\hat{\rho}_N(0)\hat{N}_{\alpha}^{\dagger}
\end{matrix}\Bigg) \;.
\end{split}
\end{equation}

It should be emphasized here that in the result, the evolution of the subsystem $\hat{\rho}_U$ is completely unitary. Therefore, encoding in a vector that belongs to the subspace $\mathcal{H}_U$ makes the transmission  through the environment docoherence-free by "preserving" the information. In principle, it means that the correction codes are no longer needed. \\

This method of encoding has been demonstrated experimentally \cite{Kwiat2000} and is still rapidly developing, in the area of potential use for quantum computers, which obviously cannot be fully isolated from the environment. \\

\hsection{Indistinguishability of photons}
\hnote{Indistinguishability of photons}
\label{sec:ch2-indistinguishability}
The indistinguishability of photons is a concept that may be approached in many different ways relating to photons' degree of freedom that are considered. Here, it is assumed that the source generates indistinguishable photons if photons from successive pulses interfere with each other via \textbf{HOM effect} (see \textit{Section \ref{sec:notes-HOM}}).\\

\begin{figure}[b!]
\centering
\includegraphics[width=0.7\linewidth]{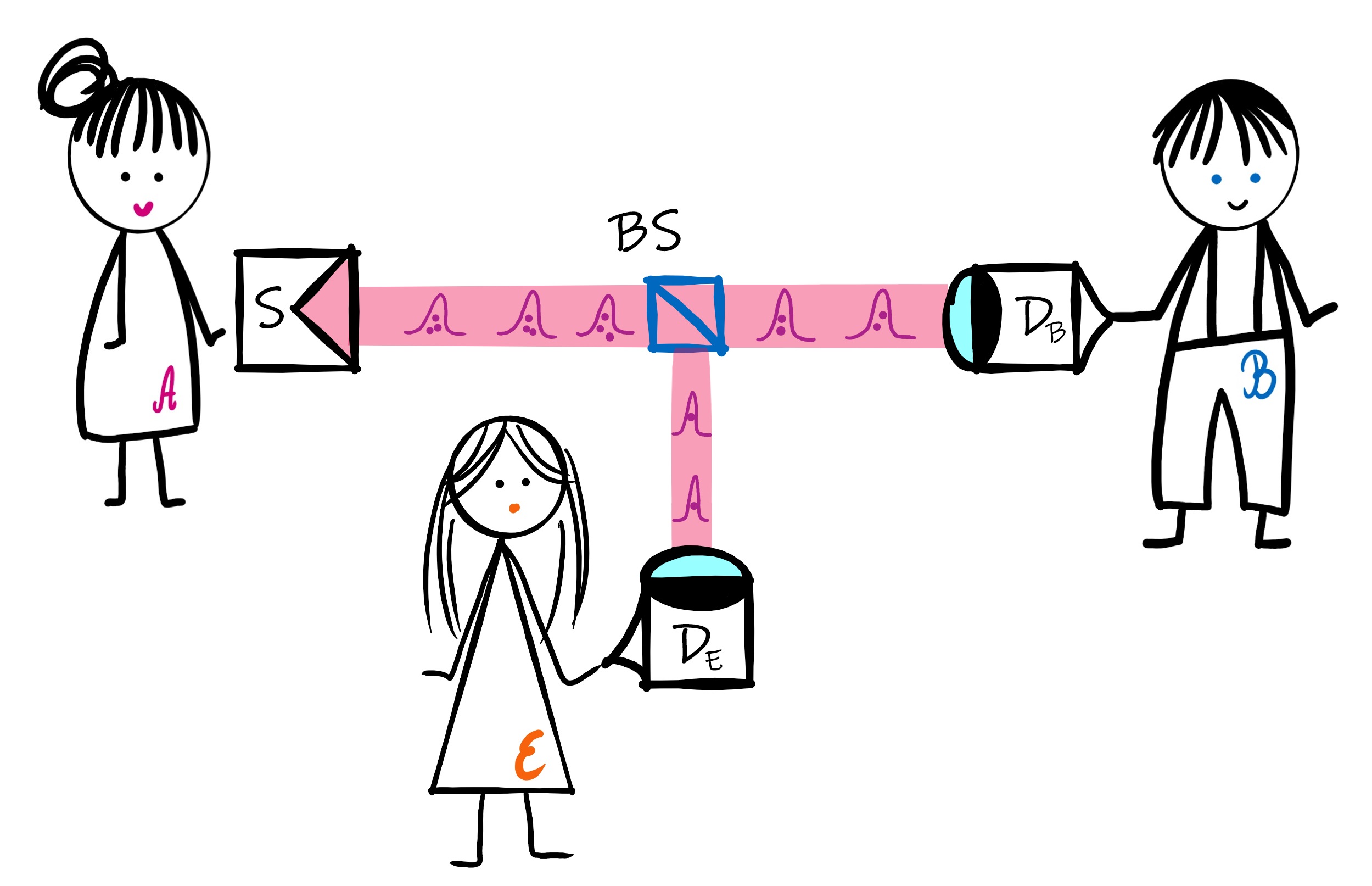}
\caption{Scheme of the beam-splitting attack. Symbols: BS -- beam splitter, S -- photon source, D -- detector, A, B, E, - Alice, Bob and Eve, respectively.}
\label{pic:splittattAB}
\end{figure}

\hsection{Second-order coherence function}
\hnote{Second-order coherence function}
\label{sec:ch2-g2}
One of the most important single-photon source parameters is the photon statistics that the source generates. In many quantum communication protocols it is crucial to send one photon at a time. More photons in a pulse may decrease the safety of 
communication by allowing the third party to eavesdrop the channel by, for example the photon splitting attack. The two communicating parties, Alice and Bob, when sending the light pulses, may be eavesdropped on by Eve, who puts a beam splitter into  their channel. If the pulses contain lots of photons, some of them can be taken out by Eve, without it being noticed by Alice or Bob. The idea of this type of attack is presented in Fig.~\ref{pic:splittattAB}.\\

To prevent such situations, it is essential to use pure single-photon sources. As it is known, photon probability distribution function and the fluctuations of the photon number depend on the type of source. To classify different light sources, the second-order correlation function $g^{(2)}(\tau)$, introduced in 1963 by American theoretical physicist Roy J. Glauber \cite{Glauber1963}, can be used. \\

The $g^{(2)}(\tau)$ function may be used to describe either quantum or classical light sources. Since the first experimental attempts were made in Hanbury-Brown Twiss (HBT) experiment scheme (see \textit{Section \ref{sec:notes-HBT}}), let us first define its classical interpretation. \\
The classical second-order intensity correlation function, which describes the correlation between two signals with temporal separation $\tau=t_2-t_1$ coming from one light source, takes a form:
\begin{equation}
g_{classical}^{(2)}(\tau)=\frac{\langle I(t)I(t+\tau)\rangle}{\langle I(t)\rangle^2}= \frac{\langle E^*(t)E(t)E^*(t+\tau)E(t+\tau)\rangle}{\langle E^*(t)E(t)\rangle^2}\; ,
\label{eq:g2class}
\end{equation}
where $I(t) \propto E^*(t)E(t)$, the intensity $I(t)$ is proportional to the amplitude of electromagnetic field $E(t)$ and $\langle ::\rangle$ denotes the averaging operation. \\
A mathematical formalism that originates form the second quantization of an electromagnetic field \cite{book:Scully, book:Fox, book:Loudon} allows the transformation from classical field quantities into its quantum mechanical equivalent operators. Therefore, the electric field $E(t)$ of a mode $k$ can be rewritten as:
\begin{equation}
E(t)\;\rightarrow\;\hat{E}_k(t)=\hat{E}^+_k(t)+\hat{E}^-_k(t) \;.
\end{equation}
The two parts $\hat{E}_k^+$,$\hat{E}_k^-$ are proportional to annihilation $\hat{a}_k$ and creation $\hat{a}^\dagger_k$ operators: 
\begin{equation}
\begin{split}
\hat{E}^+_k(t) \; \propto \; \hat{a}_k \exp{\Big(-i\big(\omega_k\cdot t-\vec{k}\cdot\vec{r} \big)\Big)} \;,\\
\hat{E}^+_k(t) \; \propto \; \hat{a}_k^\dagger \exp{\Big(i\big(\omega_k\cdot t-\vec{k}\cdot\vec{r} \big)\Big)} \;.
\end{split}
\end{equation}
Using these equations, Eq.~\ref{eq:g2class} may be rewritten as:
\begin{equation}
g^{(2)}_{quantum}(\tau)=\frac{\langle\hat{E}^-_k(t)\hat{E}^+_k(t)\hat{E}^-_k(t+\tau)\hat{E}^+_k(t+\tau)\rangle}{\langle \hat{E}^-_k (t)\hat{E}^+_k(t)\rangle^2}\; ,
\end{equation}
and for $\tau=0$ it finally takes a form:
\begin{equation}
g^{(2)}_{quantum}(0)=\frac{\langle \hat{a}_k^\dagger   \hat{a}_k^\dagger\hat{a}_k\hat{a}_k \rangle}{\langle \hat{a}_k^\dagger\hat{a}_k\rangle^2}.
\end{equation}
Then, using the photon number operator $\hat{n}=\hat{a}^\dagger\hat{a}$ as the substitution and considering its eigenvalues \cite{book:Fox, book:Scully}, while minding the commutation relation, we get:
\begin{equation}
\begin{split}
g^{(2)}_{quantum}(0)\,&=\,\frac{\langle \sqrt{n}\sqrt{n-1}\sqrt{n-1}\sqrt{n}\rangle}{\langle \sqrt{n}\rangle^2}\,=\,\frac{\langle n(n-1)\rangle}{\langle n\rangle^2}\\
&=\,\frac{\langle n^2 \rangle-\langle n \rangle}{\langle n \rangle^2}\,=\,\frac{(\Delta n)^2+\langle n\rangle^2-\langle n \rangle}{\langle n \rangle^2}\\
&=\,1+\frac{(\Delta n)^2-\langle n \rangle}{\langle n \rangle^2},
\end{split}
\end{equation}
where $\langle n \rangle$ is an average photon number and $(\Delta n)^2$ is its variance. \\

Now, we can compute the value of second-order coherence $g^{(2)}(0)$ for different types of light, with known statistics. The results are collected in Tab. \ref{tab:g2}. \\

\begin{table}[h!]
\centering
\begin{tabular}{r|c|c}
\hline
\textbf{light source} & \textbf{variance $(\Delta n)^2$} & \textbf{$g^{(2)}(0)$}\\
\hline\hline
thermal source & $\langle n^2\rangle+\langle n\rangle$ & $2$ \\
\hline
coherent source & $\langle n\rangle$ & $1$ \\
\hline
\multirow{2}{*}{Fock source} & \multirow{ 2}{*}{0} & $0$, for $n=0$ \\
 &  & $1-\frac{1}{n}$, for $n\geq1$ \\
\hline
\end{tabular}
\caption{Values of second-order coherence function $g^{(2)}(0)$ for different light sources.}
\label{tab:g2}
\end{table}

\begin{figure}[b!]
\centering
\includegraphics[width=0.6\linewidth]{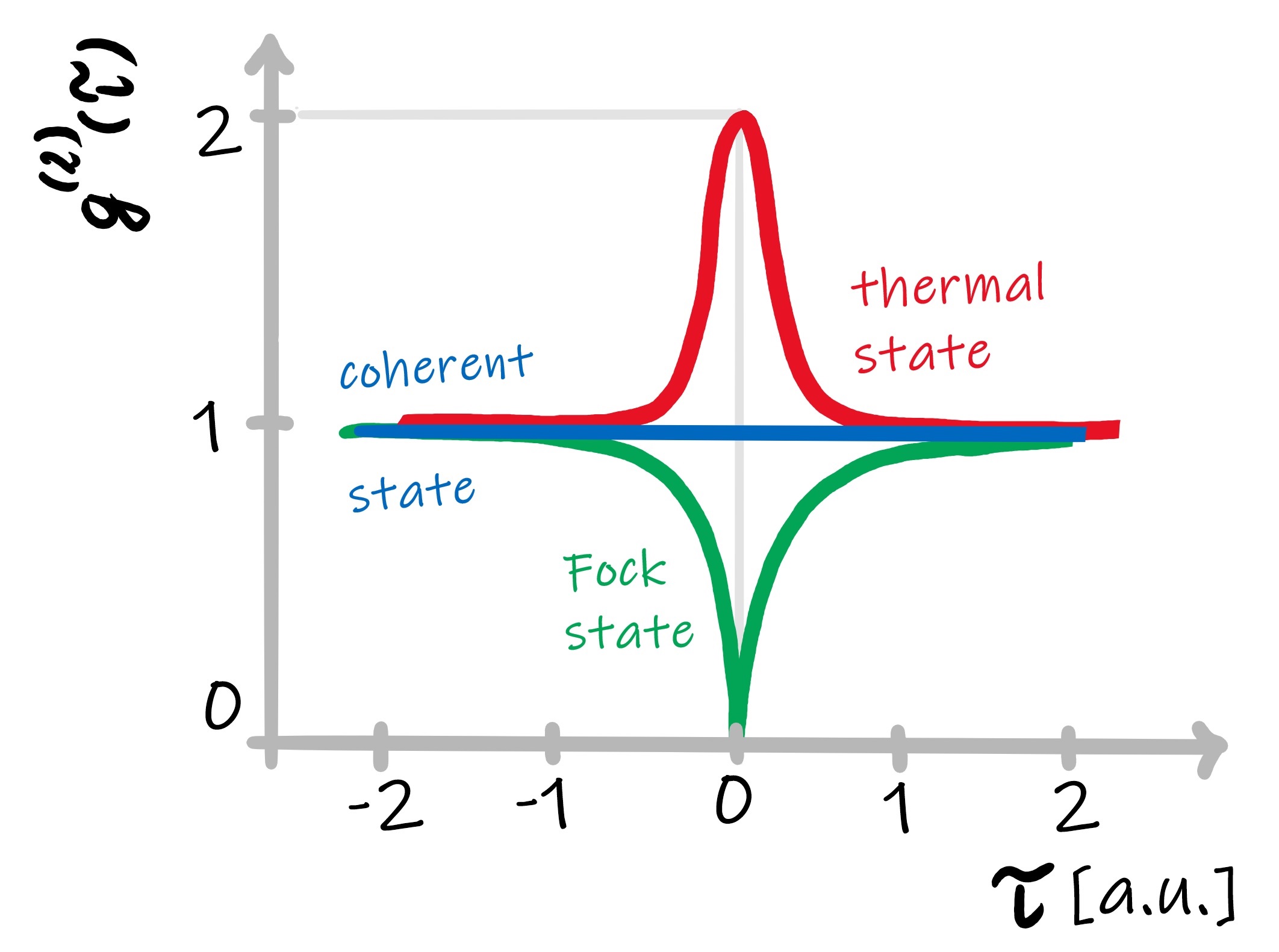}
\caption{The shapes of second-order coherence functions $g^{(2)}(\tau)$ for photons emitted by different light sources.}
\label{pic:g2res}
\end{figure}

Despite the fact that the value of $g^{(2)}(0)$ is what characterizes the "singleness" of the photon sources, the shape of $g^{(2)}(\tau)$ functions, which are schematically presented in Fig.~\ref{pic:g2res}, also reveals some information.\\

For example, for coherent light sources, the function remains  flat. It means that emitted photons are completely uncorrelated. The correlation between photon emissions can be seen in the $g^{(2)}(\tau)$ shape for thermal sources. Such sources have higher probability of generating more than one photon at a time. This effect is called \textit{photon bunching} \cite{book:Loudon}. Unlike thermal sources, the value of second-order coherence function for Fock sources is always less than 1 ($g^{(2)}(0)<1$). It means that the probability of multi-photon generation is reduced. This is the \textit{anti-bunching effect} that occurs only if the light source is non-classical \cite{book:Loudon}. \\
Non-classical light sources with the average photon number equal to 1 ($\langle n\rangle=1$), and thus $g^{(2)}(0)=1$, are the perfect single-photon sources. \\

\hsubsection{Hanbury-Brown-Twiss experiment}
\hnote{Hanbury-Brown-Twiss experiment}
\label{sec:notes-HBT}

In 1954, two British astronomers, Robert Hanbury Brown and Richard Q. Twiss proposed an experiment, which was later named the HBT experiment, that  allows for measuring of the angular size of stars \cite{HBT1954}. The setup was based on intensity inteferometer. Two telescopes, followed by detectors, were pointed at the same star. Then, correlation between electronic signals from both detectors was checked. As it turned out, the correlation changed with the change in telescope view position. When increasing the distance, the correlations dropped down. The reason of such behavior is that the statistics for a light produced by a star is thermal. It means that the probability of multi-photon emission is high; thus, such source creates correlation in photon arrival time \cite{book:Scully,book:Loudon}. \\

\begin{figure}[t!]
\centering
\includegraphics[width=0.7\linewidth]{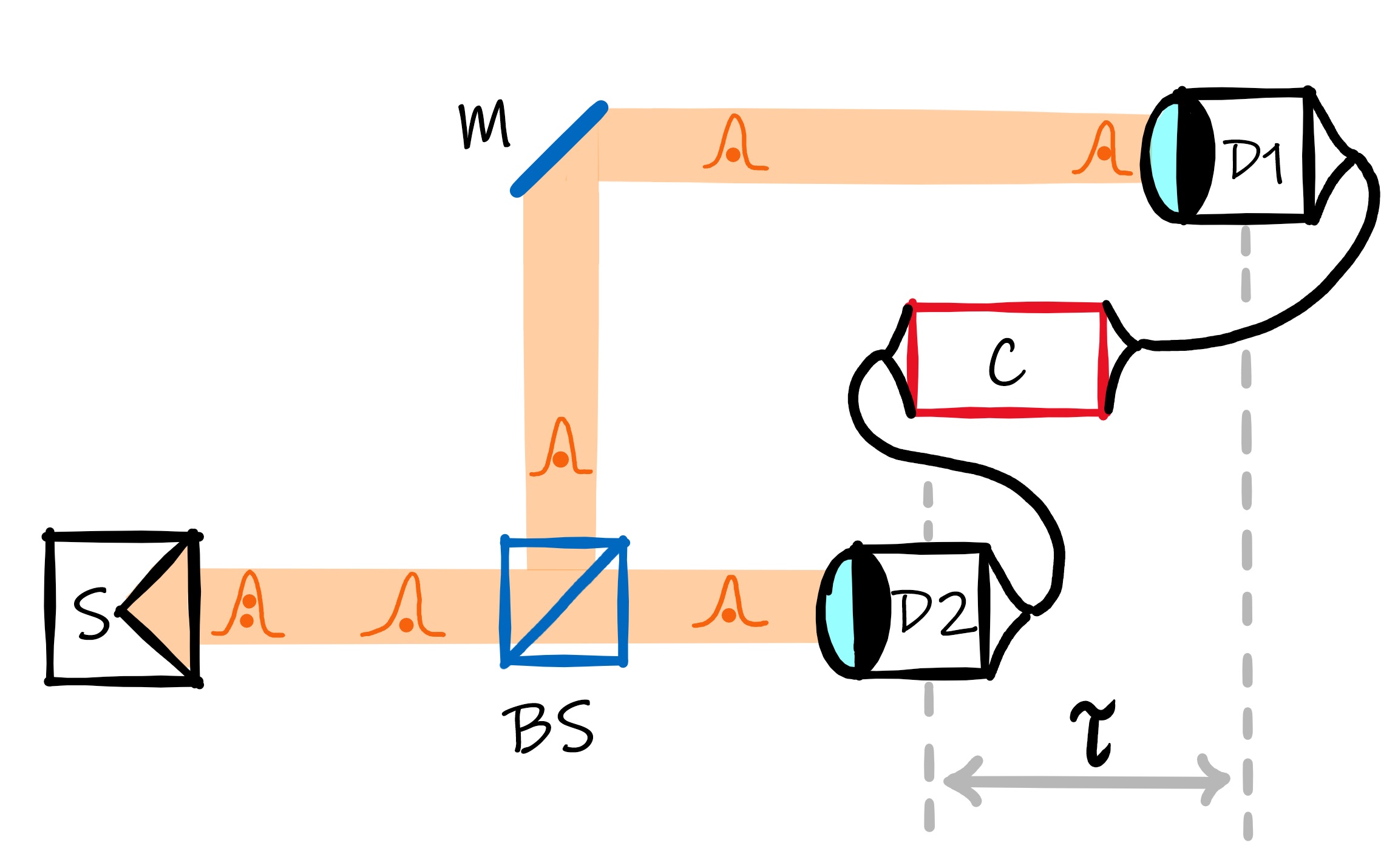}
\caption{HBT experimental setup. Symbols: S -- photon source, BS -- beam splitter, D -- detectors, C -- coincidence correlator, $\tau$ -- delay time between detectors.}
\label{pic:hbtset}
\end{figure}

Nowadays, the HBT experiment is a tool, which allows for the characterization of a light source. The setup, the scheme of which is presented in Fig.~\ref{pic:hbtset}, is  still used to measure correlations in photon arrival time. However, instead of changing the detectors' position, the time differences between detection are collected. If a light source generates single photons, each photon is observed by only one detector, so the anti-correlation effect is perceived (see \textit{Section \ref{sec:ch2-g2}}). \\

\hsection{A guide to SNLO software}
\hnote{A guide to SNLO software}
\label{sec:ch5-snlo}
SNLO \cite{snlo} is a free software developed by Dr. Arleen Smith, which includes data (and its references) for more than 50 nonlinear crystals. Thus, it is widely used around the globe when designing quantum optical experiments. \\

The software consists of various modules that allow theoretical examination of second-order nonlinear processes, such as second-harmonic generation, three-wave mixing, and spontaneous parametric down-conversion in a chosen crystal. The functions of all modules along with examples are extensively described in the manual; however, in this section, only modules greatly important for designing SPDC sources are presented. \\

\begin{figure}[t!]
\centering
\includegraphics[width=0.85\linewidth]{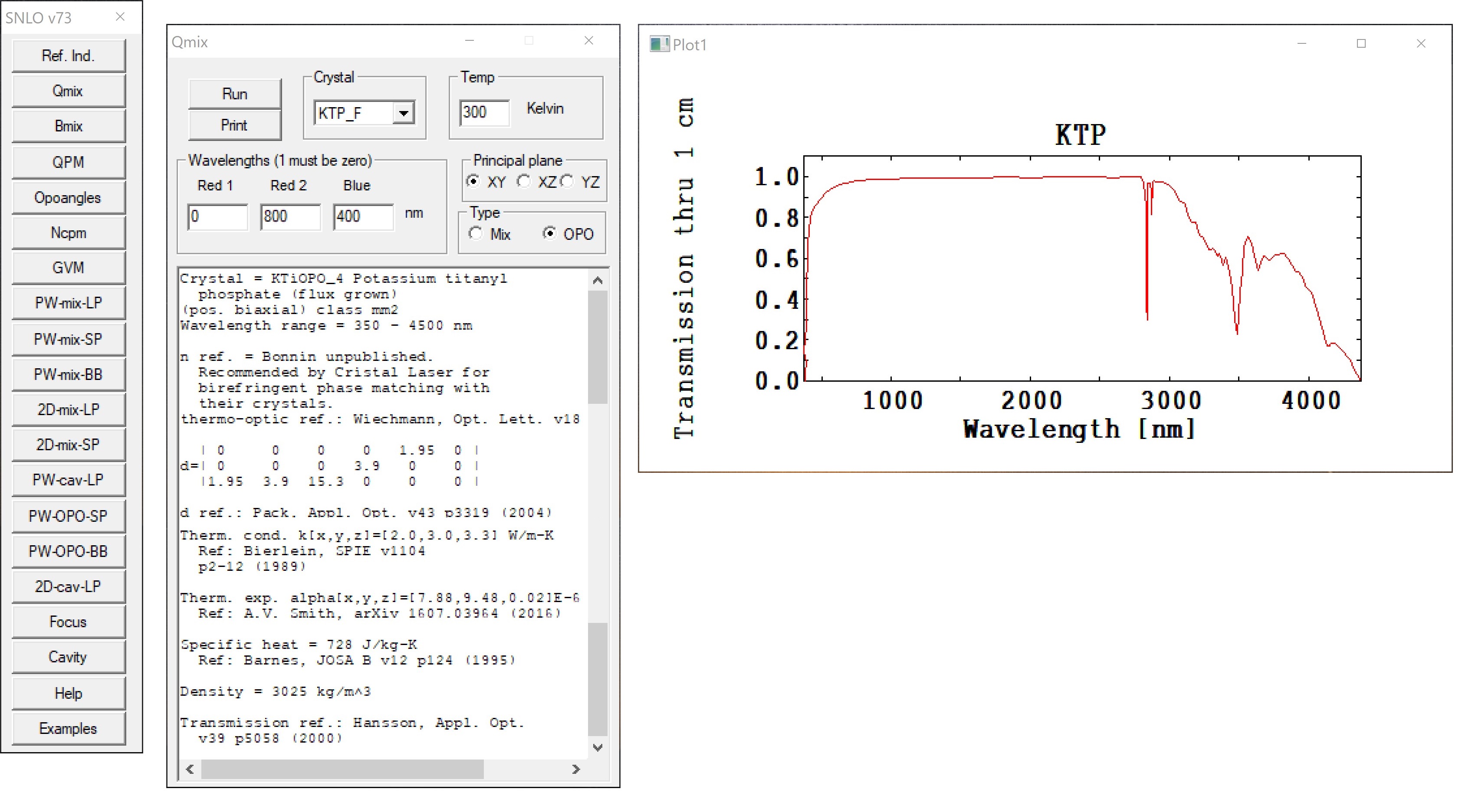}
\caption{Crystal specification and transmission function. Data for \textbf{KTP} crystal presented as an example.}
\label{pic:snlo-qmix}
\end{figure}

General parameters of a crystal, such as refractive index, can be checked in \textbf{\textsl{Ref. Ind.}} module (see software menu on the left hand side of Fig.~\ref{pic:snlo-qmix}). It also allows for manual calculation of phase matching and group velocities for specific beam wavelength and its propagation angle in a given crystal. \\

\begin{figure}[b!]
\centering
\includegraphics[width=0.85\linewidth]{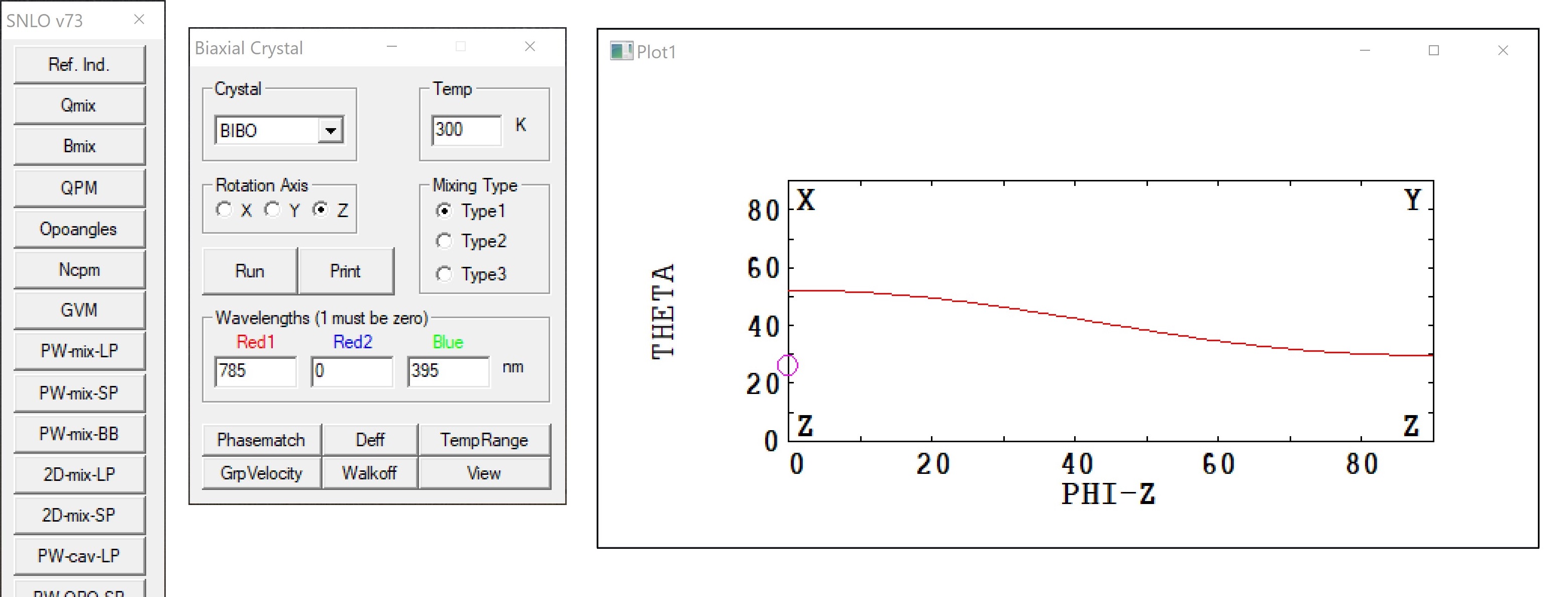}
\caption{Phase-matching conditions in biaxial nonlinear crystal. Data for \textbf{BiBO} crystal presented as an example.}
\label{pic:snlo-bc-pm}
\end{figure}

In next module, \textbf{\textsl{Qmix}}, clicking on \textsl{Run} button, offers quick check of phase-matching conditions, assuming the collinear propagation inside a crystal. It also shows the transmission function through the crystal and other parameters. Among them the most useful one is effective nonlinear coefficient (tensor in some cases). As mentioned in \textit{Section \ref{sec:ch3-eff}}, these two parameters are helpful when estimating the conversion efficiency of a crystal. \\

\begin{figure}[t!]
\centering
\includegraphics[width=0.9\linewidth]{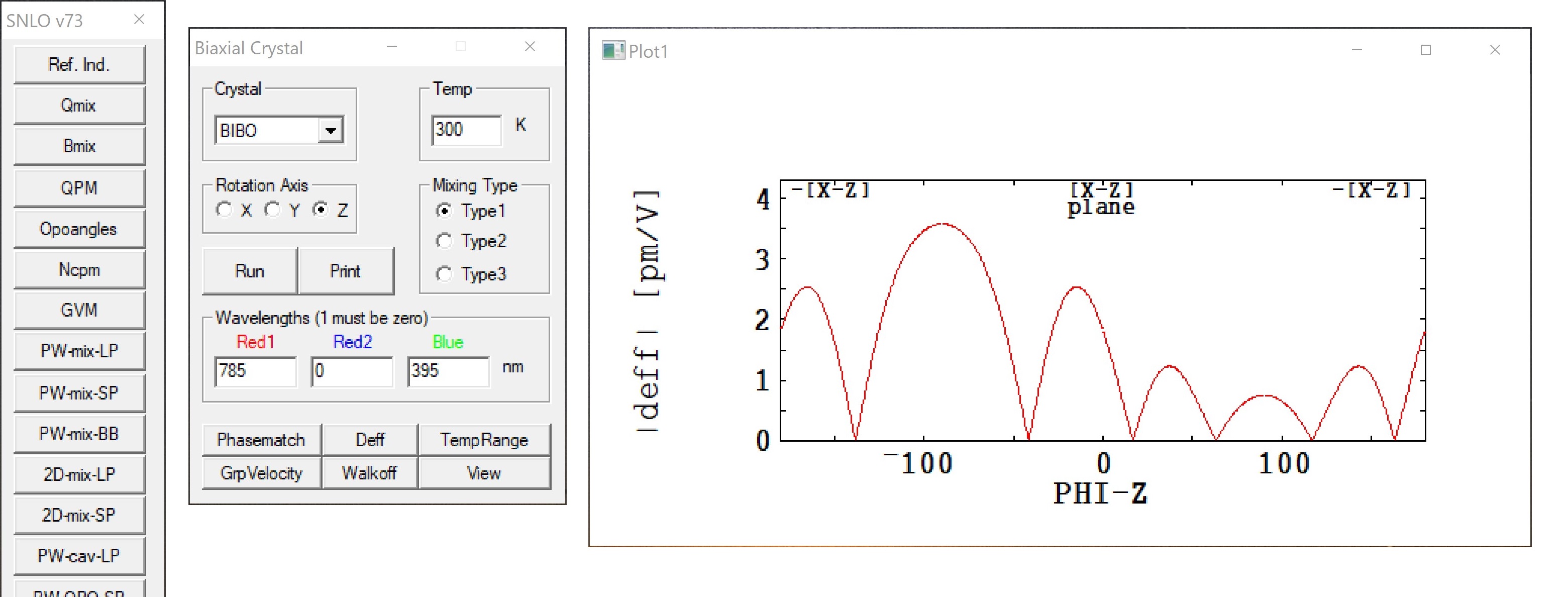}
\caption{Effective nonlinear coefficient and its crystal axis angular dependence. Data for \textbf{BiBO} crystal presented as an example.}
\label{pic:snlo-deff}
\end{figure}

Another module, \textbf{\textsl{Bmix}}, is utilized to examine the beam propagation inside a biaxial nonlinear crystals, when a pump beam is not collinear to the crystal axes. The software generates the phase-matching curves for given wavelengths (\textsl{Run} button, results presented in Fig.~\ref{pic:snlo-bc-pm}) and its dependence on temperature (\textsl{TempRange} button). It also shows the changes of effective nonlinearity coefficient (\textsl{Deff} button) in a function of  cut angle of a crystal, as depicted in Fig.~\ref{pic:snlo-deff}. Other parameters that can be obtained using these modules are the group velocities (\textsl{GrpVelocity} button) for given wavelengths and polarization walk-offs (\textsl{Walkoff} button), both dependent on crystal's cut angle. \\
The \textsl{View} button function shows all data in text format.\\

\begin{figure}[b!]
\centering
\includegraphics[width=0.9\linewidth]{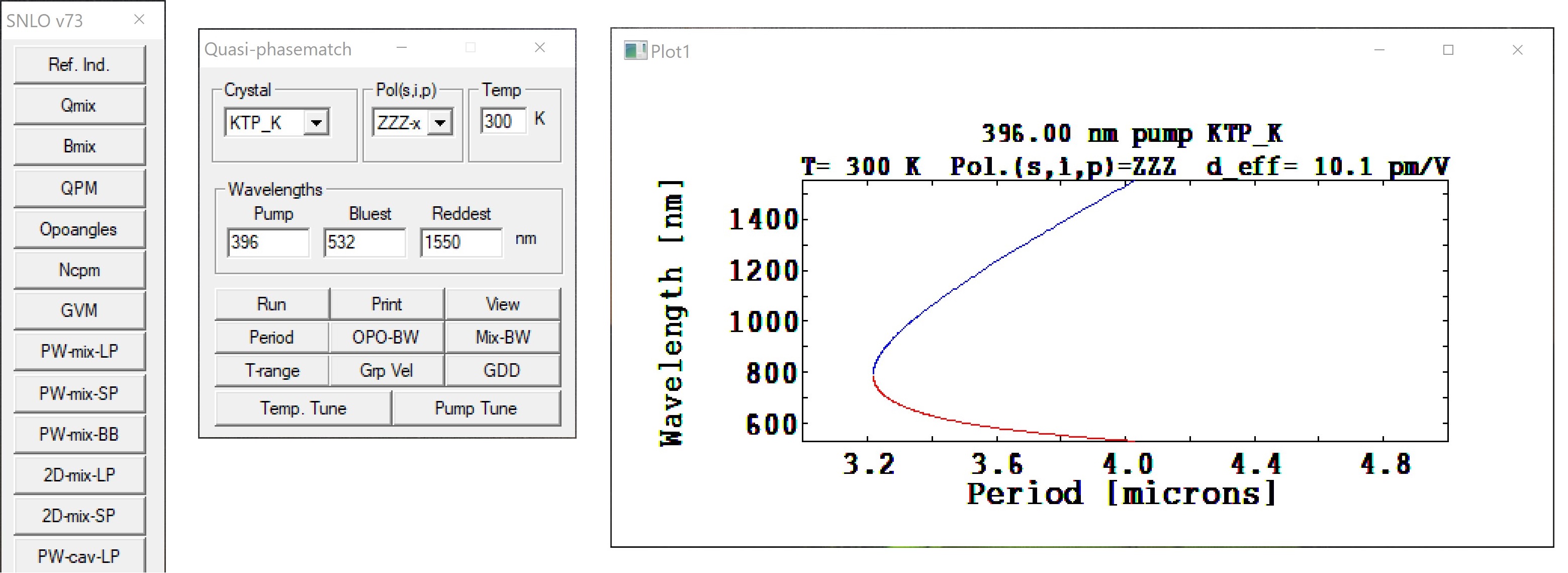}
\caption{Quasi-phase-matching conditions in a function of crystal poling period. Data for \textbf{PPKTP} crystal presented as an example.}
\label{pic:snlo-qpm2}
\end{figure}

\begin{figure}[t!]
\centering
\includegraphics[width=0.9\linewidth]{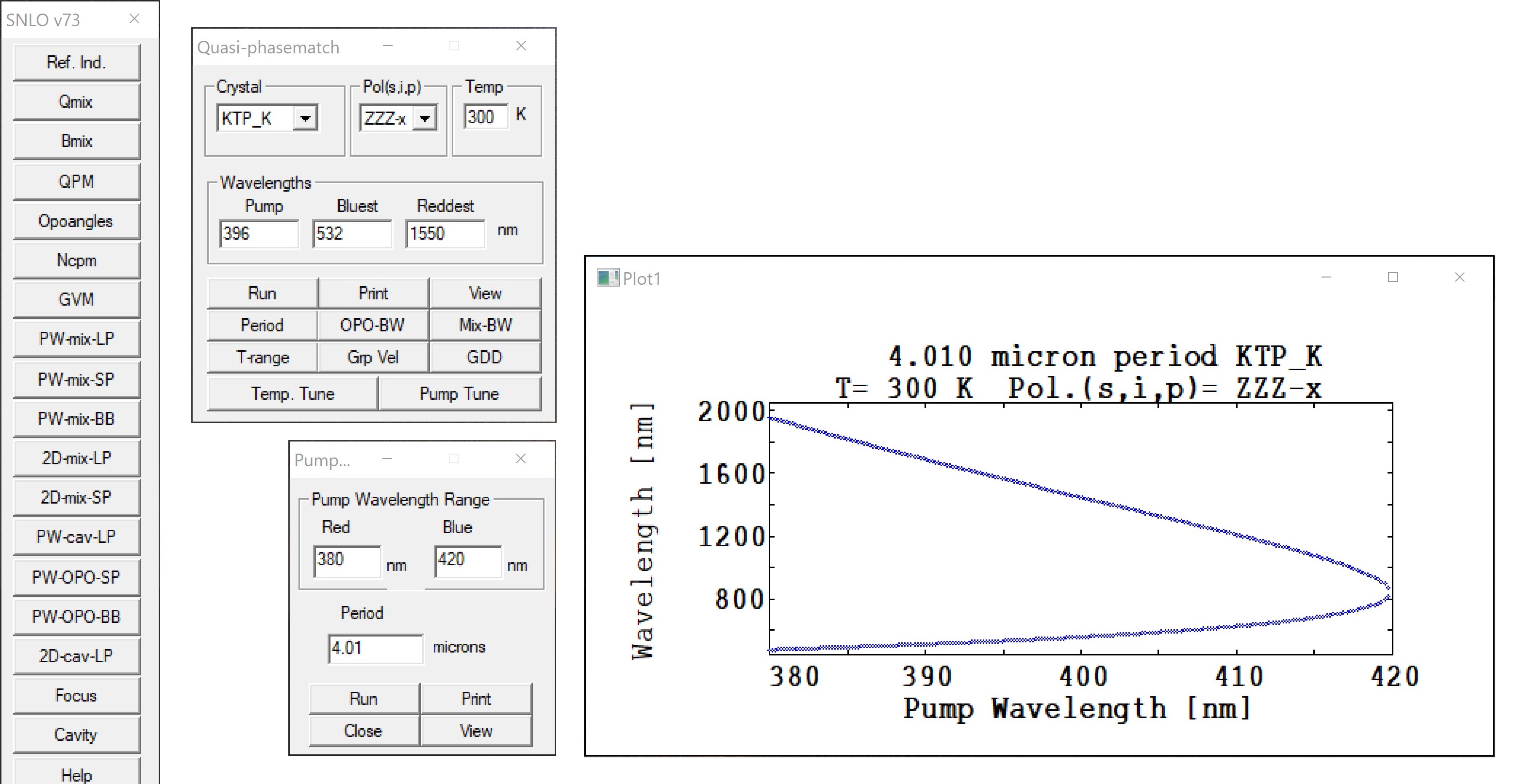}
\caption{Generated wavelengths dependence on pump wavelength for specified poling period of a crystal. Data for \textbf{PPKTP} crystal presented as an example.}
\label{pic:snlo-pump}
\end{figure}

\begin{figure}[h!]
\centering
\includegraphics[width=0.9\linewidth]{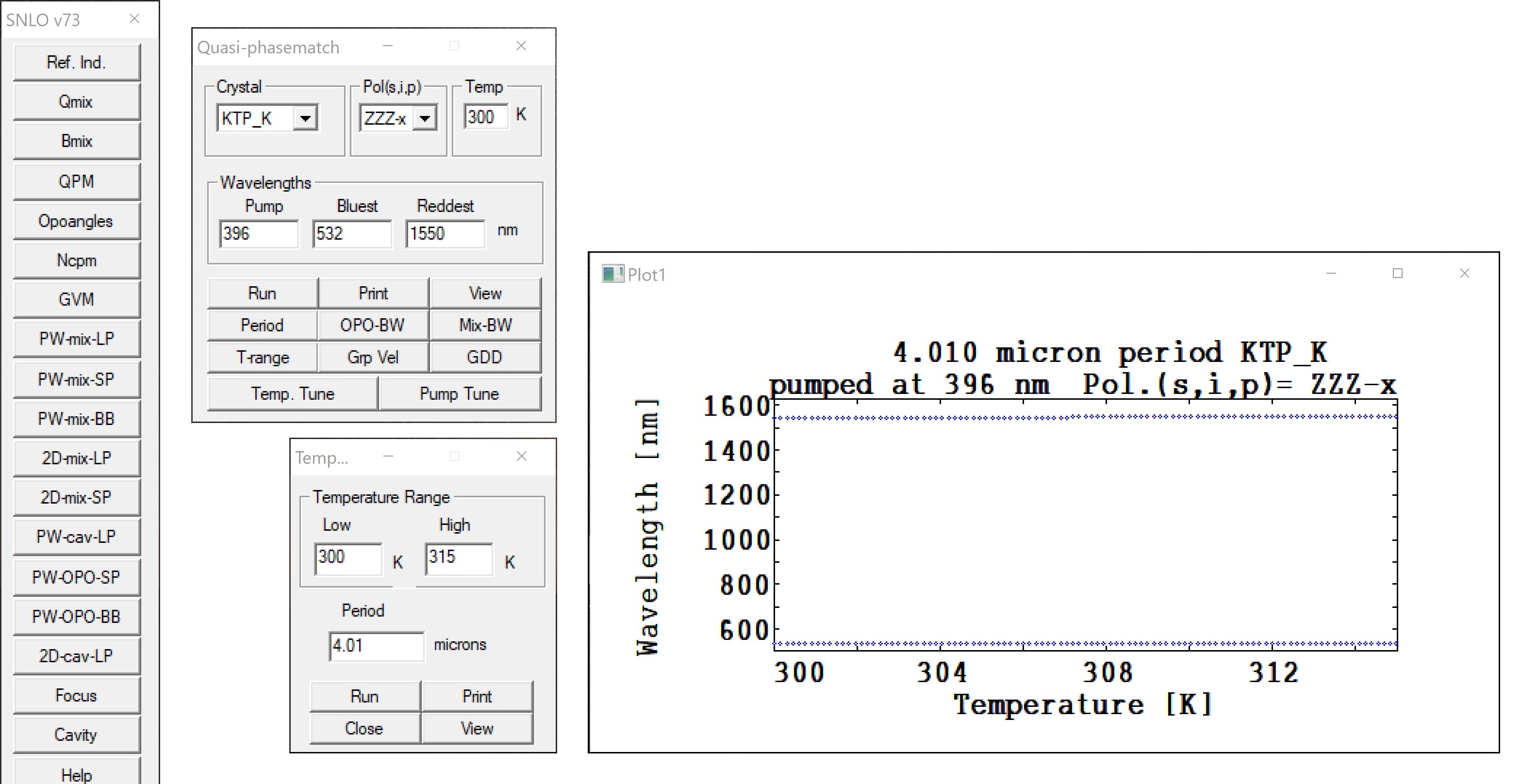}
\caption{Temperature dependence of generated wavelengths for specified poling period of a crystal. Data for \textbf{PPKTP} crystal presented as an example.}
\label{pic:snlo-temp}
\end{figure}

The last most important module to describe, in the context of designing a single-photon source, is \textbf{\textsl{QPM}} that returns both, the properties of a chosen material and a beam, in case of collinear propagation inside it. \\
As presented in Fig.~\ref{pic:snlo-qpm2}, when clicking on the \textsl{Run} button, the quasi-phase-matching conditions are calculated in a given crystal and its temperature, for specific wavelengths, pump, signal, and idler in case of SPDC process. \\
Next, as showed in Fig.~\ref{pic:snlo-pump}, by clicking \textsl{Pump Tune} button, for a crystal with a chosen period, the quasi-phase-matching conditions are calculated in a function of pump beam central wavelength. Similarly, by clicking \textsl{Temp. Tune} button, the quasi-phase matching conditions are calculated in a function of a crystal temperature, which is visualized in Fig.~\ref{pic:snlo-temp}. \\
The \textsl{View} button always shows plotted data in text mode. \\

Even though the SNLO software is quite handy, due to its main assignment as the software for the designing of  the optical parametric oscillators (OPO) \cite{book:Boyd,book:Dmitriev, book:Nikogosyan}, it has some limitations. For instance, in case of designing the SPDC source that is based on periodically poled crystal, sometimes it is necessary to apply noncollinear beam paths. Such situation is helpful when the beams of generated photons are to be separated while finding the proper dichroic mirror is a problem (for example, the central wavelengths of converted photons are too close to each other, the dichroic mirror has big losses or generates the noisy fluorescence). 
In these circumstances, it is necessary to perform one's own calculations on assumptions that cannot be made using the software. It is possible, however, to extract the Sellmeier equation \cite{book:Boyd,book:Dmitriev, book:Nikogosyan} that describes the chromatic dispersion inside a specific crystal and is all that is needed for such calculations. \\
Similar work was done in Ref. \cite{Andrzej2018}, where the difference between results from SNLO and own simulation is presented. \\

\FloatBarrier
\hsection{Quantum state tomography}
\hnote{Quantum state tomography}
\label{sec:ch5-qst}
As it was mentioned in \textit{Section \ref{sec:ch3-qst}}, \textbf{quantum state tomography} is a method that allows for density matrix reconstruction based on multiple measurements on identical quantum states. The number of measurements depends on the dimension of Hilbert space to which the discussed matrix belongs. \\
According to this fact, a set of measurement operators ${\hat{M}_k}$, which complements a \textbf{positive operator-valued measure (POVM)} \cite{PST, Toninelli2019}, is constructed. The $\hat{M}_k$ are \textit{the positive semi-definite operators} which form a basis, thus:
\begin{equation}
\sum_{i=k}^K \hat{M}_k\,=\, \mathbb{1} \; ,
\end{equation}
for K-dimensional Hilbert space. \\
Using this method with regard to \textit{Born's rule} \cite{Born1926, James2001}, the probability $p(k)$ of getting an outcome $k$ when measuring the state described by $\hat{\rho}$ is equal to:
\begin{equation}
p(k)=Tr\big(\hat{M}_k \hat{\rho} \big)\; .
\end{equation}
As a consequence, for perfect detection, an expected number of counts that should be registered is equal to:
\begin{equation}
n_k^E(\hat{\rho})\,=\,N\cdot p(k)\,=\, N \cdot Tr\big(\hat{M}_k \hat{\rho} \big)\; ,
\end{equation} 
where $N$ is an average number of
photons involved in the measurement. \\
Unfortunately, due to losses and detection imperfections, there are no perfect measurements, therefore real detected counts are equal to $n_k^M$ and, in general, are different than expected, so $n_k^M\,\neq\,n_k^E$.\\

\hsubsection{Methods of parameter estimation}
\label{sec:ch5-est}
Since the measured values vary from those which are expected, to tie them together, and later reconstruct the density matrix, it is necessary to use some kind of \textbf{estimation}. There are many methods of parameters' estimation useful in this task. Here, only a few will be shortly described. Nonetheless, the foundation of each of them is some kind of mathematical optimization problem in which an extremum of a given function $f(\hat{\rho})$ is sought. \\

The simplest method of estimation is the method of \textbf{least squares (LS)} \cite{Jakub2021, Acharaya2019}, where the minimum value of a function $\mathscr{F}_{LS}(\hat{\rho})$, defined as:
\begin{equation}
\mathscr{F}_{LS}(\hat{\rho})\,=\, \sum_{k=1}^K\big(n_k^E(\hat{\rho})-n_k^M \big)^2\; ,
\end{equation}
where $K$ is the the number of measurement operators $\hat{M}_k$ involved in a particular scenario. \\
The LS method utilizes the linear dependence between estimated and measured counts. Despite the fact that it translates into a linear regression problem which is relatively easy computationally, the LS method is not very accurate \cite{Acharaya2019}. This results in recovering density matrices that not always produce a physical quantum state. Nonetheless, it can be the first step towards other more precise methods. \\

Another estimation method is \textbf{minimum chi-squared }\cite{Berkson1980, Jack2009}. Usually, it is used under assumption that the estimated and measured data are different due to statistical fluctuations from limited numbers of observations. \\
The most popular form of $\chi^2(\hat{\rho})$ function for minimalization procedure is:
\begin{equation}
\chi^2(\hat{\rho})=\sum_{k=1}^M\bigg(\frac{n_k^M-n_k^E(\hat{\rho})}{\sqrt{n_k^M}} \bigg)^2 \; ,
\end{equation}
where $K$, just like before, is the the number of measurement operators $M_k$. Despite that fact, many other variants of $\chi^2(\hat{\rho})$ function may be used \cite{Berkson1980}. \\

The last mentioned here, and probably the most used and significantly more accurate method is \textbf{maximum likelihood estimation (MLE)} \cite{Acharaya2019, Artur2021}. The parameters are fitted by maximizing a likelihood function $\mathscr{L}(\hat{\rho})$, which may take a form \cite{Ikuta2017}:
\begin{equation}
\mathscr{L}(\hat{\rho})=\sum_{k=1}^K \bigg(\frac{\big( n_k^M-n_k^E(\hat{\rho})\big)^2}{n_k^E(\hat{\rho})} + ln\, n_k^E(\hat{\rho})\bigg) \; ,
\end{equation}
where $K$, similarly to previously described functions, is the the number of measurement operators $\hat{M}_k$. \\

The exact shape of likelihood function depends on an assumed probability distribution of parameters that result in a set of photon counts $n_k^M$ \cite{James2001, Acharaya2019, Ikuta2017}. \\
In some cases, the likelihood function may be supplemented with an error function $f_e(k)$ in the following form:
\begin{equation}
\mathscr{L}(\hat{\rho})=\sum_{k=1}^K \bigg(\frac{\big(( n_k^M-f_e(k))-n_k^E(\hat{\rho})\big)^2}{n_k^E(\hat{\rho})} + ln\, n_k^E(\hat{\rho})\bigg) \; ,
\label{eq:mle-ef}
\end{equation} 
where the form of an error function is determined by the specific properties of experimental setup when measuring with operator $\hat{M}_k$. The form of likelihood function from Eq. \ref{eq:mle-ef} was proposed in Ref. \cite{Artur2022new} and will be elaborated on further in  \textit{Section \ref{sec:ch5-expexamp}}.\\

\hsubsection{Density matrix reconstruction algorithm}
\label{sec:ch5-alg}

The density matrix reconstruction algorithms are based on one kind of estimation function $f(\hat{\rho})$, usually one from described earlier in \textit{Section \ref{sec:ch5-est}}. All of the algorithms are parametrized by $\hat{\rho}$ density matrix elements. To simplify the computation, a specific form of density matrix $\hat{\rho}$ is assumed. \\

In general, all density matrices may be factorized using \textit{Cholesky decomposition} \cite{PST2005, James2001}:
\begin{equation}
\hat{\rho}=\frac{T^{\dagger} T}{Tr(T^{\dagger} T)}\; ,
\label{eq:rho}
\end{equation}
where $T$ denotes a lower triangular matrix and $Tr(\cdot)$ is a trace operator. What is important, only diagonal elements of $T$ matrix are the real numbers, the rest of them may be complex ones.
Therefore, for any $K$-dimensional system, the number of parameters that are needed to be found is equal to $K^2$ -- $K$ real parameters on diagonal and $2(K(K-1)/2)$ real parameters in lower triangle, where complex numbers may occur. \\

Thanks to such factorization, the problem of estimation of $\hat{\rho}$ translates to the estimation of $T$ elements, through which the number of parameters needed to be found is reduced. This small change allows for considerably quicker computations. \\

Depending on the function used for estimation of $\hat{\rho}(T)$ parameters, the condition for finding an extremum of such a function is $f'\big(\hat{\rho}(T)\big)=0$. The extremum is searched iteratively. The consecutive steps of an algorithm can be written as follows:

\begin{quote}
\begin{enumerate}
\item Choose some initial values for matrix $T$.
\item Calculate $f'(T)$.
\item Check if $|f'(T)-\epsilon|=0$, for any earlier  defined numerical accuracy $\epsilon$.
\begin{enumerate}
\item If YES, go to point 4.
\item If NO, change values slightly for elements of matrix $T$ and go back to point 2.
\end{enumerate}
\item Calculate $\hat{\rho}$ using Eq. \ref{eq:rho} and end the task.
\end{enumerate}
\end{quote} 

Presented algorithm can be implemented relatively easily. Moreover, for many types of computing software, such as, for example, \textit{Wolfram Mathematica} or \textit{Matlab}, similar algorithms are already built-in. \\

\hsubsection{Experimental examples}
\label{sec:ch5-expexamp}
The examples presented here focus on polarization-encoded photon states. Of course, there are many different ways to encode qubits (see \textit{Section \ref{sec:ch1-quantnet}}). Nonetheless, probably the easiest way to do that experimentally is to exploit the photon's polarization degree of freedom. Moreover, examples shown below were chosen to be pictorial and as intuitive as possible.\\
 
\hspace{1.4cm}\textbf{Single-qubit tomography}\\

\begin{figure}[b!]
\centering
\includegraphics[width=0.8\linewidth]{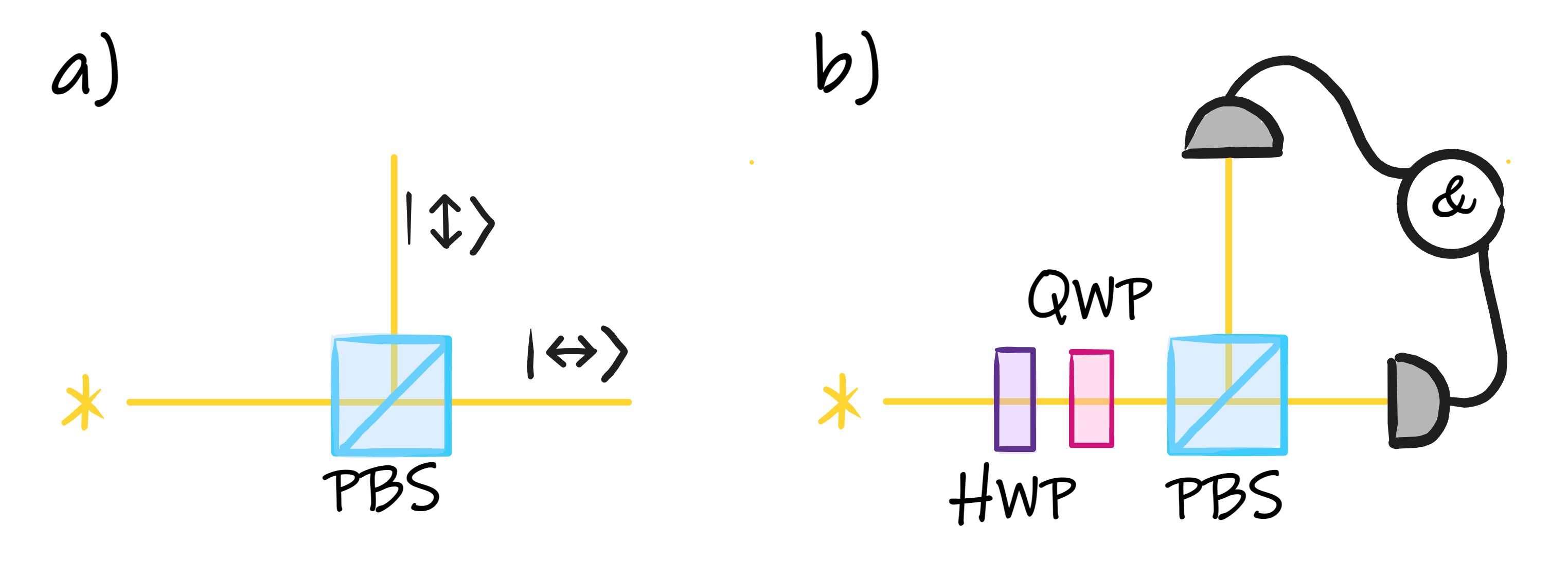}
\caption{a) The working principle of polarizing beamsplitter. b)Scheme of the setup for quantum-state tomography for photonic qubit. Symbols: $PBS$ -- polarizing beamsplitter, $HWP$ -- half-wave plate, $QWP$ -- quarter-wave plate, $D$ -- single-photon detector.}
\label{pic:pqst}
\end{figure}

As described in \textit{Section \ref{sec:ch5-pauli}}, density matrix of a polarization qubit state can be expanded using Pauli matrices (see Eq. \ref{eq:paulimat}):
\begin{equation}
\hat{\rho}=\frac{1}{2}\big(\mathds{1}+\vec{b}\cdot\vec{\sigma}\big) \; ,
\label{eq:blochvec2}
\end{equation}
where $\vec{b}$ stands for Bloch vector, and $\vec{\sigma}=(\sigma_x,\sigma_y,\sigma_z)\big)$ . Due to that, in order to define photon state, it is enough to find elements of $\vec{b}$. \\
The experimental setup that allows for finding these elements is presented in Fig. \ref{pic:pqst}. It consists of three optical elements that influence polarization of a beam -- half-wave plate (HWP), quarter-wave plate (QWP), and  polarizing beamsplitter (PBS). \\

The PBS is the element that splits the photon beam into two, one is in $|\udarr\rangle$, another is in $|\lrarr\rangle$ polarization state. It means that PBS projects an unknown polariazation state on the states from the $\{|\udarr\rangle,|\lrarr\rangle\}$ basis. \\
The use of single PBS would be enough if the photon state also belonged to one of basis states. Mostly, however, it is not. Therefore, it is necessary to transform other polarization states into the basis one, for example:
\begin{equation}
\begin{split}
\big\{|\circright\rangle,|\ruarr\rangle\big\}&\mapsto|\lrarr\rangle \; , \\ 
\big\{|\circleft\rangle,|\rdarr\rangle\big\}&\mapsto|\udarr\rangle \; .
\end{split}
\end{equation}
The HWP and QWP are optical elements which are the perfect solution for this task. Their Jones matrices (see \textit{Section \ref{sec:ch1-pol}}) are included in Tab. \ref{tab:polmat}. Fortunately, these matrices can be also factorized using Pauli matrices, in such a case one gets:
\begin{equation}
\begin{split}
HWP(\phi)\,&=\, sin(2\phi)\sigma_x\,+\,cos(2\phi)\sigma_z \; , \\
QWP(\phi)\,=\, \frac{1}{2}\big((1-i)\mathds{1}+&2(1-i)cos(\phi)sin(\phi)\sigma_x+(1+i)cos(2\phi)\sigma_z\big) \; .
\end{split}
\end{equation}

Using waveplates along with PBS effectively projects states to the basis ones, which is enough to measure in all directions.  \\

\begin{table}[h!]
\centering
\begin{tabular}{x{2.5cm}|x{1cm}|x{1cm}}
\hline
\textbf{measurement}& \textbf{HWP} & \textbf{QWP}\\
\hline\hline
$\sigma_x$&$22.5^o$&$0^o$\\
$\sigma_y$&$0^o$&$45^o$\\
$\sigma_z$&$0^o$&$0^o$\\
\hline
\end{tabular}
\end{table}
The table above presents proper angles of waveplates for $M_k$ measurements. It should be noted here that a measurement that characterizes the intensity of a beam is also needed. Therefore, the number of $M_k$ measurements that allow one to fully reconstruct density matrix of single qubit is equal to $4$.\\ 

\vspace{0.5cm}
\hspace{1.4cm}\textbf{Two-qubit tomography}\\

The two-qubit quantum-state tomography has similar procedure to its single counterpart. It is also based on the rotations of waveplates. However, the amount of settings is larger. The minimum number of $M_k$ measurements is equal to $16$. \\
All combinations of waveplates angles as well as the polarization state on which they project are collected in Tab. \ref{tab:polqst}. 

\begin{figure}[b!]
\centering
\includegraphics[width=0.6\linewidth]{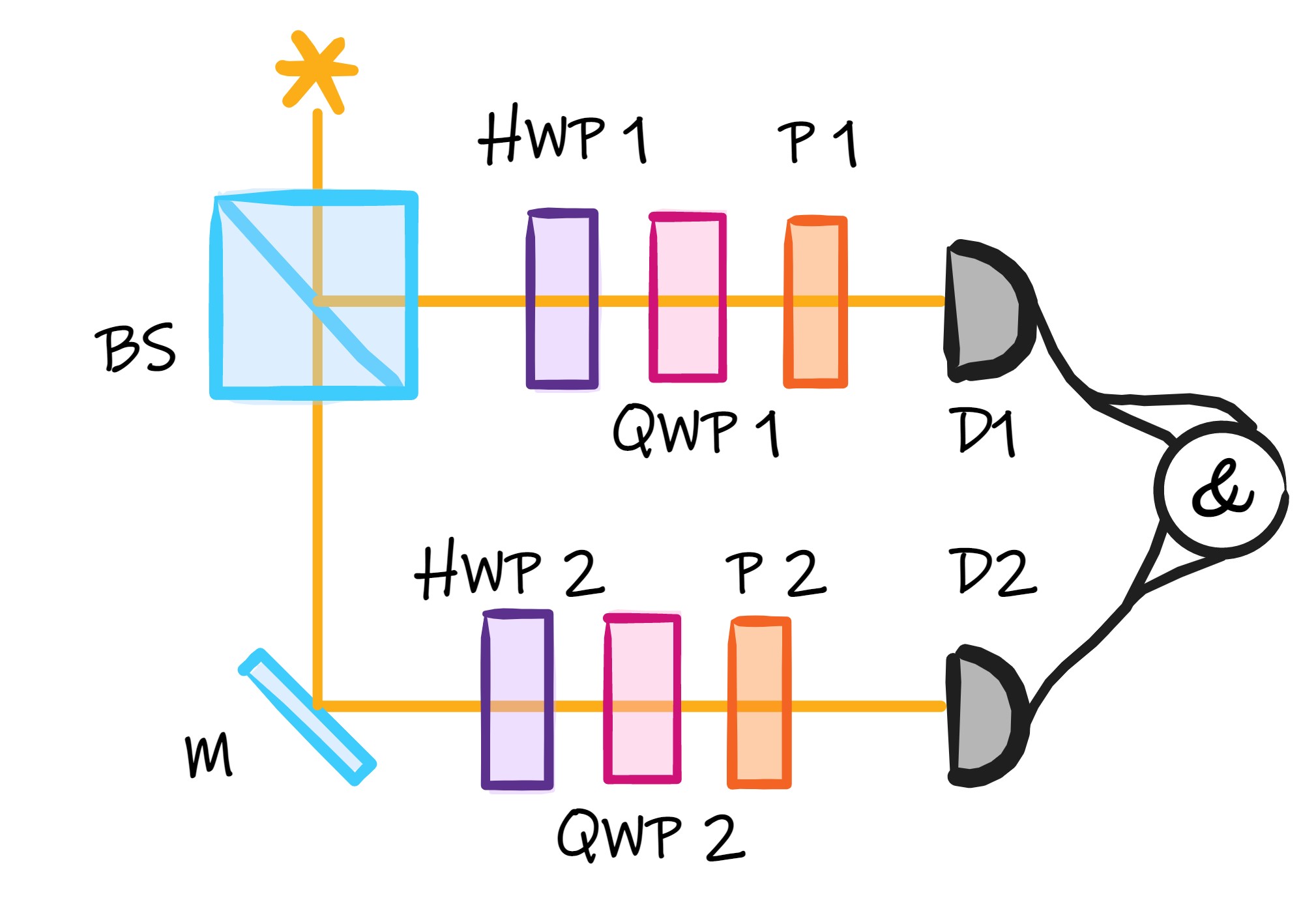}
\caption{Scheme of the setup for two-qubit quantum state tomography. A part of the experimental setup of S2 source (presented in Fig. \ref{pic:s2}).}
\label{pic:tqst}
\end{figure}

\begin{table}[t!]
\centering
\begin{tabular}{x{0.05\linewidth}|x{0.15\linewidth}|x{0.15\linewidth}|x{0.08\linewidth}|x{0.08\linewidth}|x{0.08\linewidth}|x{0.08\linewidth}}
\hline
\textbf{no.}&\textbf{state on channel 1} & \textbf{state on channel 2}& \textbf{HWP1 angle}& \textbf{QWP1 angle}&\textbf{HWP2 angle}& \textbf{QWP2 angle}\tabularnewline
\hline \hline
1&$|\lrarr\rangle$&$|\lrarr\rangle$&$45^o$&$0^o$&$45^o$&$0^o$\\
2&$|\lrarr\rangle$&$|\udarr\rangle$&$45^o$&$0^o$&$0^o$&$0^o$\\
3&$|\udarr\rangle$&$|\udarr\rangle$&$0^o$&$0^o$&$0^o$&$0^o$\\
4&$|\udarr\rangle$&$|\lrarr\rangle$&$0^o$&$0^o$&$45^o$&$0^o$\\
5&$|\circright\rangle$&$|\lrarr\rangle$&$22.5^o$&$0^o$&$45^o$&$0^o$\\
6&$|\circright\rangle$&$|\udarr\rangle$&$22.5^o$&$0^o$&$0^o$&$0^o$\\
7&$|\ruarr\rangle$&$|\udarr\rangle$&$22.5^o$&$45^o$&$0^o$&$0^o$\\
8&$|\ruarr\rangle$&$|\lrarr\rangle$&$22.5^o$&$45^o$&$45^o$&$0^o$\\
9&$|\ruarr\rangle$&$|\circright\rangle$&$22.5^o$&$45^o$&$22.5^o$&$0^o$\\
10&$|\ruarr\rangle$&$|\ruarr\rangle$&$22.5^o$&$45^o$&$22.5^o$&$45^o$\\
11&$|\circright\rangle$&$|\ruarr\rangle$&$22.5^o$&$0^o$&$22.5^o$&$45^o$\\
12&$|\lrarr\rangle$&$|\ruarr\rangle$&$45^o$&$0^o$&$22.5^o$&$45^o$\\
13&$|\udarr\rangle$&$|\ruarr\rangle$&$0^o$&$0^o$&$22.5^o$&$45^o$\\
14&$|\udarr\rangle$&$|\circleft\rangle$&$0^o$&$0^o$&$22.5^o$&$90^o$\\
15&$|\lrarr\rangle$&$|\circleft\rangle$&$45^o$&$0^o$&$22.5^o$&$90^o$\\
16&$|\circright\rangle$&$|\circleft\rangle$&$22.5^o$&$0^o$&$22.5^o$&$90^o$\\
\hline
\end{tabular}
\caption{Measurement table for two-qubit quantum state tomography.}
\label{tab:polqst}
\end{table}

Of course, the choice of different basis states than the ordinary ones $\{|\udarr\rangle,|\lrarr\rangle\}$ will cause the change of the polarization states on detection channels, therefore this measurement table is only one example.\\

The experimental setup that allows for such measurements is shown in Fig. \ref{pic:tqst}. As it can be noticed, the polarization elements are similar to those in Fig. \ref{pic:pqst}, except that PBS has been changed to polarizer (P). The downside of this approach is the loss of the count rate on detectors, because the photons that are projected into the orthogonal state to which the polarizer is transmissive are lost. On the other hand, this architecture of the setup is slightly more experimentally friendly due to the fact that here are only two arms with two couplers, instead of four. \\
It should also be mentioned that the role of a beamsplitter (BS), which is not the polarizing one here, is only to separate the qubits before measuring them. It is made possible here by reversed Hong-Ou-Mandel effect (see \textit{Section \ref{sec:ch3-qst}} and \textit{\ref{sec:notes-HOM})}. \\

Assuming that described measurements are done, the next step is to compute the relevant function portrayed earlier (\textit{Section \ref{sec:ch5-est}}) with proper coincidence counts. \\
The results obtained from S2 source, which are presented and elaborated on more in \textit{Section \ref{sec:ch3-qst}}, were calculated using a modified MLE algorithm. The modification includes the error function, shown in Eq. \ref{eq:mle-ef}. The error function, which was used was defined as:
\begin{equation}
f_e(k) = a \sqrt{c_k}+b \; ,
\end{equation}
where $c_k$ is the single count corresponding to the measurement $M_k$ and $a,b>0$. The origin of such an error function was an observation on the behavior of single-photon detectors when illuminating them with different intensity of single-photon beams.  \\

\begin{figure}[t!]
\centering
\includegraphics[width=0.95\linewidth]{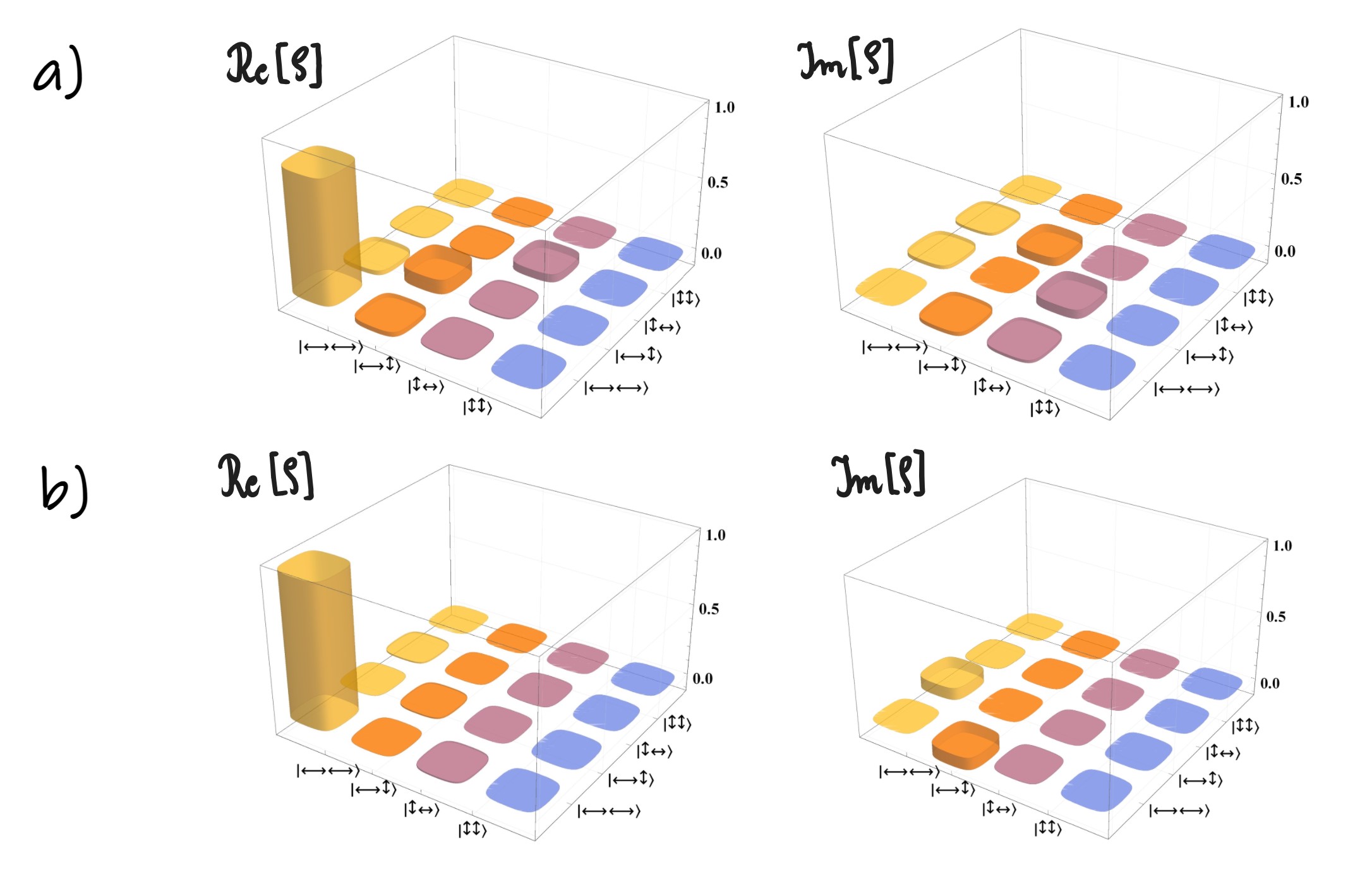}
\caption{The real (left-hand side) and imaginary (right-hand side) part of density matrices of a test  quantum state. Comparison on estimation methods used for reconstruction: a) standard MLE, b) MLE supplied with error function. Algorithm was performed by dr Artur Czerwiński.}
\label{pic:qstcomp}
\end{figure}

Although this method differs from the standard
approach, it gives nicer results \cite{Artur2022}. The comparison of density matrices reconstructed using MLE with and without error function for the test measurement performed on a source that generates $|\lrarr\lrarr\rangle$ polarization state is presented in Fig. \ref{pic:qstcomp}. \\

The comparison of calculated parameters (see \textit{Section \ref{sec:ch5-fidelity}} and \textit{\ref{sec:ch5-conc}}) for the density matrix  of the state generated by S2 source is collected in the table below. \\

\begin{table}[h!]
\centering
\begin{tabular}{x{0.18\linewidth}|x{0.2\linewidth}|x{0.3\linewidth}}
\hline
\textbf{parameter} & \textbf{MLE recontruction} & \textbf{MLE with $f_e(k)$ reconstruction} \\ \hline\hline
fidelity & $0.675233$ & $0.972216$ \\
concurrence & $0.140303$ & $0.985286$ \\
\hline
\end{tabular}
\end{table}
\hsection{Fidelity of a quantum state}
\hnote{Fidelity of a quantum state}
\label{sec:ch5-fidelity}

Although the copying of quantum states is proven as unfeasible by quantum mechanics (see \textit{Section \ref{sec:notes-noclo}}), sometimes it is useful to check the similarity between two states. For example, the similarity between experimentally generated and theoretically predicted state may be verified. 
A parameter that enables such verification is called \textbf{fidelity} \cite{Jozsa1994}, and is defined as:
\begin{equation}
F(\hat{\rho}, \hat{\sigma})=\Big( Tr\big(\sqrt{\sqrt{\hat{\rho}}\hat{\sigma}\sqrt{\hat{\rho}}}\big)\Big)^2 \; ,
\end{equation}
where $\hat{\rho}$ and $\hat{\sigma}$ are the density operators of two quantum states and $Tr$ is the trace operator.
It should be noted here that for pure states represented as $\hat{\rho}=|\psi_{\rho}\rangle\langle\psi_{\rho}|$ and $\hat{\sigma}=|\psi_{\sigma}\rangle\langle\psi_{\sigma}|$, fidelity is simply the squared overlap between the states, i.e., $F(\hat{\rho},\hat{\sigma})=|\langle\psi_{\rho}|\psi_{\sigma}\rangle|^2$. \\

It is worth to mention that fidelity may be used as an indicator of entanglement of a quantum state when it is calculated between selected state from Bloch basis (see \textit{Section \ref{sec:ch5-blochbasis}}) and any measured or chosen one. \\
Value of fidelity is included in the range of $[0,1]$. The closer to $1$, the more entangled a quantum state is; thus, for the maximally entangled state it is equal to $1$. \\

\hsection{Concurrence}
\hnote{Concurrence}
\label{sec:ch5-conc}

\textbf{Concurrence} is a value that quantifies the absolute amount of entanglement for any given qubit state\cite{Hill1997,James2001,Artur2022}. It is defined as:
\begin{equation}
C(\rho)=max\{0, \lambda_1-\lambda_2-\lambda_3-\lambda_4\} \; ,
\end{equation}
where $\lambda_i$ are  arranged in decreasing order eigenvalues of the R-matrix, defined as:
\begin{equation}
R(\rho)=\sqrt{\sqrt{\rho}\big((\sigma_y\otimes\sigma_y)\rho^*(\sigma_y\otimes\sigma_y)\big)\sqrt{\rho}} \; ,
\end{equation}
where $\rho^*$ is the complex conjugate of $\rho$ and $\sigma_y$ is the Pauli matrix (see \textit{Section \ref{sec:ch5-pauli}}). \\

Similarly to fidelity, for any density matrix $\rho$, the concurrence takes values from the range  of $[0,1]$, for the maximally entangled state it is equal to $1$, whereas for separable ones it is equal to $0$. \\

Moreover, the definition of concurrence may be generalized to measure the amount of entanglement in multiparticle pure states \cite{Bhaskara2017}. \\
\hsection{Visibility of a satellite}
\hnote{Visibility of a satellite}
\label{sec:ch4-sattime}

\begin{figure}[t!]
\centering
\includegraphics[width=0.65\linewidth]{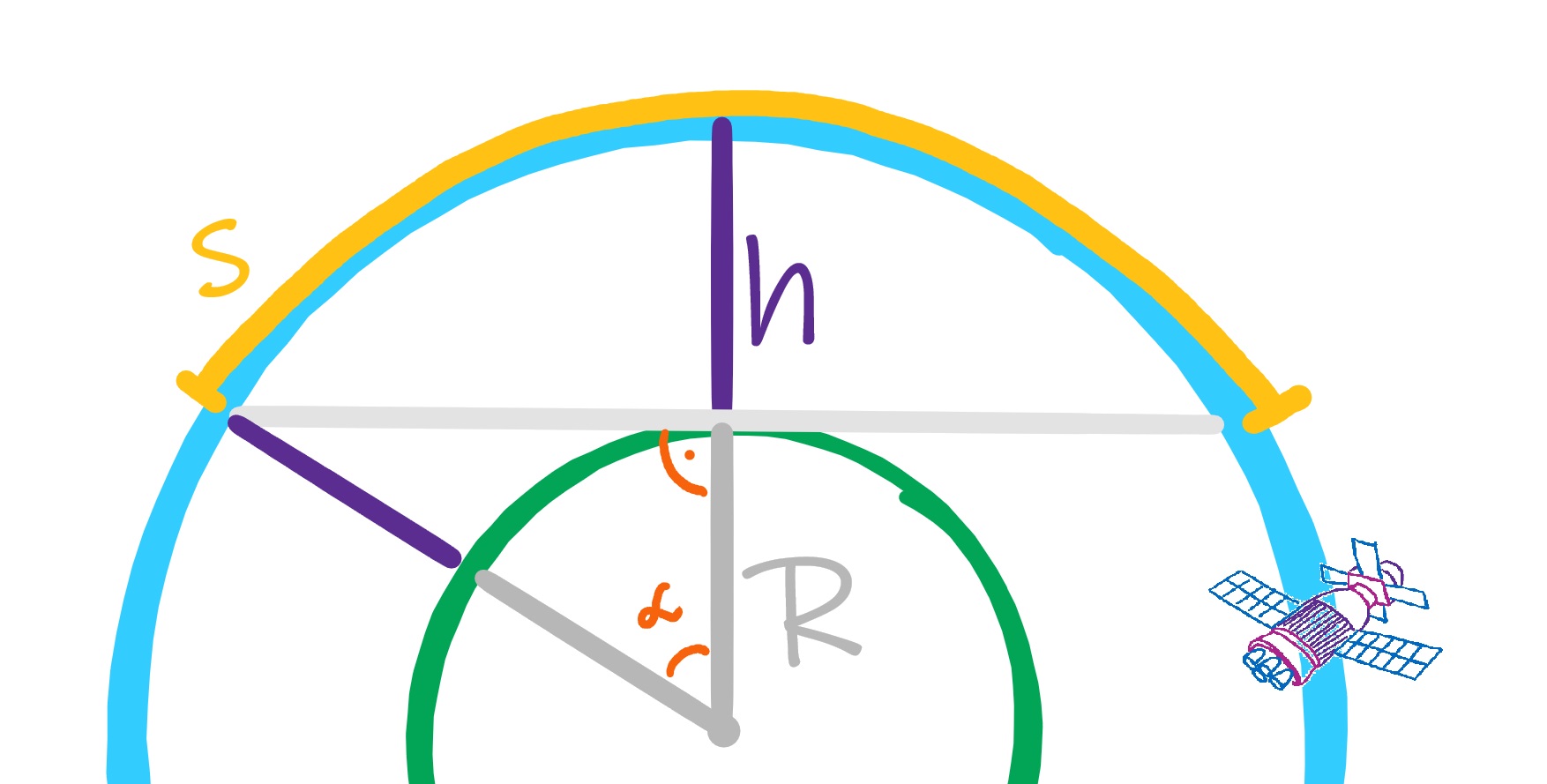}
\caption{The geometrical scheme of satellite's circular orbit that crosses the zenith point above the observer. Symbols: $h$ -- the altitude, $R$ -- Earth radius, $s$ -- the segment of the circle on which the satellite is visible for the ground station, $\alpha$ -- angle that is correlated with the segment $s$.}
\label{pic:sattimeorb}
\end{figure}

The period of time in which it is possible to communicate directly with a satellite is important when choosing the transmission protocol. Assuming that the satellite is in a specific circular orbit with the altitude $h$ crossing the zenith (as shown in Fig.~\ref{pic:sattimeorb}), the observer sees the satellite only as it traverses the segment of the circle $s$, which determines the time of the communication. \\

Let us consider a satellite with mass $m$ orbiting the Earth (with mass $M$), which moves in circular motion. The centripetal force $F$ acting upon the satellite is given by:
\begin{equation}
F=\frac{m\,v^2}{R+h}\, ,
\end{equation}
where $h$ is the altitude of the satellite, $R$ is the Earth radius, $v$ is the speed of the satellite motion. Since this force, as well as the satellite's motion, is the result of the gravitational force $F_G$ that attracts the satellite:
\begin{equation}
F_G=G\frac{m\,M}{(R+h)^2}\, ,
\end{equation}
these two equations can be equated, so:
\begin{equation}
\frac{m\,v^2}{R+h}=G\frac{m\,M}{(R+h)^2}\, ,
\end{equation}
which gives
\begin{equation}
v=\sqrt{\frac{G\,M}{R+h}}\, ,
\end{equation}
where $G$ is a gravitational constant. Next, taking an elementary relation such as $v=s/t$, where $s$ is the segment of the circle, marked in Fig.~\ref{pic:sattimeorb}, the time $t$, throughout which there is possibility of the communication with satellite, is be obtained:
\begin{equation}
t=s\cdot\sqrt{\frac{R+h}{G\,M}}\, . \\
\end{equation}

At last, the length of segment $s$ can be determined, using the expression :
\begin{equation}
s=\frac{\alpha}{2\pi}\,2\pi(R+h)
\end{equation}
and trigonometric relation $cos\,\frac{\alpha}{2}=\frac{R}{R+h}$, so:
\begin{equation}
s=(R+h)\,arccos\Big(\frac{R}{R+h}\Big)\, ,
\end{equation}
and therefore the time $t$ is equal to:
\begin{equation}
t=\frac{(R+h)^{3/2}}{(G\,M)^{1/2}}arccos\Big(\frac{R}{R+h}\Big)\, . \\
\label{eq:sattime}
\end{equation}

\begin{figure}[t!]
\centering
\includegraphics[width=0.75\linewidth]{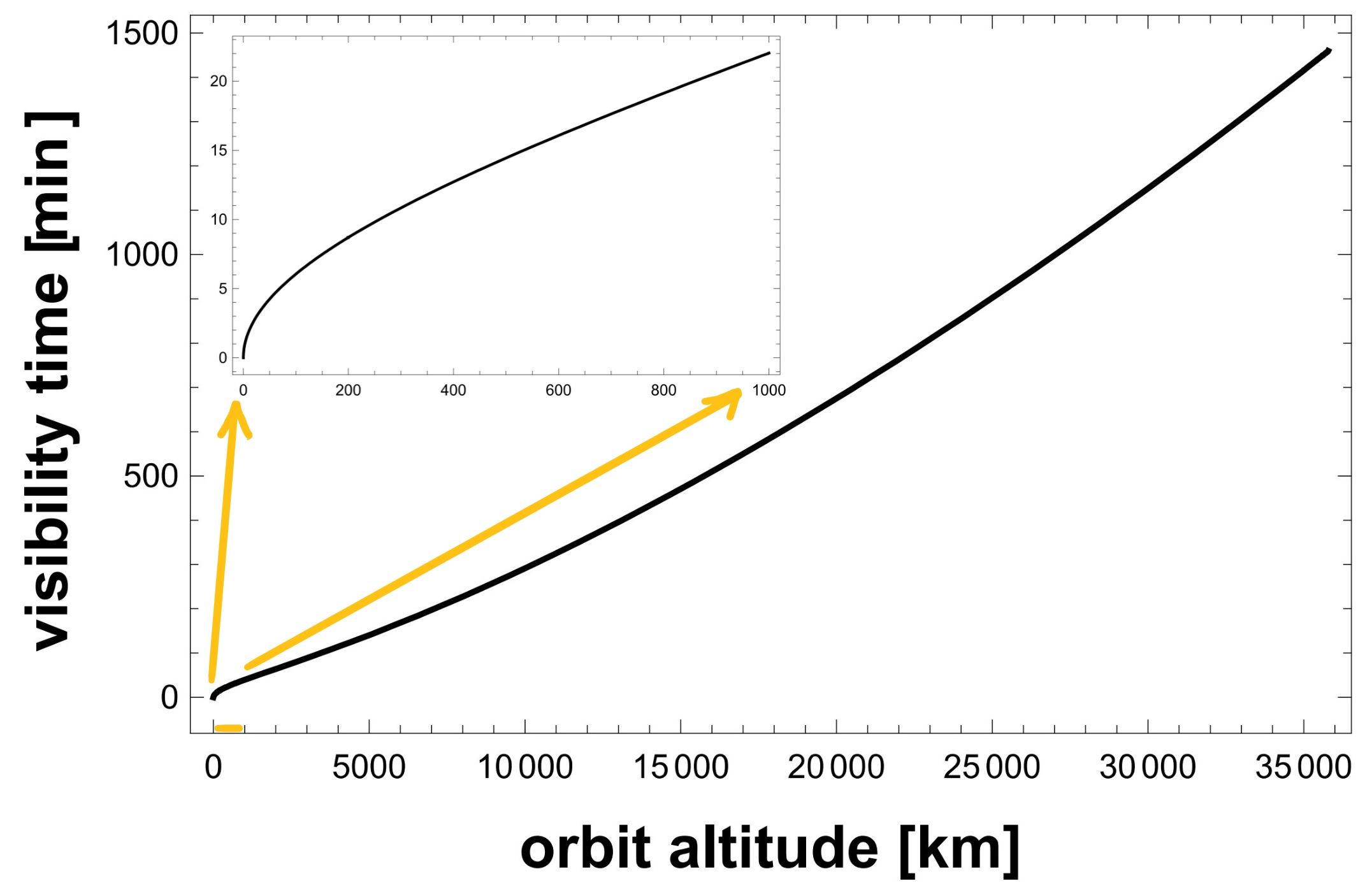}
\caption{The relation between visibility time of a satellite and its altitude, described with Eq.~\ref{eq:sattime}}
\label{pic:sattimeres}
\end{figure}
\FloatBarrier

Fig.~\ref{pic:sattimeres} presents the relation between the altitude  $h$ and visibility time $t$ of a satellite,  during which the communication with a ground station is possible. \\

\begin{small}

\end{small}

\cleartorecto
\onecolumn

\thispagestyle{empty}

\mbox{}\vspace{50mm}

\begin{quotation}
\begin{minipage}{0.8\linewidth}
\noindent {\large Dearest Readers},\\

\noindent It has been said that, everything has an end. It seems this particular adventure has inevitably approached the end. The Author of this very publication wants to give her sincere thanks to you, Kind Readers, for your time and dedication to exploring this not always perfect work. She also hopes that you had the most delightful and fruitful time while reading it, despite all shortcomings and inconveniences. \\

\noindent Fear not, Gentle Readers, and accept the most sincere of apologies if there was any misunderstanding or information that misled you on a way to acquiring the knowledge. With a certain probability, the mastering of quantum mechanical theory is the journey that never really ends. Therefore the Author wishes all the luck to you, herself, and everyone interested, on the road through endless spaces of Science. \\

\noindent Yours sincerely,\\
Marta Misiaszek-Schreyner

\end{minipage}
\end{quotation}

\end{document}